\documentclass[a4paper,12pt]{article}
\pdfoutput=1
\usepackage{jheppub}
\usepackage[utf8]{inputenc}
\usepackage{url}
\usepackage{mathrsfs}  
\usepackage{pdflscape}
\usepackage{refcount}
\usepackage{booktabs}
\usepackage{todonotes}
\usepackage{enumitem}
\usepackage{adjustbox}
\usepackage{afterpage}
\usepackage{mathdots}
\usepackage{ytableau, youngtab} 

\usepackage{amsfonts}
\usepackage{bbm}

\usepackage{pdfpages}
\usepackage{caption}
\usepackage{subcaption}

\usepackage{tikz}
\usetikzlibrary{positioning}
\usetikzlibrary{arrows}
\usetikzlibrary{decorations.pathreplacing}
\usetikzlibrary{shapes.geometric}
\usetikzlibrary{calc}
\usetikzlibrary{decorations.pathmorphing}
\usetikzlibrary{shapes.misc}

\usetikzlibrary{positioning,trees,decorations.pathmorphing,decorations.markings,decorations.pathreplacing,calc,shapes,patterns,arrows,chains,arrows.meta,fit,fadings,decorations.markings,graphs,graphs.standard,quotes,plotmarks,calligraphy,decorations.pathreplacing}
\usetikzlibrary{intersections, calc}
\usetikzlibrary{shapes.geometric}
\usetikzlibrary{matrix, arrows.meta}
\usetikzlibrary{shapes.geometric}
\usetikzlibrary{decorations.pathmorphing}
\usetikzlibrary{shapes.misc}
\usetikzlibrary{arrows.meta}
\usetikzlibrary{angles, quotes}

\definecolor{goodgreen}{RGB}{55,169,49}

\definecolor{darkyellow}{RGB}{230,170,10}

\definecolor{brightyellow}{RGB}{255,240,190}

\tikzset{flavour/.style={draw=none,minimum size=0.3mm,fill=white, regular polygon,regular polygon sides=4,draw}}
\tikzset{gaugeBig/.style={inner sep=1mm,draw=none,fill=white,minimum size=2mm,circle, draw}}
\tikzset{bd/.style={circle, draw=black, inner sep=0pt, fill=black, minimum size=2mm}}
\tikzset{wd/.style={circle, draw=black, inner sep=0pt, fill=white, minimum size=2mm}}
\tikzset{Dynkin/.style={circle, draw=black, inner sep=0pt, fill=white, minimum size=2mm}}
\tikzstyle{ligne}=[draw, very thick] 
\tikzstyle{gridline}=[draw, gray] 
\tikzset{gauge/.style={circle, draw,inner sep=2.5pt}}
\tikzset{gaugeo/.style={circle, draw,inner sep=2.5pt,fill=orange}}
\tikzset{gaugec/.style={circle, draw,inner sep=2.5pt,fill=cyan}}
\tikzset{gauger/.style={circle, draw,inner sep=2.5pt,fill=red}}
\tikzset{gaugeb/.style={circle, draw,inner sep=2.5pt,fill=blue}}
\tikzset{gaugeg/.style={circle, draw,inner sep=2.5pt,fill=green}}
\tikzset{gaugem/.style={circle, draw,inner sep=2.5pt,fill=magenta}}
\tikzset{hasse/.style={circle, fill,inner sep=2pt}}
\tikzset{shrinky/.style={circle, fill,inner sep=1pt}}
\tikzset{sized/.style={circle, draw, inner sep=1.5pt}}
\tikzset{seven/.style={circle, draw,inner sep=3pt}}

\tikzset{dotto/.style={circle, orange, draw,inner sep=1.5pt,fill=orange}}
\tikzset{dottp/.style={circle, purple, draw,inner sep=1.5pt,fill=purple}}
\tikzset{dottc/.style={circle, cyan, draw,inner sep=1.5pt,fill=cyan}}
\tikzset{dottr/.style={circle, red, draw,inner sep=1.5pt,fill=red}}
\tikzset{dottb/.style={circle, blue, draw,inner sep=1.5pt,fill=blue}}
\tikzset{dottg/.style={circle, green, draw,inner sep=1.5pt,fill=green}}
\tikzset{dottm/.style={circle, magenta, draw,inner sep=1.5pt,fill=magenta}}

\tikzset{redgauge/.style={draw=none,minimum size=0.4cm,fill=red,circle, draw}}
\tikzset{gauge3/.style={draw=none,minimum size=0.4cm,fill=white,circle, draw}}
\tikzset{bluegauge/.style={draw=none,minimum size=0.4cm,fill=blue,circle, draw}}
\tikzset{redflavor/.style={draw=none,minimum size=0.6cm,fill=red, regular polygon,regular polygon sides=4,draw}}
\tikzset{blueflavor/.style={draw=none,minimum size=0.6cm,fill=blue, regular polygon,regular polygon sides=4,draw}}
\tikzset{flavour2/.style={draw=none,minimum size=0.6cm,fill=white, regular polygon,regular polygon sides=4,draw}}
\tikzset{rede/.style={line width=0.5mm,red}}
\tikzset{bluee/.style={line width=0.5mm,blue}}
\tikzset{pinkline/.style={line width=0.5mm,purple}}
\pgfdeclarelayer{edgelayer}
\pgfdeclarelayer{nodelayer}

\DeclareMathOperator{\U}{U}
\DeclareMathOperator{\SU}{SU}
\DeclareMathOperator{\SO}{SO}
\DeclareMathOperator{\SL}{SL}

\DeclareMathOperator{\USp}{USp}
\DeclareMathOperator{\tr}{Tr}

\usepackage{epsfig}
\usepackage{amsmath}
\usepackage{amssymb}
\usepackage{amsfonts}
\usepackage{amsxtra}
\usepackage{amsthm}
\usepackage{mathrsfs}
\usepackage{makeidx}
\usepackage{graphics}
\usepackage{dsfont}
\usepackage{mathtools}
\usepackage{graphicx}
\usepackage{placeins}
\usepackage{bm}
\usepackage[capitalise]{cleveref}
\usepackage{empheq}
\usepackage{colortbl}
\usepackage{xcolor}
\usepackage{titlesec}
\usepackage{longtable}
\usepackage{hhline}
\usepackage{float}
\usepackage{color}
\usepackage{tikz}
\usepackage{xfrac}
\usepackage{footnote}
\usepackage{rotating}
\usepackage{lscape}
\usepackage{makecell}
\usepackage{environ}
\usepackage{tabularx}
\usepackage{subfiles}
\usepackage{ytableau}
\usepackage{tikz-3dplot}
\usepackage{slashed}
\usepackage{pifont}
\usepackage{multirow}
\usepackage{mdframed}
\usepackage{bbm}
\usepackage[normalem]{ulem}
\usepackage{arydshln}
\usepackage{enumitem} 
\usepackage{fancybox}

\usetikzlibrary{positioning,trees,decorations.pathmorphing,decorations.markings,decorations.pathreplacing,calc,shapes,patterns,arrows,chains,arrows.meta,fit,fadings,decorations.markings,graphs,graphs.standard,quotes,plotmarks,calligraphy,decorations.pathreplacing}

\usetikzlibrary{positioning}
\usetikzlibrary{chains}
\usetikzlibrary{arrows, arrows.meta ,fit,decorations.pathreplacing}
\tikzstyle{every picture}+=[remember picture]
\tikzstyle{na} = [baseline]
\tikzstyle{ligne}=[draw, thick]

\usetikzlibrary{arrows, decorations.markings, calc, fadings, decorations.pathreplacing, patterns, decorations.pathmorphing, positioning}
\tikzset{>={Latex[width=1.5mm,length=1.5mm]}}
\tikzset{bd/.style={circle, draw=black, inner sep=0pt, fill=black, minimum size=1.2mm}}
\tikzset{bld/.style={circle, draw=blue, inner sep=0pt, fill=blue, minimum size=1.2mm}}
\tikzset{wd/.style={circle, draw=black, inner sep=0pt, fill=white, minimum size=1.2mm}}
\tikzset{rd/.style={circle, draw=red, inner sep=0pt, fill=red, minimum size=.9mm}}
\tikzset{wrd/.style={circle, draw=red, inner sep=0pt, fill=white, minimum size=.9mm}}
\usetikzlibrary{graphs,graphs.standard,quotes}

\def\node#1#2{\overset{#1}{\underset{#2}{{\color{gray} \bullet}}}}

\def\node#1#2{\overset{#1}{\underset{#2}{\circ}}}

\tikzstyle{every picture}+=[remember picture]
\tikzstyle{na} = [baseline=-.5ex]

\providecommand{\abs}[1]{\lvert#1\rvert}

\newcommand{\eg}{e.g. }

\newcommand{\ie}{i.e. }

\numberwithin{equation}{section}
\newcommand{\bes}[1]{\begin{equation} \begin{split} #1\end{split} \end{equation}}

\newcommand{\be}{\begin{equation}} \newcommand{\ee}{\end{equation}}
\newcommand{\bea}{\begin{equation} \begin{aligned}} \newcommand{\eea}{\end{aligned} \end{equation}}

\def\tilde{\widetilde}

\def\hat{\widehat}

\def\bar{\overline}

\def\rt2{\sqrt{2}}

\def\tr{\mathop{\rm tr}}

\def\abs#1{\left|#1\right|}

\def\fa{\mathfrak{a}}
\def\fb{\mathfrak{b}}

\def\CF{{\cal F}}

\def\CH{{\cal H}}
\def\CI{{\cal I}}

\def\CM{{\cal M}}
\def\CN{{\cal N}}
\def\CO{{\cal O}}

\def\CQ{{\cal Q}}
\def\CR{{\cal R}}
\def\CS{{\cal S}}

\def\1{{\ds 1}}

\newcommand{\cO}{\mathcal{O}}

\newcommand{\fm}{\mathfrak{m}}
\newcommand{\fn}{\mathfrak{n}}

\newcommand{\fz}{\mathfrak{z}}

\newcommand{\fy}{\mathfrak{y}}
\newcommand{\fq}{\mathfrak{q}}
\newcommand{\fc}{\mathfrak{c}}
\newcommand{\fl}{\mathfrak{l}}

\def\SO{\mathrm{SO}}

\def\O{\mathrm{O}}
\def\SU{\mathrm{SU}}
\def\SL{\mathrm{SL}}

\def\Spin{\mathrm{Spin}}
\def\Pin{\mathrm{Pin}}
\def\Usp{\mathrm{\USp}}

\def\su{\mathfrak{su}}

\def\so{\mathfrak{so}}

\def\usp{\mathfrak{usp}}
\def\fp{\mathfrak{p}}

\def\fy{\mathfrak{y}}
\def\fx{\mathfrak{x}}

\def\repa{\raise4pt\hbox{$\square$}\mkern-14mu\raise-4pt\hbox{$\square$}}
\def\repab{\overline{\raise4pt\hbox{$\square$}\mkern-14mu\raise-4pt\hbox{$\square$}\mkern-1mu}}

\def\smileface{\ensuremath{\hbox{\large$\bigcirc$}\mkern-15mu\raise-1pt\hbox{\scriptsize$\smallsmile$}%
\mkern-10mu\raise4pt\hbox{..}\mkern4mu}}
\def\frownface{\ensuremath{\hbox{\large$\bigcirc$}\mkern-15mu\raise-1pt\hbox{\scriptsize$\smallfrown$}%
\mkern-10mu\raise4pt\hbox{..}\mkern4mu}}

\newcommand{\ba}{\begin{array}}
\newcommand{\ea}{\end{array}}
\newcommand{\bi}{\begin{itemize}}
\newcommand{\ei}{\end{itemize}}
\def\vec#1{\bm{#1}}
\def\bea#1\eea{\allowdisplaybreaks \begin{align}#1\end{align}}
 \newcommand{\ben}{\begin{enumerate}}
\newcommand{\een}{\end{enumerate}}
\newcommand{\bean}{\begin{eqnarray*}}
\newcommand{\eean}{\end{eqnarray*}}
\newcommand{\eref}[1]{(\ref{#1})}

\newcommand{\PE}{\mathop{\rm PE}}

\newcommand{\tQ}{\widetilde{Q}}

\newcommand{\BC}{\mathbb{C}}

\newcommand{\BZ}{\mathbb{Z}}

\newcommand{\diag}{\mathrm{diag}}

\newcommand{\Sym}{\mathrm{Sym}}

\definecolor{light-gray}{gray}{0.5}
\definecolor{new-green}{rgb}{0,0.7,0.3}
\definecolor{cerulean}{rgb}{0.0, 0.48, 0.65}
\definecolor{claret}{rgb}{0.50, 0.09, 0.20}
\definecolor{darkred}{rgb}{0.7, 0.11, 0.11}
\definecolor{scarlet}{rgb}{1.0, 0.13, 0.0}
\definecolor{orange-red}{rgb}{1.0, 0.27, 0.0}
\definecolor{blue-green}{rgb}{0.0, 0.5, 0.65}
\definecolor{green-red}{rgb}{0.5, 0.65, 0.0}

\newcommand{\blue}{\color{blue}}

\newcommand{\red}{\color{red}}

\newcommand{\violet}{\color{violet}}

\newcommand{\cerulean}{\color{cerulean}}

\newcommand{\orangered}{\color{orange-red}}

\def\aup#1 {\overset{#1}{\uparrow} \, \overset{\tilde{#1}}{\downarrow}}

\tikzset{snake it/.style={decorate, decoration={snake, amplitude=.4mm, segment length=2mm,
                       post length=0mm,pre length=0mm}}}

\def\u{\mathfrak{u}}
\DeclareMathAlphabet{\mymathds}{U}{BOONDOX-ds}{m}{n}
\tikzstyle{double_border} = [draw, double, double distance=1pt]

\colorlet{verdescuro}{green!50!black}

\theoremstyle{plain}

\title{Limits of the Superconformal Index and the Moduli Space of 3d $\mathcal{N}=3$ Theories}
\author[a]{Riccardo Comi,}
\author[b,c]{Sebastiano Garavaglia,}
\author[b,c]{William Harding,}
\author[c,d]{\\ and Noppadol Mekareeya}
\affiliation[a]{Abdus Salam Centre for Theoretical
Physics, Imperial College London, London SW7 2AZ, UK}
\affiliation[b]{Dipartimento di Fisica, Universit\`a di Milano-Bicocca, Piazza della Scienza 3, I-20126 Milano, Italy}
\affiliation[c]{INFN, sezione di Milano-Bicocca,
Piazza della Scienza 3, I-20126 Milano, Italy}
\affiliation[d]{Department of Physics, Faculty of Science, Chulalongkorn University, Phayathai Road, Pathumwan, Bangkok 10330, Thailand}
\emailAdd{rcomi@ic.ac.uk} \emailAdd{s.garavaglia18@campus.unimib.it}
\emailAdd{w.harding@campus.unimib.it}
\emailAdd{n.mekareeya@gmail.com}

\abstract{
We compute the Hilbert series of three-dimensional $\mathcal{N}=3$ quiver gauge theories by taking a specific limit of the superconformal index. Our approach introduces auxiliary fugacities associated with symmetries which, while not present in the full theory, arise as effective symmetries on specific branches of the moduli space. By evaluating the index in a limit governed by these parameters, we successfully isolate the Hilbert series of the desired branches. We validate our results against the literature and provide several new extensions. We focus primarily on linear and circular quivers with unitary gauge groups, which originate from Type IIB brane configurations involving generic $(p,q)$ fivebranes. We further generalise this approach to star-shaped and orthosymplectic $\mathcal{N}=3$ quivers. Finally, we investigate the geometric branches of affine Dynkin quivers, demonstrating agreement with known results, while offering new predictions for unexplored cases.
}

\begin{document}
\maketitle

\section{Introduction}
A fundamental feature of supersymmetric (SUSY) theories is the structure of their SUSY-preserving vacua, commonly referred to as the moduli space. The study of these spaces is of central interest, as they reveal many intrinsic properties of the underlying theories. In 3d theories, the moduli space is parametrised by operators constructed from elementary fields—known as mesonic operators—as well as by monopole operators, which may be dressed by elementary fields. Semiclassically, these monopole operators can be understood as the insertion of point-like defects in the path integral, such that the magnetic flux through a sphere enclosing the point is non-vanishing. 

In 3d theories, another remarkable phenomenon occurs: mirror symmetry \cite{Intriligator:1996ex}. This infrared (IR) duality relates pairs of 3d theories such that the mesonic operators of one theory are exchanged with the (dressed) monopole operators of the dual theory. This leads to a plethora of consequences encoded in the structure of the moduli space, which consequently exhibits a rich variety of features.

A standard strategy for studying the moduli space of 3d theories involves brane configurations \cite{Hanany:1996ie}, a method that has proven extremely successful for 3d $\mathcal{N}=4$ theories. These techniques have been extended to brane setups involving generic $(p,q)$ fivebranes \cite{Marino:2025uub,Marino:2025ihk}, which are expected to engineer 3d $\mathcal{N}=3$ theories \cite{Kitao:1998mf}. In this work, we provide a field-theoretic analysis of the moduli space of 3d $\mathcal{N}=3$ theories.\footnote{An approach using a mixture of field-theory and brane techniques has been already proposed for Abelian $\mathcal{N}=3$ theories in \cite{Assel:2017eun}.}

A similar field-theoretic approach for $\mathcal{N}=4$ theories consists of taking suitable limits of the superconformal index \cite{Bhattacharya:2008zy,Bhattacharya:2008bja,Kim:2009wb,Imamura:2011su,Kapustin:2011jm,Dimofte:2011py,Aharony:2013dha,Aharony:2013kma} to extract the Hilbert series of the two main branches: the Higgs and the Coulomb branch \cite{Razamat:2014pta}. This strategy exploits a specific property of $\mathcal{N}=4$ theories: the R-symmetry group is $\SU(2)_H \times \SU(2)_C$, where operators belonging to the Higgs or Coulomb branches transform non-trivially under only one of the two $\SU(2)$ subgroups. One can thus take a limit that effectively projects onto the desired operators to obtain the Hilbert series.\footnote{It is worth mentioning that a limit of the superconformal index for certain 3d $\CN=2$ theories was also studied in \cite{Hanany:2015via, Benvenuti:2025odh}.}

In $\mathcal{N}=3$ theories, the picture becomes more complex. The moduli space does not generally consist of only two main branches, but is instead expected to be the intersection of multiple branches. These branches are readily enumerated in the brane picture, where each corresponds to the motion of D3-branes along $(p,q)$ fivebranes of identical type. Consequently, for a given type of $(p,q)$ fivebrane, the number of branches is simply the number of pairs of such fivebranes between which D3-branes can be suspended. However, counting these branches from a field-theoretic perspective is significantly more difficult, as it requires precise knowledge of the relations among the operators parametrising the branches. Since these are typically dressed monopole operators, their relations are intrinsically non-perturbative. A second complication arises because the R-symmetry group of $\mathcal{N}=3$ theories is only $\SU(2)$. Consequently, the $\mathcal{N}=4$ approach cannot be directly extended, as there is no clear $\SU(2)$ subgroup decomposition to distinguish operators belonging to different branches.

In this work, we propose a strategy to overcome these challenges. While our approach takes inspiration from brane analysis, it is grounded in field-theoretic considerations. We demonstrate the effectiveness of this method through a variety of examples that replicate known results and provide Hilbert series for branches of theories where a brane construction is not readily available. 

Our strategy is as follows: given a 3d $\mathcal{N}=3$ theory, we provide a prescription to identify the dressed monopole operators that give rise to the branches of the theory, clarifying when these operators belong to the same or distinct branches. Once the operators are identified, the task reduces to a counting problem that accounts for the relations among them. To achieve this, we analyse the superconformal index, which is expected to encode the full operator spectrum and their non perturbative relations. For generic $\mathcal{N}=3$ theories, the superconformal index depends solely on a fugacity $x$ associated with the Cartan subgroup of the $\SU(2)$ R-symmetry, on top of fugacities for global symmetries that we will neglect for the moment. To extract the Hilbert series of a given branch of the moduli space from the index, we introduce an {\it auxiliary axial fugacity} $a$. While $a$ is not associated with a global symmetry of the full theory, it becomes a well-defined symmetry when we restrict our attention to a specific branch where vacuum expectation values of the operators that explicitly break such a symmetry vanish. 

By carefully introducing this fugacity, the operators (which include dressed monopoles) are graded by a combination of $x$ and $a$. We then extract the Hilbert series by selecting operators with a specific weight—specifically, where the power of $a$ is twice the power of $x$. This procedure isolates the operators parametrising the branch of interest while projecting out others. The resulting Hilbert series can then be tested against known descriptions, such as \cite{Cremonesi:2016nbo, Assel:2017eun}, and with the magnetic quivers recently presented in \cite{Marino:2025uub, Marino:2025ihk}. This allows us to propose that this limiting procedure is generally applicable to $\CN=3$ theories.

The theories for which we develop this prescription primarily consist of quiver theories with unitary gauge groups in linear or circular configurations. These are of particular interest due to their realisation via Type IIB brane setups preserving six supercharges. We then generalise the prescription to orthosymplectic quivers and unitary star-shaped theories, the latter of which lack a standard Hanany-Witten brane construction. Additionally, we study the geometric branch of a class of affine Dynkin quivers. In certain cases, these theories are realised by an M2-brane probing $\mathbb{C}^2/\Gamma$ singularities, where $\Gamma$ is a discrete subgroup of $\SU(2)$ following an ADE classification. For these theories, the geometric branch corresponds precisely to the space transverse to the M2-brane. We prescribe a strategy to compute this branch, replicating expected results and extending the analysis to theories with generic Chern-Simons levels. Finally, we provide quantitative comments on the geometric branch limit for cases involving multiple M2-branes.

The paper is organised as follows. In Section \ref{sec:generalprescrition}, we propose the prescription to identify the branches of $\mathcal{N}=3$ quivers and compute the corresponding Hilbert series as a limit of the superconformal index. In Section \ref{sec:section2}, we provide a series of examples involving Chern-Simons quivers with unitary gauge nodes. In Section \ref{sec:orthosimpex}, we consider orthosymplectic quivers. Finally, in Section \ref{sec:geometricbranch}, we propose a method to compute the geometric branch for a set of affine Dynkin quivers.

\section{The prescription} \label{sec:generalprescrition}
The moduli space of 3d $\mathcal{N}=3$ theories consists of an intersection of multiple hyperK\"ahler cones, typically parametrised by the vacuum expectation values (VEVs) of dressed monopole operators. In this section, we propose a strategy to derive the Hilbert series of each cone by taking a specific limit of the superconformal index, following the spirit of \cite{Razamat:2014pta}. Before describing our prescription for $\mathcal{N}=3$ theories, we briefly review the well-established case of $\mathcal{N}=4$ theories.

\subsection{Moduli space of 3d $\mathcal{N}=4$ theories from the index}
In 3d $\mathcal{N}=4$ theories, there are two primary branches, each of which is a hyperK\"ahler cone. These branches are effectively distinguished by their R-symmetry. Specifically, the R-symmetry of 3d $\mathcal{N}=4$ theories is $\SU(2)_H \times \SU(2)_C$, and the chiral operators parametrising each branch transform under only one of these $\SU(2)$ subgroups. 

The superconformal index of 3d $\mathcal{N}=4$ theories typically depends on two fugacities, $\mathcal{I}(x,a)$. These are associated, respectively, with the $\U(1)_R$ subgroup of the R-symmetry (the diagonal combination of the two Cartan generators of $\SU(2)_H \times \SU(2)_C$) and its commutant $\U(1)_A$ (the anti-diagonal combination). While the index also depends on additional fugacities associated with global symmetries, we neglect them for the present discussion. We work in the convention where a hypermultiplet carries charge $1/2$ under $\U(1)_R$ and charge $+1$ under $\U(1)_A$. 

The strategy proposed in \cite{Razamat:2014pta} consists of first redefining the fugacities as
\begin{equation}\label{N=4fugred}
    x = c h ~,\quad  a = \left( \frac{h}{c} \right)^{1/2} ~,
\end{equation}
so that the pair $(h,c)$ is associated with the $\SU(2)_H$ and $\SU(2)_C$ subgroups of the R-symmetry, respectively. The Hilbert series of the two branches is then obtained as follows:
\begin{equation}\label{N=4limitforhs}
\begin{split}
     H[\text{Higgs}](t) &= \lim_{c \to 0} \mathcal{I}(x,a) \big|_{\eqref{N=4fugred},h=t}~, \\
     H[\text{Coulomb}](t) &= \lim_{h \to 0} \mathcal{I}(x,a) \big|_{\eqref{N=4fugred},c=t}~.
\end{split}
\end{equation}
These limits are guaranteed to be well-defined if the theory satisfies a ``goodness" condition in the sense of \cite{Gaiotto:2008ak}. This condition ensures that no chiral monopole operator in the low-energy theory has a conformal dimension (computed from the UV R-symmetry group) that violates the unitarity bound.

To illustrate the strategy and set the notation for later convenience, let us consider the example of $\mathcal{N}=4$ $\U(N)$ SQCD with $F$ flavours. The superconformal index reads
\begin{equation}
\begin{split}
    \mathcal{I}(x;a;w;\vec{v}) = & \frac{1}{N!} \sum_{\{m^{(\alpha)}\} \in \mathbb{Z}^N} \oint \prod_{\alpha=1}^N \frac{du^{(\alpha)}}{2 \pi i u^{(\alpha)}} w^{\sum_{\alpha=1}^N m^{(\alpha)}} Z^{U(N)}(x;\{u^{(\alpha)}\};\{m^{(\alpha)}\}) \\
    & \qquad \quad \times \prod_{\alpha,\beta=1}^N Z^1_{\text{chir}}(x;a^{-2} u^{(\alpha)}(u^{(\beta)})^{-1}; m^{(\alpha)}-m^{(\beta)}) \\
    & \qquad \quad \times \prod_{s=\pm1}\prod_{\alpha = 1}^N \prod_{\rho=1}^F Z^{1/2}_{\text{chir}}(x;a(u^{(\alpha)} v_\rho^{-1})^s; sm^{(\alpha)}) ~,
\end{split}
\end{equation}
where the contributions of the vector and chiral multiplets are
\begin{equation}\label{indexvecchircontrib}
\begin{split}
    & Z^{\U(N)}_{\text{vec}}(x;\{u^{(\alpha)}\};\{m^{(\alpha)}\}) \\
    & = \prod_{1 \leq \alpha < \beta \leq N} x^{-|m^{(\alpha)}-m^{(\beta)}|} \prod_{s = \pm1} \left(1-(-1)^{m^{(\alpha)}-m^{(\beta)}}x^{|m^{(\alpha)}-m^{(\beta)}|} ({u^{(\alpha)}})^s ({u^{(\beta)}})^{-s} \right) ~, \\
    &Z^{R}_{\text{chir}}(x;f;m) = \left( f x^{R-1} \right)^{-\frac{|m|}{2}} \prod_{j=1}^\infty \frac{1-(-1)^m x^{2-R+|m|+2j}f^{-1}}{1-(-1)^m x^{R+|m|+2j}f} ~.
\end{split}
\end{equation}
To extract the Hilbert series, we compute the limits as in \eqref{N=4limitforhs} by first redefining the fugacities according to \eqref{N=4fugred}:
\begin{equation}\label{N=4indexch}
\begin{split}
    &\mathcal{I}(c;h;w;\vec{v}) \\&= \frac{1}{N!} \sum_{\{m^{(\alpha)}\} \in \mathbb{Z}^N} \oint \prod_{\alpha=1}^N \frac{du^{(\alpha)}}{2 \pi i u^{(\alpha)}} w^{\sum_{\alpha=1}^N m^{(\alpha)}} c^{2 \Delta(\{m^{(\alpha)}\})}\\
    & \qquad \times \prod_{1\leq \alpha \neq \beta \leq N} \left(1-(-1)^{m^{(\alpha)}-m^{(\beta)}} (h c)^{|m^{(\alpha)}-m^{(\beta)}|} \frac{u^{(\alpha)}}{u^{(\beta)}} \right) \\
    & \qquad \times \prod_{\alpha,\beta=1}^N \prod_{j=1}^\infty \frac{1-(-1)^{|m^{(\alpha)}-m^{(\beta)}|} h^{2+|m^{(\alpha)}-m^{(\beta)}|+2j}
    c^{|m^{(\alpha)}-m^{(\beta)}|+2j} (u^{(\alpha)})^{-1} u^{(\beta)}}{1-(-1)^{|m^{(\alpha)}-m^{(\beta)}|} h^{|m^{(\alpha)}-m^{(\beta)}|+2j}
    c^{2+|m^{(\alpha)}-m^{(\beta)}|+2j}u^{(\alpha)} (u^{(\beta)})^{-1}} \\
    & \qquad \times \prod_{s=\pm1}\prod_{\alpha = 1}^N \prod_{\rho=1}^F \prod_{j=1}^\infty \frac{1-(-1)^{m^{(\alpha)}} h^{1+|m^{(\alpha)}|+2j} c^{2+|m^{(\alpha)}|+2j} (u^{(\alpha)} v_\rho^{-1})^s }{1-(-1)^{m^{(\alpha)}} h^{1+|m^{(\alpha)}|+2j} c^{|m^{(\alpha)}|+2j} (u^{(\alpha)} v_\rho^{-1})^s} ~.
\end{split}
\end{equation}
Here, the $c$ factor in the first line is obtained by collecting all exponential prefactors from the definition of the vector and chiral multiplets in \eqref{indexvecchircontrib}. The function $\Delta(\{m^{(\alpha)}\})$ in the exponent represents the dimension of the bare BPS monopole operator for a given flux:
\begin{equation}
    \Delta(\{m^{(\alpha)}\}) = -\sum_{1 \leq \alpha < \beta \leq N} |m^{(\alpha)} - m^{(\beta)}| + \frac{F}{2} \sum_{\alpha=1}^N |m^{(\alpha)}| ~.
\end{equation}

\paragraph{Higgs branch limit:}
We begin with the Higgs branch limit. To obtain this Hilbert series, we take $c \to 0$. Each flux sector contributes a term weighted by $c^{\Delta(\{m^{(\alpha)}\})}$. Crucially, if we assume the goodness condition $F \geq 2N$, then $\Delta(\{m^{(\alpha)}\}) \geq 1$ for any flux $\{m^{(\alpha)}\} \neq \mathbf{0}$. Consequently, only the zero-flux sector contributes, while any non-zero flux contribution is suppressed. This is consistent with the expectation that the Higgs branch Hilbert series is parametrised solely by mesonic operators rather than monopoles. 

The limit $c \to 0$ is well-defined, with no divergent terms, and yields
\begin{equation} \label{MolienQCD}
\begin{split}
    H[\text{Higgs}](t;\{v^{(\rho)}\}) = & \frac{1}{N!} \oint \prod_{\alpha=1}^N \frac{du^{(\alpha)}}{2 \pi i u^{(\alpha)}} \prod_{1 \leq \alpha \neq \beta \leq N} \left( 1 - \frac{u^{(\alpha)}}{u^{(\beta)}} \right) \\
    & \times \PE\left[t \sum_{s=\pm 1} \sum_{\alpha=1}^N \sum_{\rho}^F  \left(\frac{u^{(\alpha)}}{v^{(\rho)}} \right)^s - t^2 \left(\sum_{i=1}^n \sum_{\alpha,\beta=1}^N \frac{u^{(\alpha)}}{u^{(\beta)}} \right)\right]~,
\end{split}
\end{equation}
where we use the asymptotic behaviour of the various terms in the index. Explicitly, for chiral multiplets, they are as follows:
\begin{equation} \label{chirallimc}
\scalebox{0.85}{$
\begin{split}
    \lim_{c \to 0} Z^{1/2}_{\text{chir}}(x; a f;m) \big|_{\eqref{N=4fugred},h=t} &\sim (c f^{-1})^{\frac{|m|}{2}} \left(\frac{1}{1-t f}\right)^{\delta_{m,0}} = \left( c f^{-1} \right) ^{\frac{|m|}{2}} \PE\left[\delta_{m,0} t f\right] ~, \\
    \lim_{c \to 0} Z^{1}_{\text{chir}}(x; a^{-2} f; m)\big|_{\eqref{N=4fugred},h=t} &\sim \left(\frac{t}{c} f^{-1} \right)^{\frac{|m|}{2}} \left(1-t^2f^{-1} \right)^{\delta_{m,0}} = \left(\frac{t}{c} f^{-1}\right)^{\frac{|m|}{2}} \PE\left[- \delta_{m,0} t^2 f^{-1}\right] ~, 
\end{split}
$}
\end{equation}
while the asymptotic behaviour for a vector multiplet is
\begin{equation}
\begin{split}
    &\lim_{c \to 0} Z^{\U(N)}_{\text{vec}}(x;\{ u^{(\alpha)} \},\{m^{(\alpha)}\})\big|_{\eqref{N=4fugred},h=t} \\
    & \sim \prod_{1 \leq \alpha < \beta \leq N} (tc)^{-|m^{(\alpha)}-m^{(\beta)}|} \prod_{s = \pm1} \left(1-(u^{(\alpha)})^s (u^{(\beta)})^{-s} \right)^{\delta_{m^{(\alpha)},m^{(\beta)}}} ~.
\end{split}
\end{equation}
The exponential prefactors reconstruct the $c^{\Delta(\{m^{(\alpha)}\})}$ term in \eqref{N=4indexch}, while the remaining terms form the integrand of \eqref{MolienQCD}, which is evaluated at zero magnetic flux. The resulting integral expression correctly provides the Hilbert series of the Higgs branch of SQCD.

\paragraph{Coulomb branch limit:}
To compute the Coulomb branch Hilbert series, we take the limit $h \to 0$. In this case, in principle, every magnetic flux contributes:
\begin{equation} \label{N=4cbhs}
\begin{split}
    \lim_{h \to 0} \mathcal{I}(c=t;h;w;\vec{v}) =& H[\text{Coulomb}](t;w) \\
    = &\sum_{\{m^{(\alpha)}\} \in \mathbb{Z}^N} t^{\Delta(\{m^{(\alpha)}\})} \frac{1}{N!} \oint \prod_{\alpha=1}^N \frac{du^{(\alpha)}}{2 \pi i u^{(\alpha)}} \frac{1}{(1-t^2)^N} \\
    &\qquad \times \prod_{1 \leq \alpha \neq \beta \leq N}
    \left( \frac{1-u^{(\alpha)}(u^{(\beta)})^{-1}}{1-t^2 u^{(\alpha)}(u^{(\beta)})^{-1}} \right)^{\delta_{m^{(\alpha)},m^{(\beta)}}} ~.
\end{split}
\end{equation}
The result is interpreted as a sum over the magnetic charge lattice, where $t$ is raised to the dimension of the monopole operator associated with a lattice point, weighted by the integral evaluating to the dressing (or $P$-) factor \cite{Cremonesi:2013lqa}. To compute the limit in \eqref{N=4cbhs}, we use the leading-order contribution of each term. For chiral multiplets, they are given by
\begin{equation} \label{chirallimh}
\begin{split}
    \lim_{h \to 0} Z^{1/2}_{\text{chir}}(x; a f;m) \big|_{\eqref{N=4fugred},c=t} &\sim \left(t f^{-1} \right)^{\frac{|m|}{2}} ~, \\
    \lim_{h \to 0} Z^{1}_{\text{chir}}(x; a^{-2} f; m)\big|_{\eqref{N=4fugred},c=t} &\sim \left(\frac{h}{t} f^{-1} \right)^{\frac{|m|}{2}} \left( \frac{1}{1-t^2 f} \right)^{\delta_{m,0}}~, 
\end{split}    
\end{equation}
while for the vector multiplet, the leading order term reads
\begin{equation}
\begin{split}
    &\lim_{h \to 0} Z^{\U(N)}_{\text{vec}}(x;\{ u^{(\alpha)} \},\{m^{(\alpha)}\})\big|_{\eqref{N=4fugred},c=t} \\
    & \sim \prod_{1 \leq \alpha < \beta \leq N} (ht)^{-|m^{(\alpha)}-m^{(\beta)}|} \prod_{s = \pm1} \left(1-(u^{(\alpha)})^s (u^{(\beta)})^{-s}\right)^{\delta_{m^{(\alpha)},m^{(\beta)}}} ~.
\end{split}    
\end{equation}
As before, the exponential prefactors form the $t^{\Delta(\{m^{(\alpha)}\})}$ contribution, with the rest constituting the integrand in \eqref{N=4cbhs}.

\subsection{The prescription for 3d $\mathcal{N}=3$ theories} \label{sec:N=3presc}
Let us now turn our attention to 3d $\mathcal{N}=3$ theories. These theories differ substantially from their $\mathcal{N}=4$ counterparts. A primary distinction is that the moduli space generally consists of an intersection of multiple hyperK\"ahler cones. These cones cannot be easily distinguished by R-symmetry, as in the $\mathcal{N}=4$ case, because the R-symmetry is reduced to a single $\SU(2)_R$. Furthermore, each cone is typically parametrised by dressed monopole operators, implying that the branches exhibit a mixture of the qualitative features characteristic of both the Higgs and Coulomb branches of $\mathcal{N}=4$ theories.

Since we lack a clear distinction between the various cones—which, in $\mathcal{N}=4$ theories, is provided by the two $\SU(2)$ subgroups of the R-symmetry—the strategy to extract the Hilbert series must be adapted. Although the superpotential does not permit a global $\U(1)_A$ axial symmetry, we nonetheless introduce an \textit{auxiliary} fugacity $a$ by assigning specific charges (positive, negative, or zero) to each hypermultiplet.

The logic of this strategy is that this new fugacity allows us to grade the operators residing on a specific branch in such a way that we can perform a limit that projects onto the operators of a single branch, analogous to the $\mathcal{N}=4$ case. While this procedure should be regarded as an effective strategy that reproduces the correct counting formula for the operators on a branch, it successfully passes a number of non-trivial tests and hints at a deeper physical interpretation.

The prescription consists of two steps. The first is the identification of the operators that can simultaneously acquire a VEV and thus form a branch of the moduli space. The second step is the practical implementation, which involves introducing the auxiliary axial fugacity $a$ and assigning charges to every field. To extract the Hilbert series, we perform the redefinition
\begin{equation}\label{N=3reparam}
    x = \mathfrak{z} t ~,\quad a = \mathfrak{z}^{-1/2}t^{1/2}\quad \text{or equivalently} \quad 
 \fz= a^{-1}x^{1/2}~, \quad t = a x^{1/2}~,
\end{equation}
and then compute the following limit of the superconformal index:
\begin{equation}\label{N=3hslimit}
    H(t) = \lim_{\mathfrak{z} \to 0} \CI(x;a) \big|_{\eqref{N=3reparam}} ~.
\end{equation}

\subsubsection*{Criteria for goodness and badness} 
This limit is well-defined, provided that potential divergences are under control. These divergences can originate from several sources. First, for a 3d $\mathcal{N}=3$ theory to avoid divergences, each local $\U(N)$ node with Chern-Simons (CS) level $k \neq 0$ and $F$ fundamental hypermultiplets must satisfy the condition
\begin{equation} \label{N=3good}
    k + F \geq N~,
\end{equation}
which can be derived from the Giveon-Kutasov duality \cite{Giveon:2008zn}. When this bound is saturated, the gauge theory confines to a trivial theory. For $k+F=N-1$ the theory exibits chiral symmetry breaking and, finally, for $k+F<N-1$ there is a runaway potential. It is important to emphasise that this condition differs significantly from the standard $\mathcal{N}=4$ goodness condition \cite{Gaiotto:2008ak}, namely, if gauge nodes with $k=0$ are present, they should satisfy
\begin{equation} 
    F \geq 2N ~,
\end{equation}
as violation in that case leads to the presence of monopole operators falling below the unitarity bound.

Once the local goodness conditions are satisfied for every node, further conditions ensure the unitarity bound is not violated for the dressed monopole operators parametrising the branches. This requirement is analogous to the criteria in \cite{Gaiotto:2008ak}. For example, consider the following $\mathcal{N}=3$ theory:

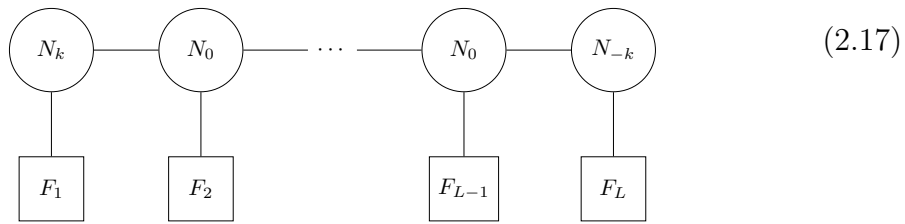
\begin{equation}
\scalebox{0.85}{
\begin{tikzpicture}[baseline,font=\footnotesize,
    circ/.style={circle, draw, minimum size=1.3cm},
    sq/.style={rectangle, draw, minimum size=1cm},
    node distance=1cm
]

    \node[circ] (n1) {$N_{k}$};
    \node[circ, right=of n1] (n2) {$N_{0}$};
    \node[right=of n2] (n3) {$\cdots$};
    \node[circ, right=of n3] (n4) {$N_{0}$};
    \node[circ, right=of n4] (n5) {$N_{-k}$};

    \node[sq, below=of n1] (s1) {$F_1$};
    \node[sq, below=of n2] (s2) {$F_2$};
    \node[sq, below=of n4] (s4) {$F_{L-1}$};
    \node[sq, below=of n5] (s5) {$F_L$};

    \draw (n1) -- (n2);
    \draw (n2) -- (n3);
    \draw (n3) -- (n4);
    \draw (n4) -- (n5);

    \draw (n1) -- (s1);
    \draw (n2) -- (s2);
    \draw (n4) -- (s4);
    \draw (n5) -- (s5);

\end{tikzpicture}}
\end{equation}
In this theory, composed of a sequence of $L$ $\U(N)$ gauge nodes with the depicted CS-levels, the lowest-dimensional dressed monopole operator arises from turning on a minimal flux $\{m^{(\alpha)}\} = (1,0,\ldots,0)$ for each node. To ensure gauge invariance, it must be dressed $k$ times with all the bifundamentals connecting the nodes. Since the bare monopole has dimension $\frac{1}{2}\sum_{i=1}^L F_i+(L-1)(N-1)-L(N-1)$, the resulting dressed monopole has dimension:
\bes{
    \frac{1}{2}(L-1)|k|+\frac{1}{2}\sum_{i=1}^L F_i -(N-1)~.
}
Requiring this dimension to be larger than or equal to one yields the condition
\begin{equation}
    (L-1) |k| + \sum_{i=1}^L F_i \geq 2N ~.
\end{equation}
This is a non-local condition, as it refers to the global topology and properties of the quiver. Specifically, it depends only on the total sum of the $F_i$ values, rather than on their individual values, and on the length of the quiver. This is not the only scenario that may appear in $\mathcal{N}=3$ theories, and we do not attempt to provide a universal rule to avoid ``badness'', which can be studied case by case with a similar strategy.

\subsubsection*{Method of assigning auxiliary axial fugacities}

The class of theories for which we implement this strategy consists mainly of linear and circular quivers with unitary gauge nodes and non-zero CS-levels. As we will demonstrate, the strategy also generalises with minor modifications to star-shaped theories and orthosymplectic quivers. A particularly interesting extension is provided by $\mathcal{N}=3$ quiver theories containing $T[\U(N)]$ blocks \cite{Gaiotto:2008ak} coupled via the gauging of their global symmetries. These theories are closely connected to Type IIB brane setups involving generic $(p,q)$ fivebranes \cite{Kitao:1998mf, Assel:2014awa}. We will comment on these generalisations later. For now, let us consider the simpler setup of a linear or circular quiver with unitary gauge groups and CS levels. To illustrate the procedure, consider the following linear quiver:\footnote{The moduli space of this example in the Abelian case $N=1$ was studied in \cite[Section 5.1]{Assel:2017eun}.}

\begin{equation} \label{extheoryN=4}
\scalebox{0.85}{
\begin{tikzpicture}[baseline,font=\footnotesize,
    circ/.style={circle, draw, minimum size=1.3cm},
    sq/.style={rectangle, draw, minimum size=1cm},
    node distance=1cm
]

    \node[circ] (n1) {$N_{k_1}$};
    \node[circ, right=of n1] (n2) {$N_{-k_1}$};
    \node[circ, right=of n2] (n3) {$N_{k_2}$};
    \node[circ, right=of n3] (n4) {$N_{k_1-k_2}$};
    \node[circ, right=of n4] (n5) {$N_{k_2-k_1}$};
    \node[circ, right=of n5] (n6) {$N_{k_1-k_2}$};
    \node[circ, right=of n6] (n7) {$N_{0}$};

    \node[sq, below=of n4] (s1) {1};
    \node[sq, below=of n6] (s2) {1};
    \node[sq, below=of n7] (s3) {$N$};

    \draw (n1) -- (n2);
    \draw (n2) -- (n3);
    \draw (n3) -- (n4);
    \draw (n4) -- (n5);
    \draw (n5) -- (n6);
    \draw (n6) -- (n7);

    \draw (n4) -- (s1);
    \draw (n6) -- (s2);
    \draw (n7) -- (s3);

\end{tikzpicture}}
\end{equation}
It is helpful to consider this theory in $\mathcal{N}=2$ notation:
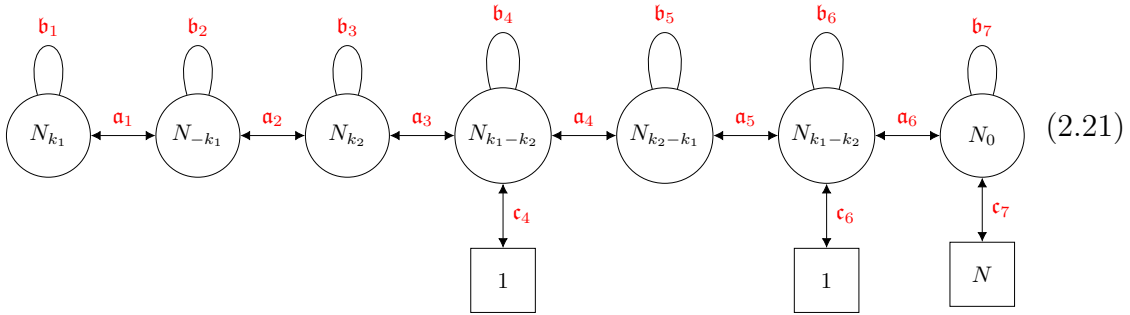
\begin{equation} \label{extheory}
\scalebox{0.85}{
\begin{tikzpicture}[baseline,font=\footnotesize,
    circ/.style={circle, draw, minimum size=1.3cm},
    sq/.style={rectangle, draw, minimum size=1cm},
    node distance=1cm,
    every loop/.style={-}
]

    \node[circ] (n1) {$N_{k_1}$};
    \node[circ, right=of n1] (n2) {$N_{-k_1}$};
    \node[circ, right=of n2] (n3) {$N_{k_2}$};
    \node[circ, right=of n3] (n4) {$N_{k_1-k_2}$};
    \node[circ, right=of n4] (n5) {$N_{k_2-k_1}$};
    \node[circ, right=of n5] (n6) {$N_{k_1-k_2}$};
    \node[circ, right=of n6] (n7) {$N_{0}$};

    \node[sq, below=of n4] (s1) {1};
    \node[sq, below=of n6] (s2) {1};
    \node[sq, below=of n7] (s3) {$N$};

    \draw[<->] (n1) -- node[above, red] {$\mathfrak{a}_1$} (n2);
    \draw[<->] (n2) -- node[above, red] {$\mathfrak{a}_2$} (n3);
    \draw[<->] (n3) -- node[above, red] {$\mathfrak{a}_3$} (n4);
    \draw[<->] (n4) -- node[above, red] {$\mathfrak{a}_4$} (n5);
    \draw[<->] (n5) -- node[above, red] {$\mathfrak{a}_5$} (n6);
    \draw[<->] (n6) -- node[above, red] {$\mathfrak{a}_6$} (n7);

    \draw[<->] (n4) -- node[right, red] {$\mathfrak{c}_4$} (s1);
    \draw[<->] (n6) -- node[right, red] {$\mathfrak{c}_6$} (s2);
    \draw[<->] (n7) -- node[right, red] {$\mathfrak{c}_7$} (s3);
    
    \path (n1) edge [loop above] node[red] {$\mathfrak{b}_1$} ();
    \path (n2) edge [loop above] node[red] {$\mathfrak{b}_2$} ();
    \path (n3) edge [loop above] node[red] {$\mathfrak{b}_3$} ();
    \path (n4) edge [loop above] node[red] {$\mathfrak{b}_4$} ();
    \path (n5) edge [loop above] node[red] {$\mathfrak{b}_5$} ();
    \path (n6) edge [loop above] node[red] {$\mathfrak{b}_6$} ();
    \path (n7) edge [loop above] node[red] {$\mathfrak{b}_7$} ();

\end{tikzpicture}}
\end{equation}
where each double-headed arrow represents a hypermultiplet, and the loops stand for the adjoint chiral multiplets belonging to the $\mathcal{N}=4$ vector multiplets. The $(\mathfrak{a}_i,\mathfrak{b}_i,\mathfrak{c}_i)$ labels denote auxiliary axial fugacities. Typically, superpotential constraints require these fugacities to be set to one. Furthermore, the supersymmetrisation of CS interactions imposes quadratic superpotential terms for the adjoint chirals of the form $\mathcal{W} \sim - \frac{k_i}{2} \phi_i^2$, see \eg~ \cite{Gaiotto:2007qi}. While one could integrate out the adjoint chirals, we retain them here as they play a key role in our analysis.

We now proceed to describe the systematic extraction of the Hilbert series for each branch of this theory.

\paragraph{Mesonic branch or the D5-branch.} In (almost) any $\mathcal{N}=3$ theory, there is a primary branch typically parametrised by mesonic operators.\footnote{With an exception in certain specific instances, such as when general $(p,q)$ fivebranes are present, this branch may be parametrised by dressed monopole operators, see \eqref{D5branchmonopole}.} As discussed later, this corresponds to the moduli space of D3-branes suspended between D5-branes, hence its name.  To compute the Hilbert series of this branch, we propose the following prescription:
\begin{itemize}
    \item \emph{Assign axial fugacities as if the theory had no CS interactions, effectively treating the theory as having $\mathcal{N}=4$ supersymmetry. The Hilbert series is then obtained by taking the Higgs branch limit of the superconformal index.}
\end{itemize}
For the example, in \eqref{extheory}, this implies fixing the axial fugacities as
\begin{equation}
    \mathfrak{a}_i = a ~,\quad \mathfrak{b}_i = a^{-2} ~,\quad \mathfrak{c}_i = a ~.
\end{equation}
The Hilbert series is then computed using the limit defined in \eqref{N=3hslimit}. 

Note that this strategy is applicable only when the starting theory satisfies the local condition $F \geq 2N$. While this condition is not strictly required for the $\mathcal{N}=3$ theory to be ``good", it ensures that the limit \eqref{N=3hslimit} yields the Molien integral associated with the theory \cite{Butti:2007jv}, which is independent of CS levels. Consequently, the D5-branch of the $\mathcal{N}=3$ theory is described by the Higgs branch of the corresponding $\mathcal{N}=4$ theory obtained by setting all CS levels to zero. 

If this prescription fails, one can employ mirror symmetry (or a more general $SL(2,\mathbb{Z})$ duality) to compute the D5-branch of the original theory as the corresponding monopole branch of the dual theory, using the prescription described below.

\paragraph{Monopole branches or the $(1,k)$-branch.}
Beyond the D5-branch, $\mathcal{N}=3$ theories possess other branches typically parametrised by the VEVs of dressed monopole operators. These branches describe the motion of D3-branes along $(1,k)$ fivebranes.\footnote{We will explicitly discuss the case of general $(p,q)$ fivebrane in Section \ref{sec:pqexample}.}

The first step is to identify the potential dressed monopole operators that contribute to a branch. The criteria for identifying such monopoles are as follows:\footnote{To establish the existence of the monopole branch discussed here, it suffices to require the presence of gauge-invariant chiral dressed operators arising from minimal magnetic fluxes. These operators are constructed solely from bare monopoles carrying a sequence of minimal fluxes, which are then dressed with chiral fields from the bifundamental hypermultiplets to ensure gauge invariance. We observe that all other dressed monopole operators are either composite or non-chiral, where the latter do not contribute to the branch of the moduli space in question. The dressed monopole operators corresponding to the geometric branch are discussed separately in Section \ref{sec:geometricbranch}.}
\begin{itemize}
    \item \emph{Each minimal sequence of consecutive gauge nodes whose CS levels sum to zero is associated with a potential dressed monopole operator.}
\end{itemize}
A sequence is ``minimal'' if it contains no proper subsequence of consecutive nodes whose CS levels also sum to zero. 
Each sequence is associated with a dressed monopole operator that can acquire a VEV. The interaction between these monopoles is determined by the following rules:
\begin{itemize}
    \item \emph{If two sequences of nodes overlap, their corresponding dressed monopoles cannot simultaneously acquire a VEV.}
    \item \emph{If two sequences are non-adjacent, the two monopoles can acquire VEVs independently, resulting in a moduli space that is the product of their individual spaces.}
    \item \emph{If two sequences are adjacent, their individual moduli spaces fuse into a larger space that is generally not a simple product.}
\end{itemize}
From these rules, we can identify the {\it maximal branches}, defined as those where the largest possible number of dressed monopole operators acquire a VEV:
\begin{itemize}
    \item \emph{The set of cones of the $\mathcal{N}=3$ theory is obtained by assigning VEVs to dressed monopole operators associated with the largest possible union of minimal sequences of gauge nodes such that the total sum of their CS-levels is zero.\footnote{For dressed elementary monopole operators, the ``total sum'' refers to $\sum_{j} k_j$ for unitary nodes with CS levels $k_j$. For orthosymplectic gauge groups, the symplectic CS levels $k'_j$ are weighted twice relative to the orthogonal levels $k_j$, \ie $\sum_j (k_j +2k'_j)$.}}
\end{itemize}
Applying this to \eqref{extheory}, we identify the minimal sequences as 
\bes{ (1,2), \, (2,3,4), \, (4,5),\,  (5,6), \, (7)~,} where gauge nodes are numbered from lefto to right. According to our rules, the monopoles for $(1,2)$ and $(2,3,4)$ cannot acquire VEVs simultaneously and thus belong to different branches. Conversely, the monopoles for $(2,3,4), (5,6),$ and $(7)$ can acquire VEVs simultaneously, forming a maximal branch. There are two additional maximal branches: one defined by $(1,2)$ and another by $(4,5)$. Together with the D5-branch, these constitute the full set of independent maximal branches for this theory.

To extract the Hilbert series for an identified branch, we use the following prescription:
\begin{itemize}
    \item \emph{For a sequence of consecutive nodes associated with monopoles that can acquire a VEV simultaneously, we assign an auxiliary $U(1)_A$ axial charge $+1$ to each bifundamental field linking two nodes within the sequence, $-1$ to any bifundamental or fundamental linked to only one node of the sequence, and zero to all other fields. Adjoint chirals are assigned non-zero charges if they join {\it any} two bifundamentals with the same axial charge (consistent with a cubic superpotential) or if the node has zero CS level. Remaining adjoint chirals are assigned zero charge, effectively becoming massive.}
\end{itemize}
In the example \eqref{extheory}, we label the three maximal branches as Branch $\text{I}$ (nodes $(1,2)$), Branch $\text{II}$ (nodes $(4,5)$), and Branch $\text{III}$ (nodes $(2,3,4), (5,6), (7)$). The axial charge assignments are summarised below:
\bes{\label{assignment123}
&\begin{tabular}{c|ccccccccc}
\hline
Branch & $\fa_1$ & $\fa_2$ & $\fa_3$ & $\fa_4$ & $\fa_5$ & $\fa_6$ & $\fc_4$ & $\fc_6$ & $\fc_7$  \\
\hline
$\text{I}$ & $a$ & $a^{-1}$ & 1 & 1 & 1 & 1& 1& 1& 1\\
$\text{II}$ & $1$ & $1$& $a^{-1}$ & $a$ & $a^{-1}$ & $1$ & $a^{-1}$ & $1$ & $1$  \\
$\text{III}$ & $a^{-1}$ & $a$ & $a$ & $a^{-1}$ & $a$ & $a^{-1}$ & $a^{-1}$ & $a^{-1}$ & $a^{-1}$  \\
\hline \hline
Branch & $\fb_1$ & $\fb_2$ & $\fb_3$ & $\fb_4$ & $\fb_5$ & $\fb_6$ & $\fb_7$  \\
\hline
$\text{I}$ & 1 & 1 &1 & 1& 1& 1& 1 \\
$\text{II}$ & 1 & 1 &1 & 1& 1& 1& 1 \\
$\text{III}$ & 1 & 1 & $a^{-2}$ & 1& 1& 1& $a^{+2}$ \\
\hline
\end{tabular} 
}
The Hilbert series is then obtained by applying the limit \eqref{N=3hslimit} to the superconformal index with these assignments. 

Before detailing the specific examples that validate our prescription, let us comment on its general consistency. For Chern-Simons theories that possess $\mathcal{N}=4$ supersymmetry, the $\SU(2)\times \SU(2)$ R-symmetry is obtained from the manifest $\mathcal{N}=3$ $\SU(2)_R$ symmetry and an extra $\U(1)_A$ axial symmetry, which provides the extra supersymmetry current leading to $\mathcal{N}=4$ supersymmetry \cite{Gaiotto:2008sd, Hosomichi:2008jd, Schnabl:2008wj, Assel:2022row, Comi:2023lfm, Li:2023ffx}. In such cases, we observe that the assigned auxiliary fugacity becomes physical, in the sense that it corresponds to a genuine global symmetry of the theory, coinciding precisely with this $\U(1)_A$. Conversely, when the theory is strictly $\mathcal{N}=3$ supersymmetric, $\U(1)_A$ is not an exact symmetry of the full theory. However, we expect that, upon restricting to a specific branch, thereby setting to zero the operators that violate this symmetry, the $\SU(2)_R$ symmetry and the auxiliary $\U(1)_A$ effectively combine into an $\SU(2) \times \SU(2)$ structure, parametrised by the fugacities $(t, \fz)$. This property is crucial for the $\fz \to 0$ limit to be well-defined. It implies that all operators not belonging to the branch are positively graded with respect to $\fz$, such that the limit simply projects out these contributions.

Further details and examples are provided in the following sections: linear quivers in Sections \ref{sec:linearex1}, \ref{sec:linearext4}, and \ref{sec:linearex2}; circular quivers in Section \ref{sec:circularex}; star-shaped examples in Section \ref{sec:starex}; orthosymplectic quivers in Section \ref{sec:orthosimpex}. The Type IIB brane setup analysis is detailed in Section \ref{sec:pqexample}, while the ``geometric branch'' found in certain circular quivers is discussed in Section \ref{sec:geometricbranch}.

\subsection{Comments on the relation with Type IIB brane setups}
Certain 3d $\CN=3$ theories are known to admit a Type IIB realisation \cite{Hanany:1996ie, Kitao:1998mf}, consisting of a stack of D3-branes intersecting a setup of fivebranes. The set of allowed fivebranes includes D5 and NS5-branes, oriented to preserve eight supercharges, as well as bound states of $p$ NS5 and $q$ D5-branes, referred to as $(p,q)$ fivebranes. The orientation of the branes is chosen such that a setup composed of generic $(p,q)$-branes preserves at least six supercharges.

The field theory residing on a stack of D3-branes intersecting a generic fivebrane setup—whether linear or circular—can be constructed via the diagonal gauging of atomic QFT blocks, each corresponding to a single fivebrane \cite{Assel:2014awa,Comi:2022aqo}. To $(0,1)$-branes (D5-branes), we associate a standard fundamental hypermultiplet, or simply a flavour. To $(1,0)$-branes (NS5-branes), we associate a bifundamental hypermultiplet.

The contribution of generic $(p,q)$-branes can be obtained by acting with an $SL(2,\mathbb{Z})$ transformation of the form $\mathrm{T}^{k_1} \mathrm{S} \mathrm{T}^{k_2} \mathrm{S} \ldots \mathrm{S} \mathrm{T}^{k_r}$ on a $(1,0)$-brane,\footnote{We adopt the following notation for the $\mathrm{S}$ and $\mathrm{T}$ generators of the $SL(2,\mathbb{Z})$ group:
\begin{equation}
    \mathrm{S} = \begin{pmatrix}
        0 & -1 \\ 1 & 0
    \end{pmatrix} ~,\quad
    \mathrm{T} = \begin{pmatrix}
        1 & 0 \\ 1 & 1
    \end{pmatrix} ~.
\end{equation}
} where the set of $k_i$ exponents is determined by the continued fraction:
\begin{equation}
    \frac{p}{q} = \frac{1}{k_1-\frac{1}{k_2-\frac{1}{\cdots-\frac{1}{k_r}}}} ~.
\end{equation}
The field theory realisation of this statement is that the contribution of a $(p,q)$-brane is generally a non-Lagrangian block that, in quiver notation, can be expressed as follows:\footnote{Note that, with respect to the notation of \cite{Assel:2014awa,Comi:2022aqo}, we adopt the opposite convention by ``acting'' with the operator on the left of the bifundamental and with its inverse on the right.}
\begin{equation} \label{pqbrane}
\scalebox{0.6}{
\begin{tikzpicture}[baseline,font=\footnotesize,
    circ/.style={circle, draw, minimum size=1.3cm},
    sq/.style={rectangle, draw, minimum size=1cm},
    node distance=1.2cm
]

    \node[sq] (n1) {$N$};
    \node[circ, right=of n1] (n2) {$N_{k_1}$};
    \node[circ, right=of n2] (n3) {$N_{k_2}$};
    \node[right=of n3] (n4) {$\cdots$};
    \node[circ, right=of n4] (n5) {$N_{k_r}$};
    \node[circ, right=of n5] (n6) {$N_{-k_r}$};
    \node[right=of n6] (n7) {$\cdots$};
    \node[circ, right=of n7] (n8) {$N_{-k_2}$};
    \node[circ, right=of n8] (n9) {$N_{-k_1}$};
    \node[sq, right=of n9] (n10) {$N$};

    \draw[blue,dashed] (n1) -- node[above, midway]{\scriptsize $T[\U(N)]$} (n2);
    \draw[blue,dashed] (n2) -- node[above, midway]{\scriptsize $T[\U(N)]$} (n3);
    \draw[blue,dashed] (n3) -- node[above, midway]{\scriptsize $T[\U(N)]$} (n4);
    \draw[blue,dashed] (n4) -- node[above, midway]{\scriptsize $T[\U(N)]$} (n5);
    \draw (n5) -- (n6);
    \draw[blue,dashed] (n6) -- node[above, midway]{\scriptsize $\bar{T[\U(N)]}$} (n7);
    \draw[blue,dashed] (n7) -- node[above, midway]{\scriptsize $\bar{T[\U(N)]}$} (n8);
    \draw[blue,dashed] (n8) -- node[above, midway]{\scriptsize $\bar{T[\U(N)]}$} (n9);
    \draw[blue,dashed] (n9) -- node[above, midway]{\scriptsize $\bar{T[\U(N)]}$} (n10);

\end{tikzpicture}}
\end{equation}
where we use the fact that a stack of D3-branes intersecting an $\mathrm{S}$ (or $\mathrm{S}^{-1}$) operator corresponds to a $T[\U(N)]$ (or $\bar{T[\U(N)]}$) theory.\footnote{A $T[\U(N)]$ theory consists of the $T[\SU(N)]$ theory with an extra BF coupling extending the $\SU(N) \times \SU(N)$ symmetry to $\U(N) \times \U(N)$. The $\bar{T[\U(N)]}$ theory employs a different choice of background configuration. We describe these theories in more detail, including their superconformal indices, in Section \ref{sec:pqexample}.} Additionally, a stack of D3-branes intersecting a $\mathrm{T}^{k}$ operator corresponds to level-$k$ Chern-Simons interactions \cite{Gaiotto:2008ak}. Note that whenever $p > 1, q \neq 0$, the corresponding field theory building block is non-Lagrangian. More precisely, it is quasi-Lagrangian, as it involves the gauging of the IR emergent symmetries of the Lagrangian $T[\U(N)]$ theory.

The theories addressed by the prescription in Section \ref{sec:N=3presc} involve brane setups with $(p,q)$-branes, where we only allow $p = 1$ and $(0,1)$-branes. For instance, the theory in \eqref{extheoryN=4} corresponds to the following brane setup:
\begin{equation} \label{exbranes}
\scalebox{0.93}{
\begin{tikzpicture}[baseline,font=\footnotesize,
    cross_node/.style={circle, draw=cyan, cross out, thick, minimum size=6pt, inner sep=0pt}
]

    \def\ystart{-1.5}
    \def\yend{1.5}

    \draw[thick] (0,0) -- node[near start, below, xshift=-2.5cm]{D3} (13.25,0);

    \draw[red, thick] (0,\ystart) -- (0,\yend) node[pos=0, below, red] {NS5};
    \draw[red, thick] (3,\ystart) -- (3,\yend) node[pos=0, below, red] {NS5};

    \draw[blue, dashed, thick] (1, \ystart) -- (2.5, \yend) node[pos=0, below, blue] {$(1, k_1)$};
    \draw[purple, dashed, thick] (4, \ystart) -- (5.5, \yend) node[pos=0, below, purple] {$(1, k_2)$};
    \draw[blue, dashed, thick] (6, \ystart) -- (7.5, \yend) node[pos=0, below, blue] {$(1, k_1)$};
    \draw[purple, dashed, thick] (8, \ystart) -- (9.5, \yend) node[pos=0, below, purple] {$(1, k_2)$};
    \draw[blue, dashed, thick] (10, \ystart) -- (11.5, \yend) node[pos=0, below, blue] {$(1, k_1)$};
    \draw[blue, dashed, thick] (12.5, \ystart) -- (14, \yend) node[pos=0, below, blue] {$(1, k_1)$};

    \node[cross_node, label={[cyan]above:D5}] at (5.8, 0.5) {}; 
    \node[cross_node] at (9.8, 0.5) {}; 
    \node[cross_node] at (11.8, 0.5) {};
    \node[] at (12.3, 0.5) {$\cdots$};
    \node[cross_node] at (12.8, 0.5) {};

\end{tikzpicture}}
\end{equation}
The prescription described in Section \ref{sec:N=3presc} admits a direct interpretation from the brane setup perspective. The branches of the theory correspond to the possible motion of D3-branes suspended on a chosen type of fivebrane. The D5-branch corresponds to the motion of D3-branes on D5-branes, while the branches typically parametrised by dressed monopole operators correspond to the motion of D3-branes on NS5 or generic $(1,k)$-branes. Each independent motion corresponds to a VEV for a dressed monopole operator. The simultaneous motion of D3-branes on two different types of fivebranes is not permitted and does not yield a branch, whereas independent motions give rise to independent branches. These observations replicate the rules established in Section \ref{sec:N=3presc}. Referring back to the example in \eqref{exbranes} (theory in \eqref{extheoryN=4}), we can immediately see that the theory should possess four branches, \ie one for each distinct type of fivebrane in the setup. Aside from the D5-branch, we can identify the dressed monopoles contributing to each branch from the setup. For example, the monopoles associated with nodes $(1,2)$ and $(4,5)$ correspond to the motion of D3-branes on NS5 and $(1,k_2)$-branes, respectively. Meanwhile, the monopoles associated with nodes $(2,3,4)$, $(5,6)$, and $(7)$ correspond to motion along $(1,k_1)$-branes. It is thus conventional to refer to branches I, II and III in \eqref{assignment123} as the NS5, $(1,k_2)$, and $(1, k_1)$ branches, respectively.

This interpretation facilitates the identification of possible branches from the brane setup and informs the prescription for assigning the axial fugacity $a$:
\begin{itemize}
    \item \emph{Once a branch corresponding to the motion of D3-branes along a chosen type of $(1,k)$-brane is selected, we assign fugacity $a^{-1}$ to each bifundamental associated with that $(1,k)$-brane,\footnote{This refers to the bifundamental hypermultiplet arising from the string stretched between the two D3-branes intersecting the $(1,k)$-brane in question.} and fugacity $a$ to each bifundamental corresponding to different $(1,\tilde{k})$-branes between any pair of the primary fivebrane type. Flavours associated with D5-branes located between the $(1,k)$-branes are assigned fugacity $a^{-1}$. Every other hypermultiplet is assigned an axial fugacity of $a^0=1$.}
\end{itemize}
This rule replicates the prescription of Section \ref{sec:N=3presc} from the point of view of the brane configuration.

Finally, we can consider more generic brane setups involving $(p,q)$-branes. While identifying the possible branches remains straightforward from the brane perspective, the field-theoretic rules in Section \ref{sec:N=3presc} must be generalised to account for $T[\U(N)]$ links. The analysis of such theories is quite complex. To provide a viable approach, we propose an example in Section \ref{sec:pqexample} demonstrating that our prescription can be generalised to these cases.

\section{Unitary quivers}\label{sec:section2}
In this section, we report examples to motivate the prescription presented in Section \ref{sec:generalprescrition} and, most importantly, to describe in detail how the computation of the branches can be performed. The examples considered in this section include quivers with unitary gauge nodes and with linear, circular or star shape. We will also include an example with $T[\U(N)]$ links, which, as commented previously, is connected to Type IIB brance setups with generic $(p,q)$ branes. Examples of quivers with orthosymplectic gauge groups are given later, in Section \ref{sec:orthosimpex}.

\subsection{Linear quiver with $\U(N)_k \times \U(N)_{-k} \times \U(N)_k$ gauge group} \label{sec:linearex1}
We consider the 3d $\CN=3$ theory described by the following quiver diagram:
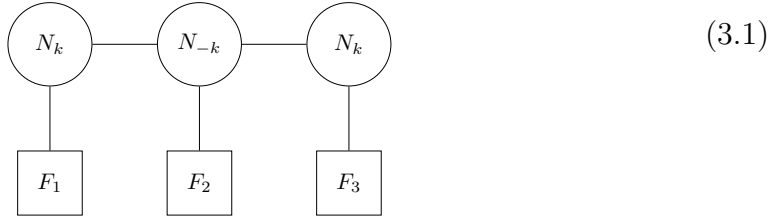
\begin{equation} \label{quivT3theory}
\scalebox{0.85}{
\begin{tikzpicture}[baseline,font=\footnotesize,
    circ/.style={circle, draw, minimum size=1.3cm},
    sq/.style={rectangle, draw, minimum size=1cm},
    node distance=1cm
]
    \node[circ] (n1) {$N_{k}$};
    \node[circ, right=of n1] (n2) {$N_{-k}$};
    \node[circ, right=of n2] (n3) {$N_{k}$};
    \node[sq, below=of n1] (s1) {$F_1$};
    \node[sq, below=of n2] (s2) {$F_2$};
    \node[sq, below=of n3] (s3) {$F_3$};
    \draw (n1) -- (n2);
    \draw (n2) -- (n3);
    \draw (n1) -- (s1);
    \draw (n2) -- (s2);
    \draw (n3) -- (s3);
\end{tikzpicture}}
\end{equation}
This theory can be realised in string theory via the brane configuration shown below:
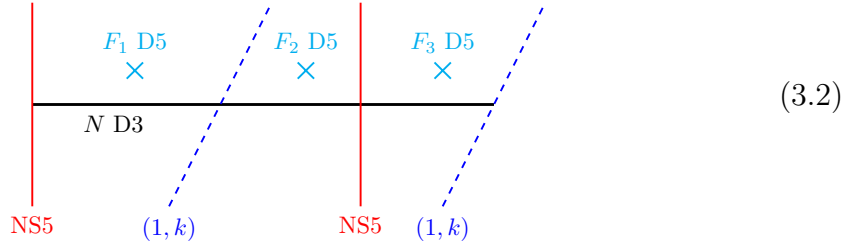
\begin{equation}
\scalebox{0.9}{
\begin{tikzpicture}[baseline,font=\footnotesize,
    cross_node/.style={circle, draw=cyan, cross out, thick, minimum size=6pt, inner sep=0pt}
]
    \def\ystart{-1.5}
    \def\yend{1.5}
    \draw[very thick] (0,0) -- node[near start, below,xshift=-0.5cm]{$N$ D3} (6.75,0);
    \draw[red, thick] (0,\ystart) -- (0,\yend) node[pos=0, below, red] {NS5};
    \draw[red, thick] (4.8,\ystart) -- (4.8,\yend) node[pos=0, below, red] {NS5};
    \draw[blue, dashed, thick] (1+1, \ystart) -- (2.5+1, \yend) node[pos=0, below, blue] {$(1, k)$};
    \draw[blue, dashed, thick] (6, \ystart) -- (7.5, \yend) node[pos=0, below, blue] {$(1, k)$};
    \node[cross_node, label={[cyan]above:$F_1$ D5}] at (1.5, 0.5) {};
    \node[cross_node, label={[cyan]above:$F_2$ D5}] at (4, 0.5) {};
    \node[cross_node, label={[cyan]above:$F_3$ D5}] at (6, 0.5) {};
\end{tikzpicture}}
\end{equation}
For the special case of $N=1$, $F_1=1$, $F_2=0$, and $F_3=1$, this model was named the $T[3]$ theory, and its moduli space was analysed in \cite[Section 4.4]{Assel:2017eun}. The magnetic quivers associated with each branch of the moduli space were proposed in \cite[Section 5.1.1]{Marino:2025uub}. In what follows, we analyse this theory by computing its superconformal index and limits thereof, as described in Section \ref{sec:generalprescrition}.

It is useful to consider the theory in 3d $\CN=2$ notation, where the field content and fugacity assignments are:
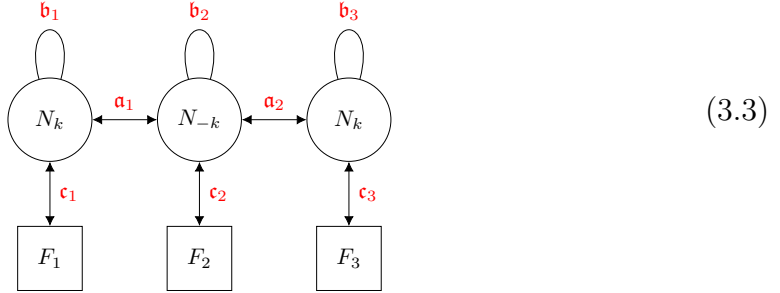
\begin{equation} \label{quivT3axial}
\scalebox{0.85}{
\begin{tikzpicture}[baseline, font=\footnotesize,
    circ/.style={circle, draw, minimum size=1.3cm},
    sq/.style={rectangle, draw, minimum size=1cm},
    node distance=1cm,
    every loop/.style={-}
]
    \node[circ] (n1) {$N_{k}$};
    \node[circ, right=of n1] (n2) {$N_{-k}$};
    \node[circ, right=of n2] (n3) {$N_{k}$};
    \node[sq, below=of n1] (s1) {$F_1$};
    \node[sq, below=of n2] (s2) {$F_2$};
    \node[sq, below=of n3] (s3) {$F_3$};
    \draw[<->] (n1) -- node[above, red] {$\fa_1$} (n2);
    \draw[<->] (n2) -- node[above, red] {$\fa_2$} (n3);
    \draw[<->] (n1) -- node[right, red] {$\fc_1$} (s1);
    \draw[<->] (n2) -- node[right, red] {$\fc_2$} (s2);
    \draw[<->] (n3) -- node[right, red] {$\fc_3$} (s3);
    \path (n1) edge [loop above] node[red] {$\fb_1$} ();
    \path (n2) edge [loop above] node[red] {$\fb_2$} ();
    \path (n3) edge [loop above] node[red] {$\fb_3$} ();
\end{tikzpicture}}
\end{equation}
The specific assignments of axial fugacities $\fa_i$, $\fb_i$, and $\fc_i$ depends on which branch we aim to compute and are assigned according to the rules in Section \ref{sec:N=3presc}. The assignment of axial charges is summarised in the following table:
\bes{ \label{branchT3label}
\begin{tabular}{c|cccccccccc}
\hline
Branch & $\fa_1$ & $\fa_2$  & $\fc_1$ & $\fc_2$ & $\fc_3$ & $\fb_1$ & $\fb_2$ & $\fb_3$  \\
\hline
NS5& $a$ & $a^{-1}$ & $a^{-1}$ & $a^{-1}$ & 1 & 1 & 1 &1 \\
$(1,k)$ & $a^{-1}$ & $a$ & $1$ & $a^{-1}$ & $a^{-1}$ & 1 & 1 &1    \\
D5 & $a$ & $a$ & $a$ & $a$ & $a$ & $a^{-2}$ & $a^{-2}$ & $a^{-2}$ \\
\hline
\end{tabular}
}
The index for the theory in \eqref{quivT3axial} is then given by
\bes{ \label{indexquivT3}
\scalebox{0.99}{$
\begin{split}
&\CI_{\eqref{quivT3axial}} = \frac{1}{(N!)^3} \sum_{\{m_1^{(\alpha)}\}, \{m_2^{(\alpha)}\}, \{m_3^{(\alpha)}\} \in \BZ^N} \oint \Bigg( \prod_{i=1}^3 \prod_{\alpha=1}^N \frac{du^{(\alpha)}_i}{2\pi i u^{(\alpha)}_i}  (u^{(\alpha)}_i)^{\kappa_i m^{(\alpha)}_i}  w_i^{m^{(\alpha)}_i} \Bigg) \\
&\qquad \times \prod_{i=1}^3  Z_{\text{vec}}^{\U(N)}(x; \{u^{(\alpha)}_i\}; \{m^{(\alpha)}_i\}) \prod_{\alpha, \beta=1}^N Z_{\text{chir}}^{1} \left(x; \fb_i u^{\alpha}_i (u^{\beta}_i)^{-1} ; m^{(\alpha)}_i - m^{(\beta)}_i  \right) \\ 
& \qquad \times \prod_{s = \pm1} \prod_{i=1}^2 \prod_{\alpha, \beta=1}^N Z_{\text{chir}}^{1/2} \left(x; \fa_i (b_i u^{(\alpha)}_i (u^{(\beta)}_{i+1})^{-1})^s; s(m^{(\alpha)}_{i} - m^{(\beta)}_{i+1}) \right) \\ 
& \qquad \times \prod_{s = \pm1} \prod_{j=1}^3 \prod_{\rho_j=1}^{F_j} \prod_{\alpha=1}^N  Z_{\text{chir}}^{1/2} \left(x; \fc_j ((f^{(\rho_j)}_j)^{-1} u^{(\alpha)}_j)^s; sm^{(\alpha)}_{j} \right)~,
\end{split}$}
}
where the Chern-Simons levels are
\bes{ \label{kappaT3}
\kappa_1 = k~, \qquad \kappa_2 = -k~, \qquad \kappa_3 = k~.
}
For definiteness, we take $k>0$ throughout this discussion. In the above, the fugacities $b_i$ are turned on only for the sake of convenience in keeping track of each bifundamental hypermultiplet, and can be absorbed in the redefinition of the fugacities for topological symmetries $w_i$. 

The goodness condition for this theory is
\begin{equation}
    F_1+F_2+k \geq \frac{N}{2} ~,\quad F_2+F_3+k \geq \frac{N}{2} ~,
\end{equation}
which ensures that no monopole operator violates the unitarity bound.

The Hilbert series for each branch is obtained by first performing the redefinition
\begin{equation}\label{ex1fugred}
    x = \fz t ~,\quad a = \fz^{-1/2}t^{1/2} ~,
\end{equation}
and then taking a specific limit of the index:
\bes{ \label{limitHS}
H_{\eqref{quivT3axial}}[\text{branch}](t;\{w_i\}; \{b_i\}; \{f_i \}) = \lim_{\fz \rightarrow 0} \CI_{\eqref{quivT3axial}} \Big|_{\eqref{branchT3label}, \, \eqref{ex1fugred}}~.
}

\subsubsection{Abelian case: $N=1$}
Let us focus on the $N=1$ case, where the index \eqref{indexquivT3} simplifies to:
\bes{
\begin{split}
\CI_{\eqref{quivT3axial}, N=1} = & \sum_{m_1,m_2, m_3 \in \BZ} \oint  \Bigg( \prod_{i=1}^3 \frac{du_i}{2\pi i u_i}  u_i^{\kappa_i m_i}  w_i^{m_i} Z_{\text{chir}}^{1} \left(x; \fb_i; 0 \right) \Bigg) \\ 
& \qquad \times \prod_{s = \pm1} \prod_{i=1}^2 Z_{\text{chir}}^{1/2} \left(x; \fa_i (b_i u_i u_{i+1}^{-1})^s; s(m_{i} - m_{i+1}) \right)  \\
&\qquad \times \prod_{j =1}^3 \prod_{\rho_j=1}^{F_j} Z_{\text{chir}}^{1/2} \left(x; \fc_j ((f_j^{(\rho_j)})^{-1} u_j)^s; sm_{j} \right) ~.
\end{split}
}
To compute the limit, we first replace the definition of the chiral multiplet contribution, given in \eqref{indexvecchircontrib}, and, for each assignment of axial fugacities in \eqref{branchT3label}, we collect an overall exponential $\fz$ factor.
A non-vanishing result in the $\fz \to 0$ limit imposes the following constraints on the magnetic fluxes $m_i$:
\bes{ \label{condflux}
\begin{tabular}{c|c|c}
\hline
Branch & Power of $\fz$ & Condition \\
\hline
NS5 & $\fz^{|m_1-m_2|+\frac{F_3}{2}|m_3|}$ & $m_1=m_2$, $m_3=0$ \\
$(1,k)$ & $\fz^{|m_2-m_3|+\frac{F_1}{2}|m_1|}$ & $m_2=m_3$, $m_1=0$ \\
D5 & $\fz^{|m_1-m_2|+|m_2-m_3|+\sum_{i=1}^3 F_i|m_i|}$ & $m_1=m_2=m_3=0$ \\
\hline
\end{tabular}
}
We now discuss the computation for each branch explicitly.

\subsubsection*{The NS5-branch}
The limit of the index for the NS5-branch is
\begin{equation}
\begin{split}
    & H[\text{NS5}](t;\{w_i\};\{b_i\}) = \lim_{\fz \to 0} \sum_{m_1,m_2,m_3 \in \mathbb{Z}} \fz^{|m_1-m_2|+\frac{F_3}{2}|m_3|} t^{2 \Delta(m_1,m_2,m_3)} \\
    &\qquad \quad \times \oint \prod_{i=1}^3 \frac{d u_i}{2\pi i u_i} u_i^{\kappa_i m_i} w_i^{m_i} PE\left[ \delta_{m_1,m_2} t \sum_{s = \pm1} (b_1 u_1 u_2^{-1})^s \right] ~.
\end{split}
\end{equation}
Here, the argument of the plethystic exponential only includes contributions from chiral fields with axial fugacity $a$, as the contribution of a chiral field with axial fugacity $a^{-1}$ becomes trivial in the $\fz \to 0$ limit (see \eqref{chirallimc} and \eqref{chirallimh}). 
The function $\Delta(m_1,m_2,m_3)$ is the conformal dimension of a bare monopole:
\bes{
\Delta(m_1, m_2, m_3) = \frac{1}{2}(F_1|m_1|+|m_1-m_2|+F_2|m_2|+|m_2-m_3|+F_3|m_3|)~.
}
In the limit $\fz \to 0$, only terms carrying zero power of $\fz$ can contribute, which imposes the condition on the magnetic fluxes: $m_1=m_2$ and $m_3=0$\footnote{As a technical aside, if $F_3$ is set to zero, then no condition on $m_3$ arises. However, the integral over $u_3$ now simply reads
\begin{equation}
    \oint \frac{d u_3 u_3^{k \, m_3}}{2 \pi i u_3} = \delta_{m_3,0}~,
\end{equation}
thus forcing $m_3$ to be zero. This argument holds for any value $F_3 \neq 0$ regardless of the condition imposed by the power of $\fz$.}. We thus obtain the following expression for the Hilbert series:
\bes{ \label{HSNSabel}
\begin{split}
&H_{\eqref{quivT3axial}}[\text{NS5}](t;\{w_{i}\};\{b_{i}\}) = \sum_{m_1=m_2 = m\in \BZ}  t^{2 \Delta(m_1 = m,\, m_2 =m, \,m_3=0)} \\
& \qquad \times \oint \frac{d u_1}{2\pi i u_1} \frac{d u_2}{2\pi i  u_2} \frac{d u_3}{2\pi i u_3} u_1^{\kappa_1 m_1} u_2^{\kappa_2 m_2} w_1^{m_1} w_2^{m_2} \PE\left[t \sum_{s=\pm 1} (b_1 u_1 u_2^{-1})^s  \right] \\
&\overset{\eqref{usefulintiden}}{=} \sum_{m\in \BZ} t^{(F_1+F_2 +1+k) |m|} \frac{1}{1-t^2} y_1^m~, \qquad y_1=b_1^{-k} w_1 w_2 \\
&= \PE\left[ t^2 + (y_1+ y_1^{-1})t^{F_1+F_2+k+1}- t^{2(F_1+F_2+k+1)}\right]~.
\end{split}
}
In the first step, we recognise that the integral over $u_3$ yields one, while the integral over $u_{1,2}$ is non-trivial and evaluates to the function $t^{k |m|}y_1^m (1-t^2)^{-1}$ upon applying identity \eqref{usefulintiden}. The $t$ factor at the numerator adds up to the conformal dimension of the bare monopole operator and represents the contribution of the dressing of bifundamental fields, so that the final power of $t$ proportional to $|m|$ is (twice) the conformal dimension of a gauge invariant dressed monopole operator. The denominator instead is analogous to a $P$-factor in the language of \cite{Cremonesi:2013lqa}.
The final expression in \eqref{HSNSabel} is the Hilbert series for the orbifold $\mathbb{C}^2/\mathbb{Z}_{F_1+F_2+k+1}$. As shown in the preceding line, this matches the Coulomb branch Hilbert series of a U(1) gauge theory with $F_1+F_2+k+1$ flavours, which is the proposed magnetic quiver for this branch \cite{Marino:2025uub}.

The elementary bare monopole operators on this branch, $V_{\pm(1,1,0)}$, have an R-charge of $\frac{1}{2}(F_1+F_2+1)$ and axial charge of $F_1+F_2+1$. The generators of this branch are
\bes{
G_0 = X_{12} X_{21}~,\quad  G_+ = X_{21}^k V_{(1,1,0)}~, \quad G_- = X_{12}^k V_{(-1,-1,0)}~,
}
which satisfy the relation
\bes{
G_+ G_- = G_0^{F_1+F_2+k+1}~,
}
where $X_{12}$ and $X_{21}$ are the chiral and antichiral fields in the bifundamental hypermultiplet connecting the first and second gauge nodes, respectively.\footnote{Here and throughout the paper, we denote by $X_{ij}$ the chiral field in the bifundamental hypermultiplet going from the $i$-th gauge node to the $j$-th gauge node. This is conventional in the literature of quiver gauge theories.}

\subsubsection*{The $(1,k)$-branch}
The result for this branch can be obtained from that of the NS5-branch by exchanging the nodes $1$ and $3$:
\bes{ \label{1kT3abel}
\begin{split}
&H_{\eqref{quivT3axial}}[(1,k)](t;\{w_{i}\},\{b_{i}\}) = \sum_{m_2=m_3 = m\in \BZ} t^{2 \Delta(m_1=0, \, m_2 = m, \, m_3 =m)} \\
& \qquad \times \oint \frac{d u_1}{2\pi i u_1} \frac{d u_2}{2\pi i u_2} \frac{d u_3}{2\pi i u_3} u_2^{\kappa_2 m_2} u_3^{\kappa_3 m_3} w_2^{m_2} w_3^{m_3}  \PE\left[t \sum_{s=\pm 1} (b_2 u_2 u_3^{-1})^s  \right] \\
&\overset{\eqref{usefulintiden}}{=} \sum_{m\in \BZ} t^{(F_2+F_3 +k+1) |m|} \frac{1}{1-t^2} y^m~, \qquad y_2=b_2^k w_2 w_3 \\
&= \PE\left[ t^2 + (y_2+ y_2^{-1})t^{F_2+F_3+k+1}- t^{2(F_2+F_3+k+1)}\right]~.
\end{split}
}
This is the Hilbert series for $\mathbb{C}^2/\mathbb{Z}_{F_2+F_3+k+1}$, matching the Coulomb branch Hilbert series of a U(1) gauge theory with $F_2+F_3+k+1$ flavours.

The generators of this branch are analogous to those of the NS5-branch, with the replacements $V_{\pm(1,1,0)} \to V_{\pm(0,1,1)}$ and $(X_{12}, X_{21}) \to (X_{32}, X_{23})$.

\subsubsection*{The D5-branch}
Finally, assuming that $F_1 \neq 0$ and $F_3 \neq0$, the Hilbert series for the D5-branch is\footnote{Notice that this assumption is sufficient, but not necessary, to have a good theory. For cases in which $F_1=0$ and/or $F_3=0$, we need an improved strategy to compute the D5-branch Hilbert series. We leave this analysis for a future work.}
\bes{ \label{HSD5T3abel}
\begin{split}
&H_{\eqref{quivT3axial}}[\text{D5}](t;b_{1,2,3}) = \oint \frac{du_1}{2\pi i u_1} \frac{d u_2}{2\pi i u_2} \frac{d u_3}{2\pi i  u_3} \\
& \qquad \quad \times \PE\left[t \sum_{s=\pm 1} \left\{ \sum_{i=1}^2 (b_i u_i u_{i+1}^{-1})^s + \sum_{j=1}^3 \sum_{\rho_j=1}^{F_j} (f^{(\rho_j)}_j u_j^{-1})^s  \right \} -3t^2 \right]~.
\end{split}
}
The term $-3t^2$ in the plethystic exponential accounts for the three adjoint chiral multiplets, which have an axial fugacity of $a^{-2}$. This result corresponds to the Higgs branch Hilbert series of the theory in \eqref{quivT3theory} with the Chern-Simons levels set to zero ($k=0$), which is the 3d $\CN=4$ theory:
\begin{equation} \label{elecquivD5}
\scalebox{0.85}{
\begin{tikzpicture}[baseline,font=\footnotesize,
    circ/.style={circle, draw, minimum size=1cm},
    sq/.style={rectangle, draw, minimum size=1cm},
    node distance=1cm
]
    \node[circ] (n1) {$1$};
    \node[circ, right=of n1] (n2) {$1$};
    \node[circ, right=of n2] (n3) {$1$};
    \node[sq, below=of n1] (s1) {$F_1$};
    \node[sq, below=of n2] (s2) {$F_2$};
    \node[sq, below=of n3] (s3) {$F_3$};
    \draw (n1) -- (n2);
    \draw (n2) -- (n3);
    \draw (n1) -- (s1);
    \draw (n2) -- (s2);
    \draw (n3) -- (s3);
\end{tikzpicture}
}
\end{equation}
The magnetic quiver for this branch is simply the mirror theory for this quiver. For example, in the special case of $F_1=1$, $F_2=0$, and $F_3=1$, \eqref{elecquivD5} is mirror dual to the magnetic quiver consisting of the $\U(1)$ gauge theory with four flavours \cite[Section 5.1.1]{Marino:2025uub}.

In summary, the results for each branch are:
\bes{
\begin{tabular}{c|c|c}
\hline
Branch & Hilbert series & Geometry \\
\hline
NS5 & \eqref{HSNSabel} & $\mathbb{C}^2/\BZ_{F_1+F_2+k+1}$ \\
$(1,k)$ & \eqref{1kT3abel} & $\mathbb{C}^2/\BZ_{F_2+F_3+k+1}$ \\
D5 & \eqref{HSD5T3abel} & Higgs branch of \eqref{elecquivD5} \\
\hline
\end{tabular}
}
These Hilbert series can also be derived using the methods presented in \cite{Cremonesi:2016nbo}; see Appendix C therein for related examples.

\subsubsection{The case of $N=2$}
We now turn to the non-Abelian case of $N=2$.

\subsubsection*{The NS5-branch}
Following a similar analysis as in the Abelian case, we compute the limit
\begin{equation}
\begin{split}
    &H[\text{NS5}](t,\{w_i\},\{b_i\}) = \lim_{\fz \to 0} \sum_{\{\{m_1^{(\alpha)}\},\{m_2^{(\alpha)}\},\{m_3^{(\alpha)}\}\in \BZ^2}\fz^{\Delta_\fz (\vec{m}_1,\vec{m}_2,\vec{m}_3)} t^{2 \Delta(\vec{m}_1,\vec{m}_2,\vec{m}_3)} \\
    &\quad \times \oint \prod_{i=1}^3 \prod_{\alpha=1,2} \frac{d u_i^{(\alpha)}}{2\pi i u_i^{(\alpha)}} \left( u_i^{(\alpha)} \right)^{\kappa_i m^{(\alpha)}} w_i^{m_i^{(\alpha)}} \prod_{1 \leq \alpha \neq \beta \leq 2} \prod_{i=1}^3 \left( 1-\frac{u_i^{(\alpha)}}{u_i^{(\beta)}} \right)^{\delta_{m_i^{(\alpha)},m_i^{(\beta)}}} \\
    &\quad \times \PE\left[t \sum_{s=\pm1} \sum_{\alpha,\beta=1}^2 \delta_{m_1^{(\alpha)},m_2^{(\beta)}} ( b_1 u_1^{(\alpha)})^s (u_2^{(\beta)})^{-s} \right] ~.
\end{split}
\end{equation}
The argument of the plethystic exponential only includes contributions from chiral fields with axial fugacity $a$. The function $\Delta(\vec{m}_1,\vec{m}_2,\vec{m}_3)$ is the conformal dimension of the bare monopole operator:
\bes{
\begin{split}
\Delta(\vec{m}_1,\vec{m}_2,\vec{m}_3) 
&= \frac{1}{2}\Big(  \sum_{j=1}^3 F_j \sum_{\alpha=1}^2|m^{(\alpha)}_j|+\sum_{\alpha,\beta=1,2} ( |m^{(\alpha)}_1-m^{(\beta)}_2| \\
&\qquad +|m^{(\alpha)}_2-m^{(\beta)}_3|) - 2\sum_{i=1}^3 |m^{(1)}_i-m^{(2)}_i| \Big)~.
\end{split}
}
The condition for a non-vanishing limit requires the exponent of $\fz$ to be zero:
\begin{equation}
\begin{split}
    &\Delta_\fz(\vec{m}_1,\vec{m}_2,\vec{m}_3) \\ 
    &= \sum_{\alpha,\beta=1,2} |m^{(\alpha)}_1-m^{(\beta)}_2|+\frac{F_3}{2} \sum_{\alpha=1}^2 | m^{(\alpha)}_3| - \sum_{i=1}^3 |m^{(1)}_i-m^{(2)}_i| = 0~.
    \end{split}
\end{equation}
We observe that the exponent of $\fz$ might be negative, which happens for $F_3 = 0$, and this could lead to divergencies in the $\fz \to 0$ limit. However, notice that the $u_3$ integral, for generic values of the magnetic fluxes and in the $\fz \to 0$ limit, reads
\begin{equation}
    \frac{1}{2} \oint \prod_{\alpha =1,2} \frac{d u_3^{(\alpha)} (u_3^{(\alpha)})^{k \,m_3^{(\alpha)}}}{2 \pi i u_3^{(\alpha)}} \prod_{s=\pm1}\left[ 1-(u_3^{(1)})^s (u_3^{(2)})^{-s} \right] = \delta_{\mathbf{m}_3,0}~,
\end{equation}
thus setting $\mathbf{m}_3$ to zero and avoiding possible negative powers of $\fz$.\footnote{This is a general feature. Whenever a gauge node sees only hypermultiplets with fugacity $a^{-1}$ or $a^0$, then its integral can be solved to effectively set its corresponding magnetic flux to zero. As expected, gauge nodes that are distant from those that are involved in the construction of dressed monopole operators do not play any role.\label{footnotedistnodes}} Then we just need to look for zero powers of $\fz$, which give a non-vanishing contribution to the limit $\fz \to 0$.
All in all, we restrict the magnetic fluxes to be of the form $\vec{m}_3 = (0,0)$ and $\vec{m}_2 = \vec{m}_3 = (\fm,\fn) \in \mathbb{Z}^2$. Furthermore, using the Weyl symmetry, we can impose $\fm \geq \fn$.

Taking the limit \eqref{limitHS}, the Hilbert series receives two distinct contributions, analogous to the examples in \cite[Appendix C]{Cremonesi:2016nbo}, namely one from fluxes where $\fm=\fn$, and another from fluxes where $\fm>\fn$:
\bes{ \label{sumtwocontrNS}
H_{\eqref{quivT3axial}}[\text{NS5}](t;w_{1,2};b_{1,2}) = H_{\fm=\fn} + H_{\fm>\fn}~.
}
For fluxes with $\fm=\fn$, each U(2) gauge group is unbroken, and the contribution is
\bes{ \label{Hm=nNS}
\scalebox{0.97}{$
\begin{split}
&H_{\fm=\fn} = \frac{1}{2^2}  \sum_{\fm \in \BZ} \oint \prod_{i=1}^2 \prod_{\alpha=1}^2 \frac{du^{(\alpha)}_i}{2\pi i u^{(\alpha)}_i} \left( \frac{u^{(1)}_1 u_1^{(2)}}{u_2^{(1)} u_2^{(2)}} \right)^{k \fm} (w_1 w_2)^{2\fm} t^{2 \Delta(\vec m_1 = \vec m_2=(\fm, \fm), \, \vec m_3 = \vec 0)} \\
& \qquad \qquad \times \prod_{1\leq \alpha \neq \beta \leq 2} \prod_{i=1}^2 \Bigg(1-\frac{u^{(\alpha)}_i}{u^{(\beta)}_i} \Bigg)  \PE\left[t \sum_{s=\pm 1} \sum_{\alpha, \beta=1,2} (b_1 u^{(\alpha)}_1)^s (u^{(\beta)}_{2})^{-s}  \right]  ~.
\end{split}
$}
}
We can simplify \eqref{Hm=nNS} by explicitly evaluating the integral using identity \eqref{U2identity}, which leads to
\bes{
H_{\fm=\fn} &= \sum_{\fm \in \BZ} \frac{t^{k(2|\fm|)} (b_1^{-k} w_1 w_2)^{2\fm}}{(1-t^2)(1-t^4)} \times t^{(F_1 + F_2+2)(2|\fm|)} \\
& = \sum_{\fm \in \BZ} \frac{t^{(F_1 + F_2+k+2)(2|\fm|)}}{(1-t^2)(1-t^4)} (b_1^{-k} w_1 w_2)^{2\fm}~.
}
Notice that the evaluation of the integral, which is the ratio in the first line, has two contributions. The $t$ exponential factor modifies the dimension of the bare monopole operator, so that it is included the contribution from the dressing. Also, the denominator can be interpreted as a $P$-factor in the language of \cite{Cremonesi:2013lqa}; see also \eqref{PU2}.

For fluxes with $\fm>\fn$, each U(2) gauge group is broken to its maximal torus U(1)$^2$. The corresponding contribution to the Hilbert series is
\bes{ \label{Hm>nNS}
H_{\fm >\fn} = \sum_{\fm >\fn \in \BZ} \, \CH(\fm)\,  \CH(\fn) \,\, t^{2\Delta(\vec m_1 = \vec m_2 = (\fm, \fn), \vec m_3 = \vec 0)}~,
}
where $\CH(m)$ is the integral part of the $N=1$ result from \eqref{HSNSabel}:
\bes{
\CH(m) &= \oint \frac{d u_1}{2\pi i u_1} \frac{d u_2}{2\pi i u_2} u_1^{\kappa_1 m} u_2^{\kappa_2 m} w_1^m w_2^{m} \times \PE\left[t \sum_{s=\pm 1} (b_1 u_1 u_2^{-1})^s  \right] \\
&= \frac{t^{k |m|}}{1-t^2} (b_1^{-k} w_1 w_2)^m~. 
}
Notice again that the result gives a $t$ exponential factor, that accounts for the dressing of bare monopoles, and a $P$-factor.
Therefore, we have
\bes{
H_{\fm > \fn} = \sum_{\fm > \fn} \frac{t^{(F_1+F_2+k+2)(|\fm|+|\fn|)-2|\fm-\fn|}}{(1-t^2)^2} (b_1^{-k} w_1 w_2)^{\fm+\fn}~.
}
Summing the two contributions as indicated in \eqref{sumtwocontrNS} yields the total Hilbert series:
\bes{
\begin{split}
&H_{\eqref{quivT3axial}}[\text{NS5}](t;w_{1,2};b_{1,2}) \\
&\quad = \sum_{\fm \geq \fn \in \BZ} t^{(F_1+F_2+k+2)(|\fm|+|\fn|) -2|\fm-\fn|}\, y_1^{\fm+\fn} \, P_{\U(2)} (t; \fm, \fn) \\
&\quad = \PE \Big[ t^2 + t^4 + \left(y_1+ y_1^{-1}\right) (t^{F_1+F_2+k}+ t^{F_1+F_2+k+2}) \\
&\qquad \qquad - t^{2(F_1+F_2+k+1)} - t^{2(F_1+F_2+k+2)}\Big]~, \quad y_1= b_1^{-1} w_1 w_2~,
\end{split}
}
where the $P$-factor is defined as in \eqref{PU2}.
This result matches the Coulomb branch Hilbert series for the proposed magnetic quiver \cite{Marino:2025uub}, which is a U(2) gauge theory with $F_1+F_2+k+2$ flavours. Notice that the $P_{\U(2)}(t;\fm,\fn)$ factor does not depend on the specific value of $(\fm,\fn)$, but only on whether $\fm$ is equal or not to $\fn$. This agreement follows from the structure of the Coulomb branch formula in \cite{Cremonesi:2013lqa}. For $m^{(1)}=m^{(2)}$, the U(2) gauge group is unbroken, and the factor $P_{\mathrm{U}(2)}(t; \fm=\fn) = \left[(1-t^2)(1-t^4)\right]^{-1}$ is the character for the Casimir invariants. For $\fm \neq \fn$, the group is broken to U(1)$^2$, and $P_{\mathrm{U}(2)}(t; \fm \neq \fn) = (1-t^2)^{-2}$ accounts for the Casimirs of the residual symmetry.

\subsubsection*{The $(1,k)$-branch}
The analysis for this branch is identical to that of the NS5-branch, with the nodes $1$ and $3$ exchanged (and $b_1$ replaced by $b_3$).

\subsubsection*{The D5-branch}
We now compute the D5-branch, assuming $F_1 > 1$ and $F_3 > 1$.\footnote{As in the Abelian case, this condition is sufficient, but not necessary, for the theory to be good. We will however leave the analysis for more general values of the flavours to a future work.}
As in the Abelian case, the magnetic fluxes are constrained to be $\vec{m}_1 = \vec{m}_2 = \vec{m}_3 = \vec{0}$, meaning that monopole operators do not contribute to this branch. 
The resulting Hilbert series is therefore the Higgs branch series of the corresponding theory with zero Chern-Simons levels ($k=0$):
\begin{equation} \label{elecquivD5N=2}
\scalebox{0.85}{
\begin{tikzpicture}[baseline,font=\footnotesize,
    circ/.style={circle, draw, minimum size=1.3cm},
    sq/.style={rectangle, draw, minimum size=1cm},
    node distance=1cm
]
    \node[circ] (n1) {$2$};
    \node[circ, right=of n1] (n2) {$2$};
    \node[circ, right=of n2] (n3) {$2$};
    \node[sq, below=of n1] (s1) {$F_1$};
    \node[sq, below=of n2] (s2) {$F_2$};
    \node[sq, below=of n3] (s3) {$F_3$};
    \draw (n1) -- (n2);
    \draw (n2) -- (n3);
    \draw (n1) -- (s1);
    \draw (n2) -- (s2);
    \draw (n3) -- (s3);
\end{tikzpicture}}
\end{equation}
The magnetic quiver for this branch is simply the mirror theory for this quiver.

\subsection{The $T_{(p,q)}[3]$ theory and its generalisation} \label{sec:pqexample}
Let us consider the following brane system:
\begin{equation}
\scalebox{0.9}{
\begin{tikzpicture}[baseline,font=\footnotesize,
    cross_node/.style={circle, draw=cyan, cross out, thick, minimum size=6pt, inner sep=0pt}
]

    \def\ystart{-1.5}
    \def\yend{1.5}

    \draw[very thick] (0,0) -- node[near start, below, xshift=-0.5cm]{$N$ D3} (6.75,0);

    \draw[red, thick] (0,\ystart) -- (0,\yend) node[pos=0, below, red] {NS5};
    \draw[red, thick] (4.5,\ystart) -- (4.5,\yend) node[pos=0, below, red] {NS5};


    \draw[blue, dashed, thick] (1+1, \ystart) -- (2.5+1, \yend) node[pos=0, below, blue] {$(2 \fp, 2 \fp \fq-1)$};

    \draw[blue, dashed, thick] (6, \ystart) -- (7.5, \yend) node[pos=0, below, blue] {$(2 \fp, 2 \fp \fq-1)$};

    \node[cross_node, label={[cyan]above:D5}] at (1.5, 0.5) {};
    \node[cross_node, label={[cyan]above:D5}] at (6, 0.5) {};

\end{tikzpicture}}
\end{equation}
This system and the corresponding field theory, known as the $T_{(p,q)}[3]$ theory, were considered in \cite[Section 7]{Assel:2017eun}. We take $p=2 \fp$ and $q = 2 \fp \fq -1$, such that $\fp, \fq \in \BZ_{\geq 1}$. We define the generators of $\SL(2,\BZ)$ as follows:
\bes{
\mathrm{S} = \begin{pmatrix} 0 & -1 \\ 1 & 0 \end{pmatrix}~, \qquad 
\mathrm{T} = \begin{pmatrix} 1 & 0 \\ 1 & 1 \end{pmatrix}~,
}
where $\mathrm{S}^2=-1$ and $(\mathrm{S} \mathrm{T})^3=1$. We define 
\bes{ \label{defJ}
J = -\mathrm{T}^{\fq} \mathrm{S} \mathrm{T}^{2 \fp}~. 
}
Since $J_k \begin{pmatrix} 1 & 0 \end{pmatrix}^t = \begin{pmatrix} 2\fp & 2\fp \fq-1 \end{pmatrix}^t$, it follows that each $(2 \fp,2 \fp \fq-1)$-brane can be regarded an NS5-brane between a $J$ wall on the left and a $J^{-1}$ wall on the right (see \cite{Assel:2014awa}). Note that the $\mathrm{T}^\kappa$ wall gives rise to the gauge group $\U(N)_\kappa$, the $\mathrm{S}$ wall gives rise to the $T[\U(N)]$ theory, and the $\mathrm{S}^{-1}$ wall gives rise to the $\bar{T[\U(N)]}$ theory.\footnote{As defined in \cite{Gaiotto:2008ak}, the $T[\U(1)]$ theory is the supersymmetric completion of the Chern-Simons coupling $\frac{1}{2\pi} \int F \wedge dW$ between two $\U(1)$ gauge fields $F$ and $W$, where each of them resides in an $\CN=4$ vector multiplet. The $T[\U(N)]$ theory is defined as the product of the $T[\SU(N)]$ theory and the $T[\U(1)]$ theory, in such a way that the $\su(N) \times \su(N)$ global symmetry of $T[\SU(N)]$ is promoted to $\u(N) \times \u(N)$ with the mixed Chern-Simons term due to the $T[\U(1)]$ theory. In terms of the index, if we denote by $\{ (f_1, f_2, \ldots, f_N), (m_1, m_2, \ldots, m_N) \}$ the flavour fugacities and corresponding magnectic fluxes, and by $\{ (w_1, w_2, \ldots, w_N), (n_1, n_2, \ldots, n_N) \}$ the topological fugacities and corresponding magnectic fluxes, then the $T[\U(1)]$ contribution to the index is $(f_1 f_2 \cdots f_N)^{n_N} (w_1 w_2 \cdots w_N)^{m_N}$. The $\bar{T[\SU(N)]}$ theory can be contain from $T[\U(N)]$ by setting $f_j \rightarrow f_j^{-1}$ and $m_j \rightarrow -m_j$, for all $j=1, \ldots, N$.} The corresponding theory is therefore
\begin{equation} \label{quivpqbrane}
\scalebox{0.85}{
\begin{tikzpicture}[baseline,font=\footnotesize,
    circ/.style={circle, draw, minimum size=1.3cm},
    sq/.style={rectangle, draw, minimum size=1cm},
    node distance=1.2cm
]

    \node[circ] (n1) {$N_{\fq}$};
    \node[circ, right=of n1] (n2) {$N_{2\fp}$};
    \node[circ, right=of n2] (n3) {$N_{-2\fp}$};
    \node[circ, right=of n3] (n4) {$N_{-\fq}$};
    \node[circ, right=of n4] (n5) {$N_{\fq}$};
    \node[circ, right=of n5] (n6) {$N_{2\fp}$};

    \node[sq, below=of n1] (s1) {1};
    \node[sq, below=of n5] (s2) {1};

    \draw[blue,dashed] (n1) -- node[above, midway]{\scriptsize $T[\U(N)]$} (n2);
    \draw (n2) -- (n3);
    \draw[blue,dashed] (n3) --node[above, midway]{\scriptsize $\bar{T[\U(N)]}$} (n4);
    \draw (n4) -- (n5);
    \draw[blue,dashed] (n5) --node[above, midway]{\scriptsize $T[\U(N)]$} (n6);

    \draw (n1) -- (s1);
    \draw (n5) -- (s2);

\end{tikzpicture}}
\end{equation}

\subsubsection{The Abelian case}
Let us focus on the case of $N=1$. In order to compute the limit of the index, we assign the following axial fugacities to the chiral fields in the quiver.
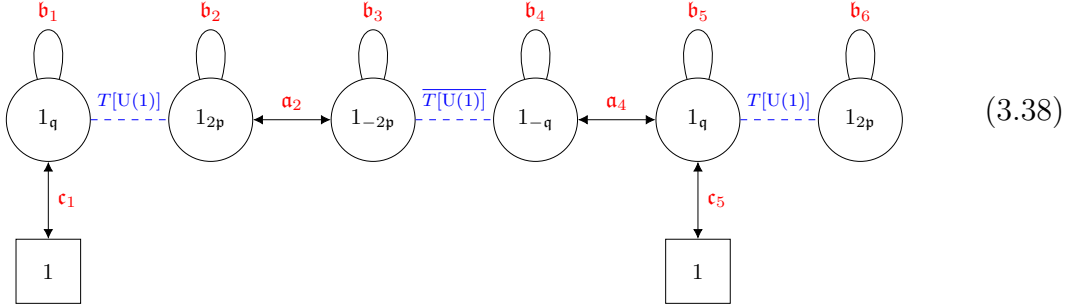
\begin{equation} \label{quivpqaxialabel}
\scalebox{0.85}{
\begin{tikzpicture}[baseline,font=\footnotesize,
    circ/.style={circle, draw, minimum size=1.3cm},
    sq/.style={rectangle, draw, minimum size=1cm},
    node distance=1.2cm,
    every loop/.style={-}
]

    \node[circ] (n1) {$1_{\fq}$};
    \node[circ, right=of n1] (n2) {$1_{2\fp}$};
    \node[circ, right=of n2] (n3) {$1_{-2\fp}$};
    \node[circ, right=of n3] (n4) {$1_{-\fq}$};
    \node[circ, right=of n4] (n5) {$1_{\fq}$};
    \node[circ, right=of n5] (n6) {$1_{2\fp}$};

    \node[sq, below=of n1] (s1) {1};
    \node[sq, below=of n5] (s2) {1};

    \draw[blue,dashed] (n1) -- node[above, midway]{\scriptsize $T[\U(1)]$} (n2);
    \draw[<->] (n2) --  node[above, red] {$\mathfrak{a}_2$} (n3);
    \draw[blue,dashed] (n3) --node[above, midway]{\scriptsize $\bar{T[\U(1)]}$} (n4);
    \draw[<->] (n4) -- node[above, red] {$\mathfrak{a}_4$} (n5);
    \draw[blue,dashed] (n5) --node[above, midway]{\scriptsize $T[\U(1)]$} (n6);

    \draw[<->] (n1) -- node[right, red] {$\mathfrak{c}_1$} (s1);
    \draw[<->] (n5) -- node[right, red] {$\mathfrak{c}_5$} (s2);

    \path (n1) edge [loop above] node[red] {$\mathfrak{b}_1$} ();
    \path (n2) edge [loop above] node[red] {$\mathfrak{b}_2$} ();
    \path (n3) edge [loop above] node[red] {$\mathfrak{b}_3$} ();
    \path (n4) edge [loop above] node[red] {$\mathfrak{b}_4$} ();
    \path (n5) edge [loop above] node[red] {$\mathfrak{b}_5$} ();
    \path (n6) edge [loop above] node[red] {$\mathfrak{b}_6$} ();

\end{tikzpicture}}
\end{equation}
The index of \eqref{quivpqaxialabel} is then
\bes{
&\CI_{\eqref{quivpqaxialabel}} = \sum_{m_1, \ldots, m_6 \in \BZ} \oint  \Bigg( \prod_{i=1}^6 \frac{du_i}{2\pi i u_i}  u_i^{\kappa_i m_i}  w_i^{m_i} Z_{\text{chir}}^{1} \left(x; \fb_i; 0 \right) \Bigg)  \\ 
& \times \prod_{s = \pm1} \prod_{i=2,3} Z_{\text{chir}}^{1/2} \left(x; \fa_i (u_i u_{i+1}^{-1})^s; s m_{i} - s m_{i+1} \right) \prod_{j =1,5} Z_{\text{chir}}^{1/2} \left(x; \fc_j (f^{-1}_j u_j)^s; sm_{j} \right) \\
& \times ( u_1^{m_2} u_2^{m_1})(u_3^{-m_4} u_4^{-m_3})(u_5^{m_6} u_6^{m_5})~,
}
where
\bes{
\kappa_1 = \fq~, \,\,\,\, \kappa_2 = 2\fp~, \,\,\,\, \kappa_3 = -2\fp~, \,\,\,\, \kappa_4 = -\fq~, \,\,\,\, \kappa_5 = \fq~, \,\,\,\, \kappa_6 = 2\fp~,
}
and for each branch the axial fugacities are as follows:
\bes{ \label{axialfugTpq3ab}
&\begin{tabular}{c|ccccccccccc}
\hline
Branch & $\fa_2$ & $\fa_4$  & $\fc_1$ & $\fc_5$ & $\fb_1$ & $\fb_2$ & $\fb_3$ & $\fb_4$ & $\fb_5$ & $\fb_6$  \\
\hline
NS5& $a$ & $a^{-1}$ & $a^{-1}$ & 1 & 1 & 1 &1& 1 & 1 & 1 \\
$(2 \fp, 2 \fp \fq-1)$ & $a^{-1}$ & $a$ & $1$ & $a^{-1}$ & $1$ & 1 &1 & 1 & 1 & 1 \\
D5 & $a$ & $a$ & $a$ & $a$ & $a^{-2}$ & $a^{-2}$ & $1$& $1$ & $a^{-2}$ & $1$ \\
\hline
\end{tabular} 
}
We can now analyse each branch separately.

\subsubsection*{The NS5-branch} 
On the right hand side of \eqref{limitHS}, we find that the power of $\fz$ is $|m_2-m_3|+\frac{1}{2}|m_5|$.  Thus, a non-vanishing result in the limit $\fz \rightarrow 0$ restricts 
\bes{
m_2=m_3 = \fm \in \BZ ~, \qquad m_5=0~. 
}
The dimension of the monopole operators is
\bes{
\Delta(m_1, \ldots, m_5) = \frac{1}{2}( |m_1| +|m_2-m_3| + |m_4-m_5| + |m_5|)~,
}
and the limit of the index for the NS5-branch reads
\bes{ \label{HSNSabelpq1}
\begin{split}
&H_{\eqref{quivpqbrane}}[\text{NS5}](t; \{ w_{i} \}; \{b_{i}\}) =  \sum_{m_1, \fm, m_4\in \BZ}  t^{2 \Delta(m_1, \fm, \fm, m_4,0)}  w_1^{m_1} (w_2 w_3)^\fm w_4^{m_4}   \\
& \times \left( \prod_{j=1}^5 \oint \frac{d u_j}{2\pi i u_j} \right) u_1^{m_1 \fq+\fm} u_2^{2 \fm \fp+m_1} u_3^{-2 \fm \fp-m_4}  u_4^{-m_4 \fq-\fm} \PE\left[t \sum_{s=\pm 1} (b_2 u_2 u_3^{-1})^s  \right]~.
\end{split}
}
The integrals over $u_1$ and $u_4$ impose the following conditions:
\bes{
\fm = -m_1 \fq ~, \,\, \fm = -m_4 \fq \quad \Rightarrow \quad m_1 = m_4 = \fn~, \,\, \fm = - \fn \fq~,
}
whereas the integral over $u_6$ and the magnetic flux $m_6$ are neglected for the reason explained in footnote \ref{footnotedistnodes}. The expression in \eqref{HSNSabelpq1} reduces to
\bes{
&H_{\eqref{quivpqbrane}}[\text{NS5}](t; \{ w_{i} \}; \{b_{i}\}) =  \sum_{\fn \in \BZ}  t^{2 \Delta(\fn, -\fn \fq, -\fn \fq, \fn,0)}  \left(\frac{w_1 w_4}{w_2^\fq w_3^\fq} \right)^{\fn}  \\
& \,\,\,\, \times \left( \prod_{j=2}^3 \oint \frac{d u_j}{2\pi i u_j} \right)  u_2^{-(2 \fp\fq-1)\fn} u_3^{(2 \fp \fq-1)\fn}  \PE\left[t \sum_{s=\pm 1} (b_2 u_2 u_3^{-1})^s  \right]~,
}
from which we see that the ``effective Chern-Simons levels'' for the second and third nodes are now $-(2\fp\fq-1)$ and $(2\fp \fq-1)$, respectively.  To form the gauge invariant quantities, we need to dress bare the monopole operator $V_{(1,-\fq,-\fq,1,0,0)}$ associated with $\fn=1$ with $X_{23}$ and $V_{(-1,\fq,\fq,-1,0,0)}$ associated with $\fn=-1$ with $X_{32}$. Similarly to \eqref{HSNSabel}, we obtain
\bes{
H_{\eqref{quivpqbrane}}[\text{NS5}](t; \{ w_{i} \}; \{b_{i}\})
&\overset{\eqref{usefulintiden}}{=}\sum_{\fn \in \BZ} t^{|\fn|[1+1 +(2 \fp \fq-1)]} \frac{1}{1-t^2} y^{\fn} ~, \quad y = b_2^{2 \fp \fq-1} \frac{w_1 w_4}{w_2^\fq w_3^\fq} \\
&= \PE \left[t^2 + (y+y^{-1}) t^{2\fp \fq+1} - t^{2(2\fp \fq+1)}  \right] ~.
}
This is the Hilbert series of $\BC^2/\BZ_{2\fp\fq+1}$, in agreement with \cite[(7.1)]{Assel:2017eun}. This space can be realised as the Coulomb branch of the magnetic quiver described by the $\U(1)$ gauge theory with $2\fp\fq+1$ flavours \cite[(5.12)]{Marino:2025uub}.  In summary, this branch is generated by
\bes{
G_0= X_{23} X_{32}~, \quad G_+ = V_{(1,-\fq,-\fq,1,0,0)} X_{23}^{2 \fp \fq-1}~, \quad G_- = V_{-(1,-\fq,-\fq,1,0,0)} X_{32}^{2 \fp \fq-1}~,
}
satisfying $G_0^{4\fp\fq+2} =G_+G_-$.

\subsubsection*{The $(2\fp, 2\fp \fq-1)$-branch} 
In this case, we have $\fz^{\frac{1}{2}|m_1| + |m4 - m5|}$. In the limit of $\fz \rightarrow 0$, a non-vanishing result imposes the conditions
\bes{
m_4=m_5 = \fm \in \BZ~, \qquad m_1=0~.
}
The limit \eqref{limitHS} of the index in this case reads
\bes{
\scalebox{0.9}{$
\begin{split}
&H_{\eqref{quivpqbrane}}[(2\fp, 2\fp \fq-1)](t; \{ w_{i} \}; \{b_{i}\}) =  \sum_{m_3, \fm, m_6 \in \BZ}  t^{2 \Delta(0, m_2, m_3,\fm , \fm)}  w_3^{m_3} (w_4 w_5)^\fm w_6^{m_6}   \\
& \times \left( \prod_{j=1}^6 \oint \frac{d u_j}{2\pi i u_j} \right) u_1^{m_2} u_2^{2 m_2 \fp} u_3^{-2 m_3 \fp-\fm}  u_4^{-m_3-\fq \fm} u_5^{m_6+\fq \fm} u_6^{2 m_6 \fp+\fm} \PE\left[t \sum_{s=\pm 1} (b_4 u_4 u_5^{-1})^s  \right]~.
\end{split}$}
}
The integrations over $u_1$, $u_2$, $u_3$, $u_6$ impose
\bes{
m_2 = 0~, \quad m_3=\fn~, \quad m_6=\fn~, \quad \fm = -2 \fp \fn~,
}
from which the expression above simplifies to
\bes{
&H_{\eqref{quivpqbrane}}[(2\fp, 2\fp \fq-1)](t; \{ w_{i} \}; \{b_{i}\}) =  \sum_{\fn \in \BZ}  t^{2 \Delta(0,0,\fn, -2\fp \fn, -2\fp\fn)}  \left(\frac{w_3 w_6}{w_4^{2\fp} w_5^{2\fp}} \right)^{\fn}  \\
& \,\,\,\, \times \left( \prod_{j=4}^5 \oint \frac{d u_j}{2\pi i u_j} \right)  u_4^{(2 \fp\fq-1)\fn} u_5^{-(2 \fp \fq-1)\fn}  \PE\left[t \sum_{s=\pm 1} (b_4 u_4 u_5^{-1})^s  \right]~.
}
We thus obtain
\bes{
&H_{\eqref{quivpqbrane}}[(2\fp, 2\fp \fq-1)](t; \{ w_{i} \}; \{b_{i}\}) \\
&\overset{\eqref{usefulintiden}}{=}\sum_{\fn \in \BZ} t^{|\fn|[1+2\fp +(2 \fp \fq-1)]} \frac{1}{1-t^2} y^{\fn} ~, \qquad y = b_4^{-(2 \fp \fq-1)} \frac{w_3 w_6}{w_4^{2\fp} w_5^{2\fp}} \\
&= \PE \left[t^2 + (y+y^{-1}) t^{2\fp( \fq+1)} - t^{4\fp( \fq+1)}  \right] ~,
}
which is the Hilbert series of $\BC^2/\BZ_{2\fp(\fq+1)}$, in agreement with \cite[(7.1)]{Assel:2017eun}. This space can be realised as the Coulomb branch of the magnetic quiver described by the $\U(1)$ gauge theory with $2\fp(\fq+1)$ flavours \cite[(5.13)]{Marino:2025uub}. 
In particular, this branch is generated by
\bes{
G_0= X_{45} X_{54}~, \,\, G_+ = V_{(0,0,1,-2\fp,-2\fp,1)} X_{45}^{2 \fp \fq-1}~, \,\, G_- = V_{-(0,0,1,-2\fp,-2\fp,1)} X_{54}^{2 \fp \fq-1}~,
}
satisfying $G_0^{8\fp\fq+8} =G_+G_-$. As usual, we denote by $X_{45}$ and $X_{54}$ the chiral and antichiral fields in the bifundamental hypermultiplet connecting the gauge nodes 4 and 5. 

\subsubsection*{The D5-branch} 
In this case, we have $\fz^{|m_1|+|m_2-m_3|+|m_4-m_5|+|m_5|}$. In the limit $\fz \rightarrow 0$, a non-vanishing result imposes the conditions
\bes{
m_1 = 0~, \quad m_2=m_3 =\fm~, \quad m_4=m_5=0~.
}
The limit \eqref{limitHS} of the index in this case is given by
\bes{
\scalebox{0.9}{$
\begin{split}
&H_{\eqref{quivpqbrane}}[\text{D5}](t; \{ w_{i} \}; \{b_{i}\}) =  \sum_{\fm, m_6 \in \BZ}  t^{2 \Delta(0, \fm, \fm, 0, 0)}  (w_2 w_3)^\fm w_6^{m_6}   \\
& \qquad \times \left( \prod_{j=1}^6 \oint \frac{d u_j}{2\pi i u_j} \right) u_1^{\fm} u_2^{2 \fm \fp} u_3^{-2 \fm \fp} u_4^{-\fm} u_5^{m_6} u_6^{2 m_6 \fp} \\
& \qquad \times \PE\left[-3t^2 + t \sum_{s=\pm 1} \{ (f_1 u_1^{-1})^s + (u_2 u_3^{-1})^s + (u_4 u_5^{-1})^s +(f_5 u_5^{-1})^s  \}  \right]~,
\end{split}$}
}
where the integral over $u_6$ sets $m_6=0$. We thus obtain
\bes{ \label{HSTpq3D5branch}
\scalebox{0.9}{$
\begin{split}
&H_{\eqref{quivpqbrane}}[\text{D5}](t; \{ w_{i} \}; \{b_{i}\}) =  \sum_{\fm \in \BZ}  (w_2 w_3)^\fm   \left( \prod_{j=1}^5 \oint \frac{d u_j}{2\pi i u_j} \right) u_1^{\fm} u_2^{2 \fm \fp} u_3^{-2 \fm \fp} u_4^{-\fm} \\
& \qquad \times \PE\left[-3t^2 + t \sum_{s=\pm 1} \{ (f_1 u_1^{-1})^s + (b_2 u_2 u_3^{-1})^s + (b_4 u_4 u_5^{-1})^s +(f_5 u_5^{-1})^s  \}  \right] \\
&\overset{\eqref{usefulintiden}}{=} \sum_{\fm \in \BZ} \left( \prod_{j=1, 5} \oint \frac{d u_j}{2\pi i u_j} \right) u_1^{\fm} u_5^{-\fm} \times t^{2  \fp |\fm|} (b_2^{-2 \fp} w_2 w_3)^\fm \times t^{ |\fm|} b_4^\fm \\
& \qquad \times \PE\left[-t^2 + t \sum_{s=\pm 1} \{ (f_1 u_1^{-1})^s  +(f_5 u_5^{-1})^s  \}  \right] \\
&= \sum_{\fm \in \BZ} \frac{t^{(2\fp+3) |\fm|} }{1-t^2} y^\fm = \PE\left[ t^2 + \left(y+ y^{-1}   \right) t^{2\fp+3} -t^{4\fp+6}\right]~, \qquad y = f_1 b_2^{-2\fp} b_4 f_5^{-1} w_2 w_3~.
\end{split}$}
}
This is the Hilbert series of $\BC^2/\BZ_{2\fp+3}$, in agreement with \cite[(7.1)]{Assel:2017eun}. This space can be realised as the Coulomb branch of the magnetic quiver described by the $\U(1)$ gauge theory with $2\fp+3$ flavours \cite[(5.14)]{Marino:2025uub}. 

We point out that, in contrast to the previous cases in which the quivers contain only fundamental and bifundamental hypermultiplets, the D5-branch in this theory is generated also by dressed monopole operators. From \eqref{HSTpq3D5branch}, we see that such generators are
\bes{ \label{D5branchmonopole}
G_+  = V_{(0,1,1,0,0,0)} \tQ_1 X_{32}^{2\fp} X_{45} Q_5~, \quad G_+  = V_{-(0,1,1,0,0,0)} Q_1 X_{23}^{2\fp} X_{54} \tQ_5~.
}
The other generators can be taken as one among the following combinations: $Q_1 \tQ_1$, $X_{23}X_{32}$, $X_{45}X_{54}$ and $Q_5 \tQ_5$, depending on which adjoint chiral field we turn on in \eqref{quivpqaxialabel}. In particular, the choice given in the last line \eqref{axialfugTpq3ab} sets $Q_1 \tQ_1 =0 $, $X_{23}X_{32} =0 $, $X_{45}X_{54}=Q_5 \tQ_5$, and so the aforementioned generator can be taken as 
\bes{
G_0 = X_{45}X_{54}=Q_5 \tQ_5~.
}
These generators satisfy the relation $G_0^{4\fp+6}  = G_+G_-$.

\subsubsection{The non-Abelian case} \label{sec:nonabelianpq}
Let us now generalise the discussion above and analyse the limit of the index for theories with $(p,q)$-branes and non-Abelian nodes. For simplicity, we focus on the $N = 2$ case, which corresponds to a Chern-Simons theory with interpolating $T[\U(2)]$ and $\bar{T[\U(2)]}$ SCFTs. Specifically, let us consider the following brane system:
\begin{equation} \label{branesetTU2}
\begin{tikzpicture}[baseline=0,font=\footnotesize,
    cross_node/.style={circle, draw=cyan, cross out, thick, minimum size=6pt, inner sep=0pt}
]

    \def\ystart{-1.5}
    \def\yend{1.5}

    \draw[very thick] (0,0) -- node[near start, below, xshift=-0.5cm]{2 D3} (5.5,0);

    \draw[red, thick] (0,\ystart) -- (0,\yend) node[pos=0, below, red] {NS5};
    \draw[red, thick] (5.5,\ystart) -- (5.5,\yend) node[pos=0, below, red] {NS5};


    \draw[blue, dashed, thick] (1+1, \ystart) -- (2.5+1, \yend) node[pos=0, below, blue] {$(2 \fp, 2 \fp \fq-1)$};

    \node[cross_node, label={[cyan]above:$L$ D5}] at (1.5, 0.5) {};
    \node[cross_node, label={[cyan]above:$R$ D5}] at (4, 0.5) {};

\end{tikzpicture}
\end{equation}
where the associated quiver theory reads
\begin{equation} \label{quivpqbraneTU2}
\begin{tikzpicture}[baseline,font=\footnotesize,
    circ/.style={circle, draw, minimum size=1cm},
    sq/.style={rectangle, draw, minimum size=0.8cm},
    node distance=1cm
]

    \node[circ] (n1) {$2_{\fq}$};
    \node[circ, right=of n1] (n2) {$2_{2\fp}$};
    \node[circ, right=of n2] (n3) {$2_{-2\fp}$};
    \node[circ, right=of n3] (n4) {$2_{-\fq}$};

    \node[sq, left=of n1] (s1) {$L$};
    \node[sq, right=of n4] (s2) {$R$};

    \draw[blue,dashed] (n1) -- node[above, midway]{\scriptsize $T[\U(2)]$} (n2);
    \draw (n2) -- (n3);
    \draw[blue,dashed] (n3) --node[above, midway]{\scriptsize $\bar{T[\U(2)]}$} (n4);

    \draw (n1) -- (s1);
    \draw (n4) -- (s2);

\end{tikzpicture}
\end{equation}
in which the flavour and topological symmetries of the $T[\U(2)]$ theory are coupled to $\CN=3$ $\U(2)$ vector multiplets with Chern-Simons levels $\fq$ and $2 \fp$, respectively. Similarly, the flavour and topological symmetries of the $\bar{T[\U(2)]}$ theory are gauged with Chern-Simons levels $- 2 \fp$ and $-\fq$, respectively.
We can assign axial fugacities to the various chiral fields appearing in quiver \eref{quivpqbraneTU2}, which can be depicted in 3d $\CN=2$ notation as follows:
\begin{equation} \label{quivpqTU2axialabel}
\begin{tikzpicture}[baseline,font=\footnotesize,
    circ/.style={circle, draw, minimum size=1cm},
    sq/.style={rectangle, draw, minimum size=0.8cm},
    node distance=1.2cm,
    every loop/.style={-}
]

    \node[circ] (n1) {$2_{\fq}$};
    \node[circ, right=of n1] (n2) {$2_{2\fp}$};
    \node[circ, right=of n2] (n3) {$2_{-2\fp}$};
    \node[circ, right=of n3] (n4) {$2_{-\fq}$};

    \node[sq, left=of n1] (s1) {$L$};
    \node[sq, right=of n4] (s2) {$R$};

    \draw[blue,dashed] (n1) -- node[above, midway]{\scriptsize $T[\U(2)]$} (n2);
    \path (n1) --  node[below, red] {$\mathfrak{a}_1$} (n2);
    \draw[<->] (n2) --  node[above, red] {$\mathfrak{a}_2$} (n3);
    \draw[blue,dashed] (n3) --node[above, midway]{\scriptsize $\bar{T[\U(2)]}$} (n4);
    \path (n3) --  node[below, red] {$\mathfrak{a}_3$} (n4);

    \draw[<->] (n1) -- node[above, red] {$\mathfrak{c}_1$} (s1);
    \draw[<->] (n4) -- node[above, red] {$\mathfrak{c}_4$} (s2);

    \path (n1) edge [loop above] node[red] {$\mathfrak{b}_1$} ();
    \path (n2) edge [loop above] node[red] {$\mathfrak{b}_2$} ();
    \path (n3) edge [loop above] node[red] {$\mathfrak{b}_3$} ();
    \path (n4) edge [loop above] node[red] {$\mathfrak{b}_4$} ();

\end{tikzpicture}
\end{equation}
The index of theory \eref{quivpqTU2axialabel}, compatible with the axial charge assignments, is then given by
\bes{ \label{indtheorypq}
\scalebox{0.9}{$
\begin{split}
&\CI_{\eref{quivpqTU2axialabel}} = \frac{1}{2^4} \sum_{\{m^{(\alpha)}_1\} \in \BZ^2} \sum_{\{m^{(\alpha)}_2\} \in \BZ^2} \sum_{\{m^{(\alpha)}_3\} \in \BZ^2} \sum_{\{m^{(\alpha)}_4\} \in \BZ^2} \oint \Bigg( \prod_{i=1}^4 \prod_{\alpha=1}^2 \frac{du^{(\alpha)}_i}{2\pi i u^{(\alpha)}_i}  (u^{(\alpha)}_i)^{\kappa_i m^{(\alpha)}_i}  w_i^{m^{(\alpha)}_i} \Bigg) \\
&\qquad \quad \times \prod_{i=1}^4  Z_{\text{vec}}^{\U(2)}\left(x; \{u^{(\alpha)}_i\}; \{m^{(\alpha)}_i\}\right) \left[\prod_{\alpha, \beta=1}^2 Z_{\text{chir}}^{1} \left(x; \fb_i u^{\alpha}_i (u^{\beta}_i)^{-1} ; m^{(\alpha)}_i - m^{(\beta)}_i  \right)\right]^{\epsilon} \\ 
& \qquad \quad \times \prod_{s = \pm1} \prod_{\alpha, \beta=1}^2 Z_{\text{chir}}^{1/2} \left(x; \fa_2 \left(u^{(\alpha)}_2 (u^{(\beta)}_{3})^{-1}\right)^s; s\left(m^{(\alpha)}_{2} - m^{(\beta)}_{3}\right) \right) \\ 
& \qquad \quad \times \prod_{s = \pm1} \prod_{j=\{1,4\}} \prod_{\rho_j=1}^{F_j} \prod_{\alpha=1}^2  Z_{\text{chir}}^{1/2} \left(x; \fc_j \left(u^{(\alpha)}_j (f^{(\rho_j)}_j)^{-1}\right)^s; sm^{(\alpha)}_{j} \right) \\ & \qquad \quad \times Z_{T[\U(2)]} \left(x; \fa_1 |u^{(\alpha)}_1; m^{(\alpha)}_1 | u^{(\alpha)}_2; m^{(\alpha)}_2\right) Z_{\bar{T[\U(2)]}} \left(x; \fa_3 |u^{(\alpha)}_3; m^{(\alpha)}_3 | u^{(\alpha)}_4; m^{(\alpha)}_4\right)~,
\end{split}$}
}
where
\bes{ \label{k1k2pq}
\kappa_1 = \fq, \,\,\,\, \kappa_2 = 2\fp, \,\,\,\, \kappa_3 = -2\fp, \,\,\,\, \kappa_4 = -\fq~.
}
For definiteness, we take $\fp, \fq>0$. We denote by $\{u^{(\alpha)}_i, m^{(\alpha)}_i\}$, with $\alpha = 1,2$, the fugacities and magnetic fluxes for the $i$-th $\U(2)$ gauge node, where $i = 1, \ldots, 4$, and with $f^{(\rho_j)}_j$ the fugacities for the $j$-th flavour node, with $\rho_1 = L$ and $\rho_2 = R$. The index contribution due to to the $T[\U(2)]$ theory is 
\bes{ \label{indexTU2}
&Z_{T[\U(2)]} (x; y | \vec f; \vec m_f | \vec w; \vec n_w) \\&= \sum_{l_z \in \BZ} \oint \frac{dz}{2 \pi i z} {\violet \left(f_1 f_2\right)^{{n_w}_2} w_2^{{m_f}_1+{m_f}_2}} \left(\frac{w_1}{w_2}\right)^{l_z} z^{{n_w}_1-{n_w}_2} \\& \qquad \qquad \times Z^{\U(1)}_{\text{adj}} (x; y) Z^{\U(1) \times \U(2)}_{\text{hyper}} (x; y | z; l_z| \vec f; \vec m_f)~,
}
where the term highlighted in {\violet violet} is the index of the $T[\U(1)]$ theory, and we denote with $(z, l_z)$, $(\vec f, \vec{m}_f)$ and $(\vec w, \vec{n}_w)$ the fugacities and magnetic fluxes for the $\U(1)$ gauge group, the flavour and the topological symmetry, respectively. Furthermore, the adjoint chiral field in the $\U(1)$ vector multiplet and the hypermultiplet in the bifundamental representation of $\U(1) \times \U(2)$ contribute as
\begin{subequations}  
\begin{align}
\begin{split} \label{adjU1}
Z^{\U(1)}_{\text{adj}} (x; y) &= Z_{\text{chir}}^{1} \left(x; y^{-2}; 0 \right)~,
\end{split} \\
\begin{split} \label{hyperU1U2}
Z^{\U(1) \times \U(2)}_{\text{hyper}} (x; y | z; l_z| \vec f; \vec m_f) &= \prod_{s = \pm1} \prod_{j=1}^2 Z_{\text{chir}}^{1/2} \left(x; y; z^s f_j^{-s}; s l_z - s {m_f}_j \right)~,
\end{split}
\end{align}
\end{subequations}
respectively. The index of the $\bar{T[\U(2)]}$ theory is then defined as
\bes{
Z_{\bar{T[\U(2)]}} (x; y | \vec f; \vec m_f | \vec w; \vec n_w) = Z_{T[\U(2)]} (x; y | \vec f^{-1}; -\vec m_f | \vec w; \vec n_w)~,
}
with $\{\vec f^{-1}, - \vec m_f\} = \{ (f_1^{-1}, f_2^{-1}), (-{m_f}_1, - {m_f}_2) \}$. 

Subsequently, we focus on the NS5-branch, in which case the axial fugacities and the parameter $\epsilon$, which appears in the second line of \eqref{indtheorypq}, are assigned as follows:
\bes{ \label{axialfTU2quiv}
&\begin{tabular}{c|ccccccccccc}
\hline
Branch & $\fa_1$ & $\fa_2$  & $\fa_3$ & $\fc_1$ & $\fc_4$ & $\fb_1$ & $\fb_2$ & $\fb_3$ & $\fb_4$ & $\epsilon$ \\
\hline
NS5& $a$ & $a$ & $a^{-1}$ & $a^{-1}$ & $a^{-1}$ & 1 &1& 1 & 1 & 0 \\
\hline
\end{tabular} 
}
The limit of the index in question is
\bes{
H_{\eqref{quivpqbraneTU2}}[\text{NS5}](t) = \lim_{\fz \rightarrow 0} \CI_{\eref{quivpqTU2axialabel}} \Big|_{\eqref{axialfTU2quiv}, \, \eqref{N=3reparam}}~.
}

\subsubsection*{Limit of the $T[\U(2)]$ index}
Let us consider the $\fz \rightarrow 0$ limit of the index of the $T[\U(2)]$ theory, depicted as 
\begin{equation} \label{quivTU2a}
\begin{tikzpicture}[baseline,font=\footnotesize,
    circ/.style={circle, draw, minimum size=1.3cm},
    sq/.style={rectangle, draw, minimum size=1cm},
    node distance=2cm, scale=0.8
]

    \node[sq, label=below:{$\red \vec f, \vec m_f$}] (s1) {2};
    \node[sq, label=below:{$\red \vec w, \vec n_w$}, right=of s1] (s2) {2};

    
     Connect the circular nodes
    \draw[blue,dashed] (s1) -- node[above, midway]{\scriptsize $T[\U(2)]$} (s2);
    \path (s1) -- node[below, midway]{$\red y$} (s2);
\end{tikzpicture}
\end{equation}
whose index is given by \eref{indexTU2}, with $y$ taking values in $\{a, a^{-1}\}$. Observe that the variables $\fz$ and $t$ defined in \eqref{N=3reparam} are swapped upon sending $a \rightarrow a^{-1}$. As a consequence, the $\fz \rightarrow 0$ limit in the theory with axial fugacity $a^{-1}$ is equivalent to taking the $t \rightarrow 0$ limit in the theory with axial fugacity $a$, and then redefining $\fz = t$. Using the self-mirror property of $T[\U(2)]$, we thus obtain
\bes{
Z^{\fz \rightarrow 0}_{T[\U(2)]} (\fz; t | \vec f; \vec m_f | \vec w; \vec n_w) = Z^{t \rightarrow 0}_{T[\U(2)]} (t; \fz | \vec w; \vec n_w | \vec f; \vec m_f )~.
}
The derivation of each of such limits is presented in Appendix \ref{app:detailTU2}. We report the important results as follows:
\bes{ \label{TU2fzto0}
\scalebox{0.86}{$
\begin{split}
& Z^{\fz \rightarrow 0}_{T[\U(2)]} (\fz; t | \vec f; \vec m_f | \vec w; \vec n_w) \\
& = \delta_{{m_f}_1, {m_f}_2} \left(w_1 w_2\right)^{{m_f}_1} t^{\abs{{n_w}_1-{n_w}_2}} \\
& \quad \times \frac{\Theta({n_w}_1-{n_w}_2) \Psi^{({n_w}_1, {n_w}_2)}_{\U(2)} \left(t^2; f_1, f_2\right) + \Theta({n_w}_2-{n_w}_1-1) \Psi^{({n_w}_2, {n_w}_1)}_{\U(2)} \left(t^2; f_1, f_2\right)}{\left(1-t^2 \frac{f_1}{f_2}\right) \left(1-t^2 \frac{f_2}{f_1}\right)}
\\
& \quad + \left(1-\delta_{{m_f}_1, {m_f}_2}\right) \fz^{\abs{{m_f}_1-{{m_f}_2}}} \Bigg\{ \delta_{{n_w}_1, {n_w}_2} \left(f_1 f_2\right)^{{n_w}_1} \frac{1}{\left(1-\fz^2 \frac{w_1}{w_2}\right) \left(1-\fz^2 \frac{w_2}{w_1}\right)} \\ 
& \quad\,\,\,  \times \Big( \Theta({m_f}_1 - {m_f}_2) \Big[ \Psi^{({m_f}_1, {m_f}_2)}_{\U(2)} \left(\fz^2; w_1, w_2\right) - t^2 \Psi^{({m_f}_1 - 1, {m_f}_2 + 1)}_{\U(2)} \left(\fz^2; w_1, w_2\right) + \cO\left(\fz^2\right) \Big]\\ 
& \quad \qquad \,\,\ + \Theta({m_f}_2 - {m_f}_1 - 1) \Big[(m_{f_1} \leftrightarrow m_{f_2}) + \cO\left(\fz^2\right) \Big] \Big)\\ 
& \quad + \left(1-\delta_{{n_w}_1, {n_w}_2}\right) t^{\abs{{n_w}_1-{n_w}_2}} \sum_{i \neq j = 1}^2 \left[w_1^{{m_f}_i} \Psi^{({n_w}_1)}_{\U(1)} \left(f_i\right) \right] \left[w_2^{{m_f}_j} \Psi^{({n_w}_2)}_{\U(1)} \left(f_j\right) \right]\Bigg\}~,
\end{split}
$}
}
where we denote by $\Theta(x)$ the Heaviside step function, such that $\Theta(x) = 1$ for $x \ge 0$ and $\Theta(x) = 0$ for $x < 0$. The expressions for the $\U(1)$ and $\U(2)$ Hall-Littlewood polynomials are given by (see \cite[(B.13)]{Cremonesi:2014kwa})
\bes{
\Psi^{(n)}_{\U(1)} \left(x\right) &= x^n~, \\
\Psi^{(n_1, n_2)}_{\U(2)} \left(t; x_1, x_2\right) &= \frac{t x_1^{1+n_2} x_2^{n_1} - x_1^{n_2} x_2^{1+n_1} + x_1^{1+n_1} x_2^{n_2} - t x_1^{n_1} x_2^{1+n_2}}{x_1 - x_2}~.
}

\subsubsection*{Limit of the index of quivers with $T[\U(2)]$ building block}
We are now ready to discuss the NS5-limit of quiver \eref{quivpqTU2axialabel}, where the axial fugacity assignments are summarised in Table \eref{axialfTU2quiv}. This can be obtained starting from the expression
\bes{ \label{limpqwithTU2a}
&\frac{1}{2^4}\oint \prod_{i=1}^4 \prod_{\alpha=1}^2 \frac{du^{(\alpha)}_i}{2\pi i u^{(\alpha)}_i}  (u^{(\alpha)}_i)^{\kappa_i m^{(\alpha)}_i}  w_i^{m^{(\alpha)}_i} \times \Bigg[ \prod_{i=1}^4 Z_{\text{vec}}^{\U(2)}\left(\fz; t; \{u^{(\alpha)}_i\}; \{m^{(\alpha)}_i\}\right) \\ 
& \qquad \times \prod_{s = \pm1} \prod_{i=1}^{L} \prod_{\alpha=1}^2  Z_{\text{chir}}^{1/2} \left(t; \fz; (f^{(i)}_L)^{-s} (u^{(\alpha)}_1)^s; s m^{(\alpha)}_{1} \right)
\\ & \qquad \times \prod_{s = \pm1} \prod_{\alpha, \beta=1}^2  Z_{\text{chir}}^{1/2} \left(\fz; t; (u^{(\alpha)}_2)^{s} (u^{(\beta)}_3)^{-s}; s m^{(\alpha)}_{2} - s m^{(\beta)}_3 \right)
\\ & \qquad \times \prod_{s = \pm1} \prod_{i=1}^{R} \prod_{\alpha=1}^2  Z_{\text{chir}}^{1/2} \left(t; \fz; (f^i_R)^{-s} (u^{(\alpha)}_4)^s; s m^{(\alpha)}_{4} \right) \Bigg]_{\fz \rightarrow 0}
\\ & \qquad \times Z^{\fz \rightarrow 0}_{T[\U(2)]} (\fz; t| \{u^{(\alpha)}_1\}; \{m^{(\alpha)}_1\} | \{u^{(\alpha)}_2\}; \{m^{(\alpha)}_2\})
\\ & \qquad \times Z^{t \rightarrow 0}_{T[\U(2)]} (t; \fz| \{(u^{(\alpha)}_3)^{-1}\}; \{-m^{(\alpha)}_3\} | \{u^{(\alpha)}_4\}; \{m^{(\alpha)}_4\})~,
}
where
\begin{subequations}
\begin{align}
\begin{split}
Z_{\text{vec}}^{\U(2)}\left(\fz; t; \{u^{(\alpha)}\}; \{m^{(\alpha)}\}\right) &= \left(\fz t\right)^{-\abs{m^{(1)}-m^{(2)}}} + (-1)^{2 m^{(1)} - 2 m^{(2)}} \left(\fz t\right)^{\abs{m^{(1)}-m^{(2)}}} \\ & - (-1)^{m^{(1)} - m^{(2)}} \left[\frac{u^{(1)}}{u^{(2)}} + \frac{u^{(2)}}{u^{(1)}}\right]~,
\end{split} \\
\begin{split}
Z_{\text{chir}}^{1/2} \left(\fz; t; u; m_u \right) &= \left(\fz u^{-1}\right)^{\frac{\abs{m_u}}{2}} \\& \times \prod_{j=0}^{\infty} \frac{1-(-1)^{m_u} \fz^{2 + \abs{m_u}+2 j} t^{1 + \abs{m_u}+2 j} u^{-1}}{1-(-1)^{m_u} \fz^{\abs{m_u}+2 j} t^{1 + \abs{m_u}+2 j} u}~.
\end{split}
\end{align}
\end{subequations}
Observe that the lowest possible power of $\fz$ in \eref{limpqwithTU2a} is
\bes{
F(\fz) = &-\abs{m^{(1)}_2 - m^{(2)}_2} -\abs{m^{(1)}_3 - m^{(2)}_3} \\ &+ \abs{m^{(1)}_2 - m^{(1)}_3} + \abs{m^{(1)}_2 - m^{(2)}_3} + \abs{m^{(2)}_2 - m^{(1)}_3} + \abs{m^{(2)}_2 - m^{(2)}_3}~,
}
where, thanks to the triangle inequality, it follows that $F(\fz) \ge 0$ for all values of the magnetic fluxes,\footnote{Observe that, in the case $\abs{m^{(1)}_2 - m^{(2)}_2} \ge \abs{m^{(1)}_3 - m^{(2)}_3}$, we can use the inequality $\abs{m^{(1)}_2 - m^{(i)}_3} + \abs{m^{(2)}_2 - m^{(i)}_3} \ge \abs{m^{(1)}_2 - m^{(2)}_2}$ for $i = \{1,2\}$, which leads to $F(\fz) \ge \abs{m^{(1)}_2 - m^{(2)}_2} -\abs{m^{(1)}_3 - m^{(2)}_3} \ge 0$. Analogously, if $\abs{m^{(1)}_3 - m^{(2)}_3} \ge \abs{m^{(1)}_2 - m^{(2)}_2}$, the inequality $\abs{m^{(1)}_3 - m^{(i)}_2} + \abs{m^{(2)}_3 - m^{(i)}_2} \ge \abs{m^{(1)}_3 - m^{(2)}_3}$ ensures that $F(\fz) \ge -\abs{m^{(1)}_2 - m^{(2)}_2} +\abs{m^{(1)}_3 - m^{(2)}_3} \ge 0$.} leading to the crucial fact that there is no divergence in taking the $\fz \rightarrow 0$ limit. Moreover, if $F(\fz) > 0$, the expression \eref{limpqwithTU2a} vanishes in the $\fz \rightarrow 0$ limit, meaning that the relevant contribution is the one coming from the choice of magnetic fluxes satisfying $F(\fz) = 0$. This can happen in three different ways: 
\bi
\item when all magnetic fluxes associated with the second and third $\U(2)$ gauge nodes in \eref{quivpqbraneTU2} are equal,
\item when $\{m^{(1)}_2 = m^{(1)}_3, m^{(2)}_2 = m^{(2)}_3\}$,
\item when $\{m^{(1)}_2 = m^{(2)}_3, m^{(2)}_2 = m^{(1)}_3\}$. 
\ei
Let us discuss the various possible cases separately, using the fact that $Z_{\text{chir}}^{1/2} \left(\fz; t; u; m_u \right)$ in the $\fz \rightarrow 0$ limit reduces to $\left(\fz u^{-1}\right)^{\frac{\abs{m_u}}{2}}$ if $m_u \neq 0$, or to $\PE\left[t u\right]$ if $m_u = 0$.
\begin{enumerate}
\item \label{item1} $\{m^{(1)}_2 = m^{(2)}_2 = m^{(1)}_3 = m^{(2)}_3 = \fm\}$: the only non-trivial contributions come from the sectors with $\{m^{(1)}_1 = m^{(2)}_1 = \fl\}$, $\{m^{(1)}_4 = m^{(2)}_4 = \fn\}$, for which the expression \eref{limpqwithTU2a} simplifies to
\bes{ 
& \frac{1}{2^4} \sum_{\fl, \fm, \fn}\oint \left(\prod_{i=1}^4 \prod_{\alpha=1}^2 \frac{du^{(\alpha)}_i}{2\pi i u^{(\alpha)}_i}\right) w_1^{2 \fl} \left(w_2 w_3\right)^{2 \fm} w_4^{2 \fn} \\& \qquad \times \left(u^{(1)}_1 u^{(2)}_1\right)^{\fq \fl} \left(u^{(1)}_2 u^{(2)}_2\right)^{2 \fp \fm} \left(u^{(1)}_3 u^{(2)}_3\right)^{-2 \fp \fm} \left(u^{(1)}_4 u^{(2)}_4\right)^{-\fq \fn} \\& \qquad \times \left[\prod_{i=1}^4 \left(2 - \frac{u^{(1)}_i}{u^{(2)}_i} - \frac{u^{(2)}_i}{u^{(1)}_i}\right)\right] \times t^{2 L \abs{\fl} + 2 R \abs{\fn}}
\\ & \qquad \times \PE\left[\sum_{s = \pm 1} \sum_{\alpha, \beta = 1}^2 t (u^{(\alpha)}_2)^s (u^{(\beta)}_3)^{-s}\right]
\\ & \qquad \times \left(u^{(1)}_1 u^{(2)}_1\right)^{\fm} \left(u^{(1)}_2 u^{(2)}_2\right)^{\fl} \times \frac{1+t^2}{\left(1-t^2 \frac{u^{(1)}_1}{u^{(2)}_1}\right) \left(1-t^2 \frac{u^{(2)}_1}{u^{(1)}_1}\right)}
\\ & \qquad \times \left(u^{(1)}_3 u^{(2)}_3\right)^{-\fn} \left(u^{(1)}_4 u^{(2)}_4\right)^{-\fm} \times \frac{1+t^2}{\left(1-t^2 \frac{u^{(1)}_4}{u^{(2)}_4}\right) \left(1-t^2 \frac{u^{(2)}_4}{u^{(1)}_4}\right)}~,
}
whre the Chern-Simons levels are parametrised as in \eref{k1k2pq}. The integration over the variables $\{u^{(\alpha)}_1\}$ and $\{u^{(\alpha)}_4\}$ yields a non-zero result if $\fm = - \fq \fl = -\fq \fn$, with $\fq \in \BZ_{>0}$. Such a condition on the magnetic fluxes reduces the expression above into a single summation, for instance by expressing $\fl$ and $\fn$ as functions of $\fm$. Upon introducing the variable $w = w_1^{-1} (w_2 w_3)^{\fq} w_4^{-1}$, the computation of the residues at the poles of the gauge fugacities finally results in 
\bes{ \label{limitTU2ma}
& \sum_{\fm \in \fq \BZ} \frac{w^{\frac{2 \fm}{\fq}} t^{\frac{2 \left(-1 + 2 \fp \fq + L + R\right) \abs{\fm}}{\fq}}}{\left(1 - t^2\right) \left(1 - t^4\right)}  = \sum_{m \in \BZ} \frac{w^{2 m} t^{2 \CF \abs{m}}}{\left(1 - t^2\right) \left(1 - t^4\right)} ~,
}
where we set $m = \frac{\fm}{\fq}$, and define
\bes{ \label{defCF}
\CF = L+R+2\fp\fq-1~.
}

We will demonstrate in Appendix \ref{app:vanishfluxes} that the contributions from the magnetic fluxes satisfying $m^{(1)}_1 \neq m^{(2)}_1$ and $m^{(1)}_4 \neq m^{(2)}_4$ vanish due to the residue theorem.

\item \label{item2} $\{m^{(1)}_2 = m^{(1)}_3 = \fm_1, m^{(2)}_2 = m^{(2)}_3 = \fm_2\}$: the only non-vanishing contributions come from the sectors with $\{m^{(1)}_1 = \fl_1 \neq m^{(2)}_1 = \fl_2\}$, $\{m^{(1)}_4 = \fn_1 \neq m^{(2)}_4 = \fn_2\}$, for which the expression \eref{limpqwithTU2a} becomes
\bes{ \label{limitem2}
& \frac{1}{2^4} \sum_{\fl_1, \fl_2, \fm_1, \fm_2, \fn_1, \fn_2}\oint \left(\prod_{i=1}^4 \prod_{\alpha=1}^2 \frac{du^{(\alpha)}_i}{2\pi i u^{(\alpha)}_i}\right) w_1^{\fl_1 + \fl_2} \left(w_2 w_3\right)^{\fm_1 + \fm_2} w_4^{\fn_1 + \fn_2} \\& \qquad \times \left(u^{(1)}_1\right)^{\fq \fl_1} \left(u^{(2)}_1\right)^{\fq \fl_2} \left(\frac{u^{(1)}_2}{u^{(1)}_3}\right)^{2 \fp \fm_1} \left(\frac{u^{(2)}_2}{u^{(2)}_3}\right)^{2 \fp \fm_2} \left(u^{(1)}_4\right)^{-\fq \fn_1} \left(u^{(2)}_4\right)^{-\fq \fn_2} \\& \qquad \times \left[\left(\fz t\right)^{-\left(\abs{\fl_1-\fl_2}+2\abs{\fm_1-\fm_2}+\abs{\fn_1-\fn_2}\right)}\right] \times t^{L \left(\abs{\fl_1}+\abs{\fl_2}\right) + R \left(\abs{\fn_1}+\abs{\fn_2}\right)}
\\ & \qquad \times \fz^{2 \abs{\fm_1-\fm_2}} \PE\left[\sum_{s = \pm 1} \sum_{\alpha = 1}^2 t (u^{(\alpha)}_2)^s (u^{(\alpha)}_3)^{-s}\right]
\\ & \qquad \times \fz^{\abs{\fl_1-\fl_2}} t^{\abs{\fm_1-\fm_2}} \Big[\left(u^{(1)}_1\right)^{\fm_1} \left(u^{(2)}_1\right)^{\fm_2} \left(u^{(1)}_2\right)^{\fl_1} \left(u^{(2)}_2\right)^{\fl_2} \\ & \qquad \qquad \qquad \quad \,\,\,\,\,\,\ + \left(u^{(1)}_1\right)^{\fm_2} \left(u^{(2)}_1\right)^{\fm_1} \left(u^{(1)}_2\right)^{\fl_2} \left(u^{(2)}_2\right)^{\fl_1}\Big]
\\ & \qquad \times \fz^{\abs{\fn_1-\fn_2}} t^{\abs{\fm_1-\fm_2}} \Big[\left(u^{(1)}_3\right)^{-\fn_1} \left(u^{(2)}_3\right)^{-\fn_2} \left(u^{(1)}_4\right)^{-\fm_1} \left(u^{(2)}_4\right)^{-\fm_2} \\ & \qquad \qquad \qquad \quad \,\,\,\,\,\,\,\,\ + \left(u^{(1)}_3\right)^{-\fn_2} \left(u^{(2)}_3\right)^{-\fn_1} \left(u^{(1)}_4\right)^{-\fm_2} \left(u^{(2)}_4\right)^{-\fm_1}\Big]~.
}
Upon performing the integration over the gauge fugacities $\{u^{(\alpha)}_1\}$ and $\{u^{(\alpha)}_4\}$, these pick up residues at the poles for the following four choices of magnetic fluxes: $\{\fm_1 = - \fq \fl_1, \fm_2 = - \fq \fl_2 \lor \fm_1 = - \fq \fl_2, \fm_2 = - \fq \fl_1 \}$ and $\{\fm_1 = - \fq \fn_1, \fm_2 = - \fq \fn_2 \lor \fm_1 = - \fq \fn_2, \fm_2 = - \fq \fn_1 \}$. These conditions simplify the integral above, which can now be expressed just as a sum over two magnetic fluxes, say $\fm_1$ and $\fm_2$, as follows:
\bes{ \label{limitTU2m1m2a}
&\frac{1}{4} \sum_{\fm_1, \fm_2 \in \fq \BZ} \frac{w^{\frac{\fm_1+\fm_2}{\fq}} t^\frac{-2 \abs{\fm_1 - \fm_2} + \left(-1 + L + R + 2 \fp \fq\right) \left(\abs{\fm_1} + \abs{\fm_2}\right)}{\fq}}{\left(1 - t^2\right)^2} \\
=& \frac{1}{4} \sum_{m_1, m_2 \in \BZ} \frac{w^{m_1+m_2} t^{-2 \abs{m_1 - m_2} + \CF\left(\abs{m_1} + \abs{m_2}\right)}}{\left(1 - t^2\right)^2}~, 
} 
where we set $w = w_1^{-1} (w_2 w_3)^{\fq} w_4^{-1}$ and, in the second equality, we define $m_{\alpha} = \frac{\fm_{\alpha}}{\fq}$ for $\alpha = 1, 2$ and use \eqref{defCF}.

On the other hand, we will demonstrate in Appendix \ref{app:vanishfluxes} that the magnetic fluxes satisfying $\{m^{(1)}_1 = m^{(2)}_1 = \fl\}$, $\{m^{(1)}_4 = m^{(2)}_4 = \fn\}$ do not contribute to the $\fz \rightarrow 0$ limit of the index. 

\item \label{item3} $\{m^{(1)}_2 = m^{(2)}_3 = \fm_1, m^{(2)}_2 = m^{(1)}_3 = \fm_2\}$: this choice of magnetic fluxes is analogous to the previous case, to which is related simply by permuting $m^{(1)}_3$ and $m^{(2)}_3$. Upon setting the topological fugacities of the theory to one, the action of the $\U(2)$ Weyl group ensures that the limit of the index \eref{limpqwithTU2a} in this case is the same as \eref{limitTU2m1m2a}.
\end{enumerate}

We can finally consider the total contribution coming from \eref{limitTU2ma} and twice \eref{limitTU2m1m2a}. Taking into account the action of the $\U(2)$ Weyl group, the sum in the latter expression can be reformulated by considering an ordered set of magnetic fluxes satisfying $m_1 \ge m_2$, and thus multiplying by an overall factor of two. If we further use the definition of the $\U(2)$ dressing factor given in \eref{PU2}, the resulting total contribution can then be expressed as
\bes{ \label{CBHSquivwithTU2a}
\scalebox{0.98}{$
\begin{split}
&\eqref{limitTU2ma} + 2\eqref{limitTU2m1m2a} \\
&=\sum_{m_1 \ge m_2 \in \BZ} P_{\U(2)}(t; m_1,m_2) w^{m_1+m_2} t^{-2 \abs{m_1 - m_2} + \CF \left(\abs{m_1} + \abs{m_2}\right)} \\&=\frac{\left(1 - t^{2 \CF-2}\right) \left(1 - t^{2 \CF}\right)}{\left(1 - t^2\right) \left(1 - t^4\right) \left(1 - w^{\pm}t^{\CF-2}\right) \left(1 - w^{\pm}t^{\CF}\right)} \\ & = \PE\left[t^2 + t^4 + \left(w+\frac{1}{w}\right) t^{\CF-2} + \left(w+\frac{1}{w}\right) t^{\CF}  - t^{2 \left(\CF-1\right)} - t^{2 \CF}\right]~,
\end{split}
$}
}
where $1-w^{\pm} t$ stands for $\left(1-w t\right) \left(1- w^{-1} t\right)$. This expression matches the Coulomb branch Hilbert series of $\CN=4$ $\U(2)$ SQCD with $\CF= L + R + 2 \fp \fq - 1$ flavours. This result finds agreement with the magnetic quiver prediction for the NS5-branch reported in \cite[(5.15)]{Marino:2025uub}.\footnote{Note that the authors of that reference consider a slightly different brane setup with respect to the one depicted in \eref{branesetTU2}.} 

As we can read from \eref{CBHSquivwithTU2a}, there are six generators subject to two relations, which can be mapped to those described for 3d $\CN=4$ $\U(2)$ SQCD in \cite[around (5.4)]{Cremonesi:2013lqa} as follows.\footnote{The case we are interested in corresponds to setting $k=2$ and $n = L+R+2\fp\fq-1$ in the notation used in that reference, where $\CN=4$ $\U(k)$ SQCD with $n$ flavours is considered.} The generators of the Coulomb branch of $\U(2)$ SQCD can be built from the adjoint field $\Phi$ in the $\CN=4$ vector multiplet, which can be put in a diagonal form as $\Phi = \diag(\psi, \phi)$, and the monopole operator with $\U(2)$ magnetic fluxes $\pm (1,0)$, denoted $V_{\pm (1,0)}$, as detailed in the aforementioned reference. We tabulate such generators and the corresponding gauge invariant quantities in theory \eref{quivpqbraneTU2} below.
\bes{
\begin{tabular}{c|c}
\hline
Generators of $\U(2)$ SQCD  & Corresponding operators in \eref{quivpqbraneTU2} \\
\hline
$\tr(\Phi)$     &  $\tr(Y_0)$ \\
$\tr(\Phi^2)$   & $\tr(Y_0^2)$ \\
$V_{(1,0)}$ & $Y_+ \left[(X_{23})^1_1\right]^{2\fp\fq-1}$\\
$V_{(-1,0)}$ & $Y_- \left[(X_{32})^1_1\right]^{2\fp\fq-1}$\\
$V_{(1,0)} \phi$ & $Y_+ \left[(X_{23})^1_1\right]^{2\fp\fq-1} y$\\
$V_{(-1,0)} \phi$ & $Y_- \left[(X_{32})^1_1\right]^{2\fp\fq-1} y$\\
\hline
\end{tabular}
}
where we define $Y_{0}=X_{23} X_{32}$, which can be put in a diagonal form as $Y_0 = \diag(x,y)$, and the monopole operators $Y_{\pm}$ are those with fluxes $\pm (1,0; - \fq,0; -\fq,0; 1,0)$ under the four $\U(2)$ gauge groups. Note that the gauge charge of such bare monopole operators can be read off from the second, fifth and seventh lines of \eqref{limitem2}, namely
$(u_1^{(1)})^\fq (u_2^{(1)}/u_3^{(1)})^{-2\fp\fq} (u_4^{(1)})^{-\fq} \times (u_1^{(1)})^{-\fq}(u_2^{(1)})\times (u_3^{(1)})^{-1} (u_4^{(1)})^{\fq} = (u_2^{(1)}/u_3^{(1)})^{-2\fp\fq+1}$.\footnote{In particular, the terms $t^2$ and $t^4$ appearing in the plethystic exponential in \eref{CBHSquivwithTU2a} correspond to the generators $\tr{Y_0}$ and $\tr{Y_0^2}$, respectively. Note that the operators $\tr{Y_0^{m+2}}$, with $m > 0$, can be built from the two said generators thanks to the Cayley-Hamilton theorem as $\tr{Y_0^{m+2}} = \tr{Y_0} \tr{Y_0^{m+1}} + \frac{1}{2} \left[\tr{Y_0^2} \tr{Y_0^{m}}- (\tr{Y_0})^2 \tr{Y_0^{m}}\right]$.}

\subsection{The $T[4]$ theory} \label{sec:linearext4}
We consider the following theory
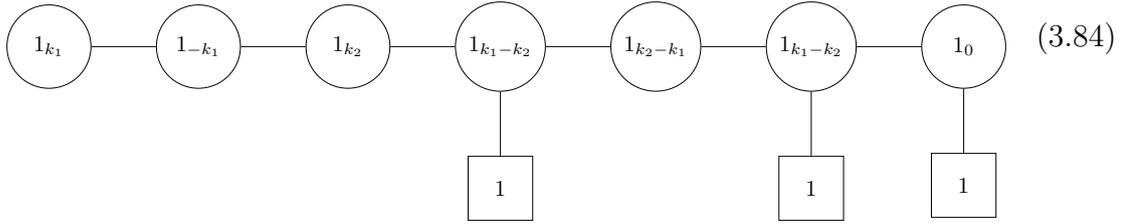
\begin{equation} \label{T4quiv}
\scalebox{0.85}{
\begin{tikzpicture}[baseline,font=\footnotesize,
    circ/.style={circle, draw, minimum size=1.3cm},
    sq/.style={rectangle, draw, minimum size=1cm},
    node distance=1cm
]

    \node[circ] (n1) {$1_{k_1}$};
    \node[circ, right=of n1] (n2) {$1_{-k_1}$};
    \node[circ, right=of n2] (n3) {$1_{k_2}$};
    \node[circ, right=of n3] (n4) {$1_{k_1-k_2}$};
    \node[circ, right=of n4] (n5) {$1_{k_2-k_1}$};
    \node[circ, right=of n5] (n6) {$1_{k_1-k_2}$};
    \node[circ, right=of n6] (n7) {$1_{0}$};

    \node[sq, below=of n4] (s1) {1};
    \node[sq, below=of n6] (s2) {1};
    \node[sq, below=of n7] (s3) {1};

    \draw (n1) -- (n2);
    \draw (n2) -- (n3);
    \draw (n3) -- (n4);
    \draw (n4) -- (n5);
    \draw (n5) -- (n6);
    \draw (n6) -- (n7);

    \draw (n4) -- (s1);
    \draw (n6) -- (s2);
    \draw (n7) -- (s3);

\end{tikzpicture}}
\end{equation}
We assume, for convenience, that $k_2 > k_1 >0$.
The corresponding brane system is
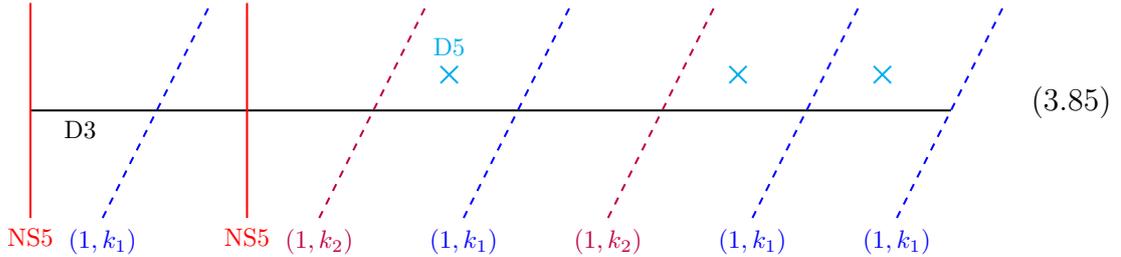
\begin{equation}
\scalebox{0.95}{
\begin{tikzpicture}[baseline,font=\footnotesize,
    cross_node/.style={circle, draw=cyan, cross out, thick, minimum size=6pt, inner sep=0pt}
]

    \def\ystart{-1.5}
    \def\yend{1.5}

    \draw[thick] (0,0) -- node[near start, below, xshift=-2.5cm]{D3} (12.75,0);

    \draw[red, thick] (0,\ystart) -- (0,\yend) node[pos=0, below, red] {NS5};
    \draw[red, thick] (3,\ystart) -- (3,\yend) node[pos=0, below, red] {NS5};


    \draw[blue, dashed, thick] (1, \ystart) -- (2.5, \yend) node[pos=0, below, blue] {$(1, k_1)$};
    
    \draw[purple, dashed, thick] (4, \ystart) -- (5.5, \yend) node[pos=0, below, purple] {$(1, k_2)$};

    \draw[blue, dashed, thick] (6, \ystart) -- (7.5, \yend) node[pos=0, below, blue] {$(1, k_1)$};
    
    \draw[purple, dashed, thick] (8, \ystart) -- (9.5, \yend) node[pos=0, below, purple] {$(1, k_2)$};
    
    \draw[blue, dashed, thick] (10, \ystart) -- (11.5, \yend) node[pos=0, below, blue] {$(1, k_1)$};

    \draw[blue, dashed, thick] (12, \ystart) -- (13.5, \yend) node[pos=0, below, blue] {$(1, k_1)$};

    \node[cross_node, label={[cyan]above:D5}] at (5.8, 0.5) {}; 
    \node[cross_node] at (9.8, 0.5) {}; 
    \node[cross_node] at (11.8, 0.5) {}; 

\end{tikzpicture}}
\end{equation}
In order to compute the limit of the index for each branch, we assign the following axial fugacities to the chiral fields in the quiver as follows.
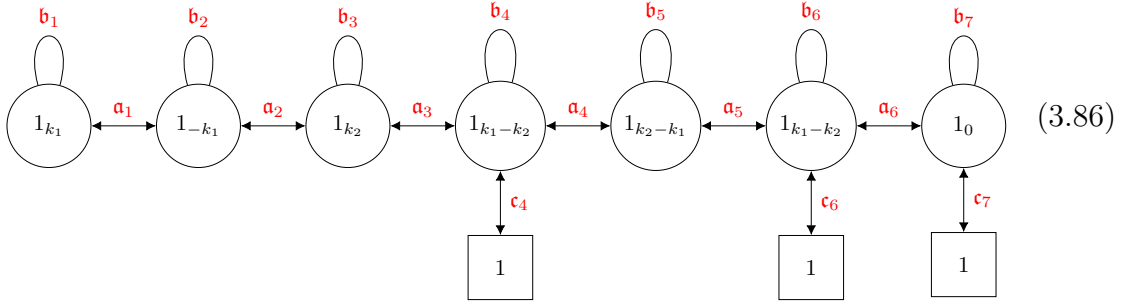
\begin{equation} \label{quivT4axial}
\scalebox{0.85}{
\begin{tikzpicture}[baseline,font=\footnotesize,
    circ/.style={circle, draw, minimum size=1.3cm},
    sq/.style={rectangle, draw, minimum size=1cm},
    node distance=1cm,
    every loop/.style={-}
]

    \node[circ] (n1) {$1_{k_1}$};
    \node[circ, right=of n1] (n2) {$1_{-k_1}$};
    \node[circ, right=of n2] (n3) {$1_{k_2}$};
    \node[circ, right=of n3] (n4) {$1_{k_1-k_2}$};
    \node[circ, right=of n4] (n5) {$1_{k_2-k_1}$};
    \node[circ, right=of n5] (n6) {$1_{k_1-k_2}$};
    \node[circ, right=of n6] (n7) {$1_{0}$};

    \node[sq, below=of n4] (s1) {1};
    \node[sq, below=of n6] (s2) {1};
    \node[sq, below=of n7] (s3) {1};

    \draw[<->] (n1) -- node[above, red] {$\mathfrak{a}_1$} (n2);
    \draw[<->] (n2) -- node[above, red] {$\mathfrak{a}_2$} (n3);
    \draw[<->] (n3) -- node[above, red] {$\mathfrak{a}_3$} (n4);
    \draw[<->] (n4) -- node[above, red] {$\mathfrak{a}_4$} (n5);
    \draw[<->] (n5) -- node[above, red] {$\mathfrak{a}_5$} (n6);
    \draw[<->] (n6) -- node[above, red] {$\mathfrak{a}_6$} (n7);

    \draw[<->] (n4) -- node[right, red] {$\mathfrak{c}_4$} (s1);
    \draw[<->] (n6) -- node[right, red] {$\mathfrak{c}_6$} (s2);
    \draw[<->] (n7) -- node[right, red] {$\mathfrak{c}_7$} (s3);
    
    \path (n1) edge [loop above] node[red] {$\mathfrak{b}_1$} ();
    \path (n2) edge [loop above] node[red] {$\mathfrak{b}_2$} ();
    \path (n3) edge [loop above] node[red] {$\mathfrak{b}_3$} ();
    \path (n4) edge [loop above] node[red] {$\mathfrak{b}_4$} ();
    \path (n5) edge [loop above] node[red] {$\mathfrak{b}_5$} ();
    \path (n6) edge [loop above] node[red] {$\mathfrak{b}_6$} ();
    \path (n7) edge [loop above] node[red] {$\mathfrak{b}_7$} ();

\end{tikzpicture}}
\end{equation}
The index of \eqref{quivT4axial} is then
\bes{
\scalebox{0.96}{$
\begin{split}
&\CI_{\eqref{quivT4axial}} = \sum_{m_1, \ldots, m_7 \in \BZ} \oint  \Bigg( \prod_{i=1}^7 \frac{du_i}{2\pi i u_i}  u_i^{\kappa_i m_i}  w_i^{m_i} Z_{\text{chir}}^{1} \left(x; \fb_i; 0 \right) \Bigg) \\ 
& \quad \times \prod_{s = \pm1} \prod_{i=1}^6 Z_{\text{chir}}^{1/2} \left(x; \fa_i (u_i u_{i+1}^{-1})^s; s m_{i} - s m_{i+1} \right) \prod_{j =4,6,7} Z_{\text{chir}}^{1/2} \left(x; \fc_j (f^{-1}_j u_j)^s; sm_{j} \right)~,
\end{split}
$}
}
where 
\bes{
\begin{array}{llll}
\kappa_1 = k_1~, &\,\,\,\, \kappa_2 = -k_1~, &\,\,\,\,\, \kappa_3 = k_2~, & \\
\kappa_4 = -(k_2-k_1) ~, &\,\,\,\,\, \kappa_5 = k_2-k_1 ~, &\,\,\,\,\, \kappa_6 = -(k_2-k_1) ~, &\,\,\,\,\, \kappa_7 = 0~,
\end{array}
}
and, for each branch, the fugacities $\fa_i$, $\fb_i$ and $\fc_i$ are given as follows:
\bes{
&\begin{tabular}{c|ccccccccc}
\hline
Branch & $\fa_1$ & $\fa_2$ & $\fa_3$ & $\fa_4$ & $\fa_5$ & $\fa_6$ & $\fc_4$ & $\fc_6$ & $\fc_7$  \\
\hline
NS5& $a$ & $a^{-1}$ & 1 & 1 & 1 & 1& 1& 1& 1\\
$(1,k_2)$ & $1$ & $1$& $a^{-1}$ & $a$ & $a^{-1}$ & $1$ & $a^{-1}$ & $1$ & $1$  \\
$(1,k_1)$ & $a^{-1}$ & $a$ & $a$ & $a^{-1}$ & $a$ & $a^{-1}$ & $a^{-1}$ & $a^{-1}$ & $a^{-1}$  \\
D5 & $a$ & $a$ & $a$ & $a$ & $a$ & $a$ & $a$ & $a$ & $a$  \\
\hline \hline
Branch & $\fb_1$ & $\fb_2$ & $\fb_3$ & $\fb_4$ & $\fb_5$ & $\fb_6$ & $\fb_7$  \\
\hline
NS5 & 1 & 1 &1 & 1& 1& 1& 1 \\
$(1,k_2)$ & 1 & 1 &1 & 1& 1& 1& 1 \\
$(1,k_1)$ & 1 & 1 & $a^{-2}$ & 1& 1& 1& $a^{+2}$ \\
D5 & $a^{-2}$ & $a^{-2}$ & $a^{-2}$ & $a^{-2}$& $a^{-2}$& $a^{-2}$& $a^{-2}$ \\
\hline
\end{tabular} 
}
The limits of the index for the NS5-branch and the D5-branch are similar to the analyses in the precedent examples. The analysis for the $(1, k_1)$-branch is slightly more intricate and we provide a detailed computation below.

\subsubsection*{The $(1, k_1)$-branch}
The power of $\fz$ on the right hand side of \eqref{limitHS} is $\fz^{|m_2-m_3|+|m_3-m_4|+|m_5-m_6|}$. A non-vanishing result in the limit $\fz \rightarrow 0$ imposes
\bes{
m_2 = m_3 = m_4 = \fm~, \quad m_5=m_6 = \fn~.
}
The limit of the index for this branch is
\bes{ \label{limitindex1k1branchT4}
\scalebox{0.9}{$
\begin{split}
&H_{\eqref{quivT4axial}}[(1,k_1)](t; \{ w_{i} \}, \{b_{i}\}) \\
&=  \sum_{m_1, \fm, \fn, m_7 \in \BZ}  t^{2 \Delta(m_1, \fm, \fm, \fm, \fn, \fn, m_7)} w_1^{m_1} (w_2 w_3 w_4)^\fm (w_5 w_6)^{\fn} w_7^{m_7}   \\
& \qquad \times \left( \prod_{j=1}^7 \oint \frac{d u_j}{2\pi i u_j} \right) u_1^{k_1 m_1} u_2^{-k_1 \fm } u_3^{k_2 \fm} u_4^{(k_1-k_2) \fm} u_5^{-(k_1-k_2) \fn} u_6^{(k_1-k_2) \fn} \\
& \qquad \times \PE\left[ t \sum_{s=\pm 1} \{  (b_2 u_2 u_3^{-1})^s + (b_3 u_3 u_4^{-1})^s +(b_5 u_5 u_6^{-1})^s  \}  \right]~,
\end{split}$}
}
where the dimension of the monopole operators is
\bes{
\Delta(m_1, \ldots, m_7) = \sum_{i=1}^6 |m_i-m_{i+1}| + |m_4| +|m_6|+|m_7|~.
}
The integral over $u_1$ sets $m_1=0$. Hence, the monopole operators on this branch possess fluxes of the form
\bes{
(0,\fm, \fm, \fm, \fn, \fn, m_7)~, \qquad \fm, \fn, m_7 \in \BZ~.
}
The expression \eqref{limitindex1k1branchT4} reduces to
\bes{
\scalebox{0.9}{$
\begin{split}
&H_{\eqref{quivT4axial}}[(1,k_1)](t; \{ w_{i} \}, \{b_{i}\}) \\
&=  \sum_{\fm, \fn, m_7 \in \BZ}  t^{|\fm-\fn|+|\fn-m_7|+2|\fm|+|\fn|+|m_7|} (w_2 w_3 w_4)^\fm (w_5 w_6)^{\fn} w_7^{m_7}   \\
& \qquad \qquad\times \left( \prod_{j=2}^6 \oint \frac{d u_j}{2\pi i u_j} \right) u_2^{-k_1 \fm } u_3^{k_2 \fm} u_4^{-(k_2-k_1) \fm} u_5^{(k_2-k_1) \fn} u_6^{-(k_2-k_1) \fn} \\
& \qquad \qquad\times \PE\left[ t \sum_{s=\pm 1} \{  (b_2 u_2 u_3^{-1})^s + (b_3 u_3 u_4^{-1})^s +(b_5 u_5 u_6^{-1})^s  \}  \right]\\
&\overset{\eqref{usefulintiden}}{=} \sum_{\fm, \fn, m_7 \in \BZ}  t^{|\fm-\fn|+|\fn-m_7|+2|\fm|+|\fn|+|m_7|} (w_2 w_3 w_4)^\fm  w_7^{m_7} \times \frac{t^{(k_2-k_1)|\fn|}}{(1-t^2)} (b_5^{-(k_2-k_1)} w_5 w_6)^\fn \\
&  \qquad \qquad\times \left( \prod_{j=2}^4 \oint \frac{d u_j}{2\pi i u_j} \right) u_2^{-k_1 \fm } u_3^{k_1 \fm} u_3^{(k_2-k_1) \fm} u_4^{-(k_2-k_1) \fm} \\
& \qquad \qquad \times \PE\left[ t \sum_{s=\pm 1} \{  (b_2 u_2 u_3^{-1})^s + (b_3 u_3 u_4^{-1})^s   \}  \right] \\
&\overset{\eqref{usefulintiden}}{=} \sum_{\fm, \fn, m_7 \in \BZ}  t^{|\fm-\fn|+|\fn-m_7|+2|\fm|+|\fn|+|m_7|}   w_7^{m_7} \times \frac{t^{(k_2-k_1)|\fn|}}{(1-t^2)} (b_5^{-(k_2-k_1)} w_5 w_6)^\fn \\
& \qquad \qquad \times \frac{t^{k_1 |\fm|} t^{(k_2-k_1) |\fm|}}{(1-t^2)^2} (b_2^{k_1} b_3^{-(k_2-k_1)} w_2 w_3 w_4)^\fm \\
&= \sum_{\fm, \fn, m_7 \in \BZ} t^{|\fm-\fn|+|\fn-m_7|+(k_2+2)|\fm|+(k_2-k_1+1)|\fn|+|m_7|} \times \frac{y_1^\fm y_2^\fn w_7^{m_7} }{(1-t^2)^3}~,
\end{split}$}
}
where
\bes{
y_1 = b_2^{k_1} b_3^{-(k_2-k_1)} w_2 w_3 w_4~, \qquad y_2 = b_5^{-(k_2-k_1)} w_5 w_6~.
}
The last line is the Coulomb branch Hilbert series of the following 3d $\CN=4$ theory:
\begin{equation} 
\scalebox{0.85}{
\begin{tikzpicture}[baseline,font=\footnotesize,
    circ/.style={circle, draw, minimum size=1cm},
    sq/.style={rectangle, draw, minimum size=1cm},
    node distance=1cm
]
    \node[circ] (n1) {$1$};
    \node[circ, right=of n1] (n2) {$1$};
    \node[circ, right=of n2] (n3) {$1$};
    \node[sq, below=of n1] (s1) {$k_2+2$};
    \node[sq, below=of n2] (s2) {$k_2-k_1+1$};
    \node[sq, below=of n3] (s3) {$1$};
    \draw (n1) -- (n2);
    \draw (n2) -- (n3);
    \draw (n1) -- (s1);
    \draw (n2) -- (s2);
    \draw (n3) -- (s3);
\end{tikzpicture}}
\end{equation}
which is the magnetic quiver for this branch, in agreement with \cite[(5.7)]{Marino:2025uub}. The fugacities for topological symmetry associated with each gauge group from left to right are $(y_1, y_2, w_7)$, respectively. The corresponding gauge magnetic fluxes are $(\fm, \fn, m_7)$, respectively.

\subsubsection*{The NS5-branch}
On this branch, we have $m_1=m_2 =\fm$, $m_3=m_4=\ldots= m_7= 0$
The limit of the index for this branch is
\bes{
\scalebox{0.99}{$
\begin{split}
&\sum_{\fm \in \BZ} \oint \frac{du_1}{2\pi i u_1} \frac{du_2}{2\pi i u_2} t^{2 \Delta(\fm, \fm, 0,0,0,0,0)} u_1^{k_1 \fm} u_2^{-k_1 \fm}  (w_1 w_2)^\fm \PE \left[t \sum_{s=\pm 1} (b_1 u_1 u_2^{-1})^s \right] \\
& \overset{\eqref{usefulintiden}}{=} \sum_{\fm \in \BZ} \frac{t^{(1+k_1)|\fm|}}{1-t^2} y^\fm  = \PE \left[t^2 + ( y +y^{-1}) t^{k_1+1} -t^{2(k_1+1)} \right]~, \quad y = b_1^{-k_1} w_1 w_2~.
\end{split}
$}
}
This is the Coulomb branch Hilbert series of the $\U(1)$ gauge theory with $k_1+1$ flavours, which is the magnetic quiver for this branch \cite[(5.4)]{Marino:2025uub}. The corresponding moduli space is $\BC^2/\BZ_{k_1+1}$.

\subsubsection*{The D5-branch}
On this branch, all of the gauge magnetic fluxes are zero: $m_1=m_2= \ldots = m_7=0$. In the limit $\fz \rightarrow 0$, the index reduces to the Molien integral that computes the Higgs branch Hilbert series of \eqref{T4quiv} with $k_1=k_2=0$:
\bes{
\oint \left( \prod_{i=1}^7 \frac{du_i}{2\pi u_i} \right) \PE\left[ -7t^2+\sum_{s = \pm 1} \left \{ \sum_{j=1}^6 (b_j u_j u_{j+1}^{-1})^s + \sum_{l=4,6,7} (f_l^{-1} u_l)^s \right \} \right]~.
}
Using \eqref{usefulintiden}, we see that the integrals over $u_1$, $u_2$ and $u_3$ are trivial. Hence, the above formula reduces to 
\bes{
\oint \left( \prod_{i=4}^7 \frac{du_i}{2\pi u_i} \right) \PE\left[ -4t^2+\sum_{s = \pm 1} \left \{ \sum_{j=4}^6 (b_j u_j u_{j+1}^{-1})^s + \sum_{l=4,6,7} (f_l^{-1} u_l)^s \right \} \right]~.
}
This is the Higgs branch Hilbert series of the following 3d $\CN=4$ theory:
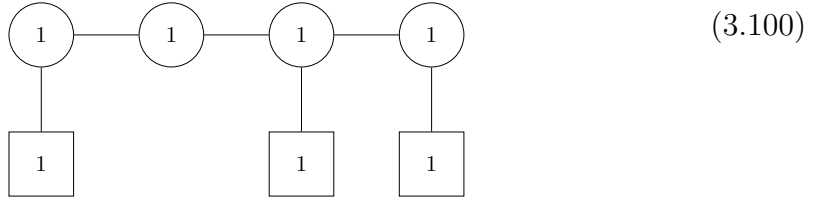
\begin{equation}
\scalebox{0.85}{
\begin{tikzpicture}[baseline,font=\footnotesize,
    circ/.style={circle, draw, minimum size=1cm},
    sq/.style={rectangle, draw, minimum size=1cm},
    node distance=1cm
]

    \node[circ] (n4) {$1$};
    \node[circ, right=of n4] (n5) {$1$};
    \node[circ, right=of n5] (n6) {$1$};
    \node[circ, right=of n6] (n7) {$1$};

    \node[sq, below=of n4] (s1) {1};
    \node[sq, below=of n6] (s2) {1};
    \node[sq, below=of n7] (s3) {1};

    \draw (n4) -- (n5);
    \draw (n5) -- (n6);
    \draw (n6) -- (n7);

    \draw (n4) -- (s1);
    \draw (n6) -- (s2);
    \draw (n7) -- (s3);

\end{tikzpicture}}
\end{equation}
The magnetic quiver for this branch is simply the mirror theory of this quiver, namely
\begin{equation}
\scalebox{0.85}{
\begin{tikzpicture}[baseline,font=\footnotesize,
    circ/.style={circle, draw, minimum size=1cm},
    sq/.style={rectangle, draw, minimum size=1cm},
    node distance=1cm
]

    \node[circ] (n6) {$1$};
    \node[circ, right=of n6] (n7) {$1$};

    \node[sq, left=of n6] (s2) {$3$};
    \node[sq, right=of n7] (s3) {$2$};

    \draw (n6) -- (n7);

    \draw (n6) -- (s2);
    \draw (n7) -- (s3);

\end{tikzpicture}}
\end{equation}
in agreement with \cite[(5.6)]{Marino:2025uub}.

\subsection{A unitary linear quiver with unequal ranks} \label{sec:linearex2}
We consider the 3d $\CN=3$ theory described by the following quiver diagram:
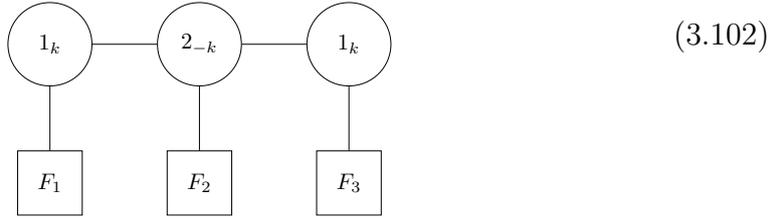
\begin{equation} \label{quivT3theoryunequalranks}
\scalebox{0.85}{
\begin{tikzpicture}[baseline,font=\footnotesize,
    circ/.style={circle, draw, minimum size=1.3cm},
    sq/.style={rectangle, draw, minimum size=1cm},
    node distance=1cm
]
    \node[circ] (n1) {$1_{k}$};
    \node[circ, right=of n1] (n2) {$2_{-k}$};
    \node[circ, right=of n2] (n3) {$1_{k}$};
    \node[sq, below=of n1] (s1) {$F_1$};
    \node[sq, below=of n2] (s2) {$F_2$};
    \node[sq, below=of n3] (s3) {$F_3$};
    \draw (n1) -- (n2);
    \draw (n2) -- (n3);
    \draw (n1) -- (s1);
    \draw (n2) -- (s2);
    \draw (n3) -- (s3);
\end{tikzpicture}}
\end{equation}
which can be realised in string theory via the brane configuration shown below:
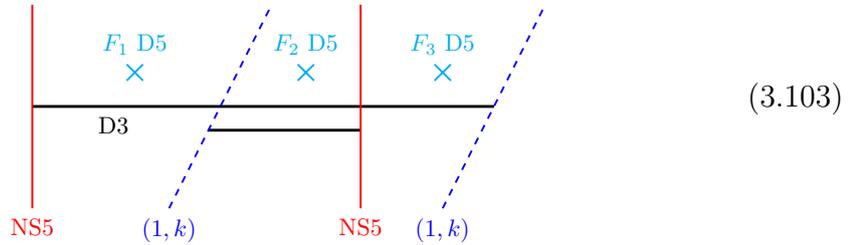
\begin{equation}
\scalebox{0.9}{
\begin{tikzpicture}[baseline,font=\footnotesize,
    cross_node/.style={circle, draw=cyan, cross out, thick, minimum size=6pt, inner sep=0pt}
]
    \def\ystart{-1.5}
    \def\yend{1.5}
    \draw[very thick] (0,0) -- node[near start, below,xshift=-0.5cm]{D3} (6.75,0);
    \draw[very thick] (2.57,-0.35) --(4.8,-0.35);
    \draw[red, thick] (0,\ystart) -- (0,\yend) node[pos=0, below, red] {NS5};
    \draw[red, thick] (4.8,\ystart) -- (4.8,\yend) node[pos=0, below, red] {NS5};
    \draw[blue, dashed, thick] (1+1, \ystart) -- (2.5+1, \yend) node[pos=0, below, blue] {$(1, k)$};
    \draw[blue, dashed, thick] (6, \ystart) -- (7.5, \yend) node[pos=0, below, blue] {$(1, k)$};
    \node[cross_node, label={[cyan]above:$F_1$ D5}] at (1.5, 0.5) {};
    \node[cross_node, label={[cyan]above:$F_2$ D5}] at (4, 0.5) {};
    \node[cross_node, label={[cyan]above:$F_3$ D5}] at (6, 0.5) {};
\end{tikzpicture}}
\end{equation}
As usual, we analyse this theory by computing its superconformal index. In order to compute the various limits of the index we assign axial fugacities to the chiral fields. In 3d $\mathcal{N}=2$ notation the field content and fugacity assignments are:
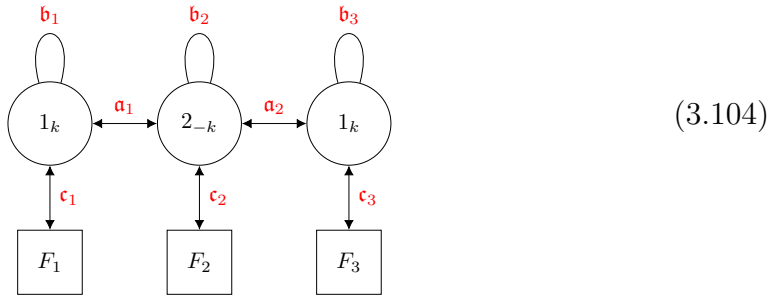
\begin{equation} \label{quivT3axialunequalranks}
\scalebox{0.85}{
\begin{tikzpicture}[baseline, font=\footnotesize,
    circ/.style={circle, draw, minimum size=1.3cm},
    sq/.style={rectangle, draw, minimum size=1cm},
    node distance=1cm,
    every loop/.style={-}
]
    \node[circ] (n1) {$1_{k}$};
    \node[circ, right=of n1] (n2) {$2_{-k}$};
    \node[circ, right=of n2] (n3) {$1_{k}$};
    \node[sq, below=of n1] (s1) {$F_1$};
    \node[sq, below=of n2] (s2) {$F_2$};
    \node[sq, below=of n3] (s3) {$F_3$};
    \draw[<->] (n1) -- node[above, red] {$\fa_1$} (n2);
    \draw[<->] (n2) -- node[above, red] {$\fa_2$} (n3);
    \draw[<->] (n1) -- node[right, red] {$\fc_1$} (s1);
    \draw[<->] (n2) -- node[right, red] {$\fc_2$} (s2);
    \draw[<->] (n3) -- node[right, red] {$\fc_3$} (s3);
    \path (n1) edge [loop above] node[red] {$\fb_1$} ();
    \path (n2) edge [loop above] node[red] {$\fb_2$} ();
    \path (n3) edge [loop above] node[red] {$\fb_3$} ();
\end{tikzpicture}}
\end{equation}
The specific assignments of axial fugacities $\fa_i$, $\fb_i$, and $\fc_i$ for each branch are summarised in the following table:
\bes{ \label{branchT3uneqranks}
\begin{tabular}{c|cccccccccc}
\hline
Branch & $\fa_1$ & $\fa_2$  & $\fc_1$ & $\fc_2$ & $\fc_3$ & $\fb_1$ & $\fb_2$ & $\fb_3$  \\
\hline
NS5& $a$ & $a^{-1}$ & $a^{-1}$ & $a^{-1}$ & 1 & 1 & 1 &1 \\
$(1,k)$ & $a^{-1}$ & $a$ & $1$ & $a^{-1}$ & $a^{-1}$ & 1 & 1 &1    \\
D5 & $a$ & $a$ & $a$ & $a$ & $a$ & $a^{-2}$ & $a^{-2}$ & $a^{-2}$ \\
\hline
\end{tabular}
}
The index for the theory in \eqref{quivT3axialunequalranks} is then given by
\bes{ \label{indexquivT3unequal}
\scalebox{0.84}{$
\begin{split}
&\CI_{\eqref{quivT3axialunequalranks}} = \frac{1}{2} \sum_{m_1,m_3 \in \BZ} \sum_{\{m^{(\alpha)}_2\} \in \BZ^2} \oint \Bigg( \prod_{\alpha=1}^2 \frac{du^{(\alpha)}_2}{2\pi i u^{(\alpha)}_2}  (u^{(\alpha)}_2)^{\kappa_2 m^{(\alpha)}_2}  w_2^{m^{(\alpha)}_2} \Bigg) \oint  \frac{du_1}{2\pi i u_1}  (u_1)^{\kappa_1 m_1}  w_1^{m_1} \\
& \qquad \quad \times \oint  \frac{du_3}{2\pi i u_3}  (u_3)^{\kappa_3 m_3}  w_3^{m_3} \times Z_{\text{vec}}^{\U(2)}\left(x; \{u^{(\alpha)}_2\}; \{m^{(\alpha)}_2\}\right) \\ & \qquad \quad \times Z_{\text{chir}}^{1} \left(x; \fb_1 ; 0  \right)Z_{\text{chir}}^{1} \left(x; \fb_3 ; 0  \right) \prod_{\alpha, \beta=1}^2 Z_{\text{chir}}^{1} \left(x; \fb_2 u^{\alpha}_2 (u^{\beta}_2)^{-1} ; m^{(\alpha)}_2 - m^{(\beta)}_2  \right) \\ 
& \qquad \quad \times \prod_{s = \pm1} \prod_{\alpha=1}^2 Z_{\text{chir}}^{1/2} \left(x; \fa_1 (b_1 u_1 (u^{(\alpha)}_{2})^{-1})^s; s m_{1} - s m^{(\alpha)}_{2} \right) \\ 
& \qquad \quad \times \prod_{s = \pm1} \prod_{\alpha=1}^2 Z_{\text{chir}}^{1/2} \left(x; \fa_2 (b_2 u^{(\alpha)}_2 (u_{3})^{-1})^s; s m^{(\alpha)}_{2} - s m_3 \right) \\ 
& \qquad \quad \times \prod_{s = \pm1} \prod_{\rho_1=1}^{F_1} Z_{\text{chir}}^{1/2} \left(x; \fc_1 ((f^{(\rho_1)}_1)^{-1} u_1)^s; sm_{1} \right)\prod_{s = \pm1} \prod_{\rho_3=1}^{F_3} Z_{\text{chir}}^{1/2} \left(x; \fc_3 ((f^{(\rho_3)}_3)^{-1} u_3)^s; sm_{3} \right)
\\ & \qquad \quad \times  \prod_{s = \pm1} \prod_{\rho_2=1}^{F_2} \prod_{\alpha=1}^2  Z_{\text{chir}}^{1/2} \left(x; \fc_2 ((f^{(\rho_2)}_2)^{-1} u^{(\alpha)}_2)^s; sm^{(\alpha)}_{2} \right) ~,
\end{split}$}
}
where the Chern-Simons levels are
\bes{ \label{kappaT3}
\kappa_1 = k~, \qquad \kappa_2 = -k~, \qquad \kappa_3 = k~.
}
For definiteness, we take $k>0$ throughout this discussion. The Hilbert series for each branch is obtained by taking a specific limit of the index:
\bes{ \label{limitHSunequalranks}
H_{\eqref{quivT3axialunequalranks}}[\text{branch}](t;\{w_i\}; \{b_i\}; \{f_i \}) = \lim_{\fz \rightarrow 0} \CI_{\eqref{quivT3axialunequalranks}} \Big|_{\eqref{branchT3uneqranks}, \, \eqref{N=3reparam}}~.
}
To compute the limit, we first examine the overall power of $\fz$ in the integrand for each branch. A non-vanishing result in the $\fz \to 0$ limit imposes the following constraints on the magnetic fluxes $m_i$:\footnote{Setting the exponent of $\fz$ to zero may yield more solutions than those reported here. However, one can check explicitly that such solutions lead to zero upon computing the limit of the index. \label{foot:moresolutions}}
\bes{ \label{condfluxT3unequal}
\scalebox{0.85}{
\begin{tabular}{c|c|c}
\hline
Branch & Power of $\fz$ & Condition \\
\hline
NS & $\fz^{|m_1-m_2^{(1)}|+|m_1-m_2^{(2)}|-|m_2^{(1)}-m_2^{(2)}|+\frac{F_3}{2}|m_3|}$ & $m_3=0~, \,\, \{m_1=m_2^{(1)} \,\, \text{or}\,\, m_1=m_2^{(2)} \}$  \\
$(1,k)$ & $\fz^{|m_3-m_2^{(1)}|+|m_3-m_2^{(2)}|-|m_2^{(1)}-m_2^{(2)}|+\frac{F_1}{2}|m_1|}$ & $m_1=0~, \,\, \{m_3=m_2^{(1)} \,\, \text{or}\,\, m_3=m_2^{(2)} \}$   \\
D5 & $\fz^{\sum_{i=1}^2(|m_1-m_2^{(i)}|+|m_3-m_2^{(i)}|)-2|m_2^{(1)}-m_2^{(2)}|} $  & $m_1=m_2^{(1)}=m_2^{(2)}=m_3=0$\\ &$^{+F_1|m_1|+F_2\left( \sum_{i=1}^2|m_2^{(i)}| \right)+F_3|m_3|}$ &  \\
\hline
\end{tabular}}
}
We now discuss the computation for each branch explicitly.
\subsubsection*{The NS5-branch}
The condition in Table \eqref{condfluxT3unequal} for a non-vanishing limit requires the exponent of $\fz$ to be zero:
\bes{\label{eq:27111}|m_1-m_2^{(1)}|+|m_1-m_2^{(2)}|-|m_2^{(1)}-m_2^{(2)}|+\frac{F_3}{2}|m_3|=0~.}
This condition restricts the magnetic fluxes to be $m_3 = 0$, $m_1 =\fm$, and $\vec{m}_2 = (\fn, \fm)$ or $\vec{m}_2 = (\fm, \fn)$, with $\fm, \fn \in \BZ$. Using the Weyl symmetry of the $\U(2)$ gauge group, we restrict our attention to the option $\fm_2 = (\fm, \fn)$. In summary, we can consider the fluxes of the form 
\bes{
m_3 =0~, \quad m_1 = \fm, \quad  \vec{m}_2 = (\fm, \fn)~.
}
Taking the limit \eqref{limitHSunequalranks}, the Hilbert series receives two distinct contributions: one from fluxes where $\fm=\fn$ and another from fluxes where $\fm\neq\fn$:
\bes{ \label{sumtwocontrNSuneq}
H_{\eqref{quivT3axialunequalranks}}[\text{NS5}](t;w_{1,2}; b_{1,2}) = H_{\fm=\fn} + H_{\fm\neq\fn}~.
}
For fluxes with $\fm=\fn$, the $\U(2)$ gauge group is unbroken, and the contribution is
\bes{ \label{Hm=nNSunequalranks}
\scalebox{0.85}{$
\begin{split}
&H_{\fm=\fn} = \frac{1}{2}  \sum_{\fm \in \BZ}  t^{2 \Delta(m_1=\fm,\   \vec m_2=(\fm, \fm), \, m_3 = 0)} \oint  \Bigg(  \prod_{\alpha=1}^2 \frac{du^{(\alpha)}_2}{2\pi i u^{(\alpha)}_2}  (u^{(\alpha)}_2)^{\kappa_2 m^{(\alpha)}_2}  w_2^{m^{(\alpha)}_2} \Bigg)_{\vec m_2=(\fm, \fm)} \\
&\,\, \times  \oint  \frac{du_1}{2\pi i u_1}   \Bigg((u_1)^{\kappa_1 m_1}  w_1^{m_1}\Bigg)_{m_1=\fm} \,\, \prod_{1\leq \alpha \neq \beta \leq 2}\Bigg(1-\frac{u^{(\alpha)}_2}{u^{(\beta)}_2} \Bigg)  \PE\left[t \sum_{s=\pm 1} \sum_{\alpha=1,2} (b_1 u_1)^2(u^{(\alpha)}_{2})^{-s}  \right]  ~,
\end{split}$}
}
where the dimension of the monopole operators is now given by
\bes{
\begin{split}
\Delta(m_1, \vec m_2, m_3)
= & \frac{1}{2} \Biggl[
    \sum_{i=1}^2 \bigl( |m_3 - m_2^{(i)}|+|m_2^{(i)}-m_1|\bigr) 
    \\ & \quad - 2|m_2^{(1)} - m_2^{(2)}| + F_1 |m_1| + F_2 \sum_{i=1}^2 |m_2^{(i)}|+ F_3 |m_3|\Biggr]~.
\end{split}
}
In fact, the only magnetic flux that contributes to \eqref{Hm=nNSunequalranks} is $\fm=0$. The expression thus reduces to
\bes{
\scalebox{0.87}{$
\begin{split}
H_{\fm=\fn} &= \frac{1}{2}  \oint    \prod_{\alpha=1}^2 \frac{du^{(\alpha)}_2}{2\pi i u^{(\alpha)}_2} \prod_{1\leq \alpha \neq \beta \leq 2}\Bigg(1-\frac{u^{(\alpha)}_2}{u^{(\beta)}_2} \Bigg) \oint  \frac{du_1}{2\pi i u_1}  \PE\left[t \sum_{s=\pm 1} \sum_{\alpha=1,2} ( b_1 u_1)^s (u^{(\alpha)}_{2})^{-s}  \right] \\
&= \frac{1}{1-t^2}~.
\end{split}$}
}
For fluxes with $\fm\neq\fn$, the U(2) gauge group is broken to $\U(1)^2$. The corresponding contribution to the limit of the index is\footnote{Note that we have an overall factor of $1$ here, in contrast to that of \eqref{indexquivT3unequal}, which is $\frac{1}{2}$. This is because the results for the allowed choices $\vec{m}_2 = (\fm,\fn)$ and $(\fn, \fm)$ are equal.}
\bes{ \label{HmneqnNSunequalranks}
\scalebox{0.92}{$
\begin{split}
H_{\fm \neq\fn} =& \sum_{\fm \neq\fn \in \BZ}  t^{2 \Delta(m_1=\fm, \, \vec m_2=(\fm, \fn), \,  m_3 = 0)}\oint  \frac{du_1}{2\pi i u_1}   \Bigg((u_1)^{\kappa_i m_1}  w_1^{m_1}\Bigg)_{\vec m_1=\fm}\\&\times \oint  \Bigg(  \prod_{\alpha=1}^2 \frac{du^{(\alpha)}_2}{2\pi i u^{(\alpha)}_2}  (u^{(\alpha)}_2)^{\kappa_i m^{(\alpha)}_2}  w_2^{m^{(\alpha)}_2} \Bigg)_{\vec m_2=(\fm, \fn)}  \PE\left[t \sum_{s=\pm 1} ( b_1 u_1)^s (u^{(1)}_{2})^{-s}  \right]~.
\end{split}
$}
}
In this case, the magnetic fluxes that contribute non-trivially to \eqref{HmneqnNSunequalranks} are those with $\fm\neq \fn=0$. We have $\Delta(m_1 =\fm,\,  \vec m_2 =(\fm,0),\,  m_3=0) = (F_1+F_2) |\fm|$. Moreover, with this magnetic flux, we have
\bes{
&\prod_{s = \pm1} \prod_{\alpha=1}^2 Z_{\text{chir}}^{1/2} \left(x; a (b_1 u_1 (u^{(\alpha)}_{2})^{-1})^s; s m_{1} - s m^{(\alpha)}_{2} \right) \Big|_{m_1=\fm,\, \vec m_2 = (\fm,0), \, \eqref{N=3reparam}} \\
&\overset{\fz \rightarrow 0}{\longrightarrow} \,\,\,\, \fz^{|\fm|} \PE\left[t \sum_{s=\pm 1} ( b_1 u_1)^s (u^{(1)}_{2})^{-s}  \right]~.
}
This explains why the argument of the plethystic exponential in \eqref{HmneqnNSunequalranks} does not depend on $u_2^{(2)}$.
Using \eqref{usefulintiden}, we obtain
\bes{
H_{\fm \neq \fn} &= \sum_{\fm \neq 0}   \frac{t^{(F_1+F_2+k)|\fm|}}{(1-t^2)} (b_1^{-k} w_1 w_2)^{\fm} \\
&= \sum_{\fm \in \BZ}   \frac{t^{(F_1+F_2+k)|\fm|}}{(1-t^2)} (b_1^{-k} w_1 w_2)^{\fm}  - \frac{1}{1-t^2}~.
}
Summing the two contributions as indicated in \eqref{sumtwocontrNSuneq} yields the total Hilbert series:
\bes{\label{eq:27112}
\begin{split}
&H_{\eqref{quivT3axialunequalranks}}[\text{NS5}](t;w_{1,2}; b_{1,2}) \\
&\quad = \sum_{\fm\in\mathbb{Z}}\frac{t^{(F_1+F_2+k)(|\fm|)}}{(1-t^2)} (b_1^{-k} w_1 w_2)^{\fm} \\
&\quad = \PE\left[t^2+t^{F_1+F_2+k}(y+y^{-1})+t^{2(F_1+F_2+k)}  \right]~, \qquad  y=b_1^{-k}w_1 w_2~.
\end{split}
}
The final expression in \eqref{eq:27112} is the Hilbert series for the orbifold $\mathbb{C}^2/\mathbb{Z}_{F_1+F_2+k}$. This matches the Coulomb branch Hilbert series of a U(1) gauge theory with $F_1+F_2+k$ flavours, which can be derived using the method described in \cite{Marino:2025ihk}.

The elementary bare monopole operators on this branch, $V_{\pm(1,(1,0),0)}$, have an R-charge of $\frac{1}{2}(F_1+F_2)$ and axial charge of $F_1+F_2$. The generators of this branch are
\bes{
G_0 = X_{12} X_{21}~,\quad  G_+ = X_{21}^k V_{(1,(1,0),0)}~, \quad G_- = X_{12}^k V_{(-1,(-1,0),0)}~,
}
which satisfy the relation
\bes{
G_+ G_- = G_0^{F_1+F_2+k}~.
}
\subsubsection*{The $(1,k)$-branch}
Since the result for this branch can be obtained from that of the NS5-branch by exchanging the nodes $1$ and $3$, we report only the final result:
\bes{ \label{eq:27113}
\begin{split}
&H_{\eqref{quivT3axialunequalranks}}[(1,k)](t;\{w_{i}\};\{b_{i}\}) \\
&=\PE\left[t^2+t^{F_3+F_2+k}(y+y^{-1})+t^{2(F_3+F_2+k)}  \right]~, \qquad y=b_2^{k}w_2 w_3~.
\end{split}
}
This is the Hilbert series for $\mathbb{C}^2/\mathbb{Z}_{F_2+F_3+k}$, matching the Coulomb branch Hilbert series of a $\U(1)$ gauge theory with $F_2+F_3+k$ flavours.

The generators of this branch are analogous to those of the NS5-branch, with the replacements $V_{\pm(1,(1,0),0)} \to V_{\pm(0,(1,0),1)}$ and $(X_{12}, X_{21}) \to (X_{32}, X_{23})$.
\subsubsection*{The D5-branch}
Finally, the Hilbert series for the D5-branch is:
\bes{ \label{eq:27114}
\begin{split}
&H_{\eqref{quivT3axialunequalranks}}[\text{D5}](t;b_{1,2,3}) = \oint \frac{du_1}{2\pi i u_1} \left( \prod_{\alpha=1}^2 \frac{d u_2^{(\alpha)}}{2\pi i  u_2^{(i)}} \right)\,\, \prod_{1\leq \alpha \neq \beta \leq 2}\Bigg(1-\frac{u^{(\alpha)}_2}{u^{(\beta)}_2} \Bigg) \,\, \frac{d u_3}{2\pi i  u_3} \\
& \times \PE\left[t \sum_{s=\pm 1} \left\{ \sum_{i=1}^2 (b_1 u_1 (u_{2}^{(i)} )^{-1})^s+ \sum_{i=1}^2 (b_2 u_3 (u_{2}^{(i)} )^{-1})^s + \sum_{\rho_j=1}^{F_1} (f^{(\rho_j)}_1 u_1^{-1})^s\right.\right.\\ &\left.\left.\qquad \quad + \sum_{i=1}^2\sum_{\rho_j=1}^{F_2} (f^{(\rho_j)}_2 (u_2^{(i)})^{-1})^s + \sum_{\rho_j=1}^{F_3} (f^{(\rho_j)}_3 u_3^{-1})^s \right \} -2t^2-t^2 \sum_{\alpha,\beta=1}^2\frac{u^{(\alpha)}_2}{u^{(\beta)}_2} \right]~.
\end{split}
}
The terms with $-t^2$ in the argument of the plethystic exponential account for the adjoint chiral multiplets of the three gauge nodes, which possess an axial fugacity $a^{-2}$. This result corresponds to the Higgs branch Hilbert series of the theory in \eqref{quivT3axialunequalranks} with the Chern-Simons levels set to zero ($k=0$), which is the 3d $\CN=4$ theory
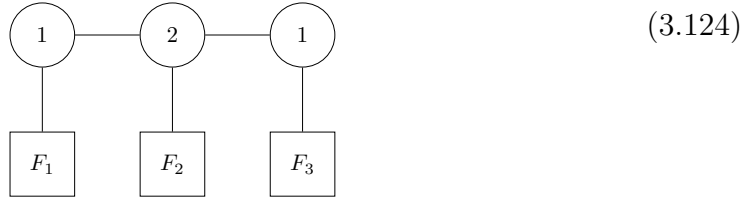
\begin{equation} \label{elecquivD51511}
\scalebox{0.85}{
\begin{tikzpicture}[baseline,font=\footnotesize,
    circ/.style={circle, draw, minimum size=1cm},
    sq/.style={rectangle, draw, minimum size=1cm},
    node distance=1cm
]
    \node[circ] (n1) {$1$};
    \node[circ, right=of n1] (n2) {$2$};
    \node[circ, right=of n2] (n3) {$1$};
    \node[sq, below=of n1] (s1) {$F_1$};
    \node[sq, below=of n2] (s2) {$F_2$};
    \node[sq, below=of n3] (s3) {$F_3$};
    \draw (n1) -- (n2);
    \draw (n2) -- (n3);
    \draw (n1) -- (s1);
    \draw (n2) -- (s2);
    \draw (n3) -- (s3);
\end{tikzpicture}}
\end{equation}
The magnetic quiver for this branch is simply the mirror theory for this quiver. For example, in the special case of $F_1=0$, $F_2=2$, and $F_3=0$, \eqref{elecquivD51511} is mirror dual to the $\U(2)$ gauge theory with four flavours.

In summary, the results for each branch are:
\bes{
\begin{tabular}{c|c|c}
\hline
Branch & Hilbert series & Geometry \\
\hline
NS5 & \eqref{eq:27112} & $\mathbb{C}^2/\BZ_{F_1+F_2+k}$ \\
$(1,k)$ & \eqref{eq:27113} & $\mathbb{C}^2/\BZ_{F_2+F_3+k}$ \\
D5 & \eqref{eq:27114} & Higgs branch of \eqref{elecquivD51511} \\
\hline
\end{tabular}
}

\subsection{A circular quiver} \label{sec:circularex}
Let us consider the following 3d $\CN=3$ theory:
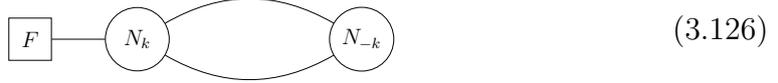
\begin{equation}
\scalebox{0.7}{ \label{quivN010}
\begin{tikzpicture}[baseline,
    gauge/.style={circle, draw, minimum size=1.2cm},
    flavor/.style={rectangle, draw, minimum size=0.8cm, label=center:$F$}
]

    \node[gauge] (g1) {$N_{k}$};
    \node[gauge, right=3cm of g1] (g2) {$N_{-k}$};

    \node[flavor, left=1cm of g1] (f) {};


    \draw (g1) to[bend left=30] (g2);
    \draw (g1) to[bend right=30] (g2);

    \draw (f) -- (g1);

\end{tikzpicture}}
\end{equation}
where, for $k=F=1$, this is the worldvolume theory on $N$ M2-branes probing the cone over $N^{0,1,0}$ \cite{Gaiotto:2009tk}. 

This theory can be realised on a system of a D3-brane on a circle, along with an NS5-brane, a $(1,k)$-brane, and $F$ D5-branes in one of the $\text{NS5}$--$(1,k)$ intervals. The index is given by
\bes{ \label{indexN010}
&\CI_{\eqref{quivN010}} = \frac{1}{(N!)^2} \sum_{\{m^{(\alpha)}_1\} \in \BZ^N} \sum_{\{m^{(\alpha)}_2\} \in \BZ^N}  \oint \Bigg( \prod_{i=1}^2 \prod_{\alpha=1}^N \frac{du^{(\alpha)}_i}{2\pi i u^{(\alpha)}_i}  (u^{(\alpha)}_i)^{\kappa_i m^{(\alpha)}_i}  w_i^{m^{(\alpha)}_i} \Bigg) \\
&\quad \times \prod_{i=1}^2  Z_{\text{vec}}^{\U(N)}(x; \{u^{(\alpha)}_i\}; \{m^{(\alpha)}_i\}) \prod_{\alpha, \beta=1}^N Z_{\text{chir}}^{1} \left(x; \fb_i u^{\alpha}_i (u^{\beta}_i)^{-1} ; m^{(\alpha)}_i - m^{(\beta)}_i  \right) \\ 
& \quad \times \prod_{s = \pm1} \prod_{i=1}^2 \prod_{\alpha, \beta=1}^N Z_{\text{chir}}^{1/2} \left(x; \fa_i (b u^{(\alpha)}_i (u^{(\beta)}_{i+1})^{-1})^s; s m^{(\alpha)}_{i} -s m^{(\beta)}_{i+1} \right) \\ 
& \quad \times \prod_{s = \pm1} \prod_{\rho=1}^{F} \prod_{\alpha=1}^N  Z_{\text{chir}}^{1/2} \left(x; \fc (f_\rho^{-1} u^{(\alpha)}_j)^s; sm^{(\alpha)}_{j} \right)~, \quad \kappa_1=-\kappa_2 = k >0~,
}
where, for each branch, $\fa_{1,2}$, $\fb_{1,2}$ and $\fc$ are as follows:
\bes{ \label{branchquivN010}
\begin{tabular}{c|ccccc}
\hline
Branch & $\fa_1$ & $\fa_2$ & $\fb_1$ & $\fb_2$  & $\fc$  \\
\hline
NS5& $a$ & $a^{-1}$ & $1$ & $1$ & $a^{-1}$  \\
$(1,k)$ & $a^{-1}$ & $a$ & $1$ & $1$ & $a^{-1}$    \\
D5 & $a$ & $a$ & $a^{-2}$ & $a^{-2}$  & $a$  \\
\hline \hline
\text{Geometric ($N=1$)} & $a$ & $a$ & $1$ & $1$ & $a^{-1}$ \\
\hline
\end{tabular}
}
We consider the limit
\bes{ \label{limquivN010}
H_{\eqref{quivN010}}[\text{branch}]= \lim_{\fz \rightarrow 0} \CI_{\eqref{quivN010}} \Big|_{\eqref{branchquivN010}, \, \eqref{N=3reparam}}~.
}
The limit of the index for the geometric branch will be discussed in Section \ref{sec:geometricbranch}. 
Using the symmetry of the theory, we see that the limits for the NS5-branch and $(1,k)$-branch are equal upon exchanging $b_1$ and $b_2$.

We will show that the NS5-branch and $(1,k)$-branch are given by $\Sym^N(\BC^2/\BZ_{k+F})$, and that the corresponding magnetic quiver is the 3d $\CN=4$ $\U(N)$ gauge theory with one adjoint hypermultiplet and $k+F$ fundamental hypermultiplets, whose Coulomb branch is the moduli space of $N$ $\U(1)$ instantons on $\BC^2/\BZ_{k+F}$.

\subsubsection{The Abelian case: $N=1$}
\paragraph{The NS5-branch.} From \eqref{limquivN010}, the power of $\fz$ for the NS5-branch limit is $|m_1-m_2|$. In order to obtain a finite non-vanishing result in the limit $\fz \rightarrow 0$, we require this to be zero, namely we have $m_1 =m_2 =\fm \in \BZ$. The dimension of the monopole operators is $\Delta(m_1, m_2) = \frac{1}{2}\left(|m_1-m_2| +F |m_1|\right)$. The limit of the index yields the Hilbert series
\bes{ \label{N=1N010}
&H_{\eqref{quivN010}}[\text{NS5}] = \\ &\sum_{\fm\in \BZ} t^{2\Delta(\fm, \fm)} \oint \frac{d u_1}{2\pi i u_1} \frac{d u_2}{2\pi i u_2} u_1^{k \fm} u_2^{-k \fm} (w_1 w_2)^\fm \PE\left[ \sum_{s = \pm 1} (b u_1 u_2^{-1})^s \right]\\
&\overset{\eqref{usefulintiden}}{=} \sum_{\fm \in \BZ} \frac{t^{k |\fm|}}{1-t^2} (b^{-k} w_1 w_2)^\fm \times t^{F|m|} = \PE \left[ t^2 +(y+y^{-1}) t^{k+F} -t^{2(k+F)}\right]~,
}
where $y= b_1^{-k} w_1 w_2$. This is the Hilbert series of $\BC^2/\BZ_{k+F}$. The result for the $(1,k)$-branch is the same.

\paragraph{The D5-branch.} For the D5-branch limit, the power of $\fz$ is $2|m_1-m_2|+F|m_1|$. Setting this to zero yields $m_1=m_2=0$. The limit of the index in this case is given by
\bes{
\scalebox{0.99}{$
\begin{split}
&H_{\eqref{quivN010}}[\text{D5}] = \\
&\oint \frac{d u_1}{2\pi i u_1} \frac{d u_2}{2\pi i u_2} \PE \left[- 2t^2 +  (b+b^{-1}) (u_1 u_2^{-1}+ u_2 u_1^{-1}) + \sum_{s=\pm 1} \sum_{\rho=1}^F (f_\rho u_1^{-1})^s \right]~.
\end{split}
$}
}
This is the Higgs branch Hilbert series of the 3d $\CN=4$ quiver \eqref{quivN010} with $k=0$. It can be identified with the moduli space of one $\SU(F)$ instanton on $\BC^2/\BZ_2$. As pointed out in \cite{Dey:2013fea, Mekareeya:2015bla}, this is isomorphic to $\BC^2/\BZ_2$ times the reduced moduli space of one $\SU(F)$ instanton on $\BC^2$, whose Hilbert series is 
\bes{
H_{\eqref{quivN010}}[\text{D5}] = \PE\left[(b^2+1+b^{-2})t^2 -t ^4\right] \times \sum_{n=0}^\infty [n,0,\ldots,0,n] t^{2n}~.
}

\paragraph{The geometric branch.} Using the result in Section \ref{sec:geometricbranch}, in particular \eqref{hsgeom}, the Hilbert series of the geometric branch is given by
\bes{ \label{geomN010N=1}
&H_{\eqref{quivN010}}[\text{geom}] \\
&=\sum_{m \in \BZ} t^{F|m|}  \oint \frac{d u_1}{2\pi i u_1} \frac{d u_2}{2\pi i u_2} w^m u_1^{k m} u_2^{-k m} \PE \left[ (b+b^{-1}) (u_1 u_2^{-1} +u_2^{-1} u_1) \right] \\
&=\sum_{m \in \BZ} t^{F|m|} \times \frac{w^m}{1-t^2} \sum_{p=0}^\infty [k |m| +2p]_{b} \, t^{k |m|+2p} \\
&= \frac{1}{1-t^2} \sum_{p=0}^\infty \sum_{m \in \BZ} [k|m|+2p]_b \, w^m \, t^{(F+k)|m|+2p}~,
}
where $[x]_b$ is the character of the $\su(2)$ representation with highest weight $x$ written in terms of $b$. This is in agreement with the unrefined Hilbert series \cite[(6.19)]{Cremonesi:2016nbo}. Moreover, for $k=F=1$, the manifest symmetry $\su(2) \oplus \u(1)$ of the theory gets enhanced to $\su(3)$, corresponding to the symmetry of the field theory dual to the $\mathrm{AdS}_4 \times N^{0,1,0}$ theory \cite{Billo:2000zr}.\footnote{If the character of the fundamental representation of $\su(3)$ is denoted by $x_1+ x_2 x_1^{-1} + x_2^{-1}$, then a fugacity map is $b = x_1 x_2^{-1/2}, \, w = x_2^{3/2}$. \label{foot:fugmapsu3su2u1}} In particular, the above expression can be rewritten as \cite[(6.21)]{Cremonesi:2016nbo}
\bes{ \label{HSminorbsu3}
H_{\eqref{quivN010}_{k=F=1}}[\text{geom}] = \sum_{n=0}^\infty [n,n]_{b,w} t^{2n}~, 
}
namely the Hilbert series of the reduced moduli space of one $\SU(3)$ instanton on $\BC^2$ \cite{Benvenuti:2010pq}, where $[n,n]_{b,w}$ is the character of the $\su(3)$ representation with highest weight $[n,n]$ in terms of $b$ and $w$.

\subsubsection{The case of $N=2$}
As we pointed out above, the results for the NS5-branch and the $(1,k)$-branch are equal. We will analyse the former for definiteness. From \eqref{limquivN010}, the power of $\fz$ for the NS5-branch is 
\bes{
\sum_{\alpha,\beta=1}^2 |m_1^{(\alpha)}-m_2^{(\beta)}| - |m^{(1)}_1-m^{(2)}_1|- |m^{(2)}_1-m^{(2)}_2|~.
}
Setting this to zero, we focus on the following solution:
\bes{
\vec m_1 = \vec m_2 = (\fm, \fn)~, \qquad \fm \geq \fn~,
}
where we use the Weyl symmetry of $\U(2)$ to arrange $\fm \geq \fn$. The dimension of the monopole operators is
\bes{ \label{dimmonN010N2}
&\Delta(\vec m_1, \vec m_2) \\
&= \frac{1}{2}\Big(  F \sum_{\alpha=1}^2|m^{(\alpha)}_1|+\sum_{\alpha,\beta=1}^2 |m^{(\alpha)}_1-m^{(\beta)}_2| - 2\sum_{i=1}^2 |m^{(1)}_i-m^{(2)}_i| \Big)~.
}
We proceed in the same way as in the previous section, namely the limit of the index receives the contributions from the cases $\fm=\fn$ and $\fm>\fn$:
\bes{ 
H_{\eqref{quivN010}}[\text{NS5}] = H_{\fm=\fn} + H_{\fm>\fn}~.
}
For fluxes with $\fm=\fn$, the limit of the index is
\bes{ 
\scalebox{0.93}{$
\begin{split}
&H_{\fm=\fn} = \frac{1}{2^2}  \sum_{\fm \in \BZ} \oint  \prod_{i=1}^2 \prod_{\alpha=1}^2 \frac{du^{(\alpha)}_i}{2\pi i u^{(\alpha)}_i} \left( \frac{u_1^{(1)}u_1^{(2)}}{u_2^{(1)}u_2^{(2)}} \right)^{k\fm} (w_1 w_2)^{2\fm} \\
& \quad \times \prod_{1\leq \alpha \neq \beta \leq 2} \prod_{i=1}^2 \Bigg(1-\frac{u^{(\alpha)}_i}{u^{(\beta)}_i} \Bigg)  \PE\left[t \sum_{s=\pm 1}\sum_{\alpha, \beta=1,2} ( b u^{(\alpha)}_1)^s (u^{(\beta)}_{2})^{-s}  \right]  t^{2 \Delta(\vec m_1 = \vec m_2=(\fm, \fm))}~,
\end{split}$}
}
where $\kappa_1 = -\kappa_2 = k >0$. This can be simplified using \eqref{U2identity} as follows:
\bes{
H_{\fm=\fn} = \sum\limits_{\fm \in \BZ} \frac{t^{k(2|\fm|)} (b_1^{-k} w_1 w_2)^{2\fm}}{(1-t^2)(1-t^4)} t^{2F|\fm|} 
= \sum\limits_{\fm \in \BZ} \frac{t^{(2F+2k)|\fm|}}{(1-t^2)(1-t^4)} (b^{-k} w_1 w_2)^{2\fm}~.
}
For fluxes with $\fm>\fn$, the limit of the index is instead
\bes{ \label{Hm>nNSN010}
H_{\fm >\fn} &= \sum_{\fm >\fn \in \BZ} \, \CH(\fm)\,  \CH(\fn) \,\, t^{2\Delta(\vec m_1 = \vec m_2 = (\fm, \fn))} \\
&= \frac{t^{k(|\fm|+|\fn|)}}{(1-t^2)^2} (b^{-k} w_1 w_2)^{\fm+\fn} \times t^{F(|\fm|+|\fn|)} = \frac{t^{(F+k)(|\fm|+|\fn|)}}{(1-t^2)^2} (b^{-k} w_1 w_2)^{\fm+\fn} ~,
}
where $\CH(m)$ is the integral part of the $N=1$ case, given in \eqref{N=1N010}:
\bes{
\CH(m) &= \frac{t^{k |m|}}{1-t^2} (b^{-k} w_1 w_2)^m~. 
}
Summing the two contributions, we obtain
\bes{
H_{\eqref{quivN010}}[\text{NS5}] &= \sum_{\fm, \fn \in \BZ} t^{(F+k)(|\fm|+|\fn|)} P_{\U(2)}(t; \fm, \fn)(b^{-k} w_1 w_2)^{\fm+\fn} \\
&=\frac{1}{2} \Big\{ \PE \left[ t^2 +(y+y^{-1}) t^{k+F} -t^{2(k+F)}\right]^2 \\
&\quad \,\,\,\, + \PE \left[ t^4 +(y^2+y^{-2}) t^{2(k+F)} -t^{4(k+F)}\right] \Big\}~,
}
where $y=b^{-k} w_1 w_2$. This is indeed the Coulomb branch Hilbert series of the 3d $\CN=4$ $\U(2)$ gauge theory with one adjoint hypermultiplet and $k+F$ fundamental hypermultiplets, which is isomorphic to $\Sym^2(\BC^2/\BZ_{k+F})$, as expected.

\subsection{A non-linear Abelian quiver} \label{sec:starex}
Let us consider the following 3d $\mathcal{N}=3$ non-linear unitary quiver:
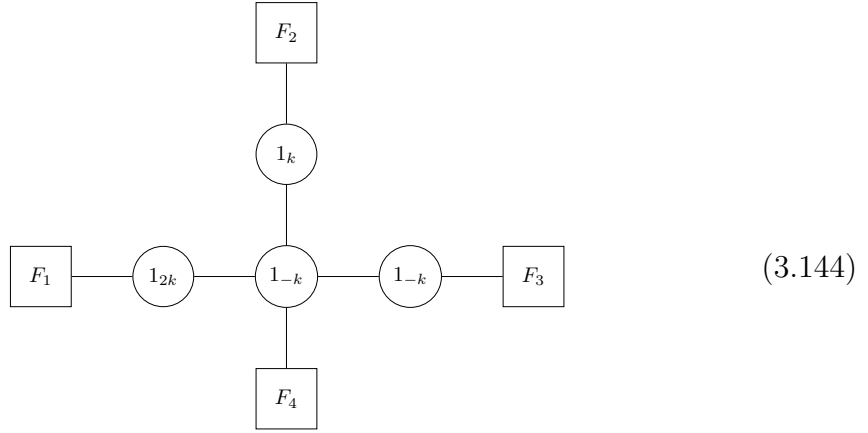
\begin{equation} \label{starshaped}
\scalebox{0.8}{
\begin{tikzpicture}[baseline,font=\footnotesize,
    circ/.style={circle, draw, minimum size=1cm},
    sq/.style={rectangle, draw, minimum size=1cm},
    node distance=1cm
]
    \node[circ] (n1) {$1_{2k}$};
    \node[circ, right=of n1] (n2) {$1_{-k}$};
    \node[circ, above=of n2] (n3) {$1_{k}$};
    \node[circ, right=of n2] (n4) {$1_{-k}$};
    \node[sq, left=of n1] (s1) {$F_1$};
    \node[sq, above=of n3] (s2) {$F_2$};
    \node[sq, right=of n4] (s3) {$F_3$};
    \node[sq,below=of n2] (s4) {$F_4$};
    \draw (n1) -- (n2);
    \draw (n2) -- (n3);
    \draw (n2) -- (n4);
    \draw (n1) -- (s1);
    \draw (n2) -- (s4);
    \draw (n3) -- (s2);
    \draw (n4) -- (s3);
\end{tikzpicture}}
\end{equation}
Note that this theory does not admit a usual brane configuration as in the other sections in this paper. Henceforth, we shall label the branches of the moduli space without referring to the types of branes. 

To compute the various limits of the index, we first assign axial fugacities to the chiral fields. In 3d $\CN=2$ notation, the field content and fugacity assignments are
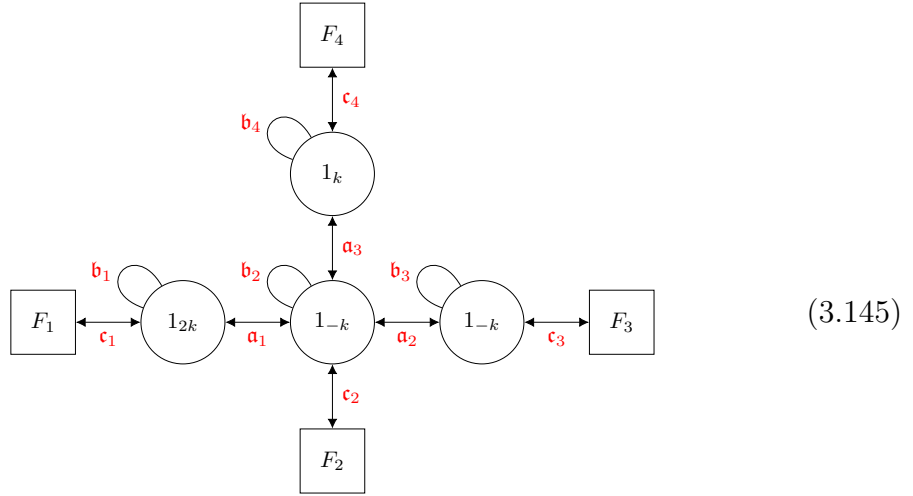
\begin{equation} \label{starshapedaxial}
\scalebox{0.85}{
\begin{tikzpicture}[baseline, font=\footnotesize,
    circ/.style={circle, draw, minimum size=1.3cm},
    sq/.style={rectangle, draw, minimum size=1cm},
    node distance=1cm,
    every loop/.style={-}
]
    \node[circ] (n1) {$1_{2k}$};
    \node[circ, right=of n1] (n2) {$1_{-k}$};
    \node[circ, right=of n2] (n3) {$1_{-k}$};
    \node[circ, above=of n2] (n4) {$1_{k}$};
    \node[sq, left=of n1] (s1) {$F_1$};
    \node[sq, below=of n2] (s2) {$F_2$};
    \node[sq, right=of n3] (s3) {$F_3$};
    \node[sq, above=of n4] (s4) {$F_4$};
    \draw[<->] (n1) -- node[below, red] {$\fa_1$} (n2);
    \draw[<->] (n2) -- node[below, red] {$\fa_2$} (n3);
    \draw[<->] (n2) -- node[right, red] {$\fa_3$} (n4);
    \draw[<->] (n1) -- node[below, red] {$\fc_1$} (s1);
    \draw[<->] (n2) -- node[right, red] {$\fc_2$} (s2);
    \draw[<->] (n3) -- node[below, red] {$\fc_3$} (s3);
    \draw[<->] (n4) -- node[right, red] {$\fc_4$} (s4);
    \path (n1) edge [out=120, in=160, looseness=5] node[red, left] {$\fb_1$} (n1);
    \path (n2) edge [out=120, in=160, looseness=5] node[red, left] {$\fb_2$} (n2);
    \path (n3) edge [out=120, in=160, looseness=5] node[red, left] {$\fb_3$} (n3);
    \path (n4) edge [out=120, in=160, looseness=5] node[red, left] {$\fb_4$} (n4);
\end{tikzpicture}}
\end{equation}
For convenience, we refer to the gauge nodes attached to $F_1$, $F_2$, $F_3$ and $F_4$ as nodes $1$, $2$, $3$ and $4$, respectively. The gauge fugacity and the corresponding magnetic flux for each of them will be denoted by $u_j$ and $m_j$ (with $j=1,2,3,4$), respectively.
Following the prescription of Section \ref{sec:N=3presc}, we can identify operators parametrising different branches of the theory. Applying the prescription for this theory, we find two maximal branches parametrised by VEVs of dressed monopole operators. Branch I involves the dressed monopole operators with minimal fluxes turned on along the horizontal line consisting of gauge nodes $1$, $2$ and $3$, whose Chern-Simons levels sum to zero. Branch II involves to the dressed monopole operators with minimal fluxes turned on along the vertical line consisting of gauge nodes $2$ and $4$, whose Chern-Simons levels also sum to zero. Then, Branch III is the mesonic branch.
The specific assignments of axial fugacities $\fa_i$, $\fb_i$, and $\fc_i$ for each branch are summarised in the following table:
\bes{ \label{branchT3abel}
\begin{tabular}{c|ccccccccccccc}
\hline
Branch & $\fa_1$ & $\fa_2$ & $\fa_3$  & $\fc_1$ & $\fc_2$ & $\fc_3$ & $\fc_4$ & $\fb_1$ & $\fb_2$ & $\fb_3$ & $\fb_4$  \\
\hline
\text{I} & $a$ & $a$ & $a^{-1}$ & $a^{-1}$ & $a^{-1}$ & $a^{-1}$ & 1 &1 &$a^{-2}$&1&1  \\
\text{II} & $a^{-1}$ & $a^{-1}$ & $a$ & 1 & $a^{-1}$ & 1 & $a^{-1}$ & 1&1&1&1 \\
\text{III} & $a$ & $a$ & $a$ & $a$ & $a$ & $a$ & $a$ & $a^{-2}$&$a^{-2}$&$a^{-2}$&$a^{-2}$  \\
\hline
\end{tabular}
}
The index for the theory in \eqref{starshapedaxial} is then given by
\bes{ \label{indexstarshaped}
\scalebox{0.89}{$
\begin{split}
&\CI_{\eqref{starshapedaxial}}  = \sum_{m_1,m_2, m_3,m_4 \in \BZ} \oint  \Bigg( \prod_{i=1}^4 \frac{du_i}{2\pi i u_i}  u_i^{\kappa_i m_i}  w_i^{m_i} Z_{\text{chir}}^{1} \left(x; \fb_i; 0 \right) \Bigg) \\ 
& \quad \times \prod_{s = \pm1} \prod_{i=1}^2 Z_{\text{chir}}^{1/2} \left(x; \fa_i (b_i u_i u_{i+1}^{-1})^s; s m_{i} -s m_{i+1} \right)  \\
&\quad \times \prod_{s=\pm 1} \prod_{j =1}^4 Z_{\text{chir}}^{1/2} \left(x; \fa_3 (b_3 u_2 u_{4}^{-1})^s; s(m_{2} - m_{4}) \right) \prod_{\rho_j=1}^{F_j}\left[Z_{\text{chir}}^{1/2} \left(x; \fc_j (f_j^{(\rho_j)})^{-s} u_j^s; sm_{j} \right)\right]~,
\end{split}$}
}
where the Chern-Simons levels are
\bes{ \label{kappaT3}
\kappa_1 = 2k~, \qquad \kappa_2 = -k~, \qquad \kappa_3 = -k~, \qquad \kappa_4 = k~.
}
We then consider the limit 
\bes{H_{\eqref{starshaped}}[\text{branch}]= \lim_{\fz \rightarrow 0} \CI_{\eqref{starshaped}} \Big|_{\eqref{branchT3abel}, \, \eqref{N=3reparam}}~.}
A non-vanishing result in the $\fz \to 0$ limit imposes the condition that the exponent of $\fz$ must be zero, so we have
\bes{ \label{condfluxstarshaped}
\scalebox{0.95}{
\begin{tabular}{c|l|l}
\hline
Branch & Exponent of $\fz$ & Condition \\
\hline
\text{I} & ${|m_1-m_2|+|m_2-m_3|+\frac{F_4}{2}|m_4|}$ & $m_1=m_2=m_3=m$~, $m_4=0$ \\
\text{II} & ${|m_2-m_4|+\frac{F_1}{2}|m_1|+\frac{F_3}{2}|m_3|}$ & $m_2=m_4$~, $m_1=m_3=0$ \\
\text{III} & $\sum_{i=1}^4F_i|m_i|+|m_1-m_2|$ & $m_1=m_2=m_3=m_4=0$ \\
& $+|m_2-m_3|+|m_2-m_4|$ & \\
\hline
\end{tabular}}
}
We now discuss the computation for each branch explicitly.
\subsubsection*{Branch I}
The limit of the index for Branch I is
\bes{ \label{Branch1starshaped}
\scalebox{0.9}{$
\begin{split}
&H_{\eqref{starshapedaxial}}[\text{I}](t;\{w_{i}\};\{b_{i}\}) = \sum_{m_1=m_2=m_3 = m\in \BZ}  t^{2 \Delta(m_1 =m_2 =m_3=m,\,m_4=0)}  \\
& \qquad \times \oint \prod_{j=1}^3 \frac{d u_j}{2\pi i u_j} u_j^{\kappa_j m_j}   w_j^{m_j}  \PE\left[t \sum_{s=\pm 1} (b_1 u_1 u_2^{-1})^s  \right] \PE\left[-t^2+t \sum_{s=\pm 1} (b_2 u_2 u_3^{-1})^s  \right] \\
&\qquad = \sum_{m\in \BZ} t^{(F_1+F_2+F_3 +1) |m|} \times \frac{t^{3k|m|}}{1-t^2} y_1^m~, \qquad y_1= b_1^{-2k}b_2^{-k} w_1 w_2 w_3 \\
&\qquad= \PE\left[ t^2 + (y_1+ y_1^{-1})t^{F_1+F_2+F_3 +1+3k}- t^{2(F_1+F_2+F_3 +1+3k)}\right]~,
\end{split}$}
}
where the result in the second equality can be obtained similarly to  \eqref{usefulintiden}.
Note that the terms inside the plethystic exponential with coefficient $t$ are precisely the contributions of chiral fields with axial fugacity $a$, whereas the term $-t^2$ comes from the adjoint chiral field with axial fugacity $a^{-2}$. The dimension of the monopole operators is given by
\bes{
&\Delta(m_1, m_2, m_3,m_4) \\
&= \frac{1}{2} \left(\sum_{i=1}^4 \left(F_i|m_i|\right)+|m_1-m_2|+|m_2-m_3|+|m_2-m_4|\right)~,
}
where, in this case, $\Delta(m_1 =m_2 =m_3=m,\,m_4=0) = (F_1+F_2+F_3+1)|m|$. The final expression in \eqref{Branch1starshaped} is the Hilbert series of $\mathbb{C}^2/\mathbb{Z}_{F_1+F_2+F_3 +1+3k}$, with generators 
\bes{
\scalebox{0.9}{$
\begin{split}
G_0 = X_{12}X_{21} = X_{23}X_{32}~, \quad G_+ = V_{(1,1,1,0)} X_{21}^{2k} X^{k}_{32}~, \quad G_- = V_{(-1,-1,-1,0)} X_{12}^{2k} X^{k}_{23}~,    
\end{split}$}
}
satisfying $G_+G_-= G_0^{2(F_1+F_2+F_3+1+3k)}$. This is, in fact, the Coulomb branch Hilbert series of the $\U(1)$ gauge theory with $F_1+F_2+F_3 +1+3k$ flavours, which can be regarded as a magnetic quiver for this branch.
\subsubsection*{Branch II}
The limit of the index for Branch II is
\bes{ \label{Branch2starshaped}
\begin{split}
&H_{\eqref{starshapedaxial}}[\text{II}](t;\{w_{i}\},\{b_{i}\}) = \sum_{m_2=m_4= m\in \BZ}  t^{2 \Delta(m_1 = 0,\, m_2 =m, \,m_3=0,\,m_4=m)}  \\
& \qquad \times \oint \frac{d u_2}{2\pi i  u_2}\oint \frac{d u_4}{2\pi i u_4} u_2^{\kappa_2 m_2}u_4^{\kappa_4 m_4}  w_2^{m_2} w_4^{m_4}  \PE\left[t \sum_{s=\pm 1} (b_3 u_2 u_4^{-1})^s  \right]\\
&\qquad \overset{\eqref{usefulintiden}}{=} \sum_{m\in \BZ} t^{(F_2+F_4 +2+k) |m|} \frac{1}{1-t^2} y_2^m~, \qquad y_2=b_3^{k} w_2 w_4\\
&\qquad= \PE\left[ t^2 + (y_2+ y_2^{-1})t^{F_2+F_4 +2+k}- t^{2(F_2+F_4 +2+k)}\right]~.
\end{split}
}
The final expression in \eqref{Branch2starshaped} is the Hilbert series of $\mathbb{C}^2/\mathbb{Z}_{F_2+F_4 +2+k}$, with generators
\bes{
G_0 = X_{24}X_{42}~,\quad G_+ = V_{(0,1,0,1)} X_{24}^k~,\quad G_- = V_{(0,-1,0,-1)} X_{42}^k ~,
}
satisfying $G_+ G_- = G_0^{2(F_2+F_4 +2+k)}$.
This is the Coulomb branch Hilbert series of a U(1) gauge theory with $F_2+F_4 +2+k$ flavours, which can be regarded as a magnetic quiver for this branch.
\subsubsection*{Branch III}
The limit of the index for Branch III is
\bes{ \label{Branch3starshaped}
\begin{split}
&H_{\eqref{starshapedaxial}}[\text{III}](t;\{w_{i}\},\{b_{i}\}) =   \oint \prod_{j=1}^4 \frac{d u_j}{2\pi i  u_j}\,\,  \PE\Big[-4t^2 + t \sum_{s=\pm 1} \Big\{(b_1 u_1 u_2^{-1})^s \\
&\qquad \qquad \qquad +(b_2 u_2 u_3^{-1})^s+(b_3 u_2 u_4^{-1})^s  + \sum_{j=1}^4 \sum_{\rho_j=1}^{F_j} (f^{(\rho_j)}_j u_j^{-1})^s  \Big\} \Bigg] ~.
\end{split}
}
The term $-4t^2$ in the plethystic exponential accounts for the four adjoint chiral multiplets, which possess axial fugacity $a^{-2}$. This result corresponds to the Higgs branch Hilbert series of the theory in \eqref{starshaped} with the Chern-Simons levels set to zero ($k=0$), which is the 3d $\CN=4$ theory
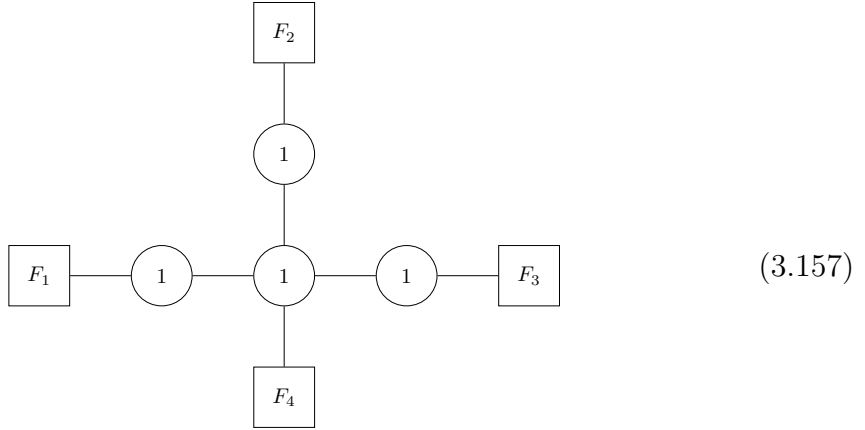
\begin{equation} 
\label{starelectquiver}
\scalebox{0.8}{
\begin{tikzpicture}[baseline, font=\footnotesize,
    circ/.style={circle, draw, minimum size=1cm},
    sq/.style={rectangle, draw, minimum size=1cm},
    node distance=1cm
]
    \node[circ] (n1) {$1$};
    \node[circ, right=of n1] (n2) {$1$};
    \node[circ, above=of n2] (n3) {$1$};
    \node[circ, right=of n2] (n4) {$1$};
    \node[sq, left=of n1] (s1) {$F_1$};
    \node[sq, above=of n3] (s2) {$F_2$};
    \node[sq, right=of n4] (s3) {$F_3$};
    \node[sq,below=of n2] (s4) {$F_4$};
    \draw (n1) -- (n2);
    \draw (n2) -- (n3);
    \draw (n2) -- (n4);
    \draw (n1) -- (s1);
    \draw (n2) -- (s4);
    \draw (n3) -- (s2);
    \draw (n4) -- (s3);
\end{tikzpicture}}
\end{equation}
It is worth discussing the mirror dual of \eqref{starelectquiver}, which serves as a magnetic quiver of this branch. Since this theory is Abelian, we can adopt the method discussed in \cite[Section 4]{deBoer:1996ck} (see also \cite{Kapustin:1999ha}, \cite[around Eq. (8)]{Tong:2000ky} and \cite[Section 6]{Cremonesi:2013lqa}), which involves the charge matrix of quiver \eqref{starelectquiver}. For simplicity, let us focus on the case of $F_1=F_2=F_3=F_4=1$. The charge matrix $\mathcal{R}$ for \eqref{starelectquiver} is then
\begin{equation}
\mathcal{R} =
\begin{pmatrix}
 1 & 0 &  0 &  0 & 1 & 0& 0\\
 0 & 1 &  0 &  0 & -1 & 1 & 1\\
 0 & 0 &  1 &  0 & 0& -1& 0\\
 0 & 0 &  0 &  1 & 0& 0& -1
\end{pmatrix} ~,
\end{equation}
where the rows correspond to each gauge group and the columns correspond to each hypermultiplet in the quiver. In particular, the entry $\CR_{ia}$ is the charge of the $a$-th hypermultiplet under the $i$-th gauge group. The charge matrix $\mathcal{S}$ for the mirror dual is given by solving the set of equations given by
\begin{equation}
    \mathcal{R}\mathcal{S}^T=0 ~.
\end{equation}
In this case, we obtain
\begin{equation}
\mathcal{S} =
\left(
\begin{array}{ccccccc}
 0 & -1 & 0 & 1 & 0 & 0 & 1 \\
 0 & -1 & 1 & 0 & 0 & 1 & 0 \\
 1 & -1 & 0 & 0 & -1 & 0 & 0 \\
\end{array}
\right) ~,
\end{equation}
which indicates that the mirror theory has three $\U(1)$ gauge groups. The matter content consists of two charge-one hypermultiplets under each gauge group, along with a (trifundamental) hypermultiplet with charge $(-1,-1,-1)$ in all gauge groups. We denote the latter by a junction in the following mirror quiver:
\begin{equation} 
\label{mirrorstarelectquiver}
\scalebox{0.8}{
\begin{tikzpicture}[baseline, font=\footnotesize,
    circ/.style={circle, draw, minimum size=1cm},
    sq/.style={rectangle, draw, minimum size=1cm},
    node distance=1cm
]
    \coordinate (n2) at (0,0);
    \node[circ, left=of n2] (n1) {$1$};
    \node[circ, above=of n2] (n3) {$1$};
    \node[circ, right=of n2] (n4) {$1$};
    \node[sq, left=of n1] (s1) {$2$};
    \node[sq, above=of n3] (s2) {$2$};
    \node[sq, right=of n4] (s3) {$2$};
    \draw (n1) -- (n2);
    \draw (n2) -- (n3);
    \draw (n2) -- (n4);
    \draw (n1) -- (s1);
    \draw (n3) -- (s2);
    \draw (n4) -- (s3);
\end{tikzpicture}}
\end{equation}
In summary, the results for each branch are as follows:
\bes{
\begin{tabular}{c|c|c}
\hline
Branch & Hilbert series & Geometry \\
\hline
\text{I} & \eqref{Branch1starshaped} & $\mathbb{C}^2/\mathbb{Z}_{F_1+F_2+F_3 +1+3k}$ \\
\text{II} & \eqref{Branch2starshaped} & $\mathbb{C}^2/\mathbb{Z}_{F_2+F_4 +2+k}$ \\
\text{III} & \eqref{Branch3starshaped} & Higgs branch of \eqref{starelectquiver} \\
\hline
\end{tabular}
}

\section{Orthosymplectic quivers} \label{sec:orthosimpex}
The main focus of this section is to generalise our prescription also in the case of three-dimensional quiver gauge theories with (special) orthogonal and symplectic gauge groups. One of the primary interesting aspects of this class of theories lies in the rich global symmetry structure associated with the $\so(N)$ gauge algebra, for which different variants of the gauge group can be considered, \eg $\SO(N)$, $\Spin(N)$, $\O(N)^\pm$ and $\Pin(N)$. These variants can be related to each other through the gauging of $\BZ_2$ symmetries \cite{Cordova:2017vab}. Explicitly, the $\SO(N)$ gauge theory possesses a $\BZ_2$ magnetic (or topological) symmetry and a $\BZ_2$ charge conjugation symmetry. Gauging the former changes the global form to $\Spin(N)$, whereas gauging the latter turns the gauge group into $\O(N)^+$.\footnote{When $N = 2 n$ is even, the charge conjugation symmetry can be regarded as the outer automorphism of the $D_n$ Dynkin diagram.} The $\O(N)^-$ variant can be reached upon gauging the diagonal combination of the said $\BZ_2$ symmetries,\footnote{Note that, in the $N=2$ case, there is no distinction between $\O(2)^+$ and $\O(2)^-$. Throughout this paper, we refer to it simply as $\O(2)$.} and, finally, the $\Pin(N)$ gauge theory arises from the sequential gauging of both the magnetic and charge conjugation symmetries.

The various global forms described above generally give rise to different physics, as the operator content of theories whose gauge group consists of distinct variants of the $\so(N)$ algebra is usually not the same. For instance, the minimal monopole operator of $\SO(N)$ with fluxes $\left(1, 0 , \ldots, 0\right)$ is even under charge conjugation and is odd under the magnetic symmetry \cite{Aharony:2013kma}.\footnote{In the $N=2$ case, neither minimal monopole operator with one unit of positive and negative $\SO(2)$ flux possesses a definite charge conjugation parity. Monopole operators which are even ({\it resp.} odd) under charge conjugation can be built from the positive ({\it resp.}) negative linear combination between the said minimal monopole operators, see \eg \cite[(2.13) and (2.14)]{Aharony:2013kma}.} As a consequence, it survives as the minimal monopole operator of the $\O(N)^+$ theory, whereas it gets projected out in both the $\Spin(N)$ and $\O(N)^-$ theories.

When the gauge theory is coupled to $N_f$ vector hypermultiplets, a non-trivial role is also played by the background magnetic fluxes for the $\usp(2 N_f)$ flavour symmetry. In particular, turning on an elementary flux $\left(1, 0, \ldots, 0\right)$ for the flavour symmetry renders the associated monopole operator odd under charge conjugation, whereas the parity under the magnetic symmetry depends on the specific values of the $\SO(N)$ fluxes. This observation becomes of essential importance when 3d $\CN=3$ and $\CN=4$ orthosymplectic quiver gauge theories are considered, in which hypermultiplets in the vector representation of an $\SO(N)$ gauge node are usually also connected to a $\USp(2 M)$ gauge node.

In the following, we analyse the limits of the superconformal index for various examples of 3d $\CN=3$ orthosymplectic theories, for which the conventions of \cite{Aharony:2013kma,Harding:2025vov} are employed. We turn on the fugacities associated with both the $\BZ_2$ magnetic and charge conjugation symmetries, denoted by $\zeta$ and $\chi$, respectively, which allow us to keep track of the various possible global forms associated with an $\so(N)$ gauge node. Considerable emphasis is especially placed on the distinction between the $\SO(N)$ and $\O(N)$ gauge groups.
\subsection{Linear quiver with $\USp(2)_{k} \times \SO(2)_{-2k} \times \USp(2)_{k}$ gauge group}
Let us consider the linear quiver
\begin{equation} \label{quivC1D1C1}
\scalebox{0.85}{
\begin{tikzpicture}[baseline,font=\footnotesize,
    circ/.style={circle, draw, minimum size=1.3cm},
    sq/.style={rectangle, draw, minimum size=1cm},
    node distance=1cm
]
    \node[circ] (n1) {{$\USp(2)_{k}$}};
    \node[circ, right=of n1] (n2) {\scalebox{0.85}{$\SO(2)_{-2k}$}};
    \node[circ, right=of n2] (n3) {$\USp(2)_{k}$};
    \node[sq, below=of n1] (s1) {$F_1$};
    \node[sq, below=of n2] (s2) {$F_2$};
    \node[sq, below=of n3] (s3) {$F_3$};
    \draw (n1) -- (n2);
    \draw (n2) -- (n3);
    \draw (n1) -- (s1);
    \draw (n2) -- (s2);
    \draw (n3) -- (s3);
\end{tikzpicture}}
\end{equation}
which can be realised through the following brane setup \cite{Feng:2000eq}:\footnote{We draw only physical branes in the brane configuration. The types of the orientifold plane for each brane interval is denoted at the bottom of the diagram.}
\begin{equation}
\scalebox{0.9}{
\begin{tikzpicture}[baseline,font=\footnotesize,
    cross_node/.style={circle, draw=cyan, cross out, thick, minimum size=6pt, inner sep=0pt}
]
    \def\ystart{-1.5}
    \def\yend{1.5}
    \draw[very thick] (0,0) -- node[near start, below,xshift=-0.5cm]{D3} (7.15,0);
    \node[green-red] at (-1,-2.35){$\O3^-$};
    \node[blue-green] at (1,-2.35){$\O3^+$};
    \node[green-red] at (3.5,-2.35){$\O3^-$};
    \node[blue-green] at (5.8,-2.35){$\O3^+$};
    \node[green-red] at (8.3,-2.35){$\O3^-$};
    \draw[red, thick] (0,\ystart) -- (0,\yend) node[pos=0, below, red] {NS5};
    \draw[red, thick] (4.8,\ystart) -- (4.8,\yend) node[pos=0, below, red] {NS5};
    \draw[blue, dashed, thick] (0.65+1, \ystart) -- (2.15+1, \yend) node[pos=0, below, blue] {$(1, 2k)$};
    \draw[blue, dashed, thick] (5.45+1, \ystart) -- (6.95+1, \yend) node[pos=0, below, blue] {$(1, 2k)$};

    \node[cross_node, label={[cyan]above:$F_1$ D5}] at (1.5, 0.5) {};
    \node[cross_node, label={[cyan]above:$F_2$ D5}] at (4, 0.5) {};
    \node[cross_node, label={[cyan]above:$F_3$ D5}] at (6, 0.5) {};
\end{tikzpicture}}
\end{equation}
For definiteness, we take $F_1 > 1$, $F_2 \ge 0$, $F_3 > 1$.\footnote{This condition ensures that the D5-branch of theory \eref{quivC1D1C1} coincides with the Higgs branch of a good 3d $\CN=4$ quiver, discussed in Section \ref{D5branchquivC1D1C1}.} In order to take the various limits of the index which reproduce the Hilbert series associated to the different branches in the moduli space of the theory, we assign axial fugacities to the chiral fields of the theory, as usual. We can easily keep track of the various choices of axial assignments by resorting to the 3d $\CN=2$ description of quiver \eref{quivC1D1C1}, namely
\begin{equation} \label{quivC1D1C1Neq2}
\scalebox{0.85}{
\begin{tikzpicture}[baseline, font=\footnotesize,
    circ/.style={circle, draw, minimum size=1.3cm},
    sq/.style={rectangle, draw, minimum size=1cm},
    node distance=1cm,
    every loop/.style={-}
]
    \node[circ] (n1) {{$\USp(2)_{k}$}};
    \node[circ, right=of n1] (n2) {\scalebox{0.85}{$\SO(2)_{-2k}$}};
    \node[circ, right=of n2] (n3) {$\USp(2)_{k}$};
    \node[sq, below=of n1] (s1) {$\SO(2F_1)$};
    \node[sq, below=of n2] (s2) {$\USp(2F_2)$};
    \node[sq, below=of n3] (s3) {$\SO(2F_3)$};
    \draw[-] (n1) -- node[above, red] {$\fa_1$} (n2);
    \draw[-] (n2) -- node[above, red] {$\fa_2$} (n3);
    \draw[-] (n1) -- node[right, red] {$\fc_1$} (s1);
    \draw[-] (n2) -- node[right, red] {$\fc_2$} (s2);
    \draw[-] (n3) -- node[right, red] {$\fc_3$} (s3);
    \path (n1) edge [loop above] node[red] {$\fb_1$} ();
    \path (n2) edge [loop above] node[red] {$\fb_2$} ();
    \path (n3) edge [loop above] node[red] {$\fb_3$} ();
\end{tikzpicture}}
\end{equation}
where the specific forms of the fugacities $\fa_i$, $\fb_i$ and $\fc_i$ which appear in the limits of the index corresponding to each possible branch of the moduli space are collected in the following table:
\bes{ \label{branchC1D1C1}
\begin{tabular}{c|cccccccccc}
\hline
Branch & $\fa_1$ & $\fa_2$  & $\fc_1$ & $\fc_2$ & $\fc_3$ & $\fb_1$ & $\fb_2$ & $\fb_3$  \\
\hline
NS5& $a$ & $a^{-1}$ & $a^{-1}$ & $a^{-1}$ & 1 & 1 & 1 & 1 \\
$(1, 2k)$& $a^{-1}$ & $a$ & 1 & $a^{-1}$ & $a^{-1}$ & 1 & 1 & 1 \\
D5 & $a$ & $a$ & $a$ & $a$ & $a$ & $a^{-2}$ & $a^{-2}$ & $a^{-2}$ \\
\hline
\end{tabular}
}
The index of the theory \eref{quivC1D1C1Neq2} reads
\bes{ \label{indexquivC1D1C1}
&\CI_{\eqref{quivC1D1C1Neq2}} = \frac{1}{4} \sum_{m_1, m_2, m_3 \in \BZ} \oint \left( \prod_{j=1}^3 \frac{du_j}{2\pi i u_j} \right) (u_1)^{2\kappa_1 m_1} (u_2)^{\kappa_2 m_2} \zeta^{m_2} (u_3)^{2\kappa_3 m_3} \\
&\qquad \qquad \qquad \times Z_{\text{vec}}^{\Usp(2)}(x; u_1; m_1)\,\, Z_{\text{vec}}^{\Usp(2)}(x; u_3; m_3) 
\,\, Z_{\text{adj-chir}}(\chi) \,\, Z_{\text{matter}}(\chi) ~,
}
where the Chern-Simons levels are chosen as
\bes{ 
\kappa_1 = k, \,\,\,\, \kappa_2 = -2 k, \,\,\,\, \kappa_3 = k~,
}
where we take $k>0$ for simplicity. Moreover, the $\USp(2)$ vector multiplet contribution reads
\bes{
Z_{\text{vec}}^{\USp(2)}(x; u; n) =  x^{-|2n|} \prod_{s=\pm 1}  (1-(-1)^{2 s n} u^{2s} x^{|2n|} )~.
}
The contribution coming from the adjoint chiral fields in the vector multiplets appearing in \eref{indexquivC1D1C1} are given explicitly by
\bes{
Z_{\text{adj-chir}}(\chi) = Z^{\USp(2)}_{\text{adj-chir}}(x; \fb_1; u_1; m_1) Z^{\SO(2)}_{\text{adj-chir}}(x; \fb_2; \chi) Z^{\USp(2)}_{\text{adj-chir}}(x; \fb_3; u_3; m_3)~,
}
with
\begin{subequations}  
\begin{align}
\begin{split} 
Z_{\text{adj-chir}}^{\USp(2)}(x; \fb; u; n) = Z_{\text{chir}}^{R=1} (x; \fb; 0) \prod_{s = \pm 1} Z_{\text{chir}}^{R=1} (x; \fb u^{2s}; 2s n )~,
\end{split}\\
\begin{split} 
Z^{\SO(2)}_{\text{adj-chir}}(x; \fb;\chi)=Z_{\text{chir}}^{R=1} (x; \chi \fb; 0)~,
\end{split}
\end{align}
\end{subequations}
whereas the matter fields in the bifundamental representation of the $\USp(2)$ and $\SO(2)$ gauge groups participate to the index \eref{indexquivC1D1C1} as
\bes{
&Z_{\text{matter}}(\chi) \\&= Z^{\USp(2)-\SO(2)}_{\text{bifund}}(x; \fa_1 | u_1; m_1 | u_2; m_2 |\chi) Z^{\USp(2)-\SO(2)}_{\text{bifund}}(x; \fa_2 | u_3; m_3 | u_2; m_2 |\chi) \\& \times Z^{\USp(2)-\SO(2 F_1)}_{\text{bifund}}(x; \fc_1 | u_1; m_1 | \vec{f_1}; \vec{0} |\chi) Z^{\USp(2 F_2)-\SO(2)}_{\text{bifund}}(x; \fc_2 | \vec{f_2}; \vec{0} | u_2; m_2 |\chi) \\&\times Z^{\USp(2)-\SO(2 F_3)}_{\text{bifund}}(x; \fc_3 | u_3; m_3 | \vec{f_3}; \vec{0} |\chi)~,
}
where
\begin{subequations}  
\begin{align}
\begin{split} \label{USp2SO2bifundchip1}
&Z^{\USp(2L)-\SO(2M)}_{\text{bifund}}(x; \fa | \{u_\alpha; n_\alpha\} | \{v_\beta; l_\beta\} |\chi=+1) \\&=  \prod_{\alpha=1}^L\prod_{\beta=1}^M \prod_{s_1, s_2=\pm 1} Z_{\text{chir}}^{1/2} (x; \fa u_\alpha^{s_1} v_\beta^{s_2};s_1 n_\alpha +s_2 l_\beta)~,
\end{split}\\
\begin{split} 
Z^{\USp(2L)-\SO(2M)}_{\text{bifund}}(x; \fa | \{u_\alpha; n_\alpha\} | \{v_\beta; l_\beta\} |\chi=-1) = \eref{USp2SO2bifundchip1} \Bigg|_{v_M^{\pm 1}=\pm1, \, l_M=0}~.
\end{split}
\end{align}
\end{subequations}
The limit of the index reproducing the Hilbert series associated to each branch reported in \eref{branchC1D1C1} is obtained, as in the unitary case, by performing the substitution \eref{N=3reparam} and, subsequently, taking the limit $\fz \rightarrow 0$:
\bes{ \label{limitHSC1D1C1}
H_{\eqref{quivC1D1C1Neq2}}[\text{branch}](t;\{w_i\}, \{b_i\}, \{f_i \}) = \lim_{\fz \rightarrow 0} \CI_{\eqref{quivC1D1C1Neq2}} \Big|_{\eqref{branchC1D1C1}, \, \eqref{N=3reparam}}~.
}
In particular, the terms contributing to the index with a positive power of $\fz$ get suppressed upon taking the above limit. In the case of the three branches we are interested in, the set of magnetic fluxes yielding a non-zero limit of the index are summarised in the table below:
\bes{ \label{condfluxC1D1C1}
\begin{tabular}{c|l|l}
\hline
Branch & Exponent of $\fz$ & Relevant fluxes \\
\hline
NS & ${|m_1\pm m_2|-2 |m_1|-2 |m_3| + F_3 |m_3|}$ & $m_1=\pm m_2~,\,\, m_3=0$\\
$(1, 2k)$ & ${|m_2\pm m_3|-2 |m_1|-2 |m_3| + F_1 |m_1|}$ & $m_1=0~,\,\, m_2=\pm m_3$\\
D5 & $|m_1\pm m_2|+|m_2\pm m_3|-4|m_1|-4 |m_3|$& $m_1=m_2=m_3=0$ \\
& $+2F_1|m_1|+2F_2 |m_2| +2F_3|m_3|$ & \\
\hline
\end{tabular} 
}
where $|m_i\pm m_j|=|m_i+m_j|+|m_i-m_j|$. Observe that, for such choices of magnetic fluxes, the exponent of $\fz$ is always zero. We now discuss the computation of the limit of index associated to each branch explicitly, and, finally, we also provide an explanation regarding the choices of magnetic fluxes yielding non-trivial contributions reported in \eref{condfluxC1D1C1}.

\subsubsection{The NS5-branch}
As anticipated in \eref{condfluxC1D1C1}, the only magnetic fluxes participating to the NS5-branch limits of the index are those satisfying the constraint
\bes{ \label{NS5fluxesC1D1C1}
m_1 = \pm m_2~,\,\,\,\,m_3 = 0~.
}
Let us start by considering the $\chi=+1$ case. When all the magnetic fluxes are set to zero, the $\fz \rightarrow 0$ limit of the index \eref{indexquivC1D1C1} takes the form
\bes{ \label{NS5limC1D1C1m1eqm2eq0}
&\frac{1}{4} \oint \left( \prod_{j=1}^3 \frac{du_j}{2\pi i u_j} \right) \left(2 - u_1^2 - \frac{1}{u_1^2}\right) \left(2 - u_3^2 - \frac{1}{u_3^2}\right) \\ & \qquad \times \PE\left[t \sum_{s_1, s_2=\pm 1} u_1^{s_1} u_2^{s_2}\right]= \frac{1}{1-t^2}~.
}
On the other hand, when $m_1 = m_2 = m$, the limit is given by
\bes{ \label{NS5limC1D1C1m1eqm2eqm}
& \frac{1}{4} \sum_{m \in \BZ_{\neq 0}} \oint \left( \prod_{j=1}^3 \frac{du_j}{2\pi i u_j} \right) \left(\frac{u_1}{u_2}\right)^{2 k m} \zeta^m \left(\fz t\right)^{-2 \abs{m}} \left(2 - u_3^2 - \frac{1}{u_3^2}\right) \\ & \qquad \qquad \times \fz^{2 \abs{m}} \PE\left[t \sum_{s=\pm 1} {u_1}^s{u_2}^{-s}\right] t^{2 \abs{m} \left(1+F_1+F_2\right)} \\ &=  \frac{1}{2} \sum_{m \in \BZ_{\neq 0}} \frac{\zeta^m t^{2 \abs{m} \left(F_1 + F_2 + k\right)}}{1-t^2}~.
}
Analogously, when $m_1 =- m_2 = -m$, the expression of the limit can be obtained from the one above by sending $u_2 \rightarrow u_2^{-1}$ and $\zeta \rightarrow \zeta^{-1}$. Taking into account that $\zeta^2 = 1$, the result of the integral turns out to be the same as the one in the third line of \eref{NS5limC1D1C1m1eqm2eqm}. The final result for the NS5-branch limit of the $\chi = + 1$ sector of the index can then be obtained by adding \eref{NS5limC1D1C1m1eqm2eq0} and twice \eref{NS5limC1D1C1m1eqm2eqm}, thus yielding
\bes{ \label{NS5limCiD1C1chiplus}
\eref{NS5limC1D1C1m1eqm2eq0}+2 \times \eref{NS5limC1D1C1m1eqm2eqm} = \sum_{m \in \BZ} P_{\SO(2)}(t;m) \zeta^m t^{2 \CF \abs{m}}~,
}
where the $\SO(2)$ dressing factor is $P_{\SO(2)} (t; m) = \PE[t^2] = \left(1-t^2\right)^{-1}$, and we define $\CF = F_1 + F_2 + k$.

Next, we can move on to the $\chi = -1$ sector. Since the $\SO(2)$ fugacity $m_2$ is set to zero for negative charge conjugation, due to \eref{NS5fluxesC1D1C1}, the only non-zero contribution to the limit is obtained by setting all of the magnetic fluxes to zero in this case. The $\fz \rightarrow 0$ limit of the index \eref{indexquivC1D1C1} can then be obtained as described in \cite{Aharony:2013kma, Harding:2025vov}, namely by substituting $u_2 \rightarrow 1$ and $u_2^{-1} \rightarrow -1$ in the expression of the plethystic exponent appearing in the integrand in \eref{NS5limC1D1C1m1eqm2eq0}. The final result is simply
\bes{ \label{NS5limCiD1C1chiminus}
 2 \times P_{\USp(2)} (t;0) - P_{\SO(2)} (t;0) = \frac{1}{1+t^2}= \frac{1}{\PE_F[t^2]}~,
}
which coincides with the dressing factor for the $\SO(2)$ gauge group with $\chi = -1$, as discussed in \cite[(5.9)]{Harding:2025vov}, where $P_{\USp(2)} (t;0) = \left(1-t^4\right)^{-1}$. Here, the fermionic plethystic exponential function $\PE_F$ is defined as in \cite[Page 10]{Feng:2007ur} and \cite[(1.1)]{Hanany:2016gbz}.

The NS5-branch limit of the index of theory \eref{quivC1D1C1}, refined with both fugacities $\zeta$ and $\chi$, can then be obtained from \eref{NS5limCiD1C1chiplus} and \eref{NS5limCiD1C1chiminus} as \cite[(5.10)]{Harding:2025vov}
\bes{ \label{NS5limc1D1C1zetachi}
&\frac{1}{2} \left[\eref{NS5limCiD1C1chiplus}+\eref{NS5limCiD1C1chiminus}\right] + \frac{1}{2} \left[\eref{NS5limCiD1C1chiplus}-\eref{NS5limCiD1C1chiminus}\right] \chi 
\\&=\frac{1-t^{4 \CF}}{\left(1-\chi t^2\right) \left(1- \zeta t^{2 \CF}\right) \left(1- \zeta \chi t^{2 \CF}\right)} = \PE\left[\chi t^2 + \left(\zeta + \zeta \chi\right) t^{2 \CF} - t^{4 \CF}\right]~.
}
This expression coincides with the Coulomb branch Hilbert series of $\SO(2)$ SQCD with $\CF$ flavours, where both $\BZ_2$ fugacities $\zeta$ and $\chi$ are turned on. This theory can thus be identified as the magnetic quiver for \eref{quivC1D1C1}.

According to the discussion in \cite[around (5.22) and (5.23)]{Cremonesi:2013lqa}, there are three generators satisfying a single relation. Explicitly, the generator at order $t^2$ can be constructed from the $\USp(2) \times \SO(2)$ bifundamental chiral fields $X_{12}$ with axial fugacity $\fa_1 = a$ in \eref{quivC1D1C1Neq2} as the Pfaffian operator $\mathrm{Pf}\left({X_{12} X_{12}}\right) \sim \epsilon^{i_1 i_2} \left(X_{12} X_{12}\right)_{i_1 i_2}$, where $i_1, i_2 = 1, 2$ are $\SO(2)$ indices. In order to identify the remaining two generators at order $t^{2 \CF}$, let us consider the bare monopole operators with magnetic fluxes $(1; \pm 1; 0)$ under the three gauge groups. These can be appropriately dressed by a combination of $2k$ chiral fields $X_{12}$ to form gauge invariant quantities. Let us denote the said gauge invariant dressed monopole operators $V_{\pm}$, where the notation indicates the magnetic flux $\pm 1$ under the $\SO(2)$ gauge group. While both operators $V_{\pm}$ are odd under the magnetic symmetry associated with the $\SO(2)$ gauge group, they lack definite parity under the charge conjugation symmetry. As pointed out in \cite[(2.14)]{Aharony:2013kma} and \cite[(6.2)]{Harding:2025vov}, it is then convenient to define the linear combinations $W_{\pm} = V_+ \pm V_-$, such that $W_+$ ({\it resp.} $W_-$) carries even ({\it resp.} odd) parity under charge conjugation, thus corresponding to the generator contributing as $\zeta t^{2 \CF}$ ({\it resp.} $\zeta \chi t^{2 \CF}$) in the plethystic exponent of \eref{NS5limc1D1C1zetachi}.

\paragraph{Gauging the charge conjugation symmetry.} If the charge conjugation symmetry of the $\SO(2)$ gauge group is gauged—replacing the $\SO(2)$ node in \eref{quivC1D1C1} with an $\O(2)$ node—the NS5-branch limit of the index is obtained by summing over $\chi = \pm 1$ and dividing by two:
\bes{ \label{NS5limc1D1C1O2}
\frac{1}{2} \left[\eref{NS5limCiD1C1chiplus}+\eref{NS5limCiD1C1chiminus}\right] = \sum_{m \in \BZ_{\ge 0}} P_{\USp(2)} (t;m) \zeta^m t^{2 \CF \abs{m}}~,
}
where we use the property $P_{\SO(2)} (t;m) = P_{\USp(2)} (t; \abs{m})$ for $m \neq 0$. This result coincides with the Coulomb branch Hilbert series of 3d $\CN=4$ $\O(2)$ SQCD with $\CF$ flavours (with vanishing background flavour fluxes) as discussed in \cite[Appendix A]{Cremonesi:2014uva}. Upon setting $\zeta = 1$, \eref{NS5limc1D1C1O2} also matches the Coulomb branch Hilbert series of 3d $\CN=4$ $\USp(2)$ SQCD with $\CF+2$ flavours, discussed in \cite[(5.12)]{Cremonesi:2013lqa}. This confirms that $\USp(2)$ SQCD is the magnetic quiver for the NS5-branch of theory \eqref{quivC1D1C1} specifically when the {\it charge conjugation symmetry is gauged}.

\paragraph{Resolving the ambiguity in magnetic quivers.} 
Let us compare our results \eqref{NS5limc1D1C1zetachi} and \eqref{NS5limc1D1C1O2} with the magnetic quivers for the NS5-branch of theory \eqref{quivC1D1C1} proposed in \cite[v1, (4.15)]{Marino:2025ihk}.\footnote{Note that the case $F_2 = 0$ is the primary focus of that reference.} In that work, both $\mathrm{SO}(2)$ SQCD with $N_f$ flavours and $\mathrm{USp}(2)$ SQCD with $N_f+2$ flavours were suggested as candidates for the magnetic quiver capturing the NS5-branch of the original theory \eqref{quivC1D1C1}. Our analysis clarifies this ambiguity: while $\mathrm{SO}(2)$ SQCD is the correct magnetic quiver for the NS5-branch of theory \eqref{quivC1D1C1}, $\mathrm{USp}(2)$ SQCD arises as the magnetic quiver only when the $\mathrm{SO}(2)_{-2k}$ gauge group of \eqref{quivC1D1C1} is replaced by $\mathrm{O}(2)_{-2k}$.

\subsubsection{The $(1, 2k)$-branch}
The magnetic fluxes contributing non-trivially to the index limit for this branch satisfy the condition
\bes{  \label{12kfluxesC1D1C1}
m_1=0~,\,\, m_2=\pm m_3~.
}
Comparing these fluxes with those in \eref{NS5fluxesC1D1C1}, and considering the axial fugacity assignments in \eref{branchC1D1C1}, it is evident that the $(1, 2k)$-branch limit can be derived directly from the NS5-branch limit via the exchange $\{u_1, m_1\} \leftrightarrow \{u_3, m_3\}$ and $F_1 \leftrightarrow F_3$. The resulting expression for the $(1, 2k)$-branch is given by \eref{NS5limc1D1C1zetachi}, with $\CF$ replaced by $\CF' = F_2 + F_3 + k$. This corresponds to the Coulomb branch Hilbert series of 3d $\CN=4$ $\SO(2)$ SQCD with $\CF'$ flavours, which we identify as the magnetic quiver for the $(1, 2k)$-branch of \eref{quivC1D1C1}. The generators follow directly from those of the NS5-branch discussed below \eref{NS5limc1D1C1zetachi} by substituting $X_{12} \leftrightarrow X_{23}$, where the latter denotes the chiral fields with axial fugacity $\fa_3 = a$ in \eref{quivC1D1C1Neq2}. Furthermore, the gauge invariant operators analogous to $V_{\pm}$ are constructed from bare monopole operators with magnetic fluxes $(0; \pm{1}; 1)$, appropriately dressed with a combination of $2k$ chiral fields $X_{23}$.

Similarly to \eref{NS5limc1D1C1O2}, gauging the charge conjugation symmetry of the $\SO(2)_{-2k}$ node in \eref{quivC1D1C1}---effectively replacing $\SO(2)_{-2k}$ with $\O(2)_{-2k}$---recovers the Coulomb branch Hilbert series of $\O(2)$ SQCD with $\CF'$ flavours (with vanishing background flavour fluxes). In particular, setting the magnetic fugacity $\zeta$ to unity yields the Coulomb branch Hilbert series of $\USp(2)$ SQCD with $\CF'+2$ flavours. Consequently, this theory is identified as the magnetic quiver for the $(1, 2k)$-branch of theory \eref{quivC1D1C1} specifically when the charge conjugation symmetry is gauged.

In this instance as well, we resolve the ambiguity noted in \cite[v1, below (4.15)]{Marino:2025ihk}, where both $\SO(2)$ SQCD with $\CF$ flavours and $\USp(2)$ SQCD with $\CF'+2$ flavours were proposed as candidate magnetic quivers for the $(1, 2k)$-branch of theory \eref{quivC1D1C1}. Specifically, we demonstrate that $\SO(2)$ SQCD is the magnetic quiver for the original theory \eref{quivC1D1C1}, whereas $\USp(2)$ SQCD is the magnetic quiver for the variant of theory \eref{quivC1D1C1} in which the $\SO(2)_{-2 k}$ gauge group is replaced by $\O(2)_{-2 k}$.

\subsubsection{The D5-branch}\label{D5branchquivC1D1C1}
The D5-branch limit of the index receives a non-trivial contribution only when the magnetic fluxes for the three gauge groups are turned off, namely
\bes{
m_1=m_2=m_3=0~.
}
When the charge conjugation fugacity is set to unity, the said limit can then be achieved by simply evaluating the integral
\bes{ \label{D5limC1D1C1chiplus}
&\frac{1}{4} \oint \left( \prod_{j=1}^3 \frac{du_j}{2\pi i u_j} \right) \left(2 - u_1^2 - \frac{1}{u_1^2}\right) \left(2 - u_3^2 - \frac{1}{u_3^2}\right) \\ & \qquad \times \PE\left[-t^2 \left\{ \left(u_1^2 + 1 + \frac{1}{u_1^2}\right) -1 + \left(u_3^2 + 1 + \frac{1}{u_3^2}\right) \right\}\right] \\ & \qquad \times \PE\left[t \sum_{s_1, s_2=\pm 1} \left\{ u_1^{s_1} u_2^{s_2} + u_2^{s_1} u_3^{s_2} \right\}\right] \\ & \qquad \times \PE\left[t \sum_{s_1, s_2=\pm 1} \left\{ \sum_{\alpha = 1}^{F_1} u_1^{s_1} (f^{(\alpha)}_1)^{s_2} + \sum_{\beta = 1}^{F_2} u_2^{s_1} (f^{(\beta)}_2)^{s_2} + \sum_{\gamma = 1}^{F_3} u_1^{s_1} (f^{(\gamma)}_3)^{s_2}\right\}\right]~,
}
which precisely coincides with the expression of the Higgs branch Hilbert series of the 3d $\CN=4$ version of theory \eref{quivC1D1C1}, with the Chern-Simons levels set to zero, when $\chi = +1$. 

Let us analyse the Higgs branch ({\it resp.} D5-branch) moment map operators of the said $\CN=4$ theory ({\it resp.} the $\CN=3$ theory \eref{quivC1D1C1}). Let us denote with $Q_1$, $Q_2$ and $Q_3$ the chiral fields carrying axial fugacity $\fc_1$, $\fc_2$ and $\fc_3$ in \eref{quivC1D1C1Neq2}. Explicitly, $Q_1$, $Q_2$ and $Q_3$ connect the three gauge groups and the $\SO(2 F_1)$, $\USp(2 F_2)$ and $\SO(2 F_3)$ flavour nodes, respectively. In addition to the $F_1 \left(2 F_1 - 1\right) + F_2 \left(2 F_2 + 1\right) + F_3 \left(2 F_3 - 1\right)$ mesons $Q_1 J Q_1$, $Q_2 Q_2$ and $Q_3 J Q_3$, where the $\SO(2)$ indices of the chiral fields $Q_2$ are contracted with a Kronecker delta and $J$ is the $\USp(2)$ invariant tensor, there are also baryons acquiring a VEV.\footnote{As explained in \cite[(2.7)]{Argyres:1996hc}, the baryons can be defined in the $\SO(N_c)$ theory with $N_f$ flavours if $2 N_f \ge N_c$. For instance, we can define baryons out of the chiral fields $Q_2$ of theory \eref{quivC1D1C1} if $F_2 \ge 1$.} These can be built using the epsilon tensor of $\SO(2)$, hence they are odd under the charge conjugation symmetry.\footnote{As a consequence, the Higgs branch of $\SO(N_c)$ SQCD with $N_f$ flavours differs from the one of $\O(N_c)^+$ SQCD with $N_f$ flavours if baryons can be constructed, \ie if $2 N_f \ge N_c$. On the other hand, if twice the number of flavours is smaller than the number of colours, there are no baryons and the charge conjugation symmetry acts trivially in the Higgs branch of the theory. Subsequently, the Higgs branch of the theory with $\SO(N_c)$ gauge group equals the one of the theory with $O(N_c)^+$ gauge group, as pointed out in \cite[Appendix B.3]{Cremonesi:2014uva}.} In this case, $F_2 \left(2 F_2 - 1\right)$ baryons can be constructed by appropriately contracting the gauge indices of $Q_2 Q_2$, and a further baryon is recovered by a gauge invariant combination of two copies of the chiral fields $Q_1$ and $Q_3$. As a result, there are $F_2 \left(2 F_2 - 1\right) + 1$ moment map operators which get projected out upon gauging the charge conjugation symmetry associated with the $\SO(2)$ gauge group. 

This can be checked explicitly by averaging the expression \eref{D5limC1D1C1chiplus} with the D5-branch limit of the index in the $\chi = -1$ sector of theory \eref{quivC1D1C1}, which reads
\bes{ \label{D5limC1D1C1chiminus}
&\frac{1}{4} \oint \left( \prod_{j=1}^3 \frac{du_j}{2\pi i u_j} \right) \left(2 - u_1^2 - \frac{1}{u_1^2}\right) \left(2 - u_3^2 - \frac{1}{u_3^2}\right) \\ & \qquad \times \PE\left[-t^2 \left\{ \left(u_1^2 + 1 + \frac{1}{u_1^2}\right) +1 + \left(u_3^2 + 1 + \frac{1}{u_3^2}\right) \right\}\right] \\ & \qquad \times \PE\left[t \sum_{s=\pm 1} \left\{ u_1^{s} + u_3^{s} \right\}\right] \times \frac{1}{\PE_{F}\left[t \sum\limits_{s=\pm 1} \left\{ u_1^{s} + u_3^{s} \right\}\right]} \\ & \qquad \times \PE\left[t \sum_{s_1, s_2=\pm 1} \left\{ \sum_{\alpha = 1}^{F_1} u_1^{s_1} (f^{(\alpha)}_1)^{s_2} + \sum_{\gamma = 1}^{F_3} u_1^{s_1} (f^{(\gamma)}_3)^{s_2}\right\}\right] \\ & \qquad \times \PE\left[t \sum_{s=\pm 1} \sum_{\beta = 1}^{F_2} (f^{(\beta)}_2)^{s}\right] \times \frac{1}{\PE_{F}\left[t \sum\limits_{s=\pm 1} \sum_{\beta = 1}^{F_2} (f^{(\beta)}_2)^{s}\right]}~,
}
where $\PE_{F}\left[t \sum_i u_i\right] = \prod_i \left(1 + t u_i\right)$.

The D5-branch limit of the index of theory \eref{quivC1D1C1} refined with the charge conjugation fugacity $\chi$ can then be obtained in a similar fashion as in \eref{NS5limc1D1C1zetachi}. Upon setting the flavour fugacities to one, this yields
\bes{
\scalebox{0.91}{$
\begin{split}
& \frac{1}{2} \left[\eref{D5limC1D1C1chiplus}+\eref{D5limC1D1C1chiminus}\right] + \frac{1}{2} \left[\eref{D5limC1D1C1chiplus}-\eref{D5limC1D1C1chiminus}\right] \chi \\&= 1 + \left\{F_1 \left(2 F_1 - 1\right) + F_2 \left(2 F_2 + 1\right) + F_3 \left(2 F_3 - 1\right) + {\cerulean \chi \left[F_2 \left(2 F_2 - 1\right) + 1\right]}\right\} t^2 + \ldots~,
\end{split}
$}
}
where the term highlighted in {\cerulean cerulean} gets projected out when the charge conjugation symmetry associated with the $\SO(2)_{-2 k}$ gauge node is gauged, in agreement with the analysis discussed above \eref{D5limC1D1C1chiminus}. As a consequence, the Hilbert series for the D5-branch of the variant of theory \eref{quivC1D1C1} in which the $\SO(2)_{-2 k}$ gauge node is replaced by $\O(2)_{-2 k}$, upon setting the flavour fugacities to one, is given by
\bes{
\scalebox{0.97}{$
\begin{split}
\frac{1}{2} \left[\eref{D5limC1D1C1chiplus}+\eref{D5limC1D1C1chiminus}\right] = 1 + \left[F_1 \left(2 F_1 - 1\right) + F_2 \left(2 F_2 + 1\right) + F_3 \left(2 F_3 - 1\right)\right] t^2 + \ldots~,
\end{split}
$}
}
where the terms at order $t^2$ are in the adjoint representation of the $\so(2 F_1) \oplus \so(2 F_2) \oplus \so(2 F_3)$ D5-branch global symmetry.

As a final observation, we point out that a magnetic quiver for the D5-branch of \eref{NS5fluxesC1D1C1} is proposed in \cite[v1, (4.17)]{Marino:2025ihk} in the case $F_2 = 0$. Such a magnetic quiver possesses an $\so(2 F_1) \oplus \so(2 F_3)$ Coulomb branch global symmetry, with $F_1 \left(2 F_1 - 1\right) + F_3 \left(2 F_3 - 1\right)$ Coulomb branch moment map operators. However, in order to match the D5-branch of theory \eref{NS5fluxesC1D1C1} with $F_2 = 0$, an extra operator is missing. This operator would get mapped to the baryon which is acted upon non-trivially by charge conjugation. As a consequence, our method clarifies that the magnetic quiver of \cite[v1, (4.17)]{Marino:2025ihk} captures the D5-branch of theory \eref{NS5fluxesC1D1C1} with $F_2 = 0$ when the central gauge node is $\O(2)_{-2 k}$, instead of $\SO(2)_{-2 k}$.
\subsubsection{Magnetic flux configurations contributing to the limits}
In order to derive the NS5-branch limit of the index \eref{indexquivC1D1C1}, we first observe that there are no non-trivial contributions if $m_3 \neq 0$. We can discard such fluxes by the following argument. Upon taking the $\fz \rightarrow 0$ limit, the integrand in principle {\it might} pick up a residue in $u_3 = 0$. Indeed, possible non-trivial contributions might arise since the gauge fugacity $u_3$ appears in the integration measure $\frac{du_3}{2\pi i u_3}$, in the Chern-Simons term $u_3^{2 k m_3}$ and in the $\USp(2)$ vector contribution, which contains a term $\fz^0 (-1)^{2 m_3} \left(u_3^2 + u_3^{-2}\right)$.\footnote{Note that, in the $\USp(2)$ vector contribution associated with the fugacity $u_3$, there are also the factors $\left(\fz t\right)^{-2 \abs{m_3}}+ (-1)^{4 m_3} \left(\fz t\right)^{2 \abs{m_3}}$, which do not depend on $u_3$. While contributing trivially to the integral if $m_3 \neq 0$, such terms are responsible for a non-zero result of the residue computation in $u_3 = 0$ if $m_3 = 0$.} In particular, a non-zero contribution arises from the residue computation if $2 k m_3 \pm 2 = 0$, which also implies $\abs{m_3} > 0$. Now, let us look at the lowest possible power of $\fz$ appearing in the integrand before taking the $\fz \rightarrow 0$ limit. There is a factor $\fz^{-2 \abs{m_1}}$ coming from the $\USp(2)$ vector contribution associated with the fugacity $u_1$, a term $\fz^{\abs{m_1+m_2}+\abs{m_1-m_2}}$ due to the chiral fields in the bifundamental representation of the $\USp(2) \times \SO(2)$ gauge groups associated with the fugacities $u_1$ and $u_2$, and, finally, there is also a factor $\CO\left(\fz^{F_3 \abs{m_3}}\right)$ coming form the chiral fields transforming in the bifundamental representation of $\USp(2) \times \SO(2 F_3)$. The overall power of $\fz$ is then $-2 \abs{m_1}+\abs{m_1+m_2}+\abs{m_1-m_2}+F_3 \abs{m_3}$, which is always larger than zero if $m_3 \neq 0$, thus vanishing in the $\fz \rightarrow 0$ limit.

Next, we also point out that a non-zero contribution coming from \eref{indexquivC1D1C1} can be extracted only if the magnetic fluxes satisfy the further condition $m_1 = \pm m_2$. In order to show the validity of this statement, let us suppose, for the sake of contradiction, that $m_1 \neq \pm m_2$. Under this hypothesis, the appearance of the gauge fugacities $u_1$ and $u_2$ in the integrand is restricted to the integration measure $\frac{du_1}{2\pi i u_1} \frac{du_2}{2\pi i u_2}$, the Chern-Simons factors $u_1 ^{2 k m_1} u_2^{2 k m_2}$, and the $\USp(2)$ vector contribution $\left(\fz t\right)^{-2 \abs{m_1}}+ (-1)^{4 m_1} \left(\fz t\right)^{2 \abs{m_1}}+ (-1)^{2 m_1} \left(u_1^2 + u_1^{-2}\right)$. The integrand picks up a residue in $u_2 = 0$ if $m_2 = 0$, which also implies, by our assumption, that $m_1 \neq 0$. On the other hand, a non-trivial contribution due to the residue in $u_1 = 0$ arises if $2 k m_1 \pm 2 = 0$, which is perfectly consistent with $m_1$ being non-zero. However, the lowest possible power of $\fz$ arising from such a non-trivial contribution to the integral \eref{indexquivC1D1C1} reads $\abs{m_1 - m_2}+\abs{m_1 + m_2} = 2 \abs{m_1}$,\footnote{This power of $\fz$ originates from the chiral fields in the bifundamental representation of the $\USp(2)$ and $\SO(2)$ gauge groups associated with the fugacities $u_1$ and $u_2$.} which is larger than zero, resulting in a trivial term upon taking the $\fz \rightarrow 0$ limit.

An analogous argument can also be applied to the $(1, 2k)$-branch limit of the index, since this is related to the NS5-branch limit discussed above by swapping $\{u_1, m_1\} \leftrightarrow \{u_3, m_3\}$ and $F_1 \leftrightarrow F_3$, as pointed out below \eref{12kfluxesC1D1C1}.

Finally, the D5-branch limit of the index \eref{indexquivC1D1C1} does not vanish only when all the magnetic fluxes are set to zero. Indeed, the triangle inequality implies that the power of $\fz$ in this particular limit, which is reported in the last row of \eref{condfluxC1D1C1}, is always larger than or equal to $-2 \abs{m_1} - 2 \abs{m_3} + 2 F_1 \abs{m_1} + 2 F_2 \abs{m_2} + 2 F_3 \abs{m_3}$, which is always larger than or equal to zero if the initial assumption that both $F_1$ and $F_3$ are larger than one. Moreover, since a positive power of $\fz$ results in a vanishing contribution upon taking the limit $\fz \rightarrow 0$, the only non-trivial contribution originates from the set of magnetic fluxes whic are all equal to zero.
\subsection{Linear quiver with $\SO(2)_{2k} \times \USp(4)_{-k} \times \SO(2)$ gauge group}
We consider the theory described by the following quiver diagram:
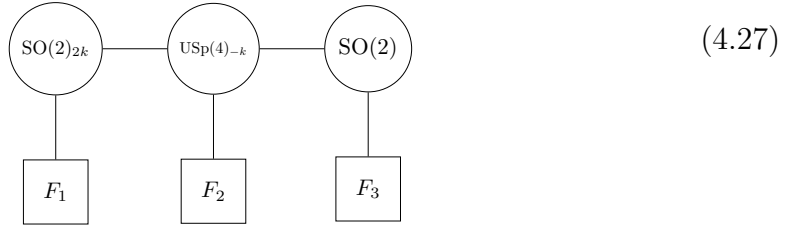
\begin{equation} \label{quivT3theoryorthopuzzle1}
\scalebox{0.85}{
\begin{tikzpicture}[baseline,font=\footnotesize,
    circ/.style={circle, draw, minimum size=1.3cm},
    sq/.style={rectangle, draw, minimum size=1cm},
    node distance=1cm
]
    \node[circ] (n1) {\scalebox{0.85}{$\SO(2)_{2k}$}};
    \node[circ, right=of n1] (n2) {\scalebox{0.7}{$\Usp(4)_{-k}$}};
    \node[circ, right=of n2] (n3) {$\SO(2)$};
    \node[sq, below=of n1] (s1) {$F_1$};
    \node[sq, below=of n2] (s2) {$F_2$};
    \node[sq, below=of n3] (s3) {$F_3$};
    \draw (n1) -- (n2);
    \draw (n2) -- (n3);
    \draw (n1) -- (s1);
    \draw (n2) -- (s2);
    \draw (n3) -- (s3);
\end{tikzpicture}}
\end{equation}
This theory can be realised in string theory via the brane configuration shown below:
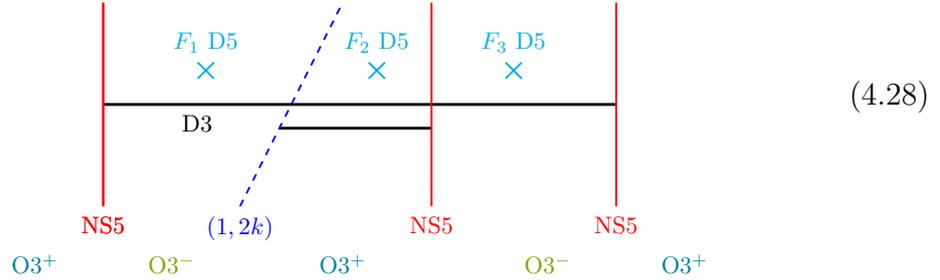
\begin{equation}
\scalebox{0.9}{
\begin{tikzpicture}[baseline,font=\footnotesize,
    cross_node/.style={circle, draw=cyan, cross out, thick, minimum size=6pt, inner sep=0pt}
]
    \def\ystart{-1.5}
    \def\yend{1.5}
    \draw[very thick] (0,0) -- node[near start, below,xshift=-0.5cm]{D3} (7.5,0);
    \draw[very thick] (2.57,-0.35) --(4.8,-0.35);
    \node[blue-green] at (-1,-2.35){$\O3^+$};
    \node[green-red] at (1,-2.35){$\O3^-$};
    \node[blue-green] at (3.5,-2.35){$\O3^+$};
    \node[green-red] at (6.5,-2.35){$\O3^-$};
    \node[blue-green] at (8.5,-2.35){$\O3^+$};
    \draw[red, thick] (0,\ystart) -- (0,\yend) node[pos=0, below, red] {NS5};
    \draw[red, thick] (4.8,\ystart) -- (4.8,\yend) node[pos=0, below, red] {NS5};
    \draw[red, thick] (0,\ystart) -- (0,\yend) node[pos=0, below, red] {NS5};
    \draw[red, thick] (7.5,\ystart) -- (7.5,\yend) node[pos=0, below, red] {NS5};
    \draw[blue, dashed, thick] (1+1, \ystart) -- (2.5+1, \yend) node[pos=0, below, blue] {$(1, 2k)$};

    \node[cross_node, label={[cyan]above:$F_1$ D5}] at (1.5, 0.5) {};
    \node[cross_node, label={[cyan]above:$F_2$ D5}] at (4, 0.5) {};
    \node[cross_node, label={[cyan]above:$F_3$ D5}] at (6, 0.5) {};
\end{tikzpicture}}
\end{equation}
Note that, for $F_1,F_2,F_3=0$, this is the $\mathcal{N}=4$ theory considered in \cite[v1, section 5]{Marino:2025ihk}. We analyse this theory by computing its superconformal index. In order to compute the various limits of the index, we assign axial fugacities to the chiral fields. In 3d $\mathcal{N}=2$ notation, the field content and fugacity assignments are
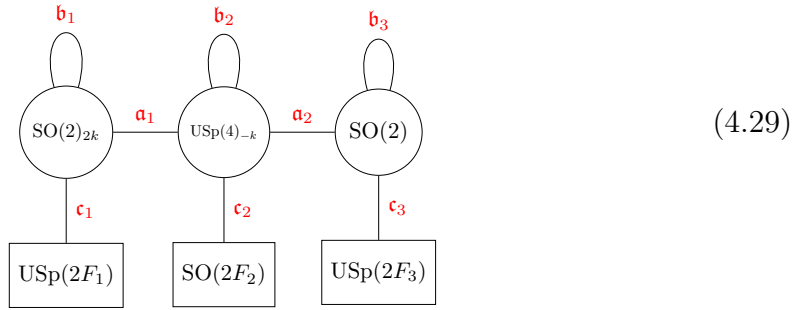
\begin{equation} \label{quivpuzzle1}
\scalebox{0.85}{
\begin{tikzpicture}[baseline, font=\footnotesize,
    circ/.style={circle, draw, minimum size=1.3cm},
    sq/.style={rectangle, draw, minimum size=1cm},
    node distance=1cm,
    every loop/.style={-}
]
    \node[circ] (n1) {\scalebox{0.85}{$\SO(2)_{2k}$}};
    \node[circ, right=of n1] (n2) {\scalebox{0.7}{$\Usp(4)_{-k}$}};
    \node[circ, right=of n2] (n3) {$\SO(2)$};
    \node[sq, below=of n1] (s1) {$\USp(2F_1)$};
    \node[sq, below=of n2] (s2) {$\SO(2F_2)$};
    \node[sq, below=of n3] (s3) {$\USp(2F_3)$};
    \draw[-] (n1) -- node[above, red] {$\fa_1$} (n2);
    \draw[-] (n2) -- node[above, red] {$\fa_2$} (n3);
    \draw[-] (n1) -- node[right, red] {$\fc_1$} (s1);
    \draw[-] (n2) -- node[right, red] {$\fc_2$} (s2);
    \draw[-] (n3) -- node[right, red] {$\fc_3$} (s3);
    \path (n1) edge [loop above] node[red] {$\fb_1$} ();
    \path (n2) edge [loop above] node[red] {$\fb_2$} ();
    \path (n3) edge [loop above] node[red] {$\fb_3$} ();
\end{tikzpicture}}
\end{equation}
The specific assignments of axial fugacities $\fa_i$, $\fb_i$, and $\fc_i$ for each branch are summarised in the following table:
\bes{ \label{branchpuzzle1}
\begin{tabular}{c|cccccccccc}
\hline
Branch & $\fa_1$ & $\fa_2$  & $\fc_1$ & $\fc_2$ & $\fc_3$ & $\fb_1$ & $\fb_2$ & $\fb_3$  \\
\hline
NS5& $a$ & $a^{-1}$ & $a^{-1}$ & $a^{-1}$ & $a^{-1}$ & 1 & 1 & $a^2$ \\
D5 & $a$ & $a$ & $a$ & $a$ & $a$ & $a^{-2}$ & $a^{-2}$ & $a^{-2}$ \\
\hline
\end{tabular}
}
The index for the theory in \eqref{quivpuzzle1} is then given by:
\bes{ \label{indexquivpuzzle1}
\scalebox{0.92}{$
\begin{split}
&\CI_{\eqref{quivpuzzle1}} = \frac{1}{8} \sum_{m_1,m_3 \in \BZ} \sum_{\{m^{(\alpha)}_2\} \in \BZ^2} \oint \Bigg( \prod_{\alpha=1}^2 \frac{du^{(\alpha)}_2}{2\pi i u^{(\alpha)}_2}  (u^{(\alpha)}_2)^{2\kappa_2 m^{(\alpha)}_2}  \Bigg) \oint  \frac{du_1}{2\pi i u_1}  (u_1)^{2\kappa_1 m_1}w_1^{m_1}  \\
&\quad \times \oint  \frac{du_3}{2\pi i u_3}w_3^{m_3} \,\, Z_{\text{vec}}^{\Usp(4)}(x; \{u^{(\alpha)}_2\}; \{m^{(\alpha)}_2\})\,\, Z_{\text{adj-chir}} \,\, Z_{\text{matter}}~, 
\end{split}$}
}
where the Chern-Simons levels are
\bes{ \label{kappaT3}
\kappa_1 = k~, \qquad \kappa_2 = -k~,
}
and we define
\bes{
\scalebox{0.8}{$
\begin{split}
Z_{\text{adj-chir}} &= \prod_{s_1,s_2=\pm1}\prod_{\alpha=1}^2 \prod_{\beta>\alpha}Z_{\text{chir}}^{1} \left(x; \fb_2 (u^{\alpha}_2)^{s_1} (u^{\beta}_2)^{s_2} ; s_1m^{(\alpha)}_2 +s_2 m^{(\beta)}_2  \right) \\ 
& \qquad \times \prod_{s=\pm1}\prod_{\alpha=1}^2 Z_{\text{chir}}^{1} \left(x; \fb_2 (u^{\alpha}_2)^{2s} ; 2sm^{(\alpha)}_2\right) Z_{\text{chir}}^{1} \left(x; \fb_2 ; 0  \right)^2 Z_{\text{chir}}^{1} \left(x; \fb_1 ; 0  \right)Z_{\text{chir}}^{1} \left(x; \fb_3 ; 0  \right)~, \\
Z_{\text{matter}} &= \Bigg[ \prod_{s_1 = \pm1} \prod_{s_2 = \pm1}\prod_{\alpha=1}^2 Z_{\text{chir}}^{1/2} \left(x; \fa_1 (b_1 u_1)^{s_1} (u^{(\alpha)}_{2})^{s_2}; s_1m_{1} +s_2 m^{(\alpha)}_{2} \right)  \\ 
& \qquad \quad \times  Z_{\text{chir}}^{1/2} \left(x; \fa_2  (u^{(\alpha)}_2)^{s_1} (b_2 u_{3})^{s_2}; s_1 m^{(\alpha)}_{2} +s_2 m_3 \right) \Bigg] \\
& \qquad \times \prod_{s_1 = \pm1}\prod_{s_2 = \pm1}  \prod_{\rho_2=1}^{F_2} \prod_{\alpha=1}^2  Z_{\text{chir}}^{1/2} \left(x; \fc_2 ( u^{(\alpha)}_2)^{s_1}(f^{(\rho_2)}_2)^{s_2}; s_1m^{(\alpha)}_{2} \right)  \\
& \qquad \times \prod_{s_1 = \pm1}\prod_{s_2 = \pm1} \prod_{\rho_1=1}^{F_1} Z_{\text{chir}}^{1/2} \left(x; \fc_1 (u_1)^{s_1}(f^{(\rho_1)}_1)^{s_2} ; s_1m_{1} \right) (1\leftrightarrow 3)~.
\end{split}$}
}
Note that $b_1$ and $b_2$ are introduced for convenience in keeping track of bifundamental hypermultiplets. They do not correspond to a symmetry of the theory and can be set to unity. Moreover, the refinement with respect to the charge conjugation symmetry fugacities $\chi_1$ and $\chi_2$ for gauge groups $\SO(2)_{2k}$ and $\SO(2)$ in \eqref{quivpuzzle1} can be done in a similar fashion to \eqref{indexquivC1D1C1}.

For definiteness, we take $k>0$ throughout this discussion. The Hilbert series for each branch is obtained by taking a specific limit of the index:
\bes{ \label{limitHSorthosymppuzzle1}
H_{\eqref{quivpuzzle1}}[\text{branch}](t;\{w_i\}; \{b_i\}; \{f_i \}) = \lim_{\fz \rightarrow 0} \CI_{\eqref{quivpuzzle1}} \Big|_{\eqref{branchpuzzle1}, \, \eqref{N=3reparam}}~.
}
To compute the limit, we first examine the overall power of $\fz$ in the integrand for each branch. A non-vanishing result in the $\fz \to 0$ limit sets the exponent of $\fz$ to zero, imposing the following constraints on the magnetic fluxes $m_i$ :
\bes{ \label{condflux2411}
\scalebox{0.85}{
\begin{tabular}{c|l|l}
\hline
Branch & Exponent of $\fz$ & Condition \\
\hline
NS5 & ${\sum_{i=1}^2\left(|m_1\pm m_2^{(i)}|-2|m_2^{(i)}|\right)-|m_2^{(1)}\pm m_2^{(2)}|}$ & $m_1=\pm m_2^{(i)}~,\,\, m_2^{(j\neq i)}=0$\\
D5 & ${\sum_{i=1}^2(|m_1\pm m_2^{(i)}|+|m_3\pm m_2^{(i)}|)} $& $m_1=m_2^{(1)}=m_2^{(2)}=m_3=0$ \\
&${+2F_1|m_1|+2F_2\sum_{i=1}^2(|m_2^{(1)}|)+2F_3|m_3|}$ &  \\
& $-2|m_2^{(1)}\pm m_2^{(2)}|-4\sum_{i=1}^2|m_2^{(i)}|$ & \\
\hline
\end{tabular} }
}
where $|m_i\pm m_j|=|m_i+m_j|+|m_i-m_j|$, and, in the first case, there is no restriction on $m_3$. Notice that the conditions in Table \ref{condflux2411} are not the only ones that can be obtained searching for solutions of the exponent of $\fz$ equal to zero.\footnote{The same argument as in Footnote \ref{foot:moresolutions} applies here.}$^,$\footnote{In this case, another subtlety arises. There are values of the fluxes for which the exponent of $\fz$ is negative. We conjecture that the integral has no residues for these values of the fluxes. This can be easily checked expanding the index up to low order and seeing that it gives zero for values of the fluxes different from the one considered. This is because monopole operators corresponding to those fluxes cannot be dressed properly and so cannot be made gauge invariant.}
We now discuss the computation for each branch explicitly.
\subsubsection*{The NS5-branch}
The condition obtained in Table \eqref{condflux2411} for the NS5-branch limit of the index is
\bes{m_1=\pm m_2^{(i)}~,\,\,\,\,m_2^{(j\neq i)}=0~, \quad i,j=1,2~.}
Using the Weyl symmetry of the $\USp(4)$ gauge group, we can restrict our attention to the option $m_1=\pm m_2^{(1)}, m_2^{(2)}=0$. Taking the limit \eqref{limitHSorthosymppuzzle1}, the Hilbert series receives the following contributions: one with $m_1 = \fm,\vec{m}_2 = ( \pm \fm, 0), m_3 =\fn$, where $\fm, \neq 0$, and the other with $m_1 = 0,\vec{m}_2 = (0, 0), m_3 =\fn$. In the first case, the signs $+$ and $-$ contribute equally to the limit of the index, so
\bes{\label{eq1121puzzle1}
H_{\eqref{quivpuzzle1}}[\text{NS5}](t;\{w_{i}\};\{b_{i}\})
=2 \times H_{m_1 = \fm,\vec{m}_2 = (\fm, 0), m_3 =\fn}+H_{m_1 = 0,\vec{m}_2 = (0, 0), m_3=\fn}~.}

For the fluxes $m_1 = \fm,\vec{m}_2 = (\fm, 0), m_3=\fn$, the $\USp(4)$ gauge group is broken to $\text{SU(2)}\times \text{U(1)}$. In this case, when both charge conjugation fugacities $\chi_1$ and $\chi_2$ are set to one, the limit of the index is\footnote{Note that we have an overall factor of $\frac{1}{4}$ here, in contrast to that of \eqref{indexquivpuzzle1}, which is $\frac{1}{8}$. This is because the results for the allowed choices $\vec{m}_2 = (\fm,0)$ and $(0, \fm)$ are equal.}
\bes{ \label{limindD1C2D1fluxesmn}
& H_{m_1 = \fm,\vec{m}_2 = (\fm, 0), m_3=\fn}
= \frac{1}{4}  \sum_{\fm \in \BZ_{\neq 0}}\sum_{\fn \in \BZ}  t^{2 \Delta(m_1=\fm,\   \vec m_2=(\fm, 0), \, m_3 = \fn)} \\
& \quad \times \oint  \frac{du_1}{2\pi i u_1} u_1^{2\kappa_1 \fm}  w_1^{\fm} \oint  \frac{du_3}{2\pi i u_3} w_3^\fn \oint  \Bigg(  \prod_{\alpha=1}^2 \frac{du^{(\alpha)}_2}{2\pi i u^{(\alpha)}_2}  (u^{(\alpha)}_2)^{2\kappa_2 m^{(\alpha)}_2}   \Bigg)_{\vec m_2=(\fm, 0)}  \\
& \quad \times \prod_{s=\pm 1}\left[1-\left(u_2^{(2)}\right)^{2s}\right]  \PE\left[t^2+t \sum_{s=\pm 1} \left( \frac{b_1 u_1}{u^{(1)}_{2}}\right)^s  \right]~,
}
where the dimension of the monopole operators is given by
\bes{
\scalebox{0.95}{$
\begin{split}
&\Delta (m_1, m_2^{(1)}, m_2^{(2)}, m_3)\\
&= \frac{1}{2} \Biggl(
    \sum_{i=1}^2 \sum_{s=\pm 1}\bigl( |m_3 +s m_2^{(i)}| +|m_1 +s m_2^{(i)}|\bigr) -2\sum_{i=1}^2 \bigl( |2m_2^{(i)}|\bigr)
    \\&\qquad -2 |m_2^{(1)} - m_2^{(2)}| - 2|m_2^{(1)} + m_2^{(2)}| + 2F_1 |m_1| + 2F_2 \sum_{i=1}^2 |m_2^{(i)}|+ 2F_3 |m_3|\Biggr) .
\end{split}$}
} 
Explicitly, for this set of magnetic fluxes, the monopole dimension reads
\bes{
&2\Delta(m_1=\fm,\   \vec m_2=(\fm, 0), \, m_3 = \fn) \\
&= |\fm-\fn|+|\fm+\fn| +|\fm|(2F_1+2F_2-4) +|\fn|(2F_3+2)~.
}
Similarly to \eqref{usefulintiden} and \eqref{U2identity}, the contribution \eref{limindD1C2D1fluxesmn} gets simplified to:
\bes{ \label{fluxconfiga}
&H_{m_1 = \fm,\vec{m}_2 = (\fm, 0), m_3 =\fn } \\
&=\frac{1}{2}\sum_{\fm \in \mathbb{Z}_{\neq 0}}\sum_{\fn \in \mathbb{Z}}\frac{t^{2k|\fm|}}{(1-t^2)^2} (b_1^{-2k} w_1)^\fm w_3^\fn \times t^{2\Delta(m_1=\fm,\   \vec m_2=(\fm, 0), \, m_3 = \fn)} \\
&= \frac{1}{2}\sum_{\fm \in \mathbb{Z}_{\neq 0}}\sum_{\fn \in \mathbb{Z}} \frac{ t^{2(F_1+F_2+k-2)|\fm|+2(F_3+1)|\fn|+|\fm+\fn|+|\fm-\fn|}}{(1-t^2)^2} (b_1^{-2k} w_1)^\fm w_3^\fn~.
}
The analogous expression with $\chi_2 = -1$ can be obtained from \eref{limindD1C2D1fluxesmn} by setting $\fn=0$, replacing the contribution $\PE[t^2]=\frac{1}{1-t^2}$ due to the adjoint chiral multiplet with $\frac{1}{PE_F[t^2]} = \frac{1}{1+t^2}$, and multiplying by an overall phase factor $\left(-1\right)^{\abs{\fm}}$ arising from the chiral fields with axial fugacity $\fa_2 = a^{-1}$ connecting the $\USp(4)_k$ gauge node with the $\SO(2)$ gauge node in \eref{quivpuzzle1}. Observe, instead, that $\chi_1$ is always equal to one when this set of magnetic fluxes is considered, since the $\SO(2)_{-2 k}$ flux $\fm$ is different from zero. Explicitly, the result refined with respect to the charge conjugation symmetry fugacities $\chi_1$ and $\chi_2$ for gauge groups $\SO(2)_{2k}$ and $\SO(2)$ in \eqref{quivT3theoryorthopuzzle1}, respectively, is
\bes{ \label{fluxconfigachi}
&H_{m_1 = \fm,\vec{m}_2 = (\fm, 0), m_3 =\fn } \\
&= \frac{1}{2}\sum_{\fm \in \mathbb{Z}_{\neq 0}} \chi_2^{\fm} \sum_{\fn \in \mathbb{Z}} \mathcal{D}(t;\chi_1=+1) \mathcal{D}(t;\chi_2) (b_1^{-2k} w_1)^\fm w_3^\fn \\
& \qquad \quad \quad \times t^{2(F_1+F_2+k-2)|\fm|+2(F_3+1)|\fn|+|\fm+\fn|+|\fm-\fn|} \Big|_{\text{$\fn=0$ if $\chi_2=-1$}}~,
}
where
\bes{ \label{DressingfactorSO2withchi}
\mathcal{D}(t; \chi) = \begin{cases}
P_{\SO(2)}(t)= \PE[t^2] = (1-t^2)^{-1}~, &\quad  \chi =1 \\
2 P_{\USp(2)}(t)- P_{\SO(2)}(t) = \frac{1}{\PE_F[t^2]}= (1+t^2)^{-1}  &\quad \chi = -1 \\
\end{cases}~,
}
defined as in \eqref{NS5limCiD1C1chiplus} and \eqref{NS5limCiD1C1chiminus}. We emphasise that, in the expression \eref{fluxconfigachi}, $\chi_1$ is always equal to $+1$ and, when $\chi_2=-1$, we have to set the flux $\fn=0$. 

For $m_1 = 0,\vec{m}_2 = (0, 0), m_3=\fn$, the $\USp(4)$ gauge group is unbroken and the corresponding contribution to the limit of the index with $\chi_1 = \chi_2 = +1$ reads
\bes{ \label{fluxconfigb}
&H_{m_1 = 0,\vec{m}_2 = (0, 0), m_3=\fn} = \frac{1}{8}  \sum_{\fn \in \BZ}  t^{2 \Delta(m_1=0,\   \vec m_2=(0, 0), \, m_3 = \fn)}\oint\frac{du_1}{2\pi i u_1} \oint  \frac{du_3}{2\pi i u_3}w_3^\fn  \\
&\,\, \times   \oint    \prod_{\alpha=1}^2 \frac{du^{(\alpha)}_2}{2\pi i u^{(\alpha)}_2} \prod_{i=1}^2\prod_{s=\pm 1}\left[1-\left(u_2^{(i)}\right)^{2s}\right] \prod_{s_1,s_2=\pm 1}\left[1-\left(u_2^{(1)}\right)^{s_1}\left(u_2^{(2)}\right)^{s_2}\right] \\& \times \PE\left[t^2+t \sum_{s1,s2=\pm 1} \left[\left( b_1^{s_1}  u_1 ^{s_1} (u^{(i)}_{2})^{s_2}\right) \right] \right]~, }
where, in this case, we have
\bes{
\Delta(m_1=0,\   \vec m_2=(0, 0), \, m_3 = \fn) = (2 F_3+4) |\fn|~.
}
When $\chi_2=-1$, the expression \eref{fluxconfigb} gets modified analogously to the discussion below \eref{fluxconfiga}: we have to set the flux $\fn=0$ and replace $\PE[t^2]=\frac{1}{1-t^2}$ with $\PE_F[t^2]=\frac{1}{1+t^2}$, as in \eqref{DressingfactorSO2withchi}. On the other hand, when $\chi_1 = -1$, we have to set $u_1 = 1$ and $u^{-1}_1 = -1$ in the contribution due to the plethystic exponential in the last line of \eref{fluxconfigb}, which then contains the terms $\prod_{i=1}^2 \prod_{s = \pm 1} \left(1-t (u^{(i)}_2)^s\right)^{-1} \left(1+t (u^{(i)}_2)^s\right)^{-1}$. As above, we obtain the following result refined with the charge conjugation fugacities $\chi_1$ and $\chi_2$:
\bes{
H_{m_1 = 0,\vec{m}_2 = (0, 0), m_3=\fn}= \sum_{\fn \in \BZ} t^{2(F_3+2)|\fn|} \mathcal{D}(t;\chi_1) \mathcal{D}(t;\chi_2) w_3^\fn~, 
}
where $\mathcal{D}(t; \chi)$ is given by \eref{DressingfactorSO2withchi}. We remark that, when $\chi_2=-1$, we have to set the flux $\fn=0$.

Summing the three contributions as indicated in \eqref{eq1121puzzle1}, we obtain
\bes{ \label{HSNS5branchpuzzle1}
&H_{\eqref{quivpuzzle1}}[\text{NS5}](t;\{w_{i}\},\{b_{i}\}) \\
&=\sum_{\fm, \fn \in \mathbb{Z}}t^{2(F_1+F_2+k-2)|\fm|+2(F_3+1)|\fn|+|\fm+\fn|+|\fm-\fn|} \\
& \qquad \quad \quad \times \mathcal{D}(t;\chi_1) \mathcal{D}(t;\chi_2) \chi_2^\fm (b_1^{-2k} w_1)^\fm w_3^\fn \Big|_{\text{$(\fm, \fn)=0$ if $(\chi_1, \chi_2)=-1$}}~,
}
which is the Coulomb branch Hilbert series of the following quiver:\footnote{Here $(\fm, \fn)$, $(\chi_1, \chi_2)$ are the quantities associated with the left and right $\SO(2)$ nodes, respectively. Note that only the phase $\chi_2^\fm$ (but not $\chi_1^{\fn}$) appears. The explanation is due to the fact that, as summarised in \eref{condflux2411}, the magnetic flux $m_3 = \fn$ associated with the rightmost $\SO(2)$ gauge node in theory \eref{quivT3theoryorthopuzzle1} is unconstrained, hence, when $\chi_2 = -1$ and, consequently, $\fn = 0$, a non-trivial phase factor $(-1)^{\abs{\fn - m^{(1)}_2}} = (-1)^{\abs{\fm}}$ can arise in the $\fz \rightarrow 0$ limit of the index due to the contribution associated with the chiral fields with axial fugacity $\fa_2 = a^{-1}$ in \eqref{quivpuzzle1}. On the other hand, the magnetic flux $m_1 = \fm$ associated with the leftmost $\SO(2)_{-2 k}$ gauge node in theory \eref{quivT3theoryorthopuzzle1} is forced to be equal to one of the two $\USp(4)$ fluxes, say $m^{(1)}_2$, with the other flux $m^{(2)}_2$ being equal to zero, see \eref{condflux2411}. Hence, upon taking the $\fz \rightarrow 0$ limit when $\chi_1 = -1$ and, consequently, $\fm = 0$, the chiral fields with axial fugacity $\fa_1 = a$ in \eqref{quivpuzzle1} only give rise to trivial phases $(-1)^{\abs{\fm-m^{(1)}_2}}=1$ and $(-1)^{\abs{\fm-m^{(2)}_2}}=1$.}
\begin{equation} \label{mquiverNSpuzzle1}
\scalebox{1}{
\begin{tikzpicture}[baseline, font=\scriptsize,
    circ/.style={circle, draw, minimum size=1cm},
    sq/.style={rectangle, draw, minimum size=1cm},
    node distance=1cm
]
    \node[circ] (n1) {\scriptsize $\SO(2)$};
    \node[circ, right=of n1] (n2) {\scriptsize $\SO(2)$};
    \node[sq, left=of n1] (s1) {$k+F_1+F_2-2$};
    \node[sq, right=of n2] (s2) {$F_3+1$};
    \draw (n1) -- (n2);
    \draw (n1) -- (s1);
    \draw (n2) -- (s2);
\end{tikzpicture}}
\end{equation}
Note that the half-hypermultiplet in the representation $[\mathbf{2}; \mathbf{2}]$ of $\SO(2) \times \SO(2)$ is exactly as described in \cite[(2.5), (2.6)]{Carta:2021dyx} with $M=1$. In particular, it contributes to the dimension of the monopole operator as $\frac{1}{2}(|\fm+\fn|+|\fm-\fn|)$. For $F_1=F_2=F_3=0$, this matches the magnetic quiver proposed in \cite[v1, (5.2)]{Marino:2025ihk} with $M=N=L=1$.

\paragraph{Further comments on the magnetic quiver.} Following the appearance of this paper on the arXiv, Ref. \cite[v2]{Marino:2025ihk} was released. The authors of that reference moved the discussion regarding theory \eref{quivT3theoryorthopuzzle1} to Section 4.2.2, where they consider the special case $F_1 = F_2 = F_3 = 0$. Consequently, they no longer provide \eref{mquiverNSpuzzle1} as a magnetic quiver for the NS5-branch of \eref{quivT3theoryorthopuzzle1}. Their revised magnetic quiver candidate is instead given by \cite[v2, (4.22)]{Marino:2025ihk}, which, for generic $F_1$, $F_2$, and $F_3$, reads
\begin{equation} \label{newmquiverNSpuzzle1}
\scalebox{1}{
\begin{tikzpicture}[baseline, font=\scriptsize,
    circ/.style={circle, draw, minimum size=1cm},
    sq/.style={rectangle, draw, minimum size=1cm},
    node distance=1cm
]
    \node[circ] (n1) {\scalebox{0.9}{$\USp(2)$}};
    \node[circ, right=of n1] (n2) {\scriptsize $\SO(2)$};
    \node[sq, left=of n1] (s1) {$k+F_1+F_2$};
    \node[sq, right=of n2] (s2) {$F_3+1$};
    \draw (n1) -- (n2);
    \draw (n1) -- (s1);
    \draw (n2) -- (s2);
\end{tikzpicture}}
\end{equation}
whose Coulomb branch Hilbert series can be obtained from \eref{HSNS5branchpuzzle1} by restricting the summation to run over $\fm \in \BZ_{\ge 0}$ and by replacing $\mathcal{D}(t;\chi_1)$ with $P_{\USp(2)} (t; \fm)$. As observed from \eref{DressingfactorSO2withchi}, this modification is exactly equivalent to summing the fugacity $\chi_1$ over $\pm 1$ and dividing by two in \eref{HSNS5branchpuzzle1}, see also the discussion around \eref{NS5limc1D1C1O2}. As a result, we conclude that \eref{newmquiverNSpuzzle1} is a magnetic quiver capturing the NS5-branch of the variant of theory \eref{quivT3theoryorthopuzzle1} in which the charge conjugation symmetry associated with the $\SO(2)_{2 k}$ node is gauged, thus turning it into an $\O(2)_{2 k}$ node.

A magnetic quiver describing the NS5-branch of theory \eref{quivT3theoryorthopuzzle1}, where the leftmost gauge node is {\it truly} $\SO(2)_{2 k}$, can also be realised starting from \eref{newmquiverNSpuzzle1} by turning on the background flux $\vec{\nu}_b= (b, 0, \ldots, 0)$, with $b \in \{0, 1 \}$, for the $\SO(2k + 2 F_1 + 2 F_2)$ flavour node. Denoting the magnetic and charge conjugation fugacities associated with the $\SO(2k + 2 F_1 + 2 F_2)$ flavour node of \eref{newmquiverNSpuzzle1} by $\tilde{\zeta}_1$ and $\tilde{\chi}_1$, respectively, the Coulomb branch Hilbert series of \eref{newmquiverNSpuzzle1} summed over $b \in \{0, 1\}$ equals the NS5-branch limit \eref{HSNS5branchpuzzle1}, provided the identification $\tilde{\zeta}_1 = \chi_1$ and $\tilde{\chi}_1 = \zeta_1 \chi_1$ holds, see \cite[v2, (4.23)]{Marino:2025ihk}. Specifically, the contribution due to $\vec{\nu}_1$ to the Coulomb branch Hilbert series of \eref{newmquiverNSpuzzle1} coincides with the terms in the NS5-branch limit of the index of \eref{quivT3theoryorthopuzzle1} that are projected out upon gauging the charge conjugation symmetry with fugacity $\chi_1$. This implies the following equality:
\bes{ \label{eqbackgroundflux}
&H_{\eqref{newmquiverNSpuzzle1}}[\text{Coulomb}](\tilde{\zeta}_1 = \chi_1, \tilde{\chi}_1 = \zeta_1 \chi_1)\Big|_{\vec{\nu}_b=\vec{\nu}_1 = (1, 0, \ldots, 0)} \\&= \eref{HSNS5branchpuzzle1}_{\chi_1} - \frac{1}{2} \Big[\eref{HSNS5branchpuzzle1}_{\chi_1 = 1} + \eref{HSNS5branchpuzzle1}_{\chi_1 = -1}\Big]_{\zeta_1 \rightarrow \zeta_1 \chi_1}~,
}
where we point out that $\zeta_1$ must be redefined as $\zeta_1 \chi_1$ in the contribution appearing inside the square brackets on the second line of the above identity.\footnote{Recall that, without turning on any background flux for the flavour symmetry, \eqref{newmquiverNSpuzzle1} is a magnetic quiver for the NS5-branch of theory \eref{quivT3theoryorthopuzzle1} with the $\SO(2)_{2 k}$ gauge node replaced by $\O(2)_{2 k}$, whose Hilbert series is given by the quantity appearing inside the square brackets in \eref{eqbackgroundflux}, divided by two. Note that this quantity does not depend on $\chi_1$, since the corresponding charge conjugation symmetry has been gauged. On the other hand, the Coulomb branch Hilbert series of \eqref{newmquiverNSpuzzle1} refined with $\tilde{\zeta}_1$ and $\tilde{\chi}_1$ contains a term $\tilde{\chi}_1^\fm = \zeta_1^\fm \chi_1^\fm$, even if $\vec{\nu}_1$ is turned off. The redefinition $\zeta_1 \rightarrow \zeta_1 \chi_1$ is thus necessary to match the two Hilbert series.} The validity of \eref{eqbackgroundflux} can be checked for generic values of the fugacities $\zeta_1$, $\zeta_2$, $\chi_1$, and $\chi_2$ by expanding both sides up to a sufficiently high order in $t$.\footnote{Actually, for the case $\chi_2 = -1$, the validity of \eref{eqbackgroundflux} can be easily shown analytically using the identity
\bes{
\sum_{\fm \in \BZ_{>0}} t^{2(F_1+F_2+k-2)|\fm|+|\fm+1|+|\fm-1|} = \sum_{\fm \in \BZ_{>0}} t^{2(F_1+F_2+k-1)|\fm|}~.
}}


\subsubsection*{The D5-branch}
Finally, the Hilbert series for the D5-branch is
\bes{ \label{eq:27114}
\scalebox{0.9}{$
\begin{split}
&H_{\eqref{quivpuzzle1}}[\text{D5}](t;b_{1,2}) = \oint \frac{du_1}{2\pi i u_1} \left(\frac{1}{8} \prod_{\alpha=1}^2 \frac{d u_2^{(\alpha)}}{2\pi i  u_2^{(i)}} \right) \oint\frac{du_3}{2\pi i u_3} \\
& \,\, \times \prod_{i=1}^2\prod_{s=\pm1}\left(1-(u_2^{(i)})^{2s}\right) \prod_{s_1,s_2=\pm1}\left(1-(u_2^{(1)})^{s_1}(u_2^{(2)})^{s_2}\right) \\
& \,\, \times \PE\Bigg[t \sum_{s_1,s_2=\pm 1} \left\{ \sum_{i=1}^2 b_1^{s_1} u_1^{s_1} (u_{2}^{(i)} )^{s_2}+ b_2^{s_2} u_3^{s_1} (u_{2}^{(i)} )^{s_2}\right\} \\
& \,\,\, \qquad + t\sum_{s_1,s_2=\pm 1}\Big\{\sum_{\rho_j=1}^{F_1} ( u_1)^{s_1}(f_1^{(\rho_1)})^{s_2}+\sum_{i=1}^2\sum_{\rho_j=1}^{F_2} ( u_2^{(i)})^{s_1}(f_2^{(\rho_2)})^{s_2}  +  \sum_{\rho_j=1}^{F_3} ( u_3)^{s_1}(f_3^{(\rho_3)})^{s_2} \Big \} \\
& \,\,\, \qquad -2t^2  -t^2 \Bigg\{\sum_{s_1,s_2=\pm 1}(u_2^{(1)})^{s_1}(u_2^{(2)})^{s_2} +\sum_{i=1}^2\sum_{s=\pm 1}(u_2^{(i)})^{2s}+2\Bigg\} \Bigg]~.
\end{split}$}
}
The coefficient of $-t^2$ inside the plethystic exponential accounts for the adjoint chiral multiplets associated with the three gauge nodes, which have an axial fugacity of $a^{-2}$. This result corresponds to the Higgs branch Hilbert series of the theory in \eqref{quivpuzzle1} with the Chern-Simons levels set to zero ($k=0$), which is the 3d $\CN=4$ theory
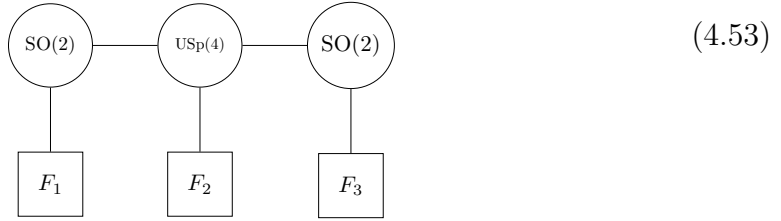
\begin{equation} \label{quivT3theoryortho}
\scalebox{0.85}{
\begin{tikzpicture}[baseline,font=\footnotesize,
    circ/.style={circle, draw, minimum size=1.3cm},
    sq/.style={rectangle, draw, minimum size=1cm},
    node distance=1cm
]
    \node[circ] (n1) {\scalebox{0.85}{$\SO(2)$}};
    \node[circ, right=of n1] (n2) {\scalebox{0.7}{$\Usp(4)$}};
    \node[circ, right=of n2] (n3) {$\SO(2)$};
    \node[sq, below=of n1] (s1) {$F_1$};
    \node[sq, below=of n2] (s2) {$F_2$};
    \node[sq, below=of n3] (s3) {$F_3$};
    \draw (n1) -- (n2);
    \draw (n2) -- (n3);
    \draw (n1) -- (s1);
    \draw (n2) -- (s2);
    \draw (n3) -- (s3);
\end{tikzpicture}}
\end{equation}
The refinement with respect to the charge conjugation symmetry fugacities is straightforward.

\subsection{Linear quiver with $\USp(2)_{k} \times \SO(5)_{-2k} \times \USp(2)$ gauge group}
We consider the 3d $\mathcal{N}=3$ theory given by the following quiver diagram:
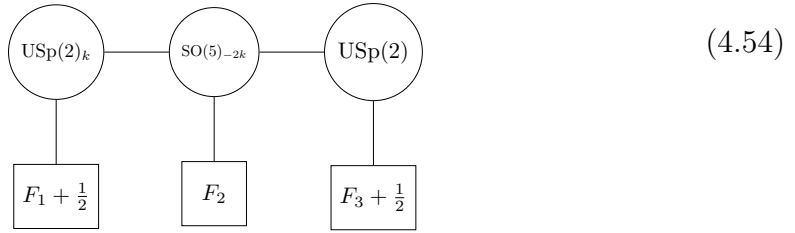
\begin{equation} \label{quivT3theoryorthopuzzle2}
\scalebox{0.85}{
\begin{tikzpicture}[baseline,font=\footnotesize,
    circ/.style={circle, draw, minimum size=1.3cm},
    sq/.style={rectangle, draw, minimum size=1cm},
    node distance=1cm
]
    \node[circ] (n1) {\scalebox{0.85}{$\USp(2)_{k}$}};
    \node[circ, right=of n1] (n2) {\scalebox{0.7}{$\SO(5)_{-2k}$}};
    \node[circ, right=of n2] (n3) {$\USp(2)$};
    \node[sq, below=of n1] (s1) {$F_1+\frac{1}{2}$};
    \node[sq, below=of n2] (s2) {$F_2$};
    \node[sq, below=of n3] (s3) {$F_3+\frac{1}{2}$};
    \draw (n1) -- (n2);
    \draw (n2) -- (n3);
    \draw (n1) -- (s1);
    \draw (n2) -- (s2);
    \draw (n3) -- (s3);
\end{tikzpicture}}
\end{equation}
which can be realised in string theory via the brane configuration shown below:
\begin{equation}
\scalebox{0.9}{
\begin{tikzpicture}[baseline,font=\footnotesize,
    cross_node/.style={circle, draw=cyan, cross out, thick, minimum size=6pt, inner sep=0pt}
]
    \def\ystart{-1.5}
    \def\yend{1.5}
    \draw[very thick] (0,0) -- node[near start, below,xshift=-0.5cm]{D3} (7.5,0);
    \draw[very thick] (2.57,-0.35) --(4.8,-0.35);
    \node[brown] at (-1,-2.35){$\tilde{\O3}^-$};
    \node[darkred] at (1,-2.35){$\tilde{\O3}^+$};
    \node[brown] at (3.5,-2.35){$\tilde{\O3}^-$};
    \node[darkred] at (6.5,-2.35){$\tilde{\O3}^+$};
    \node[brown] at (8.5,-2.35){$\tilde{\O3}^-$};
    \draw[red, thick] (0,\ystart) -- (0,\yend) node[pos=0, below, red] {NS5};
    \draw[red, thick] (4.8,\ystart) -- (4.8,\yend) node[pos=0, below, red] {NS5};
    \draw[red, thick] (0,\ystart) -- (0,\yend) node[pos=0, below, red] {NS5};
    \draw[red, thick] (7.5,\ystart) -- (7.5,\yend) node[pos=0, below, red] {NS5};
    \draw[blue, dashed, thick] (1+1, \ystart) -- (2.5+1, \yend) node[pos=0, below, blue] {$(1, 2k)$};

    \node[cross_node, label={[cyan]above:$F_1$ D5}] at (1.5, 0.5) {};
    \node[cross_node, label={[cyan]above:$F_2$ D5}] at (4, 0.5) {};
    \node[cross_node, label={[cyan]above:$F_3$ D5}] at (6, 0.5) {};
\end{tikzpicture}}
\end{equation}
Note that, for $F_1,F_2,F_3=0$, this is the $\mathcal{N}=4$ theory considered in \cite[v1, section 5]{Marino:2025ihk}. We analyse this theory by computing its superconformal index. As usual, in order to compute the various limits of the index we assign axial fugacities to the chiral fields. In 3d $\mathcal{N}=2$ notation, the field content and fugacity assignments are
\begin{equation} \label{quivpuzzle2}
\scalebox{0.85}{
\begin{tikzpicture}[baseline, font=\footnotesize,
    circ/.style={circle, draw, minimum size=1.3cm},
    sq/.style={rectangle, draw, minimum size=1cm},
    node distance=1cm,
    every loop/.style={-}
]
    \node[circ] (n1) {\scalebox{0.85}{$\USp(2)_{k}$}};
    \node[circ, right=of n1] (n2) {\scalebox{0.7}{$\SO(5)_{-2k}$}};
    \node[circ, right=of n2] (n3) {$\USp(2)$};
    \node[sq, below=of n1] (s1) {\scriptsize $\SO(2F_1+1)$};
    \node[sq, below=of n2] (s2) {\scriptsize $\USp(2F_2)$};
    \node[sq, below=of n3] (s3) {\scriptsize $\SO(2F_3+1)$};
    \draw[-] (n1) -- node[above, red] {$\fa_1$} (n2);
    \draw[-] (n2) -- node[above, red] {$\fa_2$} (n3);
    \draw[-] (n1) -- node[right, red] {$\fc_1$} (s1);
    \draw[-] (n2) -- node[right, red] {$\fc_2$} (s2);
    \draw[-] (n3) -- node[right, red] {$\fc_3$} (s3);
    \path (n1) edge [loop above] node[red] {$\fb_1$} ();
    \path (n2) edge [loop above] node[red] {$\fb_2$} ();
    \path (n3) edge [loop above] node[red] {$\fb_3$} ();
\end{tikzpicture}}
\end{equation}
where the specific assignments of axial fugacities $\fa_i$, $\fb_i$, and $\fc_i$ for each branch are summarised in the following table:
\bes{ \label{branchpuzzle2}
\begin{tabular}{c|cccccccccc}
\hline
Branch & $\fa_1$ & $\fa_2$  & $\fc_1$ & $\fc_2$ & $\fc_3$ & $\fb_1$ & $\fb_2$ & $\fb_3$  \\
\hline
NS5& $a$ & $a^{-1}$ & $a^{-1}$ & $a^{-1}$ & $a^{-1}$ & 1 & 1 & $a^2$ \\
D5 & $a$ & $a$ & $a$ & $a$ & $a$ & $a^{-2}$ & $a^{-2}$ & $a^{-2}$ \\
\hline
\end{tabular}
}
The index for the theory in \eqref{quivpuzzle2} is then given by
\bes{ \label{indexquivpuzzle2}
\scalebox{0.92}{$
\begin{split}
\CI_{\eqref{quivpuzzle2}} &= \frac{1}{32} \sum_{m_1,m_3 \in \BZ} \sum_{\{m^{(\alpha)}_2\} \in \BZ^2} \oint  \frac{du_1}{2\pi i u_1} u_1^{2\kappa_1 m_1} \Bigg( \prod_{\alpha=1}^2 \frac{du^{(\alpha)}_2}{2\pi i u^{(\alpha)}_2} (u^{(\alpha)}_2)^{2\kappa_2 m^{(\alpha)}_2} \Bigg) \frac{du_3}{2\pi i u_3} \\
  & \qquad \qquad \times \zeta^{m^{(1)}_2 + {m^{(2)}_2}}  \times Z_{\text{vec}} \,\, Z_{\text{adj-chir}} \,\, Z_{\text{matter}}~,
\end{split}$}
}
where the Chern-Simons levels are
\bes{ \label{kappaT3}
\kappa_1 = k >0 ~, \qquad \kappa_2 = -k~, 
}
and we define the following abbreviated notations:
\bes{
\scalebox{0.9}{$
\begin{split}
Z_{\text{vec}} &= Z_{\text{vec}}^{\SO(5)}(x; \{u^{(\alpha)}_2\}; \{m^{(\alpha)}_2\}; \chi)\,  Z_{\text{vec}}^{\USp(2)}(x; \{u_1\}; \{m_1\}) \, Z_{\text{vec}}^{\USp(2)}(x; \{u_3\}; \{m_3\})~,
\end{split}$}
}
\bes{
\scalebox{0.9}{$
\begin{split}
&Z_{\text{adj-chir}} = \prod_{s_1,s_2=\pm1}\prod_{1\leq\alpha < \beta \leq 2} Z_{\text{chir}}^{1} \left(x; \fb_2 (u^{\alpha}_2)^{s_1} (u^{\beta}_2)^{s_2} ; s_1m^{(\alpha)}_2 +s_2 m^{(\beta)}_2  \right) \\ 
& \qquad \qquad \times \prod_{s=\pm1}\prod_{\alpha=1}^2 Z_{\text{chir}}^{1} \left(x; \fb_2 (u^{\alpha}_2)^{s} ; sm^{(\alpha)}_2\right) Z_{\text{chir}}^{1} \left(x; \fb_2 ; 0  \right)^2 Z_{\text{chir}}^{1} \left(x; \fb_1 ; 0  \right) \\
& \qquad \qquad \times \prod_{s=\pm 1}Z_{\text{chir}}^{1} \left(x; \fb_1 u_1^{2s}; 2sm_1  \right) Z_{\text{chir}}^{1} \left(x; \fb_3 ; 0  \right)\prod_{s=\pm 1}Z_{\text{chir}}^{1} \left(x; \fb_3 u_3^{2s}; 2sm_3  \right)~,
\end{split}$}
}
\bes{
\scalebox{0.9}{$
\begin{split}
&Z_{\text{matter}} = \prod_{s_1 = \pm1} \prod_{s_2 = \pm1}\prod_{\alpha=1}^2 Z_{\text{chir}}^{1/2} \left(x; \fa_1  u_1^{s_1} (u^{(\alpha)}_{2})^{s_2}; s_1m_{1} +s_2 m^{(\alpha)}_{2} \right) \\ 
& \qquad \times \prod_{s = \pm1} Z_{\text{chir}}^{1/2} \left(x; \fa_1  \chi u_1^{s}; sm_{1}  \right) \prod_{s_1 = \pm1}\prod_{s_2 = \pm1}  \prod_{\alpha=1}^2 Z_{\text{chir}}^{1/2} \left(x; \fa_2  (u^{(\alpha)}_2)^{s_1} u_{3}^{s_2}; s_1 m^{(\alpha)}_{2} +s_2 m_3 \right) \\
& \qquad \times \prod_{s = \pm1} Z_{\text{chir}}^{1/2} \left(x; \fa_2 \chi u_3^{s}; sm_{3}  \right) \prod_{s_1 = \pm1}\prod_{s_2 = \pm1}  \prod_{\rho_2=1}^{F_2} \prod_{\alpha=1}^2  Z_{\text{chir}}^{1/2} \left(x; \fc_2 ( u^{(\alpha)}_2)^{s_1}(f_2^{\rho_2})^{s_2}; s_1m^{(\alpha)}_{2} \right) \\
& \qquad \times \prod_{s = \pm1} \prod_{\rho_2=1}^{F_2}   Z_{\text{chir}}^{1/2} \left(x;\fc_2 \chi (  f_2^{\rho_2})^{s}; 0 \right) \prod_{s_1 = \pm1}\prod_{s_2 = \pm1}  \prod_{\rho_1=1}^{F_1} Z_{\text{chir}}^{1/2} \left(x; \fc_1 (u_1)^{s_1}(f_1^{(\rho_1)})^{s_2}; s_1m_{1} \right) \\ 
& \qquad \times \prod_{s = \pm1} \prod_{s_1 = \pm1}\prod_{s_2 = \pm1}\prod_{\rho_3=1}^{F_3} Z_{\text{chir}}^{1/2} \left(x; \fc_3 (u_3)^{s_1}(f_3^{(\rho_3)})^{s_2}; s_1m_{3} \right)~,
\end{split}$}
}
with 
\bes{
\scalebox{0.95}{$
\begin{split}
Z_{\text{vec}}^{\SO(5)}(x; \vec z; \vec m; \chi)
&= x^{-|m_1|-|m_2|-|m_1- m_2|-|m_1+m_2|} \prod_{j=1}^2 \prod_{s=\pm 1} \left(1- (-1)^{s m_j} x^{|m_j|} z_j^{s} \chi \right)\\
& \quad \times \prod_{s_1, s_2 = \pm 1} \left(1- (-1)^{s_1 m_1+s_2 m_2} x^{|s_1 m_1+s_2 m_2|} z_1^{s_1} z_2^{s_2}  \right)~.
\end{split}$}
}

The Hilbert series for each branch is obtained as follows:
\bes{ \label{limitHSorthosymppuzzle2}
H_{\eqref{quivpuzzle2}}[\text{branch}](t;\{w_i\}, \{b_i\}, \{f_i \}) = \lim_{\fz \rightarrow 0} \CI_{\eqref{quivpuzzle2}} \Big|_{\eqref{branchpuzzle2}, \, \eqref{N=3reparam}}~.
}
To compute the limit, we consider the exponent of $\fz$ in the integrand for each branch. A non-vanishing result in the $\fz \to 0$ limit sets such exponents to zero, and thus imposes the following constraints on the magnetic fluxes $m_i$:
\bes{ \label{condflux4121}
\scalebox{0.8}{
\begin{tabular}{c|l|l}
\hline
Branch & Exponent of $\fz$ & Condition \\
\hline
NS & ${\sum_{i=1}^2\left(|m_1\pm m_2^{(i)}|-|m_2^{(i)}|\right)-|m_2^{(1)}\pm m_2^{(2)}|-|m_1|}$ & $m_1=\pm m_2^{(i)}~,\,\, m_2^{(j\neq i)}=0$\\
D5 & ${\sum_{i=1}^2(|m_1\pm m_2^{(i)}|+|m_3\pm m_2^{(i)}|)-2|m_2^{(1)}\pm m_2^{(2)}|} $& $m_1=m_2^{(1)}=m_2^{(2)}=m_3=0$ \\
& $-2|m_3|-2|m_1|+2F_1|m_1|+2F_2\sum_{i=1}^2(|m_2^{(1)}|)$ &  \\
& $+2F_3|m_3|-3\sum_{i=1}^2|m_2^{(i)}|$ & \\
\hline
\end{tabular} }
}
where $|m_i\pm m_j|=|m_i+m_j|+|m_i-m_j|$, and, in the first case, there is no restriction on $m_3$. 

\subsubsection*{The NS5-branch}
The condition obtained in Table \eqref{condflux4121} for the NS branch limit of the index is
\bes{m_1=\pm m_2^{(i)}~,\quad m_2^{(j\neq i)}=0~, \,\, i,j=1,2~.}
Using the Weyl symmetry of the $\SO(5)$ gauge group, we can restrict our attention to the option $m_1=\pm m_2^{(1)}, m_2^{(2)}=0$. Taking the limit \eqref{limitHSorthosymppuzzle2}, the Hilbert series receives contributions from the following sets of fluxes:
\begin{enumerate}
    \item $m_1 = 0~,\quad \vec{m}_2 = (0, 0)~, \quad m_3 =0~,$
    \item $m_1 = \fm~, \quad \vec{m}_2 = (\pm\fm, 0)~, \quad m_3 =0~,$
    \item $m_1 = \fm~, \quad \vec{m}_2 = (\pm \fm, 0)~, \quad m_3 =\fn~,$
    \item $m_1 = 0~, \quad \vec{m}_2 = (0, 0)~, \quad m_3 =\fn~,$
\end{enumerate}
with $\fm,\fn \neq  0$. Note that the contributions from the $+$ and $-$ signs in the second and third cases are equal, so the total Hilbert series is given by
\bes{\label{eq1121}
H_{\eqref{quivpuzzle2}}[\text{NS5}](t;\{w_{i}\};\{b_{i}\})
&=H_{m_1 = 0,\vec{m}_2 = (0, 0), m_3 =0}+2 \times H_{m_1 = \fm,\vec{m}_2 = (\fm, 0), m_3 =0}\\
& + 2 \times H_{m_1 = \fm,\vec{m}_2 = (\fm, 0), m_3 =\fn}+H_{m_1 = 0,\vec{m}_2 = (0, 0), m_3 =\fn}~.
}
For the fluxes $m_1 = 0,\vec{m}_2 = (0, 0), m_3=0$, all the gauge groups are unbroken. In this case, the limit of the index is
\bes{ 
& H_{m_1 = 0,\vec{m}_2 = (0, 0), m_3 =0}
= \frac{1}{32}  t^{2 \Delta(m_1=0,\   \vec m_2=(0, 0), \, m_3 = 0)}  \times \\
& \quad \times \oint  \frac{du_1}{2\pi i u_1} \oint  \frac{du_3}{2\pi i u_3} \oint    \prod_{\alpha=1}^2 \frac{du^{(\alpha)}_2}{2\pi i u^{(\alpha)}_2}   \prod_{s=\pm 1}\left[1-\left(u_1\right)^{2s}\right] \prod_{s=\pm 1}\left[1-\left(u_3\right)^{2s}\right] \\
& \quad\times \prod_{s_1,s_2=\pm 1}\left[1-(u_2^{(1)})^{s_1}(u_2^{(2)})^{s_2}\right]\prod_{s=\pm 1}\left[1-\chi (u_2^{(1)})^{s}\right]\prod_{s=\pm 1}\left[1- \chi (u_2^{(2)})^{s}\right]  \\&\quad \times \PE\left[t^2\left(u_3^2+1+\frac{1}{u_3^2}\right)+t\left(\chi \sum_{s=\pm 1} \left( u_1\right)^s+\sum_{i=1}^2\sum_{s_1,s_2=\pm 1} (u_1)^{s_1} (u^{(i)}_{2})^{s_2}      \right)  \right]~,
}
where the dimension of the monopole operators is given by
\bes{
\scalebox{0.95}{$
\begin{split}
&\Delta (m_1, m_2^{(1)}, m_2^{(2)}, m_3)\\
&= \frac{1}{2} \Biggl(
    -2|m_1|-2|m_3| +\sum_{i=1}^2 \sum_{s=\pm 1}\bigl( |m_3 +s m_2^{(i)}| +|m_1 +s m_2^{(i)}|\bigr) -2\sum_{i=1}^2 \bigl( |m_2^{(i)}|\bigr)
    \\&\qquad \,\, -2 |m_2^{(1)} - m_2^{(2)}| - 2|m_2^{(1)} + m_2^{(2)}| + 2F_1 |m_1| + 2F_2 \sum_{i=1}^2 |m_2^{(i)}|+ 2F_3 |m_3|\Biggr) ~.
\end{split}$}
} 
The contribution gets simplified to
\bes{H_{m_1 = 0,\vec{m}_2 = (0, 0), m_3 =0} = P_{\USp(2)}(t; 0) \times \PE[t^4] = \frac{1}{(1-t^4)^2}~,}
where $P_{\USp(2)}(t; 0) = \PE[t^4] =\frac{1}{1-t^4}$ is the contribution of $\tr(\phi^2)$, where $\phi$ is the adjoint chiral field in the vector multiplet of the rightmost $\USp(2)$ gauge group, and the other $\PE[t^4] = \frac{1}{1-t^4}$ factor is the contribution of the gauge invariant quantity $\epsilon^{\alpha_1 \alpha_2}\epsilon^{\alpha_3 \alpha_4} \delta_{i_1 i_3} \delta_{i_2 i_4} X^{i_1}_{\alpha_1}X^{i_2}_{\alpha_2}X^{i_3}_{\alpha_3}X^{i_4}_{\alpha_4}$, where $X^i_\alpha$ is the bifundamental chiral field in $\SU(2)_k \times \SO(5)_{-2k}$ with $i=1, 2, \ldots, 5$ and $\alpha=1, 2$.

For $m_1 = 0,\vec{m}_2 = (0, 0), m_3=\fn$, the $\SO(5)$ and the first $\USp(2)$ gauge group are unbroken, while the last $\USp(2)$ group is broken to $\U(1)$, and the corresponding limit of the index is
\bes{
\scalebox{1}{$
\begin{split}
&H_{m_1 = 0,\vec{m}_2 = (0,0), m_3=\fn} = \frac{1}{32}  \sum_{\fn \in \BZ_{\neq 0}}  t^{2 \Delta(m_1=0,\   \vec m_2=(0, 0), \, m_3 = \fn)}\,\, \chi^{\abs{\fn}} \\
& \quad \times \oint\frac{du_1}{2\pi i u_1}  \oint \frac{du_3}{2\pi i u_3}  \oint   \prod_{\alpha=1}^2 \frac{du^{(\alpha)}_2}{2\pi i u^{(\alpha)}_2}   \prod_{s=\pm 1}\left[1-\left(u_1\right)^{2s}\right]      \\
& \quad\times \prod_{s_1,s_2=\pm 1}\left[1-(u_2^{(1)})^{s_1}(u_2^{(2)})^{s_2}\right]\prod_{s=\pm 1}\left[1-\chi (u_2^{(1)})^{s}\right]\prod_{s=\pm 1}\left[1-\chi (u_2^{(2)})^{s}\right]  \\
&\quad \times \PE\left[t^2+t\left( \chi \sum_{s=\pm 1} \left( u_1\right)^s+\sum_{i=1}^2\sum_{s_1,s_2=\pm 1} (u_1)^{s_1} (u^{(i)}_{2})^{s_2}      \right)  \right]~, 
\end{split}$}
}
where, in this case, the dimension of the monopole operator simplifies to
\bes{
\Delta(m_1=0,\   \vec m_2=(0, 0), \, m_3 = \fn) = ( F_3+1) |\fn|~.
}
As above, we have
\bes{H_{m_1 = 0,\vec{m}_2 = (0, 0), m_3=\fn}=\sum_{\fn >0}\frac{t^{2(1+F_3)|\fn|)} \chi^{\fn}}{(1-t^2)(1-t^4)}~.
}
For $m_1 = \fm,\vec{m}_2 = (\fm, 0), m_3=0$, the $\SO(5)$ and the first $\USp(2)$ gauge group are broken, while the last $\USp(2)$ group is unbroken, and the corresponding contribution to the Hilbert series is\footnote{Note that we have an overall factor of $\frac{1}{16}$ here, in contrast to that of \eqref{indexquivpuzzle2} which is $\frac{1}{32}$. This is because the results for the allowed choices $\vec{m}_2 = (\fm,0)$ and $(0, \fm)$ are equal.}
\bes{\label{eq:15121}
&H_{m_1 = \fm,\vec{m}_2 = (\fm, 0), m_3=0} \\
&= \frac{1}{16}  \sum_{\fm \in \BZ_{\neq 0}}  t^{2 \Delta(m_1=\fm,\   \vec m_2=(\fm, 0), \, m_3 = 0)} \zeta^{\fm} \chi^{\abs{\fm}} \oint\frac{du_1}{2\pi i u_1}u_1^{2k_1\fm} \oint  \frac{du_3}{2\pi i u_3}  \\
&\quad \times   \oint  \Bigg(  \prod_{\alpha=1}^2 \frac{du^{(\alpha)}_2}{2\pi i u^{(\alpha)}_2}  (u^{(\alpha)}_2)^{2\kappa_2 m^{(\alpha)}_2}   \Bigg)_{\vec m_2=(\fm, 0)}    \prod_{s=\pm 1}\left[1-\chi (u_2^{(2)})^{s}\right] \\
&\quad \times \prod_{s=\pm 1}\left[1-\left(u_3\right)^{2s}\right] \,\, \PE\left[t^2\left(u_3^2+1+\frac{1}{u_3^2}\right)+t\left( \sum_{s=\pm 1} u_1^s (u^{(1)}_{2})^{-s}      \right)  \right]~, }
where, in this case, we have
\bes{
\Delta(m_1=\fm,\   \vec m_2=(\fm, 0), \, m_3 = 0) = (-1+F_1+F_2) |\fm|~.
}
As above, we find
\bes{\label{eq:15122}
H_{m_1 = \fm,\vec{m}_2 = (\fm, 0), m_3=0}=\frac{1}{2}\sum_{\fm >0}\frac{t^{2(-1+F_1+F_2+k)|\fm|)} \zeta^{\fm} \chi^{\fm}}{(1-t^2)(1-t^4)}~.
}
For $m_1 = \fm,\vec{m}_2 = (\fm, 0), m_3=\fn$, all gauge groups are broken to subgroups, and the corresponding limit of the index is
\bes{\label{eq:15121}
&H_{m_1 = \fm,\vec{m}_2 = (\fm, 0), m_3=\fn} = \frac{1}{16}  \sum_{\fn \in \BZ_{\neq 0}} \sum_{\fm \in \BZ_{\neq 0}} t^{2 \Delta(m_1=\fm,\   \vec m_2=(\fm, 0), \, m_3 = \fn)} \zeta^{\fm} \chi^{\abs{\fm}+\abs{\fn}}  \\
&\quad \times  \oint\frac{du_1}{2\pi i u_1}u_1^{2k_1\fm} \oint  \frac{du_3}{2\pi i u_3}  \oint  \Bigg(  \prod_{\alpha=1}^2 \frac{du^{(\alpha)}_2}{2\pi i u^{(\alpha)}_2}  (u^{(\alpha)}_2)^{2\kappa_2 m^{(\alpha)}_2}   \Bigg)_{\vec m_2=(\fm, 0)} \\
&\quad \times \prod_{s=\pm 1}\left[1-\chi (u_2^{(2)})^{s}\right] \,\, \PE\left[t^2+t\left( \sum_{s=\pm 1} u_1^s (u^{(1)}_{2})^{-s}      \right)  \right]~, }
where, in this case, the monopole dimension reduces to
\bes{
&\Delta(m_1=\fm,\   \vec m_2=(\fm, 0), \, m_3 = \fn) \\
&= \frac{1}{2}(-4|\fm|+|\fm-\fn|+|\fm+\fn|+2(F_1+F_2)|\fm|+2F_3|\fn|) ~.
}
As above, we have
\bes{\label{eq:15123}
H_{m_1 = \fm,\vec{m}_2 = (\fm, 0), m_3=\fn}=\frac{1}{2}\sum_{\fm,\fn >0}\frac{t^{2\left((-2+F_1+F_2+k)|\fm|+\frac{1}{2}|\fm-\fn|+\frac{1}{2}|\fm+\fn|+F_3|\fn|\right)} \zeta^{\fm} \chi^{\fm+\fn}}{(1-t^2)^2}~.}
Summing the four contributions as indicated in \eqref{eq1121}, we obtain
\bes{ \label{HSNS5branchpuzzle2}
&H_{\eqref{quivpuzzle2}}[\text{NS5}](t;\{w_{i}\},\{b_{i}\};\zeta, \chi) \\
&=\sum_{\fm, \fn >0}\frac{t^{2 \hat{\Delta}(\fm, \fn)} \zeta^{\fm} \chi^{\abs{\fm}+\abs{\fn}}}{(1-t^2)^2}+\sum_{\fm >0}\frac{t^{2 \hat{\Delta}(\fm, 0)} \zeta^{\fm} \chi^{\abs{\fm}}}{(1-t^2)(1-t^4)} \\
& +\sum_{\fn >0}\frac{t^{2 \hat{\Delta}(0, \fn)} \chi^{\abs{\fn}}}{(1-t^2)(1-t^4)}+\frac{1}{(1-t^4)^2} \\
&=\sum_{\fm, \fn \geq 0} t^{2\hat{\Delta}(\fm,\fn)} \zeta^{\fm} \chi^{\abs{\fm}+\abs{\fn}} P_{\USp(2)}(t; \fm) P_{\USp(2)}(t; \fn)~,
}
with
\bes{ \label{Deltahat}
\hat{\Delta}(\fm, \fn) =(-2+F_1+F_2+k)|\fm|+{\orangered \frac{1}{2}}|\fm-\fn|+{\orangered \frac{1}{2}}|\fm+\fn|+F_3|\fn|~,
}
where we highlight in {\orangered orange} the presence of factors of {\orangered $\frac{1}{2}$} in the monopole dimension. This turns out to be equal to the Coulomb branch Hilbert series of the following quiver:
\footnote{In this quiver, the magnetic symmetry associated with $\zeta$ is not manifest. We shall henceforth set it to unity in the subsequent discussion.}
\begin{equation} \label{magquiverNSpuzzle1}
\scalebox{1}{
\begin{tikzpicture}[baseline, font=\scriptsize,
    circ/.style={circle, draw, minimum size=1cm},
    sq/.style={rectangle, draw, minimum size=1cm},
    node distance=1cm
]
    \node[circ] (n1) {\scriptsize $\USp(2)$};
    \node[circ, right=of n1] (n2) {\scriptsize $\USp(2)$};
    \node[sq, left=of n1] (s1) {$\SO(2k+2F_1+2F_2)$};
    \node[sq, right=of n2] (s2) {$\SO(2F_3+4)$};
    \draw[orange-red] (n1) -- node[above]{$\frac{1}{2}$} (n2);
    \draw (n1) -- (s1);
    \draw (n2) -- (s2);
\end{tikzpicture}}
\end{equation}
where the line in {\orangered orange} with label {\orangered $\frac{1}{2}$} corresponds to the half-hypermultiplet in the representation $[\mathbf{2}; \mathbf{2}]$ of $\USp(2) \times \USp(2)$.\footnote{This should be contrasted with the line connecting two $\USp(2k)$ gauge group in quiver \cite[(2.29)]{Mekareeya:2015bla} (see also \cite[Section 4.2]{Intriligator:1997kq}, \cite[Figure 28]{Dey:2013fea}, and \cite[Figure 6]{Tachikawa:2014qaa}), whose Higgs branch describes the moduli space of instantons of $\SO(8)$ on $\BC^2/\BZ_2$. In that reference, such a line denotes the full hypermultiplet in the representation $[\mathbf{2k}; \mathbf{2k}]$ of $\USp(2k) \times \USp(2k)$. In the case of $\USp(2) \times \USp(2)$ with fluxes $\fm$ and $\fn$, such a full hypermultiplet contributes to the dimension of the monopole formula as $|\fm-\fn|+|\fm+\fn|$, without a factor of $1/2$.} This is in accordance with the factors of {\orangered $\frac{1}{2}$} in \eqref{Deltahat}. For $F_1=F_2=F_3=0$ this matches the magnetic quiver proposed in \cite[v1, (5.4)]{Marino:2025ihk} with $M=N=L=1$. Note that quiver \eqref{magquiverNSpuzzle1} can be rewritten as
\begin{equation} \label{magquiverNSpuzzle1a}
\scalebox{0.85}{
\begin{tikzpicture}[baseline, font=\scriptsize,
    circ/.style={circle, draw, minimum size=1cm},
    sq/.style={rectangle, draw, minimum size=1cm},
    node distance=1cm
]
    \node[circ] (n1) {\scriptsize $\USp(2)$};
    \node[circ, right=of n1] (n2) {\scriptsize $\USp(2)$};
    \node[sq, left=of n1] (s1) {$\SO(2k+2F_1+2F_2-1)$};
    \node[sq, right=of n2] (s2) {$\SO(2F_3+3)$};
    \node[sq, above=of n1] (s3) {$\SO(1)$};
    \node[sq, above=of n2] (s4) {$\SO(1)$};
    \draw[orange-red] (n1) -- node[above]{$\frac{1}{2}$} (n2);
    \draw (n1) -- (s1);
    \draw (n2) -- (s2);
    \draw (n1) -- (s3);
    \draw (n2) -- (s4);
\end{tikzpicture}}
\end{equation}
Gauging the charge conjugation symmetry associated with $\chi$, whose Hilbert series is $\frac{1}{2}\sum_{\chi=\pm 1} H_{\eqref{quivpuzzle2}}[\text{NS5}](t;\{w_{i}\},\{b_{i}\};\zeta=1, \chi)$, corresponds to the Coulomb branch of the following theory:
\begin{equation} \label{magquiverNSpuzzle1a}
\scalebox{0.85}{
\begin{tikzpicture}[baseline, font=\scriptsize,
    circ/.style={circle, draw, minimum size=1cm},
    sq/.style={rectangle, draw, minimum size=1cm},
    node distance=1cm
]
    \node[circ] (n1) at (0,0) {\scriptsize $\USp(2)$};
    \node[circ, right=of n1] (n2) {\scriptsize $\USp(2)$};
    \node[circ] (n3) at (1.1,1.5) {\scriptsize $\O(1)$};
    \node[sq, left=of n1] (s1) {$\SO(2k+2F_1+2F_2-1)$};
    \node[sq, right=of n2] (s2) {$\SO(2F_3+3)$};
    \draw[orange-red] (n1) -- node[above]{$\frac{1}{2}$} (n2);
    \draw (n1) -- (s1);
    \draw (n2) -- (s2);
    \draw (n1)--(n3);
    \draw (n2)--(n3);
\end{tikzpicture}}
\end{equation}

\paragraph{Further comments on the magnetic quiver.} Following the initial release of this work, an updated version of Ref.~\cite[v2]{Marino:2025ihk} appeared on the arXiv. In this revision, the discussion regarding theory \eref{quivT3theoryorthopuzzle2} was relocated to Section 4.2.2. Replacing the previous proposal \eref{magquiverNSpuzzle1}, a new candidate for the magnetic quiver capturing the NS5-branch of \eref{quivT3theoryorthopuzzle2} is presented in \cite[v2, (4.30)]{Marino:2025ihk}:\footnote{We note that \cite[v2]{Marino:2025ihk} explicitly considers the case $F_1 = F_2 = F_3 = 0$.}
\begin{equation} \label{magquiverNSpuzzle2v2}
\scalebox{1}{
\begin{tikzpicture}[baseline, font=\scriptsize,
    circ/.style={circle, draw, minimum size=1cm},
    sq/.style={rectangle, draw, minimum size=1cm},
    node distance=1cm
]
    \node[circ] (n1) {\scriptsize $\SO(3)$};
    \node[circ, right=of n1] (n2) {\scalebox{0.85}{$\USp(2)$}};
    \node[sq, left=of n1] (s1) {$\USp(2k+2F_1+2F_2-2)$};
    \node[sq, right=of n2] (s2) {$\SO(2F_3+3)$};
    \draw (n1) -- (n2);
    \draw (n1) -- (s1);
    \draw (n2) -- (s2);
\end{tikzpicture}}
\end{equation}
The Coulomb branch Hilbert series of this quiver matches that of \eref{HSNS5branchpuzzle2},\footnote{Observe that the $\SO(3)$ vector multiplet contributes $-\abs{\fm}$ to the monopole dimension in \eref{Deltahat}, whereas the $\USp(2)$ vector multiplet contributes $-2 \abs{\fm}$. However, accounting for the differing contributions from the hypermultiplets in \eref{magquiverNSpuzzle1} and \eref{magquiverNSpuzzle2v2}, the total monopole dimension of the latter equals that of the former, namely \eref{Deltahat}.} provided that the magnetic and charge conjugation fugacities, $\tilde{\zeta}$ and $\tilde{\chi}$, satisfy the identification $\tilde{\zeta} = \zeta \chi$ and $\tilde{\chi} = \chi$.\footnote{With these assignments, the fugacities appear in the Coulomb branch Hilbert series as $\tilde{\zeta}^\fm \tilde{\chi}^\fn = \zeta^\fm \chi^{\fm+\fn}$, in agreement with \eref{HSNS5branchpuzzle2}.}$^,$\footnote{In \cite[v2, (4.31)]{Marino:2025ihk}, the authors instead consider the map $\tilde{\zeta} = \zeta \chi$, $\tilde{\chi} = s$, and $\tilde{\chi}_F = s \chi$, where $\tilde{\chi}_F$ is the charge conjugation fugacity associated with the $\SO(2 F_3 + 3)$ flavour node of \eref{magquiverNSpuzzle2v2}. These fugacities contribute to the Coulomb branch Hilbert series as $\tilde{\zeta}^\fm \tilde{\chi}^\fn \tilde{\chi}_F^\fn = \zeta^\fm \chi^{\fm + \fn} s^{2 \fn}$. Since $\fn$ is an integer and $s$ is a $\BZ_2$ fugacity, this result remains consistent with \eref{HSNS5branchpuzzle2}.} A key advantage of this description is that the magnetic fugacity associated with $\zeta$ becomes manifest in \eref{magquiverNSpuzzle2v2}.

\subsubsection*{The D5-branch}
Finally, the Hilbert series for the D5-branch is
\bes{ \label{eq:27114}
\scalebox{0.83}{$
\begin{split}
&H_{\eqref{quivpuzzle2}}[\text{D5}](t;b_{1,2,3}) =\frac{1}{32} \oint \frac{du_1}{2\pi i u_1} \left( \prod_{\alpha=1}^2 \frac{d u_2^{(\alpha)}}{2\pi i  u_2^{(i)}} \right) \oint\frac{du_3}{2\pi i u_3} \\
& \,\, \times \prod_{i=1}^2\prod_{s=\pm1}\left(1-\chi (u_2^{(i)})^{s}\right) \prod_{s_1,s_2=\pm1}\left(1-(u_2^{(1)})^{s_1}(u_2^{(2)})^{s_2}\right)\prod_{s=\pm 1}(1-u_1^{2s})  (1 \leftrightarrow 3)\\
& \,\, \times \PE\Bigg[t \sum_{s_1,s_2=\pm 1} \left\{ \sum_{i=1}^2  u_1^{s_1} (u_{2}^{(i)} )^{s_2}+  u_3^{s_1} (u_{2}^{(i)} )^{s_2}\right\} +2t \chi \sum_{s=\pm 1}\left(u_1^s+u_3^s\right) +t \chi \sum_{\rho_j=1}^{F_2} \sum_{s_2=\pm 1}(f_2^{(\rho_2)})^{s_2} \\
& \,\,\, \qquad  + t\sum_{s_1,s_2=\pm 1}\Big\{\sum_{\rho_j=1}^{F_1} ( u_1)^{s_1}(f_1^{(\rho_1)})^{s_2}+(1 \leftrightarrow 3)+\sum_{i=1}^2\sum_{\rho_j=1}^{F_2} ( u_2^{(i)})^{s_1}(f_2^{(\rho_2)})^{s_2} \Big \}  \\
& \,\,\, \qquad -t^2\left(1+u_1^{-2}+u_1^{2}\right)-(1 \leftrightarrow 3)  -t^2 \Bigg(\sum_{s_1,s_2=\pm 1}(u_2^{(1)})^{s_1}(u_2^{(2)})^{s_2} +\chi \sum_{i=1}^2\sum_{s=\pm 1}(u_2^{(i)})^{s}+2\Bigg) \Bigg]~.
\end{split}$}
}
The coefficient of $-t^2$ inside the plethystic exponential accounts for the adjoint chirals in the vector multiplets of the three gauge nodes, which have an axial fugacity of $a^{-2}$. This result corresponds to the Higgs branch Hilbert series of the theory in \eqref{quivpuzzle2} with the Chern-Simons levels set to zero ($k=0$), which is the 3d $\CN=4$ theory
\begin{equation} \label{quivT3theoryortho}
\scalebox{0.85}{
\begin{tikzpicture}[baseline,font=\footnotesize,
    circ/.style={circle, draw, minimum size=1.3cm},
    sq/.style={rectangle, draw, minimum size=1cm},
    node distance=1cm
]
    \node[circ] (n1) {\scalebox{0.85}{$\USp(2)$}};
    \node[circ, right=of n1] (n2) {\scalebox{0.7}{$\SO(5)$}};
    \node[circ, right=of n2] (n3) {$\USp(2)$};
    \node[sq, below=of n1] (s1) {$F_1+\frac{1}{2}$};
    \node[sq, below=of n2] (s2) {$F_2$};
    \node[sq, below=of n3] (s3) {$F_3+\frac{1}{2}$};
    \draw (n1) -- (n2);
    \draw (n2) -- (n3);
    \draw (n1) -- (s1);
    \draw (n2) -- (s2);
    \draw (n3) -- (s3);
\end{tikzpicture}}
\end{equation}
Gauging charge conjugation symmetry by summing $\chi$ over $\pm 1$ and dividing the result by two amounts to replacing $\SO(5)$ by $\O(5)^+$.

\section{Geometric branch} \label{sec:geometricbranch}

In this section, we describe how to generalise the prescription to study the geometric branches of affine-shaped quivers. 

\subsection{Circular Abelian quivers: affine $A_N$ quivers}
Let us consider the following theory:
\begin{equation} \label{CSquivflv}
\scalebox{0.8}{
\begin{tikzpicture}[baseline,
    gauge/.style={circle, draw, minimum size=1.2cm},
    flavor/.style={rectangle, draw, minimum size=0.8cm}
]

    \def\radius{2.5cm}

    \node[gauge] (g1) at (-60:\radius) {$1_{k_1}$};
    \node[gauge] (g2) at (0:\radius)   {$1_{k_2}$};
    \node[gauge] (g3) at (60:\radius)  {$1_{k_3}$};
    \node[gauge] (g4) at (120:\radius) {$1_{k_4}$};
    \node[gauge] (g5) at (180:\radius) {$1_{k_5}$};
    \node[gauge] (gG) at (240:\radius) {$1_{k_n}$};

    \node[flavor, right=0.5cm of g1] (f1) {$f_1$};
    \node[flavor, right=0.5cm of g2] (f2) {$f_2$};
    \node[flavor, right=0.5cm of g3] (f3) {$f_3$};
    \node[flavor, left=0.5cm of g4]  (f4) {$f_4$};
    \node[flavor, left=0.5cm of g5]  (f5) {$f_5$};
    \node[flavor, left=0.5cm of gG]  (fG) {$f_n$};


    \draw (g1) to[bend right=15] (g2);
    \draw (g2) to[bend right=15] (g3);
    \draw (g3) to[bend right=15] (g4);
    \draw (g4) to[bend right=15] (g5);
    \draw (gG) to[bend right=15] (g1);

    \draw[dashed] (g5) to[bend right=15] (gG);

    \draw (g1) -- (f1);
    \draw (g2) -- (f2);
    \draw (g3) -- (f3);
    \draw (g4) -- (f4);
    \draw (g5) -- (f5);
    \draw (gG) -- (fG);
\end{tikzpicture}}
\end{equation}
where we write
\bes{ \label{defF}
F = f_1 + f_2 + \cdots + f_n~,
}
and assume that
\bes{ \label{zerosumCS}
\sum_{i=1}^n k_i =0~.
}
Let us denote the chiral multiplets in each bifundamental hypermultiplet $\CS_i$ between node $\U(1)_{k_{i}}$ and $\U(1)_{k_{i+1}}$ by $(A_i, B_i)$, and those in each fundamental hypermultiplet $\CQ_i$ of the node $\U(1)_{k_i}$ by $(Q_i, \tQ_i)$. The superpotential is
\bes{ \label{supCSquivflv}
W=\sum_{i=1}^n \phi_i (A_i B_i - B_{i-1}A_{i-1} + Q_i \tQ_i)+ \frac{1}{2} \sum_{i=1}^n k_i \phi_i^2~.
}
This theory was studied in \cite{Gaiotto:2009tk}, and the special case in which $f_i=0$ for all $i=1, \ldots, n$ was studied in \cite{Jafferis:2008qz}. Since $\phi_i$ are massive, we can integrate them out and obtain
\bes{
W_{\text{eff}} = \sum_{i=1}^n \frac{2 \pi}{k_i} (A_i B_i - B_{i-1}A_{i-1} + Q_i \tQ_i)^2~.
}
The index of this theory is given by
\bes{ \label{indexCSquivflv}
\scalebox{1}{$
\begin{split}
&\CI_{\eqref{CSquivflv}}(x; \{w_i \}; \{ v_i\};\{y_{i, \ell} \}) \\
&= \sum_{m_1, \ldots, m_n \in \BZ} \oint \frac{du_1}{2\pi u_1} \ldots \frac{du_n}{2\pi u_n} \,\, \prod_{i=1} u_i^{k_i m_i} w_i^{m_i} \\
& \quad \times \prod_{s = \pm1} \prod_{i=1}^n Z_{\text{chir}}^{1/2} \left(x; (v_i u_i u_{i+1}^{-1})^s; s(m_{i} - m_{i+1}) \right) \prod_{\ell =1}^{f_i} Z_{\text{chir}}^{1/2} \left(x; (y_{i, \ell}^{-1} u_i)^s; s m_i \right)~,
\end{split}$}
}
where the contribution of a chiral multiplet with R-charge $R$ is given by \eqref{indexvecchircontrib}.

Theory \eqref{CSquivflv} can be realised on D3-brane on a circle along with various fivebranes. There are several branches of the moduli space that can be obtained from the limit of the index, as explained in the preceding sections. In the following, we will describe another branch of the moduli space, known as the {\it geometric branch}.

\subsection*{Geometric branch}
The geometric branch is obtained by considering
\bes{ \label{defgeometricbranch}
\phi_i = \phi~, \quad \text{for all $i=1, \ldots, n$}~.
}
For simplicity, we set $f_i=0$ for all $i=1, \ldots, n$ for now. We will discuss the case in which some or all $f_i$ are non-zero later.
We also assume that $A_i$, $B_i$ are non-zero on this branch. Varying $W_{\text{eff}}$ with respect to $A_i$ (or $B_i$) and dividing by $B_i$ (or $A_i$), we obtain the $F$-terms:
\bes{ \label{F1}
& (k^{-1}_i+k^{-1}_{i+1}) A_i B_i - k^{-1}_{i+1} A_{i+1} B_{i+1} -k^{-1}_{i} A_{i-1} B_{i-1} =0~.
}
As pointed out in \cite[(2.36), (2.6)]{Jafferis:2008qz}, the $F$-terms and $D$-terms corresponding to \eqref{F1} can be written together as\footnote{Here, our matrix $\CM$ corresponds to $M^{\text{gen}}$ in \cite[(2.36)]{Jafferis:2008qz}.}
\bes{ \label{classicalFD}
0 = \sum_{j=1}^n \CM_{ij} \CS^\dagger_j \sigma_\alpha \CS_j~,
}
where $\sigma_\alpha$ are the Pauli matrices, and
\bes{
\CM_{ij} = \begin{cases} k_i^{-1}+ k_{i+1}^{-1} & \text{if $j=i$} \\ -k_{i+1}^{-1} & \text{if $j=i+1$} \\ -k_{i}^{-1} & \text{if $j=i-1$} \\ 0 & \text{otherwise}  \\  \end{cases}~.
}
As pointed out in \cite[(2.36)]{Jafferis:2008qz}, we can also write  $\CM_{ij}$ as $\CM_{ij} = \sum_a k_a^{-1} X_{ai} X_{aj}$, where the index $a$ runs over the nodes and the index $i$ over the hypermultiplets, and $X_{ai}$ is the charge matrix such that has value $1$ if $A_i$ is ingoing to the node $a$, $-1$ if it is outgoing from it, and $0$ otherwise.  The rank of the matrix $\CM$ is $n-2$, provided that the sum of the Chern--Simons levels is zero \cite{Jafferis:2008qz} as in \eqref{zerosumCS}. Indeed, 
\bes{
\CM \mathbf{\beta}^t  =0~, \quad \text{with} \,\,\,\, \beta= \begin{pmatrix} 1 & 1 & \cdots & 1 \\ p_1 & p_2 &\cdots & p_n \end{pmatrix}~,
}
where $p_i = \sum_{j=1}^i k_j$, so that $k_j = p_{j}-p_{j-1}$. In fact, $(1,p_1)$, $(1,p_2)$, $\ldots,$ $(1,p_n)$ are the fivebranes involving in the brane realisation of theory \eqref{CSquivflv} with $f_i=0$. It was shown in \cite{Jafferis:2008qz} that the geometric branch is isomorphic to the hypertoric manifold
\bes{
\mathbb{H}^n///\ker(\beta)~.
}
The authors of \cite{Gaiotto:2009tk} generalised this result to the case in which some or all of the $f_i$ are non-zero. In the corresponding brane configurations, there are $F$ D5-branes partitioned as in \eqref{defF}. In which case, the {\it quantum corrected} geometric branch is described by
\bes{ \label{geomflvquiv}
\mathbb{H}^{n+1}///\ker(\tilde{\beta})~, \quad \text{with} \,\,\,\, \tilde{\beta}= \begin{pmatrix} 1 & 1 & \cdots & 1 & 0  \\ p_1 & p_2 &\cdots & p_n & F \end{pmatrix}~.
}
The set of holomorphic functions on \eqref{geomflvquiv} contains chiral fields $A_i$ and $B_i$ appropriately dressed with such monopole operators. 

\subsubsection*{The Hilbert series as a limit of the index}
The Hilbert series of the geometric branch of this quiver can be computed from the limit of the index of the theory with the following axial fugacity assignments:
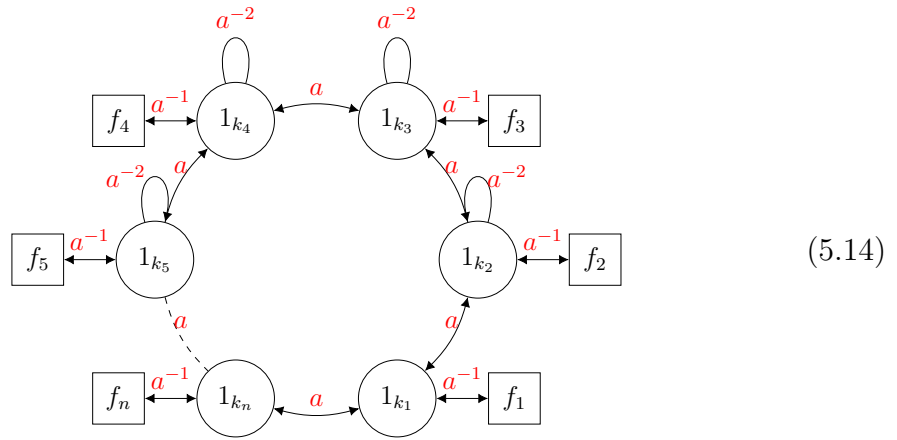
\begin{equation} \label{CSquivflvgeoma}
\scalebox{0.85}{
\begin{tikzpicture}[baseline,
    gauge/.style={circle, draw, minimum size=1.2cm},
    flavor/.style={rectangle, draw, minimum size=0.8cm}
]

    \def\radius{2.5cm}

    \node[gauge] (g1) at (-60:\radius) {$1_{k_1}$};
    \node[gauge] (g2) at (0:\radius)   {$1_{k_2}$};
    \node[gauge] (g3) at (60:\radius)  {$1_{k_3}$};
    \node[gauge] (g4) at (120:\radius) {$1_{k_4}$};
    \node[gauge] (g5) at (180:\radius) {$1_{k_5}$};
    \node[gauge] (gG) at (240:\radius) {$1_{k_n}$};

    \node[flavor, right=0.8cm of g1] (f1) {$f_1$};
    \node[flavor, right=0.8cm of g2] (f2) {$f_2$};
    \node[flavor, right=0.8cm of g3] (f3) {$f_3$};
    \node[flavor, left=0.8cm of g4]  (f4) {$f_4$};
    \node[flavor, left=0.8cm of g5]  (f5) {$f_5$};
    \node[flavor, left=0.8cm of gG]  (fG) {$f_n$};

    \path (g2) edge [loop above, draw, looseness=8] node[right, red] {$a^{-2}$} (g2);
    \path (g3) edge [loop above, draw, looseness=8] node[above, red] {$a^{-2}$} (g3);
    \path (g4) edge [loop above, draw, looseness=8] node[above, red] {$a^{-2}$} (g4);
    \path (g5) edge [loop above, draw, looseness=8] node[left, red] {$a^{-2}$} (g5);


    \draw[<->] (g1) to[bend right=15] node[above, red] {$a$} (g2);
    \draw[<->] (g2) to[bend right=15] node[above, red] {$a$} (g3);
    \draw[<->] (g3) to[bend right=15] node[above, red] {$a$} (g4);
    \draw[<->] (g4) to[bend right=15] node[above, red] {$a$} (g5);
    \draw[<->] (gG) to[bend right=15] node[above, red] {$a$} (g1);

    \draw[dashed] (g5) to[bend right=15] node[above, red] {$a$} (gG);

    \draw[<->] (g1) -- node[above, red] {$a^{-1}$} (f1);
    \draw[<->] (g2) -- node[above, red] {$a^{-1}$} (f2);
    \draw[<->] (g3) -- node[above, red] {$a^{-1}$} (f3);
    \draw[<->] (g4) -- node[above, red] {$a^{-1}$} (f4);
    \draw[<->] (g5) -- node[above, red] {$a^{-1}$} (f5);
    \draw[<->] (gG) -- node[above, red] {$a^{-1}$} (fG);
\end{tikzpicture}}
\end{equation}
where the quiver is depicted in the 3d $\CN=2$ notation, and each bidirectional arrow between the circular nodes denotes the chiral multiplets $A_i$ and $B_i$, whereas that between a circular node and a square node represents $Q_i$ and $\tQ_i$.

The corresponding index \eqref{indexCSquivflv}  is
\bes{ \label{indexwitha}
\scalebox{0.9}{$
\begin{split}
&\CI_{\eqref{CSquivflvgeoma}}(x; a; w, v) \\
&= \left[ Z_{\text{chir}}^{1} \left(x; a^{-2}; 0 \right) \right]^{n-2}  \sum_{m \in \BZ} \,\, \oint \frac{du_1}{2\pi u_1} \cdots \frac{du_n}{2\pi u_n} \,\, \prod_{i=1} u_i^{k_i m} w_i^{m} \\
& \quad \times \prod_{s = \pm1} \prod_{i=1}^n Z_{\text{chir}}^{1/2} \left(x; a (b u_i u_{i+1}^{-1})^s; s(m_{i} - m_{i+1}) \right) \prod_{\ell =1}^{f_i} Z_{\text{chir}}^{1/2} \left(x; a^{-1} (u_i)^s; s m_i \right)~.
\end{split}$}
}
where we identify $u_{n+1} \equiv u_1$. The factor $\left[ Z_{\text{chir}}^{1} \left(x; a^{-2}; 0 \right) \right]^{n-2}$ is the contribution of the Lagrange multipliers (or ``flipping fields'') that impose the $(n-2)$ independent relations $0 = \sum_{j=1}^n \CM_{ij} \CS^\dagger_j \sigma_\alpha \CS_j$, where we recall that $(n-2)$ is equal to the rank of the matrix $\CM$. We denote these $(n-2)$ Lagrange multipliers in \eqref{CSquivflvgeoma} by the loops at every circular node, except at $\U(1)_{k_1}$ and $\U(1)_{k_n}$.

Let us substitute \eqref{N=3reparam} in the index \eqref{indexwitha}.  The power of $\fz$ is 
\bes{
|m_1-m_2| +|m_2-m_3| +\ldots+ |m_{n-1}-m_n|~.
}
To obtain a finite result in the limit $\fz \rightarrow 0$ (while keeping $t$ fixed), the above expression must be zero. This puts the following constraint on the magnetic fluxes:
\bes{ \label{fluxgeombr}
m_1= m_2= \ldots= m_n \equiv m \in \BZ~,
}
in agreement with \eqref{defgeometricbranch}. The dimension of the monopole operator is
\bes{
&\Delta(m_1, \ldots, m_n) \\
&=\frac{1}{2} \left(|m_1-m_2| +|m_2-m_3| +\ldots+ |m_{n-1}-m_n|+\sum_{i=1}^n f_i |m_i| \right)~,
}
where, with the flux configuration \eqref{fluxgeombr}, we have
\bes{ \label{dimmongeom}
\Delta(m, \ldots, m)= F|m| ~.
}
In the limit $\fz \rightarrow 0$, the index becomes the following Hilbert series:
\bes{ \label{hsgeom}
H(t, v, w) &= \text{PE}\left[-(n-2)t^2\right] \sum_{m \in \mathbb{Z}} w^m t^{F|m|}~  \\
& \quad \times \left( \prod_{i=1}^{n} \oint \frac{du_i}{2\pi i u_i} u_i^{k_i m} \right) \text{PE} \left[ t \sum_{i=1}^{n} \left( b \frac{u_i}{u_{i+1}} + b^{-1} \frac{u_{i+1}}{u_i} \right) \right]~,
}
where $w=\prod_{j=1}^n w_j$ is the fugacity for the diagonal combination of the topological symmetries associated to each gauge group. This is in agreement with \cite[(6.18)]{Cremonesi:2016nbo}. Note that $b$ and $w$ correspond to the two Cartan elements of the isometry of the geometric branch. The terms in the plethystic exponential in the second line are the contributions of $A_i$ and $B_i$, and their associated flavour fugacities are $v$ and $v^{-1}$. The factor $t^{F|m|}$ is the dimension of the monopole operator given by \eqref{dimmongeom}. The factor $\text{PE}\left[-(n-2)t^2\right]$ corresponds to the $(n-2)$ independent $F$-terms given by the rank of the matrix $\CM$ discussed above. Observe that the chiral fields $Q_i$ and $\tQ_i$ do not appear in the chiral ring. Only the sum $F$ of the number of flavours contributes to the dimension of the monopole operators.

When $k_i=0$ for all $i=1, \ldots, n$, the theory has a manifest $\CN=4$ supersymmetry, and its geometric branch is simply $\BC^2/\BZ_n \times \BC^2/\BZ_{F}$, since the theory can be realised on a single M2-brane probing such a space in this case \cite{Porrati:1996xi}. Other explicit examples can be found in \cite{Cremonesi:2016nbo}.

\subsection{Comments on general non-Abelian circular quivers}\label{rank2affine}
In the previous subsection, we have considered the case where all gauge groups of \eqref{CSquivflvgeoma} are Abelian. We now analyse the theories with non-Abelian gauge groups.
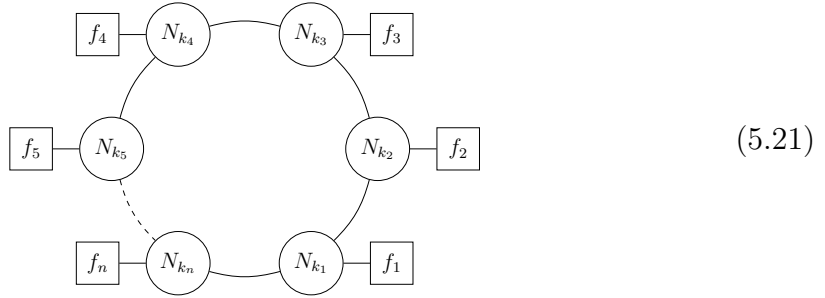
\begin{equation} \label{CSquivflvnonab}
\scalebox{0.7}{
\begin{tikzpicture}[baseline,
    gauge/.style={circle, draw, minimum size=1.2cm},
    flavor/.style={rectangle, draw, minimum size=0.8cm}
]

    \def\radius{2.5cm}

    \node[gauge] (g1) at (-60:\radius) {${N}_{k_1}$};
    \node[gauge] (g2) at (0:\radius)   {${N}_{k_2}$};
    \node[gauge] (g3) at (60:\radius)  {${N}_{k_3}$};
    \node[gauge] (g4) at (120:\radius) {${N}_{k_4}$};
    \node[gauge] (g5) at (180:\radius) {${N}_{k_5}$};
    \node[gauge] (gG) at (240:\radius) {${N}_{k_n}$};

    \node[flavor, right=0.5cm of g1] (f1) {$f_1$};
    \node[flavor, right=0.5cm of g2] (f2) {$f_2$};
    \node[flavor, right=0.5cm of g3] (f3) {$f_3$};
    \node[flavor, left=0.5cm of g4]  (f4) {$f_4$};
    \node[flavor, left=0.5cm of g5]  (f5) {$f_5$};
    \node[flavor, left=0.5cm of gG]  (fG) {$f_n$};


    \draw (g1) to[bend right=15] (g2);
    \draw (g2) to[bend right=15] (g3);
    \draw (g3) to[bend right=15] (g4);
    \draw (g4) to[bend right=15] (g5);
    \draw (gG) to[bend right=15] (g1);

    \draw[dashed] (g5) to[bend right=15] (gG);

    \draw (g1) -- (f1);
    \draw (g2) -- (f2);
    \draw (g3) -- (f3);
    \draw (g4) -- (f4);
    \draw (g5) -- (f5);
    \draw (gG) -- (fG);
\end{tikzpicture}}
\end{equation}
To begin, we consider the special case where all the CS-levels are set to zero, namely $k_i = 0 \, \forall \, i=1,\ldots,n $, and take the values fo the $f_i$ parameters to be generic as long as $F = \sum_{i=1}^n f_i \neq 0$ (the case of $F=0$ is commented in \cite{circularpaper}). This is the case of a 3d $\mathcal{N}=4$ theory whose brane setup consist of only NS5 and D5 branes. This provides a simpler setup to study the problem and obtain a result that, as we will comment later in the subsection, holds also for generic $k_i$ levels.

We focus on the case where all gauge groups are $\U(2)$. The index reads
\begin{equation}\label{indcircN=4}
\scalebox{0.88}{$
\begin{split}
    & \mathcal{I}_{N=2} (x; \{w_i\},\{v_i\},\{y_{i,l}\}) \\
    & = \frac{1}{(2!)^n} \sum_{\{m_1^{(\alpha)}\},\ldots,\{m_n^{(\alpha)}\} \in \mathbb{Z}^2} \oint \prod_{\alpha=1}^2 \frac{d u_1^{(\alpha)} }{2 \pi i u_1^{(\alpha)}} \ldots \frac{d u_n^{(\alpha)} }{2 \pi i u_n^{(\alpha)}} \prod_{i=1}^n w_i^{\sum_{\alpha=1}^2 m_i^{(\alpha)}}
    \frac{1}{[Z^{1}_{\text{chir}}(x;a^{-2},0) ]^2} \\
    & \times \prod_{i=1}^n Z^{\U(2)}_{\text{vec}} (x;\{u_i^{(\alpha)}\};\{m_i^{(\alpha)}\}) \prod_{\alpha,\beta=1}^2 Z^{1}_{\text{chir}}(x; a^{-2} u_i^{(\alpha)} (u_i^{(\beta)})^{-1}; m_i^{(\alpha)}-m_i^{(\beta)} ) \\
    &  \times \prod_{s = \pm 1}\prod_{i=1}^n \prod_{\alpha,\beta=1}^2 Z^{1/2}_{\text{chir}}\left(x; a (b u_i^{(\alpha)})^s (u_{i+1}^{(\beta)})^{-s}; s m_i^{(\alpha)}-s m_{i+1}^{(\beta)}\right) \\
    & \times \prod_{s=\pm 1} \prod_{i=1}^n \prod_{\rho=1}^F \prod_{\alpha=1}^2 Z^{1/2}_{\text{chir}} (x; a^{-1}(f_\rho^{-1} u_i^{\alpha})^{s}; s m_i^{(\alpha)})~,
\end{split}$}
\end{equation}
with periodic parameters such that $\vec{u}_{n+1} \equiv \vec{u}_1$ and $\vec{m}_{n+1} \equiv \vec{m}_1$. Notice that, in the second line, we divide by the contribution of two chiral multiplets with axial charge $-2$. This is conventional and is part of the prescription that we propose to compute the geometric branch. The computation that we will present will depend very little on this factor.
It is also part of the prescription that we assign axial charge $-1$ to all the adjoint chiral fields and flavours while we assign charge $+1$ to each bifundamental. This choice is instead crucial for the computation that will follow.

The geometric branch is expected to arise as the limit of the index
\begin{equation}
    H_{\eqref{indcircN=4}}[\text{geom}](t) = \lim_{\fz \rightarrow 0} \CI_{\eqref{indcircN=4}} \Big|_{\eqref{N=3reparam}}~.
\end{equation}
In the limit of $\fz \to 0$ only the flux sectors carrying zero power of $\fz$ contribute. The power of $\fz$ reads
\begin{equation}
- \sum_{i=1}^n \sum_{\alpha>\beta}^2 |m_i^{(\alpha)} - m_i^{(\beta)}| + \sum_{i=1}^n \sum_{\alpha, \beta = 1}^2 |m_i^{(\alpha)} - m_{i+1}^{(\beta)}| ~,
\end{equation}
which is a non-negative quantity for any set of fluxes $\textbf{m}_i$ and is zero if and only if $\textbf{m}_1 = \ldots = \textbf{m}_n \equiv \bf{m}$.\footnote{Indeed the fluxes can be equal up to permutations of their components. However, in the following, we pick a fixed Weyl chamber such that $m_i^{(\alpha)} \geq m_i^{(\beta)}$ if $\alpha < \beta$ for any $i=1,\ldots,n$. Therefore, we do not need to take into account permutations in the computation that follows. } Since all the magnetic fluxes are identified, the limit does not depend on the FI parameters of all the nodes but only on their diagonal combination $w = \prod_{i=1}^n w_i$.
The limit yields
\begin{equation}\label{geombrcirclim}
\begin{split}
    & H_{\eqref{indcircN=4}}(t;w;b) = \sum_{m_1 > m_2} t^{F (|m_1|+|m_2|)} w^{m_1+m_2} A_1(t,b)^2 + \sum_{m} t^{2F |m|} w^{2m} \frac{A_2(t,b)}{1-t^2}~,
\end{split}
\end{equation}
where the first contribution comes from the situation where $m^{(1)}=m_1 \neq m^{(2)}=m_2$, thus each $\U(2)$ gauge group is broken to $\U(1)\times \U(1)$, while the second contribution comes from the case $m^{(1)}=m^{(2)} =m$, which preserves each $\U(2)$ gauge group. In \eqref{geombrcirclim}, we defined the contribution
\begin{equation} \label{defAN}
\begin{split}
    A_N(t,b) = & \frac{1}{(N!)^n} \oint \prod_{\alpha=1}^N \frac{d u_1^{(\alpha)} }{2 \pi i u_1^{(\alpha)}} \cdots \frac{d u_n^{(\alpha)} }{2 \pi i u_n^{(\alpha)}} \prod_{1 \leq \alpha \neq \beta \leq N} \prod_{i=1}^n \left( 1 - \frac{u_i^{(\alpha)}}{u_i^{(\beta)}} \right) \\
    & \times \PE\left[t \sum_{s=\pm 1} \sum_{\alpha,\beta=1}^N \sum_{i=1}^n  b \left(\frac{u_i^{(\alpha)}}{u_{i+1}^{(\beta)}} \right)^s - t^2 \left( {
    \red -1} + \sum_{i=1}^n \sum_{\alpha,\beta=1}^N \frac{u_i^{(\alpha)}}{u_i^{(\beta)}} \right)\right]~,
\end{split}
\end{equation}
in a such way that $A_1(t,b)$ is the Hilbert series of $\mathbb{C}^2/\mathbb{Z}_n$:
\bes{
A_1(t, b) = H_{\BC^2/\BZ_n}(t,b) = \PE[ t^2 +(b+b^{-1}) t^n - t^{2n} ]~.
}
The term ${\red -1}$, highlighted in {\red red}, deserves some explanations as follows. First of all, its presence renders $A_1(t,b)$ the Hilbert series of $\mathbb{C}^2/\mathbb{Z}_n$, as it should be for the Higgs branch of a Kronheimer-Nakajima quiver. Secondly, upon substituting \eqref{defAN} into \eqref{geombrcirclim}, we see that each term in the latter contains the factor $\PE\left[2\, t^2\right] = \frac{1}{(1-t^2)^2}$, which is in agreement with the limit of the term $1/[Z^{1}_{\text{chir}}(x;a^{-2},0) ]^2$ in \eqref{indcircN=4}.

For $N >1$ the quantity $A_N$ is related to $A_1$. For example, the relations enjoyed by $A_{2,3}$ are given by
\begin{equation}\label{circmolienrelation}
\scalebox{0.95}{$
\begin{split}
    & A_2(t,b) = \frac{1}{2(1-t^2)} \times \left[ (1-t^2)^2A_1(t,b)^2 + (1-t^4) \times A_1(t^2,b^2) \right] \,, \\
    & A_3(t,b) = \frac{1}{6(1-t^2)} \big[ (1-t^2)^3 \times A_1(t,b)^3 + 2(1-t^4)(1-t^2) \times A_1(t,b)A_1(t^2,b^2) \\
    & \qquad\qquad\qquad\qquad \, + 3(1-t^6) \times A_1(t^3,b^3) \big] ~.
\end{split}
$}
\end{equation}
These relations, and those for higher $N$, have been found first in \cite{circularpaper} (see the analogous relations also in \cite[(2.1), (2.5)]{Hanany:2012dm}), from the analysis of the $F=0$ case.

Going back to \eqref{geombrcirclim}, we perform a couple of manipulations. For the first term, we split the sum into two independent summations over $m_1$ and $m_2$, which requires to introduce the missing contributions of $m_1=m_2$ in the sum. By doing that we are led to
\begin{equation}
\begin{split}
 H_{\eqref{indcircN=4}}(t;w;b) &= \frac{1}{2} (1-t^2)^2 \left( \frac{1}{1-t^2} \sum_{m} t^{F|m|} w^m \right)^2 A_1(t,b)^2  \\
    & \quad + \sum_m t^{2F|m|} w^{2m} \left[ \frac{A_2(t,b)}{1-t^2} - \frac{1}{2} A_1(t,b)^2 \right]~,
\end{split}
\end{equation}
where, in the first line, we also multiplied and divided by $(1-t^2)^2$, so that we recognise that the sum yields the Hilbert series of $\mathbb{C}^2/\mathbb{Z}_F$, namely
\bes{
H_{\BC^2/\BZ_F}(t;w) = \frac{1}{1-t^2} \sum_{m} t^{F|m|} w^m= \PE\left[ t^2+(w+w^{-1}) t^{F} -t^{2F}\right]~.
}
We then use the relation for $A_2$ in \eqref{circmolienrelation} to get
\begin{equation}
\scalebox{0.87}{$
\begin{split}
    & H_{\eqref{indcircN=4}}(t;w;b) \\
    &= \frac{1}{2} (1-t^2)^2 \left( \frac{1}{1-t^2} \sum_{m} t^{F|m|} w^m \right)^2 A_1(t,b)^2 \\
    & \quad\,\,  + \sum_m \frac{1}{1-t^4} t^{2F|m|} w^{2m} \left[ \frac{(1-t^2)^2 A_1(t,b)^2+(1-t^4)A_1(t^2,b^2)}{2(1-t^2)^2} - \frac{1}{2} A_1(t,b)^2 \right] \\
    & = \frac{1}{2} (1-t^2)^2 \left( \frac{1}{1-t^2} \sum_{m} t^{F|m|} w^m \right)^2 A_1(t,b)^2  + \frac{1}{2} (1-t^4)^2 \sum_{m} \frac{1}{1-t^4}t^{2F|m|} w^{2m} A_1(t^2,b^2) \\
    & = \frac{1}{2} (1-t^2)^2 \Big[ H_{\BC^2/\BZ_{F}}(t,w)^2 H_{\BC^2/\BZ_{n}}(t,b)^2 + \frac{(1+t^2)^2}{(1-t^2)^2} H_{\BC^2/\BZ_{F}}(t^2,w^2) H_{\BC^2/\BZ_{n}}(t^2,b^2)  \Big]~.
 \end{split}$}
\end{equation}
We then recognise the contribution of the Hilbert series of $\mathbb{C}^2/\mathbb{Z}_n \times \mathbb{C}^2/\mathbb{Z}_F$ to write
\bes{\label{geombranchcircres}
    &H_{\eqref{indcircN=4}}(t;w;b) \\
    &= \frac{1}{2} (1-t^2)^2 \left[ H_{\mathbb{C}^2/\mathbb{Z}_n \times \mathbb{C}^2/\mathbb{Z}_F}(t;w;b)^2 + \frac{(1+t^2)^2}{(1-t^2)^2}H_{\mathbb{C}^2/\mathbb{Z}_n \times \mathbb{C}^2/\mathbb{Z}_F}(t^2;w^2;b^2)  \right]~.
}
Notice that the factor $\frac{(1+t^2)^2}{(1-t^2)^2} = \PE[4t^2-2t^4]$ is a deviation from the formula being equal to the symmetric product of two $\mathbb{C}^2/\mathbb{Z}_n \times \mathbb{C}^2/\mathbb{Z}_F$ spaces (up to a prefactor), each of which is the geometric branch in the Abelian case. In particular, the formula for the Hilbert series for such a symmetric product is
\bes{\label{Sym2ZnZF}
    & H_{\Sym^2(\mathbb{C}^2/\mathbb{Z}_n \times \mathbb{C}^2/\mathbb{Z}_F)}(t;w;b) \\
    &= \frac{1}{2} \left[ H_{\mathbb{C}^2/\mathbb{Z}_n \times \mathbb{C}^2/\mathbb{Z}_F}(t;w;b)^2
    +H_{\mathbb{C}^2/\mathbb{Z}_n \times \mathbb{C}^2/\mathbb{Z}_F}(t^2;w^2;b^2)  \right]~.
}
Using relations of the form in \eqref{circmolienrelation}, it is quite straightforward to generalise this result to the case of $\U(N)$ gauge groups to find a similar deviation from the symmetric product of $N$ $\mathbb{C}^2/\mathbb{Z}_n \times \mathbb{C}^2/\mathbb{Z}_F$ spaces.

To conclude, we now comment on the generalisation of this result for the case of circular quivers with non-zero CS levels, thus to brane setups including other type of fivebranes besides D5 and NS5. The steps leading to \eqref{geombrcirclim} are identical, the difference is that the $A_N$ contribution now depends on the values of the magnetic fluxes due to the presence of CS levels. This makes the computation quite more difficult and we will not attempt to carry it out. However, we propose that the formula in \eqref{geombranchcircres} should hold for any quiver by replacing $\mathbb{C}^2/\mathbb{Z}_n \times \mathbb{C}^2/\mathbb{Z}_F$ with the geometric branch of the Abelian case:
\begin{equation}\label{geombranchcircfinal}
    H_{N=2}(t;w;b) = \frac{1}{2}(1-t^2)^2\left[ H_{N=1}(t;w;b)^2 + \frac{(1+t^2)^2}{(1-t^2)^2}H_{N=1}(t^2;w^2;b^2) \right] ~.
\end{equation}
A direct check of this proposal is given in the example discussed below.

\subsubsection{Circular quiver \eqref{quivN010}} \label{sec:N010geomN=2}
In this subsection, we consider the geometric branch of the circular quiver \eqref{quivN010} with $N=2$ and $k=F=1$. This can be computed in a similar way to \eqref{indexN010} and \eqref{branchquivN010}, but with the following modifications. First of all, the factor 
\bes{
\prod_{i=1}^2 \prod_{\alpha, \beta=1}^2 Z_{\text{chir}}^{1} \left(x; \fb_i u^{\alpha}_i (u^{\beta}_i)^{-1} ; m^{(\alpha)}_i - m^{(\beta)}_i  \right)
}
in \eqref{indexN010} should be further divided by $Z_{\text{chir}}^{1} \left(x; \fb_i ; 0  \right)^2$ to remove the trace of the adjoint chiral multiplet for each gauge group. Secondly, the parameters $\fb_i$ should be taken as $\fb_i = a^{-2}$ (with $i=1, 2$) for the geometric branch, whereas the other parameters are taken as the last line of \eqref{branchquivN010}.  These modifications are parts of the prescription and are necessary to obtain the limit of the index that contains a contribution from the global symmetry current $\su(3)$, which is a global symmetry for the theory corresponding to $N^{0,1,0}$.

On the geometric branch, the magnetic fluxes of each gauge node are equal and take the form
\bes{
\vec m_1 = \vec m_2 = (\fm, \fn)~,
}
where, as usual, we can use the $\U(2)$ Weyl group to set $\fm \geq \fn$.  The limit of the index receives two contributions, one from $\fm=\fn$ and the other from $\fm >\fn$:
\bes{ 
H_{\eqref{quivN010}_{N=2,k=F=1}}[\text{geom}] = H_{\fm=\fn} + H_{\fm>\fn}~.
}
Similarly to \eqref{Hm=nNS}, for fluxes with $\fm=\fn$, the limit of the index is
\bes{  \label{HmeqngeomN010N=2}
\scalebox{0.88}{$
\begin{split}
&H_{\fm=\fn} = \frac{1}{2^2}  \sum_{\fm \in \BZ} \oint  \Bigg(  \prod_{i=1}^2 \prod_{\alpha=1}^2 \frac{du^{(\alpha)}_i}{2\pi i u^{(\alpha)}_i}  (u^{(\alpha)}_i)^{\kappa_i m^{(\alpha)}_i}  w_i^{m^{(\alpha)}_i} \Bigg)_{\vec m_1 = \vec m_2=(\fm, \fm)}  t^{2F|m|} \\
& \times \prod_{1\leq \alpha \neq \beta \leq 2} \prod_{i=1}^2 \Bigg(1-\frac{u^{(\alpha)}_i}{u^{(\beta)}_i} \Bigg)  \PE\left[t \sum_{s, \sigma=\pm 1}\sum_{\alpha, \beta=1}^2 \left(b^{\sigma}  \frac{ u^{(\alpha)}_1}{u^{(\beta)}_{2}}\right)^s - t^2 \left( {\blue -2}+  \sum_{i=1}^2 \sum_{\alpha, \beta=1}^2\frac{u^{(\alpha)}_i}{u^{(\beta)}_i}\right)\right]~,
\end{split}$}
}
where $\kappa_1 = -\kappa_2 = k >0$, and the term ${\blue -2}$ in {\blue} takes into account of the tracelessness of the adjoint chiral multiplet for each gauge group. This is in agreement with the discussion below \eqref{defAN}. The power $2F|m|$ of $t$ in the first line comes from two times \eqref{dimmonN010N2} with $\fm=\fn$. For reference, we report the result for the unrefined Hilbert series for $H_{\fm=\fn}$ (with $b=w_1=w_2=1$) as follows: 
\bes{1 + 4 t^2 + 18 t^4 + 46 t^6 + 109 t^8+\ldots~.
}
For fluxes with $\fm>\fn$, similarly to \eqref{Hm>nNS}, the limit of the index is
\bes{ \label{HmneqngeomN010N=2}
H_{\fm >\fn} &= \sum_{\fm >\fn \in \BZ} \, \CH(\fm)\,  \CH(\fn) \,\, t^{F (|\fm| + |\fn|)} \\
&= \sum_{\fm >\fn} \, \frac{(w_1 w_2)^{\fm+\fn}}{(1-t^2)^2} \sum_{p_1, p_2=0}^\infty [k |\fm| +2p_1]_{b} [k |\fn| +2p_2]_{b} \, t^{(k+F) (|\fm|+ |\fn|) +2p_1+2p_2} ~,
}
where $\CH(m)$ is the integral part of the $N=1$ case, given in the third line of \eqref{geomN010N=1} with $w= w_1 w_2$:
\bes{
\CH(m) &= \frac{(w_1 w_2)^m}{1-t^2} \sum_{p=0}^\infty [k |m| +2p]_{b} \, t^{k |m|+2p}~,
}
and the power $F(|\fm|+|\fn|)$ of $t$ in the first line comes from two times \eqref{dimmonN010N2} with $\fm>\fn$.
For reference, we report the result for $H_{\fm>\fn}$ (with $b=w_1=w_2=1$) as follows: 
\bes{
4 t^2 + 30 t^4 + 132 t^6 + 409 t^8+ \ldots~.
}
Summing \eqref{HmeqngeomN010N=2} and \eqref{HmneqngeomN010N=2}, we obtain the following result written in terms of the character of $\su(3)$ representations as in \eqref{HSminorbsu3}, with the subscripts $b$ and $w$ omitted for brevity:
\bes{ \label{quivN010geom}
H_{\eqref{quivN010}}[\text{geom}](t;b;w) = 1 &+[1, 1] t^2 +(2 [2, 2] + 2 - [1, 1])t^4 \\
& + (2 [3, 3] + [4, 1] + [1, 4] - [3, 0] - [0, 3]) t^6+\ldots~,
}
where we use the fugacity map as stated in Footnote \ref{foot:fugmapsu3su2u1}. Due to the fact that this result contains terms with negative coefficients, the limit of the index in this case cannot be interpreted as a Hilbert series, since the latter counts gauge invariant quantities parametrising the moduli space.  Note that we can also derive \eref{quivN010geom} directly from formula \eqref{geombranchcircfinal} simply by taking $H_{N=1}(t; w; b)$ to be \eqref{HSminorbsu3}. This confirms the consistency between the analysis in this subsection and that in the previous one. 

It is instructive to compare this with the Hilbert series of the second symmetric power of the closure of the minimal nilpotent orbit of $\su(3)$, which is the expected geometric branch of this theory:
\bes{ \label{quivN010geomsym}
&H\left[\Sym^2(\bar{\min\, \su(3)})\right](t;b;w) = \frac{1}{2} \left[\sum_{n=0}^\infty [n,n]_{b,w} t^{2n} + \sum_{n=0}^\infty [n,n]_{b^2,w^2} t^{4n} \right]\\
&= 1 +[1, 1] t^2 +(2 [2, 2] + [1, 1]+1)t^4 \\
& \quad \,\,\,\,\, + (2 [3, 3] + [4, 1] + [1, 4] +2[2,2]+ [3,0] + [0, 3]+[1,1]) t^6+\ldots~,
}
where we drop the subscripts $b,w$ in the last equality.

\subsection{Affine $\mathcal{N}=4$ $D_N$ quiver} 
The prescription for studying the geometric branch can also be extended to general affine quivers, beyond the $A_N$ type. In particular, in this section, we study the geometric branch of the $\mathcal{N}=4$ $D_N$ affine quiver with an arbitrary number of flavours attached to each node of the quiver. The theory is realised by the following quiver:
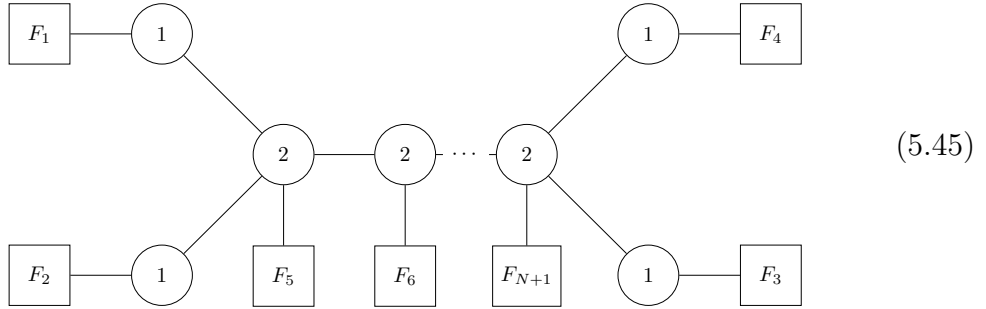
\begin{equation} \label{affined4}
\scalebox{0.8}{
\begin{tikzpicture}[baseline,font=\footnotesize,
    circ/.style={circle, draw, minimum size=1cm},
    sq/.style={rectangle, draw, minimum size=1cm},
    node distance=1cm
]
    \node[circ] (m1) at (0,2) {$1$};
    \node[circ] (m2) at (0,-2)  {$1$};
    \node[circ] (m3) at (8,2)  {$1$};
    \node[circ] (m4) at (8,-2)  {$1$};
    \node[circ] (n1) at (2,0)  {$2$};
    \node[circ] (n2) at (4,0) {$2$};
    \node[] (n3) at (5,0) {$\cdots$};
    \node[circ] (n4) at (6,0) {$2$};
    \node[sq, left=of m1] (s1) {$F_1$};
    \node[sq,left=of m2] (s4) {$F_2$};
    \node[sq, right=of m3] (s2) {$F_4$};
    \node[sq, right=of m4] (s3) {$F_{3}$};
    \node[sq,below=of n1] (s5) {$F_5$};
    \node[sq,below=of n2] (s6) {$F_6$};
    \node[sq,below=of n4] (s7) {$F_{N+1}$};
    \draw (m1) -- (n1);
    \draw (m2) -- (n1);
    \draw (m3) -- (n4);
    \draw (m4) -- (n4);
    \draw (n1) -- (n2);
    \draw (n2) -- (n3);
    \draw (n3) -- (n4);
    \draw (m1) -- (s1);
    \draw (m2) -- (s4);
    \draw (m3) -- (s2);
    \draw (m4) -- (s3); 
    \draw (n1) -- (s5);
    \draw (n2) -- (s6);
    \draw (n4) -- (s7);
\end{tikzpicture}}
\end{equation}
For convenience, we refer to the gauge nodes attached to $F_j$ as the $j$-th gauge node. As usual, we denote by $X_{ij}$ the chiral field in the bifundamental hypermultiplet going from the $i$-th gauge node to the $j$-th gauge node.  The theory can be obtained by a single M2-brane probing $\BC^2/\hat{D}_N \times \BC^2/\BZ_{F}$; see \cite{Porrati:1996xi}, with $F$ given by
\begin{equation}
    F = F_1 + F_2 + F_3 + F_4 + 2\sum_{i=5}^{N+1} F_i ~.
\end{equation}
Hence, $\BC^2/\hat{D}_N \times \BC^2/\BZ_{F}$ is the expected geometric branch of the theory. We confirm this prediction with the explicit computation of the Hilbert series.
To compute the various limits of the index, we first assign axial fugacities to the chiral fields. In 3d $\CN=2$ notation, the field content and fugacity assignments are
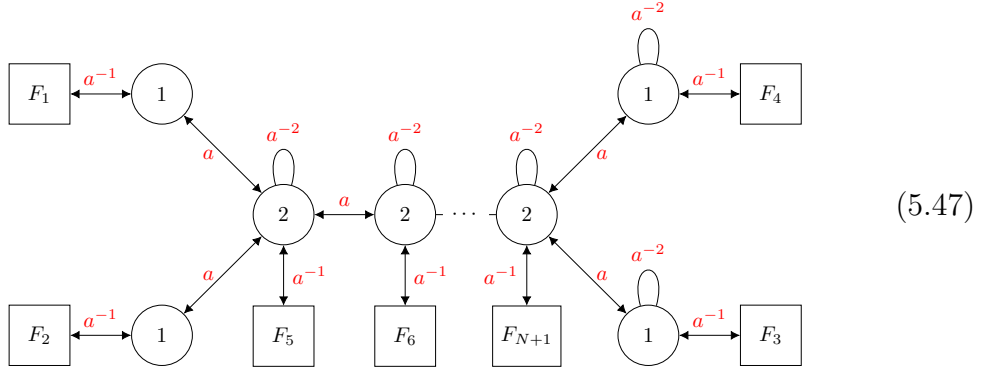
\begin{equation} \label{affineD4axial}
\scalebox{0.8}{
\begin{tikzpicture}[baseline,font=\footnotesize,
    circ/.style={circle, draw, minimum size=1cm},
    sq/.style={rectangle, draw, minimum size=1cm},
    node distance=1cm,
    every loop/.style={-}
]
    \node[circ] (m1) at (0,2) {$1$};
    \node[circ] (m2) at (0,-2)  {$1$};
    \node[circ] (m3) at (8,2)  {$1$};
    \node[circ] (m4) at (8,-2)  {$1$};
    \node[circ] (n1) at (2,0)  {$2$};
    \node[circ] (n2) at (4,0) {$2$};
    \node[] (n3) at (5,0) {$\cdots$};
    \node[circ] (n4) at (6,0) {$2$};
    \node[sq, left=of m1] (s1) {$F_1$};
    \node[sq, left=of m2] (s4) {$F_2$};
    \node[sq, right=of m3] (s2) {$F_4$};
    \node[sq, right=of m4] (s3) {$F_{3}$};
    \node[sq,below=of n1] (s5) {$F_5$};
    \node[sq,below=of n2] (s6) {$F_6$};
    \node[sq,below=of n4] (s7) {$F_{N+1}$};
    \draw[<->] (m1) -- node[left, red] {$a$} (n1);
    \draw[<->] (m2) -- node[left, red] {$a$} (n1);
    \draw[<->] (m3) -- node[right, red] {$a$} (n4);
    \draw[<->] (m4) -- node[right, red] {$a$} (n4);
    \draw[<->] (n1) -- node[above, red] {$a$} (n2);
    \draw[] (n2) -- (n3);
    \draw[] (n3) -- (n4);
    \draw[<->] (m1) -- node[above, red] {$a^{-1}$} (s1);
    \draw[<->] (m2) -- node[above, red] {$a^{-1}$} (s4);
    \draw[<->] (m3) -- node[above, red] {$a^{-1}$} (s2);
    \draw[<->] (m4) -- node[above, red] {$a^{-1}$} (s3);
    \draw[<->] (n1) -- node[right, red] {$a^{-1}$} (s5);
    \draw[<->] (n2) -- node[right, red] {$a^{-1}$} (s6);
    \draw[<->] (n4) -- node[left, red] {$a^{-1}$} (s7);
    \path (m3) edge [loop above] node[red] {$a^{-2}$} ();
    \path (m4) edge [loop above] node[red] {$a^{-2}$} ();
    \path (n1) edge [loop above] node[red] {$a^{-2}$} ();
    \path (n2) edge [loop above] node[red] {$a^{-2}$} ();
    \path (n4) edge [loop above] node[red] {$a^{-2}$} ();
\end{tikzpicture}}
\end{equation}
where we omit two adjoint chirals in the vector multiplets of two $\U(1)$ gauge groups, as part of the prescription to compute the limit of the index for the geometric branch. 
As the result will not depend on the specific choice of the values $F_i$, let us consider the simplified case where $F_1=F$, while every other $F_i =0$. The index for the theory \eref{affineD4axial} is given by
\bes{ \label{indexaffined4}
\scalebox{0.95}{$
\begin{split}
    & \CI_{\eqref{affineD4axial}}(x;\{y_i\},\{w_i\}) = \\
    &= \frac{1}{2^{N-3}} \sum_{m_1,\ldots,m_4 \in \BZ} \sum_{\vec{n}_1, \ldots, \vec{n}_{N-3} \in \mathbb{Z}^2} \oint \prod_{i=1}^4 \frac{d u_i w_i^{m_i}}{2 \pi i u_i} \prod_{i=5}^{N+1} \prod_{\alpha=1}^2 \frac{d v^{(\alpha)}_i w_{i}^{n_i^{(\alpha)}}}{2 \pi i v^{(\alpha)}_i} Z^{\U(N)}_{\text{vec}}(x;\vec{v}_i;\vec{n}_i) \\
    &\quad \times \left[Z^1_{\text{chir}}(x;a^{-2};0)\right]^{2} \prod_{i=5}^{N+1} \prod_{\alpha,\beta=1}^2 Z^1_{\text{chir}}(x;a^{-2} v_i^{(\alpha)} (v_i^{(\beta)})^{-1}; n_i^{(\alpha)}-n_i^{(\beta)}) \\
    &\quad \times \prod_{i=5}^{N} \prod_{\alpha,\beta=1}^2 \prod_{s = \pm1} Z^{1/2}_{\text{chir}}\left(x; a (v_i^{(\alpha)})^s (v_{i+1}^{(\beta)})^{-s}; s n_i^{(\alpha)}-s n_{i+1}^{(\beta)}\right) \\
    &\quad \times \prod_{i=1}^2 \prod_{\alpha=1}^2 \prod_{s=\pm1} \Bigg[Z^{1/2}_{\text{chir}}\left(x; a u_i^s (v_5^{(\alpha)})^{-s}; s m_i-s n_5^{(\alpha)}\right) \\ & \qquad \qquad \qquad \,\,\,\, \times Z^{1/2}_{\text{chir}}\left(x; a u_{i+2}^s (v_{N+1}^{(\alpha)})^{-s}; s m_{i+2}-s n_{N+1}^{(\alpha)}\right) \Bigg] \\
    &\quad \times \prod_{\rho=1}^F \prod_{s=\pm1} Z^{1/2}_{\text{chir}}\left(x; a^{-1} u_1^s (y^{(\rho)})^{-s}; s m_1\right)~,
\end{split}
$}
}
where, in the third line, the factor $\left[Z^1_{\text{chir}}(x;a^{-2};0)\right]^{2}$ corresponds to two chiral multiplets with R-charge one and fugacity $a^{-2}$ under the two $\U(1)$ gauge groups attached to $F_3$ and $F_4$.
To compute the geometric branch, as usual, we redefine
\begin{equation}
    x=\fz t~,\quad a=\fz^{-1/2}t^{1/2} ~,
\end{equation}
and then perform the $\fz \to 0$ limit.
A non-vanishing result in the $\fz \to 0$ limit imposes the condition that the exponent of $\fz$ must be zero. In this case, the exponent of $\fz$ is given by
\begin{equation}
\begin{split}
    & - \sum_{i=5}^{N+1} |n_i^{(1)}-n_i^{(2)}| + \sum_{i=5}^{N} \sum_{\alpha,\beta=1}^2 |n_i^{(\alpha)}-n_{i+1}^{(\beta)}| +  \\
    & + \sum_{i=1}^2 \sum_{\alpha=1}^2 ( |m_i - n_3^{(\alpha)}| + |m_{i+1} - n_{N+1}^{(\alpha)}|) + F |m_1| ~.
\end{split}
\end{equation}
In the limit, the following conditions are imposed on the fluxes in the theory:
\bes{ \label{fluxaffineD}
m_i = m ~,\quad \vec{n}_i = (m,m) ~.
}
The limit of the index for the geometric branch is then given by
\begin{equation} \label{geometricBaffineD4}
\begin{split}
    &H_{\eqref{affined4}}[\text{geom}](t; y) = \\
    &= \sum_{m \in \mathbb{Z}} \frac{t^{F |m|} y^m}{1-t^2} \left( \frac{1}{2} \right)^{N-3} \oint \prod_{i=1}^4 \frac{du_i}{2 \pi i u_i} \prod_{i=5}^{N+1} \prod_{\alpha=1}^2 \frac{dv_i^{(\alpha)}}{2\pi i v_i^{(\alpha)}} (1-t^2)^{2} \\
    &\quad \times \prod_{i=5}^{N+1} \prod_{s=\pm1} \left(1- (v_i^{(1)})^s (v_i^{(2)})^{-s} \right) \prod_{\alpha, \beta = 1, 2} \PE \left[ -t^2 (v_i^{(\alpha)})^s (v_i^{(\beta)})^{-s} \right] \\
    &\quad \times \prod_{i=5}^{N} \prod_{\alpha,\beta=1}^2 \prod_{s=\pm1} \PE \left[ t (v_i^{(\alpha)})^s (v_{i+1}^{(\beta)})^{-s} \right] \\ & \quad \times \prod_{i=1}^2 \prod_{\alpha=1}^2 \prod_{s=\pm 1} \PE \left[ t u_i^s (v_5^{(\alpha)})^{-s} +t u_{i+2}^s (v_{N+1}^{(\alpha)})^{-s} \right]~,
\end{split}
\end{equation}
where we redefine the topological fugacity as
\begin{equation}
    y = w_1 w_2 w_{3} w_{4} \prod_{i=5}^{N+1} w^2_i ~.
\end{equation}
We then recognise that the integral, which does not depend on the magnetic flux, evaluates to the Hilbert series of $\mathbb{C}^2/\hat{D}_N$, while the sum over fluxes yields the Hilbert series of $\mathbb{C}^2/\mathbb{Z}_F$. Thus the geometric branch Hilbert series is that of $\mathbb{C}^2/\hat{D}_N \times \mathbb{C}^2/\mathbb{Z}_F$.

Let us discuss this in further details, focusing on the case of $N=4$ for simplicity. The generators of $\BC^2/\BZ_{F}$ are given by
\bes{
G_0 = \tr(X_{15}X_{51}) = \tr(X_{25}X_{52})~, \qquad G_\pm = V_{\pm (1,1,1,1,(1,1))} ~,   
}
satisfying $G_+G_-= G_0^{2F}$. The equalities on the left follow from the $F$-terms associated with the adjoint chiral in the vector multiplet of the $\U(2)$ gauge group, and $\tr$ denotes the trace over the fundamental representation of such a gauge group. On the other hand, the generators of $\BC^2/\hat{D}_4$, along with their relation, are explicitly given in \cite[(3.4), (3.7), Figure 3]{Lindstrom:1999pz}.

We discuss the geometric branch of $D_4$ affine Dynkin quivers with Chern-Simons levels in Appendix \ref{app:affineD4withCSlevels}. It is straightforward to generalise the above analysis to $E_N$ affine Dynkin quivers. Furthermore, we anticipate that rank-two---and more generally higher-rank---$D_N$ affine Dynkin quivers will admit a treatment analogous to that presented for the $A_N$ case in Section \ref{rank2affine}. 

\acknowledgments 
{\small
We thank Fabio Marino and Sinan Moura Soys\"uren for useful conversations. We are also grateful to Mario Francesco D'Angelo for his assistance with the computations in Appendix \ref{app:affineD4withCSlevels}. R.C. is supported by the STFC grant ST/X000575/1. N.M. gratefully acknowledges support from the Simons Center for Geometry and Physics, Stony Brook University, during the 22nd Simons Physics Summer Workshop (2025). N.M. also thanks Seoul National University for hospitality during the workshop ``Aspects of Supersymmetric QFT 2025''. Special thanks go to Carlotta Meneghini and Michele Sarzana for their warm hospitality during the completion of this project. Research of N.M. and W.H. is partially supported by the MUR-PRIN grant No. 2022NY2MXY (Finanziato dall’Unione europea – Next Generation EU, Missione 4 Componente 1 CUP H53D23001080006, I53D23001330006).} 
\appendix

\appendix
\section{Some useful identities}
In this Appendix, we collect various identities used in the main text. 

For linear quivers, we apply the following identity for the $\U(1)_\kappa \times \U(1)_{-\kappa}$ gauge group with a bifundamental hypermultiplet and gauge magnetic fluxes $(m, m)$:
\bes{ \label{usefulintiden}
\oint \frac{du}{2\pi i u} \frac{dv}{2\pi i v} w^m u^{\kappa m} v^{-\kappa m} \PE\left[t \sum_{s=\pm 1} (b u v^{-1})^s  \right]  = \frac{t^{|\kappa| |m|}}{1-t^2} (b^{-\kappa} w)^m~.
}
Note that the factor $(1-t^2)^{-1}$ is actually the $P$-factor $P_{\U(1)}(t; m)$ introduced in \cite{Cremonesi:2013lqa}.

Observe that identity \eref{usefulintiden} can be generalised, for example, to a linear quiver containing the $\U(2)_\kappa \times \U(2)_{-\kappa}$ gauge group and a bifundamental hypermultiplet. We consider the gauge magnetic fluxes $\vec m_1 = \vec m_2=(m_1, m_2)$. There are two contributions: one from the case of $m_1=m_2$ (each $\U(2)$ gauge factor is preserved), and the other from the case of $m_1 \neq m_2$ (each $\U(2)$ gauge factor is broken to $\U(1)^2$):
\bes{ \label{U2identity}
\scalebox{0.92}{$
\begin{split}
&\oint  \Bigg(  \prod_{i=1}^2 \prod_{\alpha=1}^2 \frac{du^{(\alpha)}_i}{2\pi i u^{(\alpha)}_i}  (u^{(\alpha)}_i)^{\kappa_i m^{(\alpha)}_i}  w_i^{m^{(\alpha)}_i} \Bigg)_{\substack{\vec m_1 = \vec m_2=(m_1,m_2) \\ \kappa_1 = -\kappa_2 = \kappa}} \\
&\quad \times \Bigg \{ \delta_{m_1, m_2} \cdot \frac{1}{2}\prod_{1\leq \alpha \neq \beta \leq 2} \prod_{i=1}^2 \Bigg(1-\frac{u^{(\alpha)}_i}{u^{(\beta)}_i} \Bigg)  \PE\left[t \sum_{s=\pm 1} \sum_{\alpha, \beta=1,2} (b u^{(\alpha)}_1)^s (u^{(\beta)}_{2})^{-s}  \right]\\
& \qquad \quad + (1-\delta_{m_1, m_2}) \PE\left[t \sum_{s=\pm 1} \sum_{\alpha=1,2} (b u^{(\alpha)}_1)^s (u^{(\alpha)}_{2})^{-s}  \right] \Bigg \}\\
&= t^{|\kappa| \sum_i |m_i|} \,\, P_{\U(2)}(t; m_1, m_2)\,\, (b^{-\kappa} w_1 w_2)^{\sum_{i} m_i}~,
\end{split}$}
}
where $P_{\U(2)}(t; m_1,m_2)$ is defined as in \cite[(A.2)]{Cremonesi:2013lqa}:
\bes{ \label{PU2}
P_{\mathrm{U}(2)}(t; m_1, m_2) = \begin{cases}
\left[(1-t^2)(1-t^4)\right]^{-1} & \qquad m_1=m_2 \\ 
(1-t^2)^{-2} & \qquad m_1 \neq m_2
\end{cases}~.
}
Moreover, throughout the paper, we consider various limits of the index. For convenience, we collect the contributions of a chiral field with R-charge $R$, whose expression is given by \eref{indexvecchircontrib}, to the index in these limits below:
\bes{ \scalebox{0.93}{
\begin{tabular}{l|c|c}
\hline
Chiral field & Limit $\fz \rightarrow 0$ & Limit $t \rightarrow 0$ \\
\hline
$Z^{\frac{1}{2}}_{\text{chiral}} (x; a z, m)$ & $\left(\fz z^{-1}\right)^{\frac{\abs{m}}{2}} \PE\left[\delta_{m, 0} t z\right]$ & $\left(\fz z^{-1}\right)^{\frac{\abs{m}}{2}}$ \\
$Z^{\frac{1}{2}}_{\text{chiral}} (x; a^{-1} z, m)$ & $\left(t z^{-1}\right)^{\frac{\abs{m}}{2}}$ & $\left(t z^{-1}\right)^{\frac{\abs{m}}{2}} \PE\left[\delta_{m, 0} \fz z\right]$ \\
$Z^{\frac{1}{2}}_{\text{chiral}} (x; z, m)$ & $\left(\fz t\right)^{\frac{\abs{m}}{4}} z^{-\frac{\abs{m}}{2}}$ & $\left(\fz t\right)^{\frac{\abs{m}}{4}} z^{-\frac{\abs{m}}{2}}$ \\
$Z^{1}_{\text{chiral}} (x; a^{2} z, m)$ & $\left(\fz t^{-1} z^{-1}\right)^{\frac{\abs{m}}{2}} \PE\left[\delta_{m, 0} t^2 z\right]$ & $\left(\fz t^{-1} z^{-1}\right)^{\frac{\abs{m}}{2}} \PE\left[-\delta_{m, 0} \fz^2 z^{-1}\right]$ \\
$Z^{1}_{\text{chiral}} (x; a^{-2} z, m)$ & $\left(\fz^{-1} t z^{-1}\right)^{\frac{\abs{m}}{2}} \PE\left[-\delta_{m, 0} t^2 z^{-1}\right]$ & $\left(\fz^{-1} t z^{-1}\right)^{\frac{\abs{m}}{2}} \PE\left[\delta_{m, 0} \fz^2 z\right]$ \\
$Z^{1}_{\text{chiral}} (x; z, m)$ & $z^{-\frac{\abs{m}}{2}}$ & $z^{-\frac{\abs{m}}{2}}$ \\
\hline
\end{tabular}
}
}
\section{Detailed computation of Section \ref{sec:nonabelianpq}} \label{app:detailTU2}
In this Appendix, we collect the derivations of various formulae and arguments presented in Section \ref{sec:nonabelianpq}.

\subsection{The $\fz \rightarrow 0$ limit of the index of $T[\U(2)]$}
Let us derive the $\fz \rightarrow 0$ limit of the index of the  $T[\U(2)]$ theory. We depict such a theory schematically as follows:
\begin{equation} \label{quivTU2}
\begin{tikzpicture}[baseline=0]
\node (Q1) {
\begin{tikzpicture}[font=\footnotesize,
    circ/.style={circle, draw, minimum size=1.3cm},
    sq/.style={rectangle, draw, minimum size=1cm},
    node distance=1.2cm
]

    \node[circ, label=below:{$\red z, l_z$} ] (n1) {$1$};
    \path (n1) edge [loop left] node[red] {$y^{-2}$} ();

    \node[sq, label=below:{$\red \vec f, \vec m_f$}, right=of n1] (s1) {2};


    \draw[<->] (n1) --  node[below, red] {$y$} (s1);

\end{tikzpicture}};
\node (Q2) [right=30mm of Q1] {
\begin{tikzpicture}[font=\footnotesize,
    circ/.style={circle, draw, minimum size=1.3cm},
    sq/.style={rectangle, draw, minimum size=1cm},
    node distance=1.2cm
]

    \node[sq, label=below:{$\red \vec f, \vec m_f$}] (s1) {2};
    \node[sq, label=below:{$\red \vec w, \vec n_w$}, right=of s1] (s2) {2};

    
     Connect the circular nodes
    \draw[blue,dashed] (s1) -- node[above, midway]{\scriptsize $T[\U(2)]$} (s2);
    \path (s1) -- node[below, midway]{$\red y$} (s2);

\end{tikzpicture}};
\node (Q3) [right=1mm of Q1] {$\times \,\ T[\U(1)] \,\ \equiv$};
\end{tikzpicture}
\end{equation}
where $y \in \{a, a^{-1}\}$. In terms of the variables $\fz$ and $t$, from \eref{indexvecchircontrib} and \eref{adjU1},  the contribution to the index of the adjoint chiral field in the $\U(1)$ vector multiplet takes the form
\bes{
Z^{\U(1)}_{\text{adj}} (\fz, t) = \prod_{j = 0}^{\infty} \frac{1 - \fz^{2 j} t^{2 + 2j}}{1 - \fz^{2+2 j} t^{2j}}~,
}
which admits the following behaviour in the limits $\fz \rightarrow 0$ and $t \rightarrow 0$:
\begin{subequations}  
\begin{align}
\begin{split} \label{adjU1fzto0}
Z^{\U(1)}_{\text{adj}} (\fz, t) &\overset{\fz \rightarrow 0}{=} 1 - t^2~,
\end{split} \\
\begin{split} \label{adjU1tto0}
Z^{\U(1)}_{\text{adj}} (\fz, t) &\overset{t \rightarrow 0}{=} \frac{1}{1 - \fz^2}~.
\end{split}
\end{align}
\end{subequations}
Let us also analyse the contribution coming from the hypermultiplet in the bifundamental representation of $\U(1) \times \U(2)$, which, using \eref{indexvecchircontrib} and \eref{hyperU1U2}, is given by
\bes{
\scalebox{0.93}{$
\begin{split}
Z^{\U(1) \times \U(2)}_{\text{hyper}} (\fz, t) &= \fz^{\sum_{i=1}^2 \abs{l_z - {\vec{m}_f}_i}} \\& \times \prod_{s=\pm 1} \prod_{i=1}^2 \prod_{j=0}^{\infty} \frac{1-(-1)^{s l_z - s {m_f}_i} \fz^{2 + \abs{s l_z - s {m_f}_i}+2 j} t^{1 + \abs{s l_z - s {m_f}_i}+2 j} z^{-s} f_i^s}{1-(-1)^{s l_z - s {m_f}_i} \fz^{\abs{s l_z - s {m_f}_i}+2 j} t^{1 + \abs{s l_z - s {m_f}_i}+2 j} z^{s} f_i^{-s}}~.
\end{split}
$}
}
In the limits $\fz \rightarrow 0$ and $t \rightarrow 0$, the expression above simplifies to\footnote{Observe that a single chiral field of R-charge $\frac{1}{2}$, whose expression is given in \eref{indexvecchircontrib}, behaves as $Z^{\frac{1}{2}}_{\text{chiral}} (\fz, t; z, m)\overset{\fz \rightarrow 0}{=} \delta_{m,0} \frac{1}{1-t z}$, $Z^{\frac{1}{2}}_{\text{chiral}} (\fz, t; z, m)\overset{t \rightarrow 0}{=} \left(\fz z^{-1}\right)^{\frac{\abs{m}}{2}}$, see \cite{Razamat:2014pta}.}
\begin{subequations}  
\begin{align}
\begin{split} \label{hyperU1U2fzto0}
Z^{\U(1) \times \U(2)}_{\text{hyper}} (\fz, t) \overset{\fz \rightarrow 0}{=} \delta_{{m_f}_1, {m_f}_2} \delta_{l_z, {m_f}_1} \prod_{s=\pm 1} \prod_{i=1}^2  \frac{1}{1-t z^{s} f_i^{-s}}~,
\end{split} \\
\begin{split} \label{hyperU1U2tto0}
Z^{\U(1) \times \U(2)}_{\text{hyper}} (\fz, t) \overset{t \rightarrow 0}{=} \fz^{\abs{l_z-{m_f}_1}+\abs{l_z-{m_f}_2}}~.
\end{split}
\end{align}
\end{subequations}
It follows that, in the case ${m_f}_1 = {m_f}_2$, the index of the $T[\U(2)]$ theory admits the following expression in the limit $\fz \rightarrow 0$:
\bes{
&Z_{T[\U(2)]} (\fz, t; {m_f}_1 = {m_f}_2) \\&\overset{\fz \rightarrow 0}{=} \sum_{l_z \in \BZ} \oint \frac{dz}{2 \pi i z} \left(f_1 f_2\right)^{{n_w}_2} w_2^{{m_f}_1+{m_f}_2} \left(\frac{w_1}{w_2}\right)^{l_z} z^{{n_w}_1-{n_w}_2} \\& \qquad \times \left(1-t^2\right) \delta_{{m_f}_1, {m_f}_2} \delta_{l_z, {m_f}_1} \prod_{s=\pm 1} \prod_{i=1}^2  \frac{1}{1-t z^{s} f_i^{-s}}
\\&=\oint \frac{dz}{2 \pi i z} \left(f_1 f_2\right)^{{n_w}_2} w_2^{2 {m_f}_1} \left(\frac{w_1}{w_2}\right)^{l_z} z^{{n_w}_1-{n_w}_2} \\& \qquad \times \frac{1-t^2}{\left(1-t \frac{z}{f_1}\right) \left(1-t \frac{f_1}{z}\right) \left(1-t \frac{z}{f_2}\right) \left(1-t \frac{f_2}{z}\right)}~.
}
Upon computing the residues at the poles of the gauge fugacity $z$, this yields the Hall-Littlewood formula for the Hilbert series of $T[\U(2)]$, as defined in \cite[Section 3.1.2]{Cremonesi:2014kwa},\footnote{Actually, in that reference, the Hall-Littlewood formula is provided for the $T[\SU(2)]$ theory. Here, instead, we are dealing with the $T[\U(2)]$ theory, hence an extra term $\left(w_1 w_2\right)^{{m_f}_1}$, which is absent in the Hall-Littlewood formula for $T[\SU(2)]$, appears. As a further comment, observe that we we can rewrite $\frac{1}{\left(1-t^2 \frac{f_1}{f_2}\right) \left(1-t^2 \frac{f_2}{f_1}\right)}$ in \eref{TU2fzto0eqm} as $\left(1-t^2\right)^2 K^{\U(2)}_{[1^2]}(t^2; f_1, f_2)$, where 
\bes{K^{\U(2)}_{[1^2]}(t; x_1, x_2) = \PE\left[t \chi^{\U(2)}_{\text{adj}} (x_1, x_2)\right] = \prod_{1 \le \alpha, \beta \le 2} \frac{1}{1-t x_{\alpha} x^{-1}_{\beta}}
}
is defined in \cite[(3.20)]{Cremonesi:2014kwa}.
} where we take ${n_w}_1 \ge {n_w}_2$: 
\bes{ \label{TU2fzto0eqm}
Z_{T[\U(2)]} (\fz, t; {m_f}_1 = {m_f}_2) &\overset{\fz \rightarrow 0}{=} \left(w_1 w_2\right)^{{m_f}_1} t^{{n_w}_1-{n_w}_2} \\& \,\ \times \frac{1}{\left(1-t^2 \frac{f_1}{f_2}\right) \left(1-t^2 \frac{f_2}{f_1}\right)} \Psi^{({n_w}_1, {n_w}_2)}_{\U(2)} \left(t^2; f_1, f_2\right)~,
}
where the expression for the $\U(2)$ Hall-Littlewood polynomial is given by \cite[(B.13)]{Cremonesi:2014kwa}
\bes{
\Psi^{(n_1, n_2)}_{\U(2)} \left(t; x_1, x_2\right) = \frac{t x_1^{1+n_2} x_2^{n_1} - x_1^{n_2} x_2^{1+n_1} + x_1^{1+n_1} x_2^{n_2} - t x_1^{n_1} x_2^{1+n_2}}{x_1 - x_2}~.
}
If ${n_w}_1 < {n_w}_2$, it is sufficient to swap ${n_w}_1 \leftrightarrow {n_w}_2$ in the equation \eref{TU2fzto0eqm}. Let us now consider the case ${m_f}_1 \neq {m_f}_2$ in the $\fz \rightarrow 0$ limit. Since \eref{hyperU1U2fzto0} vanishes, we have to study the leading order in $\fz$, \ie the expression with the lowest non-zero power of $\fz$ which appears in the limit of the index of the $T[\U(2)]$ theory . As can be observed from \eref{hyperU1U2tto0}, given that we sum over $l_z \in \BZ$, the leading order in $\fz$ is reached when ${m_f}_{\text{min}} \equiv \min\{{m_f}_1, {m_f}_2\} \le l_z \le \max\{{m_f}_1, {m_f}_2\} \equiv {m_f}_{\text{max}}$, for which the power of $\fz$ is ${\abs{{m_f}_1 - {m_f}_2}}$. For instance, if we focus on the contribution coming from $l_z = {m_f}_1$, the hypermultiplet in the bifundamental representation of $\U(1) \times \U(2)$ behaves as
\bes{ \label{hyperU1U2fzto0lzeqmf1}
Z^{\U(1) \times \U(2)}_{\text{hyper}} (\fz, t; l_z = {m_f}_1) \overset{\fz \rightarrow 0}{=} \left[\delta_{l_z, {m_f}_1} \frac{1}{\left(1-t \frac{z}{f_1}\right) \left(1-t \frac{f_1}{z}\right)}\right] \fz^{\abs{l_z-{m_f}_2}}~,
}
then the expression for the $T[\U(2)]$ theory in the $l_z = {m_f}_1 \neq {m_f}_2$ sector reads
\bes{ \label{TU2fzto0lzeqmf1}
\sum_{l_z \in \BZ} &\oint \frac{dz}{2 \pi i z} \left(f_1 f_2\right)^{{n_w}_2} w_2^{{m_f}_1+{m_f}_2} \left(\frac{w_1}{w_2}\right)^{l_z} z^{{n_w}_1-{n_w}_2} \\& \times \left(1-t^2\right) \delta_{l_z, {m_f}_1} \frac{1}{\left(1-t \frac{z}{f_1}\right) \left(1-t \frac{f_1}{z}\right)} \fz^{\abs{l_z-{m_f}_2}}
\\=&\oint \frac{dz}{2 \pi i z} \left(f_1 f_2\right)^{{n_w}_2} w_1^{{m_f}_1} w_2^{{m_f}_2} z^{{n_w}_1-{n_w}_2} \frac{1-t^2}{\left(1-t \frac{z}{f_1}\right) \left(1-t \frac{f_1}{z}\right)} \fz^{\abs{{m_f}_1-{m_f}_2}}
\\ =& \left[w_1^{{m_f}_1} \Psi^{({n_w}_1)}_{\U(1)} \left(f_1\right) \right] \left[w_2^{{m_f}_2} \Psi^{({n_w}_2)}_{\U(1)} \left(f_2\right) \right] \fz^{\abs{{m_f}_1-{m_f}_2}} t^{\abs{{n_w}_1-{n_w}_2}}~,
}
where $\Psi^{(n)}_{\U(1)} \left(x\right) = x^n$. This is just the product of the Hall-Littlewood formulae of two $T[\U(1)]$ theories, multiplied by an overall factor $\fz^{\abs{{m_f}_1-{m_f}_2}} t^{\abs{{n_w}_1-{n_w}_2}}$. 
Similarly, the contribution coming from the $l_z = {m_f}_2$ sector can be obtained immediately from \eref{hyperU1U2fzto0lzeqmf1} and \eref{TU2fzto0lzeqmf1} by swapping the pair $(f_1, {m_f}_1)$ with $(f_2, {m_f}_2)$. Finally, we also have to consider the ${m_f}_{\text{min}} < l_z < {m_f}_{\text{max}}$ sectors, which contribute to the limit as
\bes{ \label{TU2fzto0mfminmfmax}
\scalebox{0.86}{$
\begin{split}
&\sum_{l_z = {m_f}_{\text{min}}+1}^{{m_f}_{\text{max}}-1} \oint \frac{dz}{2 \pi i z} \left(f_1 f_2\right)^{{n_w}_2} w_2^{{m_f}_1+{m_f}_2} \left(\frac{w_1}{w_2}\right)^{l_z} z^{{n_w}_1-{n_w}_2} \left(1-t^2\right) \fz^{\abs{l_z-{m_f}_1}+\abs{l_z-{m_f}_2}}
\\ & = \delta_{{n_w}_1, {n_w}_2} \left(1-t^2\right) \fz^{\abs{{m_f}_1-{m_f}_2}} \,\ \sum_{i = 1}^{\abs{{m_f}_1-{m_f}_2}-1} \left[w_1^{{m_f}_{\text{min}}+i} \Psi^{({n_w}_1)}_{\U(1)} \left(f_1\right) \right] \left[w_2^{{m_f}_{\text{max}}-i} \Psi^{({n_w}_2)}_{\U(1)} \left(f_2\right) \right]~.
\end{split}
$}
}
The expressions \eref{TU2fzto0lzeqmf1} and \eref{TU2fzto0mfminmfmax}, together with the one for the $l_z = {m_f}_2$ contribution to the limit, can be also written in a more meaningful way by manifestly distinguishing the two cases in which the background fluxes ${n_w}_1$ and ${n_w}_2$ either differ or take the same value. In particular, in the latter case, the said expressions can be incorporated by reformulating the various terms of the type $w^{{m_f}_i}_1 w^{{m_f}_j}_2$ in terms of $\U(2)$ Hall-Littlewood polynomials as functions of $\fz^2$, $(w_1, w_2)$ and the background fluxes $({n_w}_1, {n_w}_2)$. In such a way, taking into account also \eref{TU2fzto0eqm}, we finally obtain the complete expression for the $\fz \rightarrow 0$ limit of the index of the $T[\U(2)]$ theory, which reads
\bes{ \label{TU2fzto0}
\scalebox{0.77}{$
\begin{split}
& Z^{\fz \rightarrow 0}_{T[\U(2)]} (\fz; t | \vec f; \vec m_f | \vec w; \vec n_w) = \delta_{{m_f}_1, {m_f}_2} \left(w_1 w_2\right)^{{m_f}_1} t^{\abs{{n_w}_1-{n_w}_2}} \\& \qquad \times \frac{\Theta({n_w}_1-{n_w}_2) \Psi^{({n_w}_1, {n_w}_2)}_{\U(2)} \left(t^2; f_1, f_2\right) + \Theta({n_w}_2-{n_w}_1-1) \Psi^{({n_w}_2, {n_w}_1)}_{\U(2)} \left(t^2; f_1, f_2\right)}{\left(1-t^2 \frac{f_1}{f_2}\right) \left(1-t^2 \frac{f_2}{f_1}\right)}
\\& \qquad + \left(1-\delta_{{m_f}_1, {m_f}_2}\right) \fz^{\abs{{m_f}_1-{{m_f}_2}}} \Bigg\{ \delta_{{n_w}_1, {n_w}_2} \left(f_1 f_2\right)^{{n_w}_1} \frac{1}{\left(1-\fz^2 \frac{w_1}{w_2}\right) \left(1-\fz^2 \frac{w_2}{w_1}\right)} \\ & \qquad \qquad \times \Big( \Theta({m_f}_1 - {m_f}_2) \Big[ \Psi^{({m_f}_1, {m_f}_2)}_{\U(2)} \left(\fz^2; w_1, w_2\right) - t^2 \Psi^{({m_f}_1 - 1, {m_f}_2 + 1)}_{\U(2)} \left(\fz^2; w_1, w_2\right) + \cO\left(\fz^2\right) \Big]\\ & 
\qquad \qquad \,\,\ + \Theta({m_f}_2 - {m_f}_1 - 1) \Big[\Psi^{({m_f}_2, {m_f}_1)}_{\U(2)} \left(\fz^2; w_1, w_2\right) - t^2 \Psi^{({m_f}_2 - 1, {m_f}_1 + 1)}_{\U(2)} \left(\fz^2; w_1, w_2\right) + \cO\left(\fz^2\right) \Big] \Big)\\ & 
\qquad \quad + \left(1-\delta_{{n_w}_1, {n_w}_2}\right) t^{\abs{{n_w}_1-{n_w}_2}} \sum_{i \neq j = 1}^2 \left[w_1^{{m_f}_i} \Psi^{({n_w}_1)}_{\U(1)} \left(f_i\right) \right] \left[w_2^{{m_f}_j} \Psi^{({n_w}_2)}_{\U(1)} \left(f_j\right) \right]\Bigg\}~,
\end{split}
$}
}
where we denote by $\Theta(x)$ the Heaviside step function, such that $\Theta(x) = 1$ for $x \ge 0$ and $\Theta(x) = 0$ for $x < 0$.
Let us comment on the result \eref{TU2fzto0}. As already pointed out around \eref{TU2fzto0eqm}, when the two magnetic fluxes associated with the $\U(2)$ flavour symmetry are equal, \ie ${m_f}_1 = {m_f}_2$, the $\fz \rightarrow 0$ limit of the index reproduces the Hall-Littlewood formula of the $T[\U(2)]$ theory. On the other hand, when the two magnetic fluxes ${m_f}_1$ and ${m_f}_2$ take different values, the $\fz \rightarrow 0$ limit of the index of the $T[\U(2)]$ theory yields a sum of various terms. Explicitly, when ${n_w}_1 = {n_w}_2$, these can be expressed in terms of $\U(2)$ Hall-Littlewood polynomials in the variables $\fz$, $(w_1, w_2)$ and $({n_w}_1, {n_w}_2)$, which give rise to a modified $T[\U(2)]$ Hall-Littlewood-type formula which is proportional to a positive power of $\fz$, appearing from the third to the fifth line in \eref{TU2fzto0}. On the other hand, when both ${m_f}_1 \neq {m_f}_2$ and ${n_w}_1 \neq {n_w}_2$, then the various terms appearing in the last line of the above limit of the $T[\U(2)]$ limit correspond to the product of two $T[\U(1)]$ couplings. 

\subsection{The $t \rightarrow 0$ limit of the index of $T[\U(2)]$}
Let us now consider the $t \rightarrow 0$ limit of the $T[\U(2)]$ theory. Taking into account \eref{adjU1tto0} and \eref{hyperU1U2tto0}, this reads
\bes{ \label{TU2tto0deltan1n2}
&\sum_{l_z \in \BZ} \oint \frac{dz}{2 \pi i z} \left(f_1 f_2\right)^{{n_w}_2} w_2^{{m_f}_1+{m_f}_2} \left(\frac{w_1}{w_2}\right)^{l_z} z^{{n_w}_1-{n_w}_2}  \frac{\fz^{\abs{l_z-{m_f}_1}+\abs{l_z-{m_f}_2}}}{\left(1-\fz^2\right)} 
\\&= \delta_{{n_w}_1, {n_w}_2} \sum_{l_z \in \BZ} \frac{\left(f_1 f_2\right)^{{n_w}_2} w_1^{l_z} w_2^{{m_f}_1+{m_f}_2-l_z} \fz^{\abs{l_z-{m_f}_1}+\abs{l_z-{m_f}_2}}}{\left(1-\fz^2\right)}~.
}
The infinite series in the expression above can be resummed in closed form thanks to the following identities, valid for ${m_f}_1 \ge {m_f}_2$:
\begin{subequations} 
\begin{align}
\begin{split} \label{TU2tto0sumlzlmf1}
\sum_{l_z = {m_f}_1 +1}^{+\infty} \frac{\left(\frac{w_1}{w_2}\right)^{l_z} \fz^{\abs{l_z-{m_f}_1}+\abs{l_z-{m_f}_2}}}{\left(1-\fz^2\right)} &= \frac{\left(\frac{w_1}{w_2}\right)^{1+{m_f}_1} \fz^{2+{m_f}_1-{m_f}_2}}{\left(1-\fz^2\right) \left(1-\fz^2 \frac{w_1}{w_2}\right)}~,
\end{split} \\
\begin{split} \label{TU2tto0summf1llzlmf2}
\sum_{l_z = {m_f}_2}^{{m_f}_1} \frac{\left(\frac{w_1}{w_2}\right)^{l_z} \fz^{\abs{l_z-{m_f}_1}+\abs{l_z-{m_f}_2}}}{\left(1-\fz^2\right)} &= \frac{w_1 \left(\frac{w_1}{w_2}\right)^{{m_f}_1}-w_2 \left(\frac{w_1}{w_2}\right)^{{m_f}_2}}{\left(1-\fz^2\right) \left(w_1 - w_2 \right)} \fz^{{m_f}_1 - {m_f}_2}~,
\end{split}\\
\begin{split} \label{TU2tto0summf2lz}
\sum_{l_z = -\infty}^{{m_f}_2 -1} \frac{\left(\frac{w_1}{w_2}\right)^{l_z} \fz^{\abs{l_z-{m_f}_1}+\abs{l_z-{m_f}_2}}}{\left(1-\fz^2\right)}&= \frac{\left(\frac{w_1}{w_2}\right)^{-1+{m_f}_2} \fz^{2+{m_f}_1-{m_f}_2}}{\left(1-\fz^2\right) \left(1-\fz^2 \frac{w_2}{w_1}\right)}~.
\end{split}
\end{align}
\end{subequations}
Upon summing together the contributions \eref{TU2tto0sumlzlmf1}--\eref{TU2tto0summf2lz}, the expression \eref{TU2tto0deltan1n2} for ${m_f}_1 \ge {m_f}_2$ reduces to
\bes{ \label{TU2tto0deltan1n2closed}
\delta_{{n_w}_1, {n_w}_2} \left(f_1 f_2\right)^{{n_w}_1} \fz^{{m_f}_1-{m_f}_2} \frac{1}{\left(1-\fz^2 \frac{w_1}{w_2}\right) \left(1-\fz^2 \frac{w_2}{w_1}\right)} \Psi^{({m_f}_1, {m_f}_2)}_{\U(2)} \left(\fz^2; w_1, w_2\right)~,
}
whereas, in the case ${m_f}_2 > {m_f}_1$, it is sufficient to swap ${m_f}_1 \leftrightarrow {m_f}_2$ in the expression above. Observe that this reproduces exactly \eref{TU2fzto0eqm} upon exchanging $(t; \vec f, \vec m_f)$ with $(\fz; \vec w, \vec n_w)$, in agreement with the self-mirror property of the $T[\U(2)]$ theory.\footnote{The $\fz \rightarrow 0$ and $t \rightarrow 0$ limits of the $T[\U(2)]$ theory, as defined in \eqref{N=3reparam} with axial fugacity $y = a$ in \eref{quivTU2}, coincide with the Higgs and Coulomb branch limits, respectively, introduced in \cite[(3.37)]{Razamat:2014pta}. Since the $T[\U(2)]$ SCFT enjoys self-mirror symmetry, the two limits coincide.} Since \eref{TU2tto0deltan1n2closed} vanishes when ${n_w}_1 \neq {n_w}_2$, in such a case we have to isolate the terms appearing with the lowest power of $t$, which contribute when the gauge flux $l_z$ takes one of the following values: $l_z = {m_f}_1 \neq {m_f}_2$, $l_z = {m_f}_2 \neq {m_f}_1$, and $l_z = {m_f}_1 = {m_f}_2$. For instance, in the $l_z = {m_f}_1 \neq {m_f}_2$ sector, we consider\footnote{Let us comment on the difference between the expressions \eref{hyperU1U2tto0lzeqmf1} and \eref{hyperU1U2fzto0lzeqmf1}. The quantity inside the square bracket in \eref{hyperU1U2tto0lzeqmf1} is $\frac{1 + \CO(\fz^2 t)}{1+ \CO(t)}$, which reduces to $\frac{1}{1+ \CO(t)}$ when $\fz \rightarrow 0$, which explains why the numerator is equal to one in \eref{hyperU1U2fzto0lzeqmf1}. On the other hand, in \eref{hyperU1U2tto0lzeqmf1}, since we are interested in the lowest order in $t$, the numerator and the denominator are of the type $1 + \CO(t)$, hence we have to keep them both.}
\bes{ \label{hyperU1U2tto0lzeqmf1}
Z^{\U(1) \times \U(2)}_{\text{hyper}} (\fz, t; l_z = {m_f}_1) \overset{t \rightarrow 0}{=} \left[\delta_{l_z, {m_f}_1} \frac{\left(1-\fz^2 t \frac{z}{f_1}\right) \left(1-\fz^2 t \frac{f_1}{z}\right)}{\left(1-t \frac{z}{f_1}\right) \left(1-t \frac{f_1}{z}\right)}\right] \fz^{\abs{l_z-{m_f}_2}}~,
}
then the corresponding contribution to the $t \rightarrow 0$ limit of the index of the $T[\U(2)]$ theory reads
\bes{ \label{TU2tto0lzeqmf1}
&\oint \frac{dz}{2 \pi i z} \left(f_1 f_2\right)^{{n_w}_2} w_1^{{m_f}_1} w_2^{{m_f}_2} z^{{n_w}_1-{n_w}_2} \\& \qquad \times \frac{1}{1-\fz^2} \frac{\left(1-\fz^2 t \frac{z}{f_1}\right) \left(1-\fz^2 t \frac{f_1}{z}\right)}{\left(1-t \frac{z}{f_1}\right) \left(1-t \frac{f_1}{z}\right)} \fz^{\abs{{m_f}_1-{m_f}_2}}
\\ &= \frac{\left[f_1^{{n_w}_1} \Psi^{({m_f}_1)}_{\U(1)} \left(w_1\right) \right] \left[f_2^{{n_w}_2} \Psi^{({m_f}_2)}_{\U(1)} \left(w_2\right) \right] \left(1-\fz^2 t^2\right) \fz^{\abs{{m_f}_1-{m_f}_2}} t^{\abs{{n_w}_1-{n_w}_2}}}{\left(1-t^2\right)}~.
}
Analogously, the contribution coming from the $l_z = {m_f}_2 \neq {m_f}_1$ sector can be achieved by swapping $(f_1, {m_f}_1)$ with $(f_2, {m_f}_2)$ in both \eref{hyperU1U2tto0lzeqmf1} and \eref{TU2tto0lzeqmf1}. Finally, when $l_z = {m_f}_1 = {m_f}_2$, the $\U(1) \times \U(2)$ bifundamental hypermultiplet contributes as
\bes{ \label{hyperU1U2tto0lzeqmf1eqmf2}
&Z^{\U(1) \times \U(2)}_{\text{hyper}} (\fz, t; l_z = {m_f}_1 = {m_f}_2) \\&\overset{t \rightarrow 0}{=} \delta_{{m_f}_1, {m_f}_2} \delta_{l_z, {m_f}_1} \frac{\left(1-\fz^2 t \frac{z}{f_1}\right) \left(1-\fz^2 t \frac{f_1}{z}\right) \left(1-\fz^2 t \frac{z}{f_2}\right) \left(1-\fz^2 t \frac{f_2}{z}\right)}{\left(1-t \frac{z}{f_1}\right) \left(1-t \frac{f_1}{z}\right) \left(1-t \frac{z}{f_2}\right) \left(1-t \frac{f_2}{z}\right)}
}
to the limit. For such value of the gauge flux $l_z$, the $t \rightarrow 0$ limit of the index of the $T[\U(2)]$ theory with ${n_w}_1 \ge {n_w}_2$ then reads
\bes{ \label{TU2tto0lzeqmf1eqmf2}
&\oint \frac{dz}{2 \pi i z} \left(f_1 f_2\right)^{{n_w}_2} \left(w_1 w_2\right)^{{m_f}_1} z^{{n_w}_1-{n_w}_2} \\& \qquad \times \frac{1}{1-\fz^2} \frac{\left(1-\fz^2 t \frac{z}{f_1}\right) \left(1-\fz^2 t \frac{f_1}{z}\right) \left(1-\fz^2 t \frac{z}{f_2}\right) \left(1-\fz^2 t \frac{f_2}{z}\right)}{\left(1-t \frac{z}{f_1}\right) \left(1-t \frac{f_1}{z}\right) \left(1-t \frac{z}{f_2}\right) \left(1-t \frac{f_2}{z}\right)}
\\ &= \left(w_1 w_2\right)^{{m_f}_1} t^{{n_w}_1-{n_w}_2} \frac{\left(1-\fz^2 t^2\right)}{\left(1-t^2\right) \left(1-t^2 \frac{f_1}{f_2}\right) \left(1-t^2 \frac{f_2}{f_1}\right)} \\ & \qquad \times \Big[\Psi^{({n_w}_1, {n_w}_2)}_{\U(2)} \left(t^2; f_1, f_2\right) - \fz^2 \Psi^{({n_w}_1 - 1, {n_w}_2 + 1)}_{\U(2)} \left(t^2; f_1, f_2\right) \\ & \qquad - \fz^2 t^2 \Psi^{({n_w}_1 + 1, {n_w}_2 - 1)}_{\U(2)} \left(t^2; f_1, f_2\right) + \fz^4 t^2 \Psi^{({n_w}_1, {n_w}_2)}_{\U(2)} \left(t^2; f_1, f_2\right)\Big]~,
}
whereas, for ${n_w}_2 \ge {n_w}_1$, it is sufficient to swap ${n_w}_1 \leftrightarrow {n_w}_2$ in the expression above. From \eref{TU2tto0deltan1n2closed}, \eref{TU2tto0lzeqmf1} and \eref{TU2tto0lzeqmf1eqmf2}, we can finally report a closed formula for the $t \rightarrow 0$ limit of the index of the $T[\U(2)]$ theory, which can be written compactly as
\bes{ \label{TU2tto0}
\scalebox{0.8}{$
\begin{split}
&Z^{t \rightarrow 0}_{T[\U(2)]} (\fz, t| \vec f; \vec m_f | \vec w; \vec n_w) = \delta_{{n_w}_1, {n_w}_2} \left(f_1 f_2\right)^{{n_w}_1} \fz^{\abs{{m_f}_1-{m_f}_2}} \\ & \qquad \times \frac{\Theta({m_f}_1-{m_f}_2) \Psi^{({m_f}_1, {m_f}_2)}_{\U(2)} \left(\fz^2; w_1, w_2\right) + \Theta({m_f}_2-{m_f}_1-1) \Psi^{({m_f}_2, {m_f}_1)}_{\U(2)} \left(\fz^2; w_1, w_2\right)}{\left(1-\fz^2 \frac{w_1}{w_2}\right) \left(1-\fz^2 \frac{w_2}{w_1}\right)} 
\\& \qquad + \left(1-\delta_{{n_w}_1, {n_w}_2}\right) t^{\abs{{n_w}_1-{n_w}_2}}
\Bigg\{ \delta_{{m_f}_1, {m_f}_2} \left(w_1 w_2\right)^{{m_f}_1} \frac{1}{\left(1-t^2 \frac{f_1}{f_2}\right) \left(1-t^2 \frac{f_2}{f_1}\right)} \\ & \qquad \qquad \times \Big( \Theta({n_w}_1 - {n_w}_2) \Big[\Psi^{({n_w}_1, {n_w}_2)}_{\U(2)} \left(t^2; f_1, f_2\right) - \fz^2 \Psi^{({n_w}_1 - 1, {n_w}_2 + 1)}_{\U(2)} \left(t^2; f_1, f_2\right) + \cO\left(t^2\right) \Big]\\ & 
\qquad \qquad \,\,\ + \Theta({n_w}_2 - {n_w}_1 - 1) \Big[\Psi^{({n_w}_2, {n_w}_1)}_{\U(2)} \left(t^2; f_1, f_2\right) - \fz^2 \Psi^{({n_w}_2 - 1, {n_w}_1 + 1)}_{\U(2)} \left(t^2; f_1, f_2\right) + \cO\left(t^2\right) \Big] \Big)\\ & 
\qquad \quad + \left(1-\delta_{{m_f}_1, {m_f}_2}\right) \fz^{\abs{{m_f}_1-{m_f}_2}} \sum_{i \neq j = 1}^2 \left[f_1^{{n_w}_i} \Psi^{({m_f}_1)}_{\U(1)} \left(w_i\right) \right] \left[f_2^{{n_w}_j} \Psi^{({m_f}_2)}_{\U(1)} \left(w_j\right) \right]\Bigg\}~.
\end{split}
$}
}
Observe that the limit above coincides with \eref{TU2fzto0} upon exchanging $\left(t; \vec{f}, \vec{m_f}\right)$ with $\left(\fz; \vec{w}, \vec{n_w}\right)$, in agreement with the self-mirror property of the $T[\U(2)]$ theory:
\bes{ \label{selfmirrorTU2}
Z^{\fz \rightarrow 0}_{T[\U(2)]} (\fz; t | \vec f; \vec m_f | \vec w; \vec n_w) = Z^{t \rightarrow 0}_{T[\U(2)]} (t; \fz | \vec w; \vec n_w | \vec f; \vec m_f )~.
}

\subsection{Comments on various magnetic flux configurations} \label{app:vanishfluxes}
Below \eqref{limpqwithTU2a}, we discuss the nontrivial contributions of the following configurations of magnetic fluxes:
\ben
\item \label{item1a} $\{m^{(1)}_2 = m^{(2)}_2 = m^{(1)}_3 = m^{(2)}_3 = \fm\}$: the only non-trivial contributions come from the sectors with $\{m^{(1)}_1 = m^{(2)}_1 = \fl\}$, $\{m^{(1)}_4 = m^{(2)}_4 = \fn\}$.

\item \label{item2a} $\{m^{(1)}_2 = m^{(1)}_3 = \fm_1, m^{(2)}_2 = m^{(2)}_3 = \fm_2\}$: the only non-vanishing contributions come from the sectors with $\{m^{(1)}_1 = \fl_1 \neq m^{(2)}_1 = \fl_2\}$, $\{m^{(1)}_4 = \fn_1 \neq m^{(2)}_4 = \fn_2\}$.
\item $\{m^{(1)}_2 = m^{(2)}_3 = \fm_1, m^{(2)}_2 = m^{(1)}_3 = \fm_2\}$: this choice of magnetic fluxes is analogous to case \ref{item2a}, to which is related simply by permuting $m^{(1)}_3$ and $m^{(2)}_3$. 
\een
We now provide the arguments for the trivial contribution of the magnetic fluxes $\{m^{(\alpha)}_1\}$ and $\{m^{(\alpha)}_4\}$ that do not satisfy the above conditions.

\paragraph{Point \ref{item1a}.} If one considers the monopole sectors with $m^{(1)}_1 \neq m^{(2)}_1$ instead, the only contributions of the gauge fugacities $\{u^{(\alpha)}_1\}$ appearing in the $\fz \rightarrow 0$ limit of \eref{limpqwithTU2a} are the the integration measure $\prod\limits_{\alpha=1}^2 \frac{du^{(\alpha)}_1}{2\pi i u^{(\alpha)}_1}$, the terms containing the Chern-Simons couplings $(u^{(1)}_1)^{\kappa_1 m^{(1)}_1}$ and $(u^{(2)}_1)^{\kappa_1 m^{(2)}_1}$, as well as the factor $\left(u^{(1)}_1 u^{(2)}_1\right)^{\fm}$ coming from the $T[\U(2)]$ block connecting the $\U(2)$ nodes with gauge fugacities $\{u^{(\alpha)}_1\}$ and $\{u^{(\alpha)}_2\}$. The integrand picks up a residue in $u^{(1)}_1 = 0$ for $\fm = - \kappa_1 m^{(1)}_1$, but there is no residue in $u^{(2)}_1 = 0$ since $\fm \neq - \kappa_1 m^{(2)}_1$ in these sectors. Analogous arguments also hold when $m^{(1)}_4 \neq m^{(2)}_4$.

\paragraph{Point \ref{item2a}.} The magnetic fluxes satisfying $\{m^{(1)}_1 = m^{(2)}_1 = \fl\}$, $\{m^{(1)}_4 = m^{(2)}_4 = \fn\}$ do not contribute to the $\fz \rightarrow 0$ limit of the index. This statement can be explained by focusing, for instance, just on the gauge fugacities $\{u^{(\alpha)}_1\}$, where we can assume $\fm_1 > \fm_2$. These fugacities appear in \eref{limpqwithTU2a} due to the contributions coming from the integration measure $\prod\limits_{\alpha=1}^2 \frac{du^{(\alpha)}_1}{2\pi i u^{(\alpha)}_1}$, the Chern-Simons terms $(u^{(1)}_1 u^{(2)}_1)^{\fq \fl}$, the factor $2-\frac{u^{(1)}_1}{u^{(2)}_1} - \frac{u^{(2)}_1}{u^{(1)}_1}$ associated with the leftmost $\U(2)$ vector multiplet in \eref{quivpqbraneTU2}, as well as the contribution associated with the $T[\U(2)]$ block on the left, namely $\Psi^{({\fm}_1, {\fm}_2)}_{\U(2)} \left(t^2; u^{(1)}_1, u^{(2)}_1\right) \times \PE\left[\sum\limits_{s = \pm 1} t^2 (u^{(1)}_1)^s (u^{(2)}_1)^{-s}\right]$. Upon parametrising $\fq \fl + \fm_1 = \fx$ and $\fq \fl + \fm_2 = \fy$, with $\fx > \fy$, the whole $\{u^{(\alpha)}_1\}$-dependence of the integrand in \eref{limpqwithTU2a} can be expressed compactly as
\bes{
\left(u^{(1)}_1 - u^{(2)}_1\right) \left[\frac{(u^{(1)}_1)^{\fx-1} (u^{(2)}_1)^{\fy-1}}{t^2 u^{(1)}_1 - u^{(2)}_1} + \frac{(u^{(1)}_1)^{\fy-1} (u^{(2)}_1)^{\fx-1}}{u^{(1)}_1 - t^2 u^{(2)}_1}\right]~.
}
Two different possibilities must be taken into account here.
\begin{itemize}
\item Let us consider the situation in which there are no poles in $u^{(1)}_1 = 0$ first. This happens if $\fy \ge 1$, which also implies $\fx \ge 2$. In such a case, the computation of the residue in $u^{(1)}_1 = t^2 u^{(2)}_1$ yields a contribution coming from the gauge fugacity $u^{(2)}_1$ with power $\fx + \fy - 1$, which is always positive. Consequently, there is no residue to be taken with respect to $u^{(2)}_1$. \item Next, let us focus, for instance, on the quantity 
\bes{
\frac{(u^{(1)}_1)^{\fy-1} (u^{(2)}_1)^{\fx}}{u^{(1)}_1 - t^2 u^{(2)}_1}~,
}
which possesses a pole of order $1-\fy$ in $u^{(1)}_1 = 0$. The computation of the residue at such pole yields a negative term $-t^{2 \fy -2} (u^{(2)}_1)^{\fx+\fy-1}$, which perfectly cancels the result coming from the computation of the residue in $u^{(1)}_1 = t^2 u^{(2)}_1$, that is $t^{2 \fy -2} (u^{(2)}_1)^{\fx+\fy-1}$. The total contribution then cancels, meaning that the integral vanishes also in this case.
\end{itemize}

\section{Affine $D_4$ quiver with Chern-Simons levels} \label{app:affineD4withCSlevels}
In this Appendix, we discuss some examples of the affine $D_4$ quiver with Chern-Simons levels. Let us first consider the following theory:
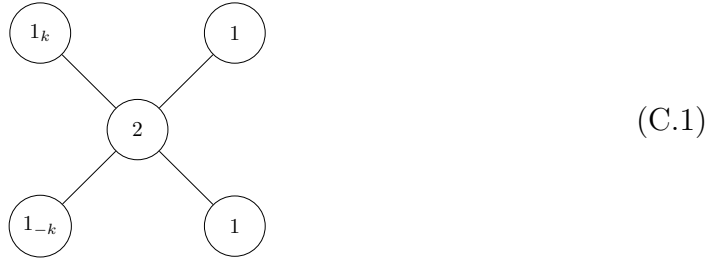
\begin{equation} \label{affined41}
\scalebox{0.8}{
\begin{tikzpicture}[baseline, font=\footnotesize,
    circ/.style={circle, draw, minimum size=1cm},
    sq/.style={rectangle, draw, minimum size=1cm},
    node distance=0.5cm, scale=0.8
]
    \node[circ] (m1) at (-2,2) {$1_{k}$};
    \node[circ] (m2) at (-2,-2)  {$1_{-k}$};
    \node[circ] (m3) at (2,2)  {$1$};
    \node[circ] (m4) at (2,-2)  {$1$};
    \node[circ] (n1) at (0,0)  {$2$};
    \draw (m1) -- (n1);
    \draw (m2) -- (n1);
    \draw (m3) -- (n1);
    \draw (m4) -- (n1);
\end{tikzpicture}}
\end{equation}
In order to compute the geometric branch limit of the index, we assign the auxiliary axial fugacities to the chiral fields as follows:
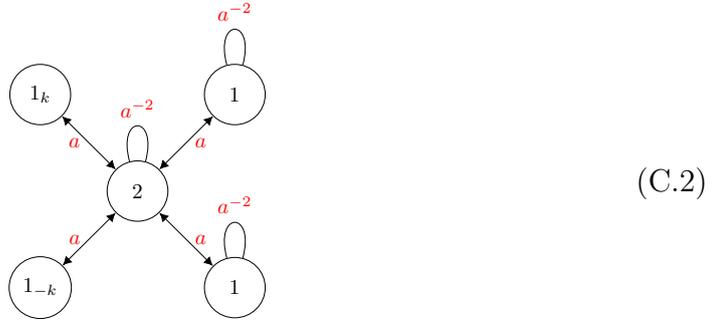
\begin{equation} \label{affineD4axial1}
\scalebox{0.8}{
\begin{tikzpicture}[baseline,font=\footnotesize,
    circ/.style={circle, draw, minimum size=1cm},
    sq/.style={rectangle, draw, minimum size=1cm},
    node distance=0.5cm,
    every loop/.style={-}, scale=0.8
]
    \node[circ] (m1) at (-2,2) {$1_{k}$};
    \node[circ] (m2) at (-2,-2)  {$1_{-k}$};
    \node[circ] (m3) at (2,2)  {$1$};
    \node[circ] (m4) at (2,-2)  {$1$};
    \node[circ] (n1) at (0,0)  {$2$};
    \draw[<->] (m1) -- node[left, red] {$a$} (n1);
    \draw[<->] (m2) -- node[left, red] {$a$} (n1);
    \draw[<->] (m3) -- node[right, red] {$a$} (n1);
    \draw[<->] (m4) -- node[right, red] {$a$} (n1);
   \path (m3) edge [loop above] node[red] {$a^{-2}$} ();
   \path (m4) edge [loop above] node[red] {$a^{-2}$} ();
    \path (n1) edge [loop above] node[red] {$a^{-2}$} ();
\end{tikzpicture}}
\end{equation}
The index is very similar to \eqref{indexaffined4}, with $F=0$ and factor $u_1^{k m_1}u_2^{-k m_2}$ added to the integrand. Upon taking the $\fz \rightarrow 0$ limit, we obtain
\bes{ \label{orbCBU2w4}
\frac{1}{k} \sum_{j=0}^{k-1} \sum_{m_1 \geq m_2 > -\infty} t^{4|m_1|+4|m_2| -2|m_1-m_2|} P_{\U(2)} (t; m_1,m_2) \left(\omega_k^j y^{1/k} \right)^{m_1+m_2}~,
}
where $\omega_k = \exp(2\pi i/k)$. This is simply the $\BZ_k$ orbifold of the Coulomb branch Hilbert series of $\U(2)$ SQCD with four flavours, where $\BZ_k$ acts in the same way as the topological symmetry associated to the $\U(2)$ gauge group.\footnote{This is very similar to the ABJM theory \cite{Aharony:2008ug}; see \cite[(4.17)]{Cremonesi:2016nbo}.} The fugacity $y$ can be mapped to the topological fugacities $w_{1,\ldots, 5}$ in \eqref{indexaffined4} as
\bes{
y = w_1 w_2 w_3 w_4 w_5^2~.
}
With the aid of \cite[(5.6)]{Cremonesi:2013lqa}, this result can be written more explicitly as
\bes{
\text{\eref{orbCBU2w4}} =\frac{1}{k} \sum_{j=0}^{k-1} \PE\left[(\omega_k^j y^{1/k}+1+\omega_k^{-j} y^{-1/k})(t^2+t^4) - t^6-t^8  \right]~.
}
For general $k$, the moment map operator contributing at order $t^2$ is $X_{15}X_{51} = X_{25}X_{52}$, where the equality follows from the $F$-term with respect to the adjoint chiral field of the central $\U(2)$ gauge group. The dressed monopole operators are $X_{51}^k X_{25}^k V_{(1,1,1,1;(1,1))}$ and $X_{15}^k X_{52}^k V_{-(1,1,1,1;(1,1))}$.\footnote{Note that the bare monopole operators $V_{\pm(1,1,1,1;(1,1))}$ have dimension zero, but they are not gauge invariant themselves due to the presence of non-trivial Chern-Simons levels. Upon dressing by the chiral fields, the dressed monopoles in question have $R$-charge $k$ and contribute at order $t^{2k}$.}  Inspired by mirror symmetry, we observe that \eqref{orbCBU2w4} can also be realised from the $\BZ_k$ orbifold of the Higgs branch of $\U(2)$ SQCD with four flavours, with two combinations of Cartan subalgebra of the $\su(4)$ flavour symmetry being gauged, namely
\bes{ \label{orbHBU2w4}
\text{\eref{orbCBU2w4}} = \frac{1}{k}\sum_{j=0}^{k-1}(1-t^2)^2\oint \frac{df_1}{2\pi i f_1} \oint \frac{df_2}{2\pi i f_2} H(t; f_1, f_2, f_3 = \omega_k^j y^{1/k} f_4, f_4)~,
}
where the Higgs branch Hilbert series of $\U(2)$ SQCD with four flavours is given by\footnote{The highest weight generating function of the Higgs branch of this theory is given in \cite[Table 4]{Hanany:2016gbz} with label $(2^2)$.}
\bes{
&H(t; f_1, f_2, f_3, f_4) \\
&= \frac{1}{2}\oint \prod_{j=1}^2\left( \frac{du_j}{2\pi i u_j} \right)\left(1-\frac{u_1}{u_2} \right) \left(1-\frac{u_2}{u_1} \right) \\
& \qquad \times \PE \Bigg[\sum_{s=\pm 1}\left(\sum_{i=1}^4 f_i^s \right)(u_1^{-s}+u_2^{-s})t - (u_1+u_2)(u_1^{-1}+u_2^{-1})t^2 \Bigg]~.
}
The next examples are the following quivers:
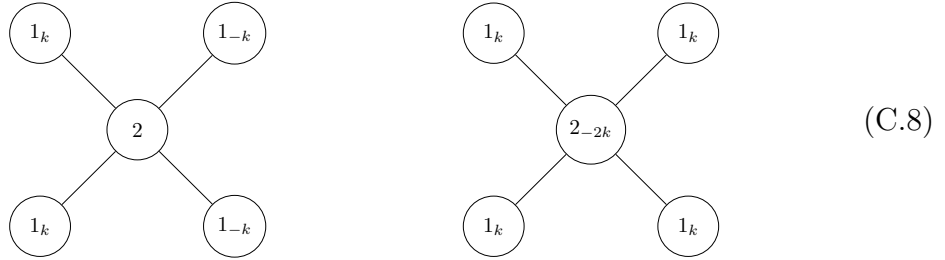
\begin{equation} \label{affined42}
\scalebox{0.8}{
\begin{tikzpicture}[baseline, font=\footnotesize,
    circ/.style={circle, draw, minimum size=1cm},
    sq/.style={rectangle, draw, minimum size=1cm},
    node distance=0.5cm, scale=0.8
]
    \node[circ] (m1) at (-2,2) {$1_{k}$};
    \node[circ] (m2) at (-2,-2)  {$1_{k}$};
    \node[circ] (m3) at (2,2)  {$1_{-k}$};
    \node[circ] (m4) at (2,-2)  {$1_{-k}$};
    \node[circ] (n1) at (0,0)  {$2$};
    \draw (m1) -- (n1);
    \draw (m2) -- (n1);
    \draw (m3) -- (n1);
    \draw (m4) -- (n1);
\end{tikzpicture}}
\qquad \qquad \qquad 
\scalebox{0.8}{
\begin{tikzpicture}[baseline, font=\footnotesize,
    circ/.style={circle, draw, minimum size=1cm},
    sq/.style={rectangle, draw, minimum size=1cm},
    node distance=0.5cm, scale=0.8
]
    \node[circ] (m1) at (-2,2) {$1_{k}$};
    \node[circ] (m2) at (-2,-2)  {$1_{k}$};
    \node[circ] (m3) at (2,2)  {$1_{k}$};
    \node[circ] (m4) at (2,-2)  {$1_{k}$};
    \node[circ] (n1) at (0,0)  {$2_{-2k}$};
    \draw (m1) -- (n1);
    \draw (m2) -- (n1);
    \draw (m3) -- (n1);
    \draw (m4) -- (n1);
\end{tikzpicture}}
\end{equation}
Although these are different theories, we will demonstrate that they have the same geometric branch. We assign the auxiliary axial fugacities to the chiral fields as follows:
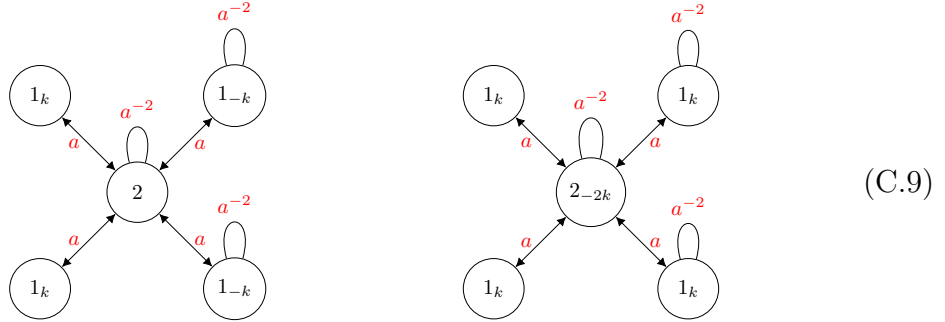
\begin{equation} \label{affineD4axial2}
\scalebox{0.8}{
\begin{tikzpicture}[baseline,font=\footnotesize,
    circ/.style={circle, draw, minimum size=1cm},
    sq/.style={rectangle, draw, minimum size=1cm},
    node distance=0.5cm,
    every loop/.style={-}, scale=0.8
]
    \node[circ] (m1) at (-2,2) {$1_{k}$};
    \node[circ] (m2) at (-2,-2)  {$1_{k}$};
    \node[circ] (m3) at (2,2)  {$1_{-k}$};
    \node[circ] (m4) at (2,-2)  {$1_{-k}$};
    \node[circ] (n1) at (0,0)  {$2$};
    \draw[<->] (m1) -- node[left, red] {$a$} (n1);
    \draw[<->] (m2) -- node[left, red] {$a$} (n1);
    \draw[<->] (m3) -- node[right, red] {$a$} (n1);
    \draw[<->] (m4) -- node[right, red] {$a$} (n1);
   \path (m3) edge [loop above] node[red] {$a^{-2}$} ();
   \path (m4) edge [loop above] node[red] {$a^{-2}$} ();
    \path (n1) edge [loop above] node[red] {$a^{-2}$} ();
\end{tikzpicture}}
\qquad \qquad \qquad  
\scalebox{0.8}{
\begin{tikzpicture}[baseline,font=\footnotesize,
    circ/.style={circle, draw, minimum size=1cm},
    sq/.style={rectangle, draw, minimum size=1cm},
    node distance=0.5cm,
    every loop/.style={-}, scale=0.8
]
    \node[circ] (m1) at (-2,2) {$1_{k}$};
    \node[circ] (m2) at (-2,-2)  {$1_{k}$};
    \node[circ] (m3) at (2,2)  {$1_{k}$};
    \node[circ] (m4) at (2,-2)  {$1_{k}$};
    \node[circ] (n1) at (0,0)  {$2_{-2k}$};
    \draw[<->] (m1) -- node[left, red] {$a$} (n1);
    \draw[<->] (m2) -- node[left, red] {$a$} (n1);
    \draw[<->] (m3) -- node[right, red] {$a$} (n1);
    \draw[<->] (m4) -- node[right, red] {$a$} (n1);
   \path (m3) edge [loop above] node[red] {$a^{-2}$} ();
   \path (m4) edge [loop above] node[red] {$a^{-2}$} ();
    \path (n1) edge [loop above] node[red] {$a^{-2}$} ();
\end{tikzpicture}}
\end{equation}
As before, the index is very similar to \eqref{indexaffined4}, where we take $F=0$ and add factor $\prod_{j=1}^2 u_j^{k m_j} \prod_{l=3}^4 u_l^{-k m_l}$ for the left quiver, $\prod_{j=1}^4 u_j^{k m_j} \prod_{\alpha=1}^2 (v_5^{(\alpha)})^{-2km_5^{(\alpha)}} $ for the right quiver. Upon taking the $\fz \rightarrow 0$ limit, we obtain a geometric branch Hilbert series similar to \eqref{orbHBU2w4}, namely
\bes{\label{geomCS2ex}
& \frac{1}{k}\sum_{j=0}^{k-1}(1-t^2)^2 \oint \frac{d u}{2\pi i u}\oint\frac{d v}{2\pi i v} \\
& \qquad \times H(t; f_1 = u , f_2 = u v, f_3 = v \omega_k^j y^{1/k} , f_4 = 1) ~.
}
For reference, we report the result for $k=1$ as follows:
\bes{
\text{\eqref{geomCS2ex}} = \PE\Big[&t^2 + \left(2y + 2 + 2y^{-1}\right) t^4 + \left(y + 1 + y^{-1}\right) t^6 \\
& - \left(y + 4 + y^{-1}\right) t^8 - \left(2y + 4 + 2y^{-1}\right) t^{10} +\ldots \Big]~.
}
We observe that, upon taking $k \rightarrow \infty$, we obtain $(1-t^2)^{-1}$ times the Hilbert series of $\BC^2/\hat{D}_4$, where the latter is given by $\PE\left[2t^4 +t^6 -t^{12} \right]$. As before, the moment map operator contributing at order $t^2$ is $X_{15}X_{51} = X_{25}X_{52}$, where the equality follows from the $F$-term with respect to the adjoint chiral field of the central $\U(2)$ gauge group. The gauge invariant dressed monopole operators are $\prod_{j=1}^2 X_{5j}^k \prod_{l=3}^4 X_{l5}^k V_{(1,1,1,1;(1,1))}$, $\prod_{j=1}^2 X_{j5}^k \prod_{l=3}^4 X_{5j}^k V_{-(1,1,1,1;(1,1))}$ for the left quiver in \eqref{affineD4axial2}, and $\prod_{j=1}^4 X_{5j}^k V_{(1,1,1,1;(1,1))}$, $\prod_{j=1}^4 X_{j5}^k V_{-(1,1,1,1;(1,1))}$ for the right quiver in \eqref{affineD4axial2}.

\bibliographystyle{JHEP}
\bibliography{bibli.bib}

\end{document}